\documentclass[12pt]{article}
\makeatletter \@addtoreset{equation}{section} \makeatother
\renewcommand{\theequation}{\thesection.\arabic{equation}}
\newcommand{\ba}{\begin{array}}
\newcommand{\ea}{\end{array}}
\newcommand{\beq}{\begin{equation}}
\newcommand{\eeq}{\end{equation}}
\newcommand{\bea}{\begin{eqnarray}}
\newcommand{\eea}{\end{eqnarray}}

\def\bce{\begin{center}}
\def\ece{\end{center}}

\def\nonu{\nonumber}

\def\pa{\partial}

\def\be{\beta}

\def\la{\lambda}

\def\eps6{{\displaystyle \mathop{\epsilon}^{6}}{}}
\def\g6{{\displaystyle \mathop{g}^{6}}{}}

\def\nab6{{\displaystyle \mathop{\nabla}^{6}}{}}

\def\0{{\sst{(0)}}}
\def\1{{\sst{(1)}}}
\def\2{{\sst{(2)}}}
\def\3{{\sst{(3)}}}
\def\4{{\sst{(4)}}}
\def\5{{\sst{(5)}}}
\def\6{{\sst{(6)}}}
\def\7{{\sst{(7)}}}
\def\8{{\sst{(8)}}}

\def\ba{\begin{array}}
\def\ea{\end{array}}
\def\beq{\begin{equation}}
\def\eeq{\end{equation}}
\def\be{\begin{equation}}
\def\ee{\end{equation}}

\def\la{\lambda}
\def\eps{\epsilon}

\def\ba{\begin{array}}
\def\ea{\end{array}}
\def\beq{\begin{equation}}
\def\eeq{\end{equation}}
\def\be{\begin{equation}}
\def\ee{\end{equation}}

\def\la{\lambda}
\def\eps{\epsilon}

\def\eps6{{\displaystyle \mathop{\epsilon}^{6}}{}}

\def\nab6{{\displaystyle \mathop{\nabla}^{6}}{}}

\newcommand{\bean}{\begin{eqnarray*}}
\newcommand{\eean}{\end{eqnarray*}}

\begin{document}
\thispagestyle{empty} \addtocounter{page}{-1}
   \begin{flushright}
\end{flushright}

\vspace*{1.3cm}

\centerline{ \Large \bf
Higher Spin Currents }
\centerline{ \Large \bf
in the  Holographic ${\cal N}=1$ Coset Minimal Model}
\vspace*{1.5cm}
\centerline{{\bf Changhyun Ahn } and {\bf Jinsub Paeng}
}
\vspace*{1.0cm}
\centerline{\it
Department of Physics, Kyungpook National University, Taegu
702-701, Korea}
\vspace*{0.8cm}
\centerline{\tt ahn@knu.ac.kr, \qquad jdp2r@knu.ac.kr
}
\vskip2cm

\centerline{\bf Abstract}
\vspace*{0.5cm}

In the ${\cal N}=1$ supersymmetric coset minimal model
based on
$(B_N^{(1)} \oplus D_N^{(1)}, D_N^{(1)})$
at level $(k,1)$ studied recently,
the standard ${\cal N}=1$ super stress tensor of spins $(\frac{3}{2},2)$
is reviewed.
By considering the stress tensor in the coset
$(B_N^{(1)}, D_N^{(1)})$
at level $k$, the higher spin-$2'$ Casimir current
was obtained previously.
By acting the above spin-$\frac{3}{2}$ current on the higher spin-$2'$
Casimir
current, its superpartner, the higher spin-$\frac{5}{2}$ current,
can be generated
and combined as the first higher spin supercurrent with spins
$(2', \frac{5}{2})$.
By calculating the operator product expansions (OPE) between the higher spin
supercurrent and itself, the next higher spin supercurrent can be generated
with spins $(\frac{7}{2},4)$.
Moreover, the other higher spin supercurrent with spins $(4',\frac{9}{2})$
can be generated by calculating the OPE between the first higher spin
supercurrent with spins $(2', \frac{5}{2})$ and the second higher spin
supercurrent with spins $(\frac{7}{2},4)$.
Finally, the higher spin supercurrent, $(\frac{11}{2},6)$, can be extracted
from the right hand side of OPE between the higher spin supercurrents,
$(2', \frac{5}{2})$ and $(4', \frac{9}{2})$.

\baselineskip=18pt
\newpage
\renewcommand{\theequation}
{\arabic{section}\mbox{.}\arabic{equation}}

\section{Introduction}

One example of the minimal model holography \cite{GG,GG1,Ahn1106,GV}
is studied in \cite{CHR}.
The gravity theory is the ${\cal N}=1$
truncation of the ${\cal N}=2$ higher spin
supergravity on $AdS_3$ space \cite{PV}.
The ${\cal N}=1$ truncation of the matter fields
provides the ${\cal N}=1$ hypermultiplet with two complex massive
scalars and two massive fermions.
The 2-dimensional dual conformal field theory
is described by the following coset minimal model
\bea
\frac{G}{H} = \frac{\widehat{SO}(2N+1)_k \oplus
\widehat{SO}(2N)_1}{\widehat{SO}(2N)_{k+1}}.
\label{coset1}
\eea
The diagonal denominator current of spin-$1$ is the sum of
two numerator currents where the fermions of spin-$\frac{1}{2}$ belong to
the second numerator factor.
The higher spin currents including the ${\cal N}=1$  super
stress tensor with spins-$(\frac{3}{2},2)$
are given by
\bea
(\frac{3}{2}, 2): \quad (2', \frac{5}{2}), \quad
(\frac{7}{2}, 4), \quad (4', \frac{9}{2}),
\quad (\frac{11}{2}, 6), \quad \cdots, \quad
(n-\frac{1}{2}, n), \quad (n', n'+\frac{1}{2}), \quad \cdots.
\label{currents}
\eea
The 't Hooft coupling constant is given by $\la = \frac{2N}{(2N+k-1)}$ in the
coset model (\ref{coset1}) and corresponds to the mass parameter in the
$AdS_3$ bulk theory.
Recently, in \cite{CV}, the ${\cal N}=1$
${\cal W}_{\infty}[\mu]$ algebra with $\mu =(1-2N)$
 by taking the field contents in (\ref{currents}) is found.

In this paper,
we would like to construct the higher spin currents for the coset model
(\ref{coset1}).
For example, in order to understand this duality, the three-point functions
can be compared to each other.
Once the higher spin currents with bosonic spins are known completely, then
in principle, the three point functions can be obtained 
\footnote{We will determine the complete expressions for the bosonic higher spin currents of spins $s=2', 4, 4'$ and $6$. Then one can proceed the previous analysis done in different coset model \cite{Ahn1202}. The zero mode for each 
bosonic current acts on the vector representation.
The zero mode of the spin-$1$ current with level $k$ in the numerator
of the coset (\ref{coset1}) 
acting on the state $|(0;v) >$ vanishes while the zero mode of the
diagonal spin-$1$ current in the denominator of the coset acting on the 
state $|(v;0)>$ vanishes. For  the former, 
the zero mode for higher spin current 
consists of multiple product of 
quadratic in the fermions of the 
numerator current (i.e. $K^a K^b(z)$ in section $2$) and for the latter,
the zero mode can be written in terms of 
the numerator current with level $k$ (i.e. $J^{AB}(z)$ in section $2$) (or 
the combination of fermions in the 
numerator current (i.e. $K^a K^b(z)$) and the numerator current with level $k$ where 
one index is fixed by $(2N+1)$ (i.e. $J^{a 2N+1}(z)$ in section $2$)).
Some identities in the trace of the 
generators of $SO(2N)$ or $SO(2N+1)$ can be used in the eigenvalue 
equations. 
One should take into account 
the fact that the first factor of the numerator group of the coset
has different from other factor groups in the coset and there exist 
spin-$\frac{1}{2}$ current in the second factor of the numerator group,
compared to the previous results in \cite{Ahn1202}.  
It would be interesting to obtain the three point 
functions from the findings of this paper explicitly. }.
The direct construction using the Jacobi identities in \cite{CV}
doesn't  tell us what the central charge is.
Only after the isomorphism between the Drinfeld-Sokolov reduction and the
coset construction is used, the central charge in the direct construction
\cite{CV} can be identified with the one in the above coset model
(\ref{coset1}) where the central charge
$c$ is equal to $c=\frac{3 N k}{(2N+k-1)}$.
Furthermore, if one considers more general coset where the second level
$1$ is replaced by an arbitrary integer $l$, then the extra current will
appear in general. The above isomorphism cannot be used in this case also.
This is one of the reasons why we are interested in the construction of
higher spin currents for the coset model explicitly.

The main procedure in this coset construction
is based on the fact that
once the lower higher spin currents are found, then
the next undetermined higher spin currents
can be generated, in principle, by calculating the operator product
expansions (OPE) between the known higher spin currents.
The lowest component spin-$2'$ in the ${\cal N}=1$ multiplet
$(2', \frac{5}{2})$ of (\ref{currents}) with known
realization of
${\cal N}=1$ stress tensor  is the fundamental higher spin current
because this current generates all the higher spin currents.
It turns out that once the spin-$2'$ current in the multiplet
$(2', \frac{5}{2})$ is found, then the spin-$\frac{3}{2}$ current in the
${\cal N}=1$ super stress tensor determines
the spin-$\frac{5}{2}$ in the above multiplet.
Then one considers the OPE between the above spin-$2'$ current and
the spin-$\frac{5}{2}$ current.
The first-order pole of this OPE determines the spin-$\frac{7}{2}$ current
in the ${\cal N}=1$ multiplet $(\frac{7}{2}, 4)$ of (\ref{currents}).
Now one can continue to calculate the OPE between the spin-$2'$ current and
the above
spin-$\frac{7}{2}$ current and it turns out that the spin-$\frac{9}{2}$
current that is the second component in $(4', \frac{9}{2})$
appears in the first-order pole of this OPE.
Furthermore, the OPE between the spin-$2'$ and the spin-$\frac{9}{2}$ current
determines the spin-$\frac{11}{2}$ current that is the first component in
$(\frac{11}{2}, 6)$.
In this way, all the half-integer spin currents can be obtained.
What about the bosonic higher spin currents?
They can be determined by calculating the OPE between
the spin-$\frac{3}{2}$ current of
${\cal N}=1$ stress tensor and any known higher spin current of half-integer
spin due to the ${\cal N}=1$ supersymmetry.

Then how one can extract the correct primary or quasi-primary
fields in the given singular
terms in the OPE?
It is known that the OPE of two quasi-primary fields, of spins $h_i$ and
$h_j$ respectively, takes the form
\cite{Bowcock,BFKNRV,BS,Nahm1,Nahm2}
\bea
&& \Phi_i(z) \; \Phi_j(w)  =  \frac{1}{(z-w)^{h_i+h_j}} \, \gamma_{ij} \nonu \\
&& +  \sum_k C_{ijk} \sum_{n=0}^{\infty} \frac{1}{(z-w)^{h_i+h_j-h_k-n}}
\left[\frac{1}{n!} \prod_{x=0}^{n-1} \frac{(h_i-h_j+h_k+x)}{(2h_k+x)}
\right]
\pa^n \Phi_k(w).
\label{PhiPhi}
\eea
$\gamma_{ij}$ corresponds to a metric on the space of quasi-primary fields.
The structure constant $C_{ijk}$
appears in the three-point function between the quasi-primary fields,
$\Phi_i(z), \Phi_j(z)$ and $\Phi_k(z)$.
The index $k$ specifies all the quasi-primary fields occurring
in the right hand side of (\ref{PhiPhi}).
The descendant fields for the quasi-primary field
$\Phi_k(w)$ of spin $h_k$ are multiple derivatives of $\Phi_k(w)$.
The relative coefficient functions
$\left[\frac{1}{n!} \prod_{x=0}^{n-1} \frac{(h_i-h_j+h_k+x)}{(2h_k+x)}
\right]$
in the descendant fields
depend on the spins
and number of derivatives.
Since the higher spin currents can be written in terms of
WZW currents, the above structure constant $C_{ijk}$ can be determined.
For the fixed $C_{ijk}$,
the relative coefficient functions in the descendant fields
using (\ref{PhiPhi})
can be obtained.
In general,
the singular terms of the OPE
are written in terms of WZW currents in complicated way.
It is quite nontrivial to rearrange those expressions in terms of
determined (and known) higher spin currents.
This rearrangement can be done using the primary or quasi-primary condition
under the spin-$2$ current  of ${\cal N}=1$ super stress tensor and
superprimary condition under the spin-$\frac{3}{2}$ current of
${\cal N}=1$ stress tensor.
The details can be seen in the next section.

In section $2$,
the fundamental OPEs between the WZW currents living in the coset model
(\ref{coset1}) are given and the spin-$2$ stress tensor and its superpartner
with
Sugawara construction are reviewed.

In section $3$,
from the observation of \cite{CHR}, the lowest higher spin
${\cal N}=1$ multiplet
$(2, \frac{5}{2})$ in (\ref{currents}) is obtained.
The additional three
higher spin supercurrents in the list (\ref{currents})  are constructed
very explicitly.

In section $4$, we summarize
what we have found in this paper and discuss the future directions.

In Appendices, some results from the detailed calculations in section $3$
are provided.

The  mathematica package by Thielemans \cite{Thielemans} is used.

\section{The GKO coset construction: Review}

For the diagonal coset model \cite{CHR}
\bea
\frac{G}{H} = \frac{\widehat{SO}(2N+1)_k \oplus
\widehat{SO}(2N)_1}{\widehat{SO}(2N)_{k+1}},
\label{coset}
\eea
the antisymmetric spin-1 fields, $J^{AB}(z)$ with level $k$, and the spin-$\frac{1}{2}$ fields, $K^{a}(z)$ with level $1$,
generate the affine Lie algebra $G = \widehat{SO}(2N+1)_k \oplus
\widehat{SO}(2N)_1$.
The index $a$  runs from 1 to $2N$ and the index $A$  runs from $1$ to
$(2N+1)$.
The number of independent fields of $J^{AB}$ and $K^a(z)$
are given by $N(2N+1)$ and $2N$ respectively.
Note that this coset model (\ref{coset}) is a little different from the ones
studied in \cite{Ahn1106,Ahn1202,CGKV,AP1301}.
The fundamental OPE of the fermion fields $K^a(z)$ is given by
\bea
&& K^{a} (z) K^{b} (w)  =
\frac{1}{(z-w)} \,  \delta^{ab} + \cdots.
\label{KKcoset}
\eea
The standard OPE of the spin-1 currents is expressed as
\bea
&& J^{AB} (z) J^{CD} (w)  =  -\frac{1}{(z-w)^2} \, k \, (-\delta^{BC}
\, \delta^{AD}+
\delta^{AC} \, \delta^{BD})
\nonu \\
&& +  \frac{1}{(z-w)} \,
  \left[  \delta^{BC} \, J^{AD}(w) +\delta^{AD} \, J^{BC}(w)-\delta^{AC}
  \, J^{BD}(w)-
\delta^{BD} \, J^{AC}(w) \right] + \cdots.
\label{JJcoset}
\eea
The diagonal
spin-1 field $J'^{ab} (z)$ with level $(k+1)$ generates
the affine Lie algebra $H =\widehat{SO}(2N)_{k+1}$. It is expressed as
\bea
J'^{ab} (z) = J^{ab}(z) +(K^{a} K^{b})(z).
\label{jprime}
\eea
Note that
we are using the double index notation for the spin-$1$ current
rather than a single index used in \cite{CHR}.
Among $(2N+1) \times (2N+1)$ matrix,
the first $(2N) \times (2N)$ matrix corresponds to
the spin-$1$ current $J^{ab}(z)$ and the remaining matrix elements
correspond to other spin-$1$ current $J^{a \,2N+1}(z)$.
The spin-2 stress energy tensor with (\ref{jprime}),
via the Sugawara construction in the coset model (\ref{coset}), is written as
\bea
T(z) & = &   -\frac{1}{4(k+2N-1)} \, (J^{AB} J^{AB}) (z)
 -\frac{1}{4(1+2N-2)} \, ((K^{a} K^{b}) (K^{a} K^{b})) (z) \nonu \\
& + &
\frac{1}{4(1+k +2N-2)} \, (J'^{ab} J'^{ab}) (z).
\label{stress1}
\eea
Since $J^{AB}(z)$ can be decomposed into $J^{a\,2N+1}(z)$, and $J^{ab}(z)$ which belongs to $SO(2N)$ subgroup,
$(J^{AB}J^{AB}) (z)$ can be written as
\bea
(J^{AB}J^{AB}) (z) = (J^{ab}J^{ab})(z) +2(J^{a\, 2N+1}J^{a\, 2N+1})(z).
\label{JAB}
\eea
Therefore, by plugging (\ref{JAB}) into  the stress tensor (\ref{stress1}),
the spin-$2$ stress tensor  can be expressed concisely as
\bea
T(z) & = &   \frac{1}{(2k + 4 N - 2)} \, \left[ K^a K^b J^{ab}
  -k K^a \pa K^a  - J^{a\, 2N+1}J^{a\, 2N+1}\right](z).
\label{stress2}
\eea
Note that there is no
$J^{ab} J^{ab}(z)$ in (\ref{stress2}).
One can easily check that there is no singular term in the OPE
between $T(z)$ and $J^{\prime ab}(w)$.
The OPE between the stress energy tensor $T(z)$ and itself, from
(\ref{KKcoset}), (\ref{JJcoset}) and (\ref{stress2}),
is given by
\bea
T (z)\; T(w)
& = & \frac{1}{(z-w)^4} \, \frac{c}{2} +\frac{1}{(z-w)^2} \, 2
T(w) +
\frac{1}{(z-w)} \, \pa T(w) +\cdots.
\label{cosettt}
\eea
The central charge in the highest singular term of
(\ref{cosettt}) is given by
\bea
c  & = &   \frac{1}{2} (2N+1)(2N) \frac{k}{(k+2N+1-2)}  +
\frac{1}{2} (2N)(2N-1) \frac{1}{(1+2N-2)} \nonu \\
&- &
\frac{1}{2} (2N)(2N-1) \frac{(1+k)}{(1+k +2N-2)}   =
\frac{3 N k}{(k+2N-1)} \leq 3N.
\label{ctilde}
\eea

The superpartner of the spin-2 current $T(z)$ is the spin-$\frac{3}{2}$ current $G(z)$ \cite{FMS}. They can be combined into a single ${\cal N}=1$ multiplet
as $(\frac{3}{2}, 2)$ in (\ref{currents}).
The OPE between $G(z)$ and itself reads as
\bea
G(z)\; G(w) = \frac{1}{(z-w)^3} \,
\frac{2}{3} c + \frac{1}{(z-w)} \, 2 T(w) +\cdots.
\label{gg}
\eea
One can construct $G(z)$ from the WZW currents
$K^a(z)$ of spin-$\frac{1}{2}$ and $ J^{AB}(z)$ of spin-$1$.
Since $K^a(z)$ has one index, $( K^a J^{a\, 2N+1})(z)$
is the only candidate for $G(z)$.
The normalization for $G(z)$ can be fixed from (\ref{gg}) and the explicit form of $G(z)$ is given by
\bea
G(z)&=& \frac{i}{\sqrt{k+2N-1}} ( K^a J^{a\, 2N+1})(z).
\label{G3}
\eea
One checks that
the OPE between $G(z)$ and the diagonal spin-$1$ current
$J^{\prime ab}(w)$ does not contain any singular terms.
As we expect, it satisfies the following OPE:
\bea
T(z) \; G(w) = \frac{1}{(z-w)^2} \, \frac{3}{2}
G(w) +\frac{1}{(z-w)} \, \pa G(w) +\cdots.
\label{tg}
\eea

The standard ${\cal N}=1$ superconformal algebra
consists of (\ref{cosettt}), (\ref{gg}) and (\ref{tg}).
In the next section, the fundamental OPEs (\ref{KKcoset}) and (\ref{JJcoset})
are used heavily and the coset stress tensor (\ref{stress2})
and its superpartner (\ref{G3}) with central charge
(\ref{ctilde})
will be used all the times.

\section{ The construction of higher spin supercurrents   }

In this section, the higher spin currents will be constructed for general $N$ explicitly from the fermion fields $K^a(z)$ of spin-$\frac{1}{2}$
and the antisymmetric spin-1 currents $J^{AB}(z)$.

\subsection{ The OPEs between the higher spin currents of
spins-$(2', \frac{5}{2})$ and itself}

$\bullet$
The OPE $O_2(z) \, O_2(w)$

Now, let us consider the stress energy tensor of the coset $\frac{\widehat{SO}(2N+1)_k }{\widehat{SO}(2N)_{k}}$  \cite{CHR}, which is
\bea
\widetilde{T}(z) & = &   -\frac{1}{4(k+2N-1)} \, (J^{AB} J^{AB}) (z)
+\frac{1}{4(k+2N-2)} \, (J^{ab} J^{ab}) (z).
\label{Ttilde1}
\eea
Note that the denominator current is simply given by
$J^{ab}(z)$.
This stress tensor
$\widetilde{T}(z)$ obeys the following OPEs from (\ref{Ttilde1}) and
(\ref{stress2}):
\bea
\widetilde{T}(z)\; \widetilde{T}(w)
& = & \frac{1}{(z-w)^4} \, \frac{\widetilde{c}}{2} +\frac{1}{(z-w)^2} \, 2
\widetilde{T}(w) +
\frac{1}{(z-w)} \, \pa \widetilde{T}(w) +\cdots,
\label{cosettt1} \\
T(z)\; \widetilde{T}(w)
& = & \frac{1}{(z-w)^4} \, \frac{\widetilde{c}}{2} +\frac{1}{(z-w)^2} \, 2
\widetilde{T}(w) +
\frac{1}{(z-w)} \, \pa \widetilde{T}(w) +\cdots,
\label{cosettt2}
\eea
where $\widetilde{c}$ is the central charge of coset $\frac{\widehat{SO}(2N+1)_k }{\widehat{SO}(2N)_{k}}$ and can be expressed as
\bea
\widetilde{c}=\frac{k N (-3 + 2 k + 2N)}{(-2 + k + 2N) (-1 + k + 2 N)}.
\label{widec}
\eea
The stress tensor $\widetilde{T}(z)$ is a quasi-primary field
under the stress tensor (\ref{stress2}).
From (\ref{cosettt}) and (\ref{cosettt2}), one can easily figure out that the following combination of $T(z)$ and $\widetilde{T}(z)$ with (\ref{ctilde})
and (\ref{widec})
gives a spin-$2'$ primary field $O_{2'}(z)$ under the stress tensor $T(z)$:
\bea
O_{2'}(z) &=& c\, \widetilde{T}(z)-\widetilde{c}\,T(z).
\label{spin2}
\eea
This is because the OPE between $T(z)$ and $O_{2'}(w)$ does not contain
the fourth-order pole.
In terms of $J^{AB}(z)$ and $K^a(z)$, the spin-$2$ current (\ref{spin2})
is expressed as \footnote{We denote $O_{2'}(z)$ as $O_2(z)$. This is O.K.
because the other spin-$2$ current is denoted by $T(z)$.}
\bea
O_{2}(z)&=&\frac{k N}{4 (-2 + k + 2 N) (-1 + k + 2 N)^2} \left[3\,J^{ab}J^{ab}+
  2 k (-3 + 2 k + 2 N)  K^a \pa K^a  \right. \nonu \\
&& \left. +2 (3 - 2 k - 2 N) K^a K^b J^{ab}- 2(-3 + k + 4 N)J^{a\, 2N+1}J^{a\, 2N+1}\right](z).
\label{O2}
\eea
Compared to (\ref{stress2}), the first term of (\ref{O2}) appears newly.

The OPE between $G(z)$ and $O_2(w)$ generates a primary spin-$\frac{5}{2}$ current $O_{\frac{5}{2}} (z)$ under the stress tensor $T(z)$ as follows:
\bea
G(z) \, O_{2}(w) = \frac{1}{(z-w)} \,O_{\frac{5}{2}}(w)+\cdots.
\label{GO2}
\eea
The explicit form of spin-$\frac{5}{2}$ current $O_{\frac{5}{2}} (z)$ is given by
\bea
O_{\frac{5}{2}}(z) &=& \frac{i k N}{2 (-2 + k + 2 N) (-1 + k + 2 N)\sqrt{k+2N-1}} \left[
  2 (-3 + 2 k + 2 N)  \pa K^a J^{a\, 2N+1} \right. \nonu \\
&& \left. -(-3 + 2 k + 2 N) K^a \pa J^{a\, 2N+1}+3K^a (J^{ab} J^{b\, 2N+1}+ J^{b\, 2N+1}J^{ab})\right](z).
\label{O5}
\eea
The OPE between $G(z)$ and $O_{\frac{5}{2}} (w)$ with (\ref{G3}) and (\ref{O5})
is described as
\bea
G(z) \, O_{\frac{5}{2}}(w) = \frac{1}{(z-w)^2} \, 4 \,O_2(w) +
\frac{1}{(z-w)} \, \pa O_2(w) +\cdots.
\label{GO5}
\eea
From (\ref{GO2}) and (\ref{GO5}), one can clearly see that the currents
$O_{\frac{5}{2}} (z)$ and $O_2(z)$ are superpartners each other.


Now let us consider the OPE between  $O_2 (z)$ and itself. From  (\ref{cosettt}),(\ref{cosettt1}), and (\ref{cosettt2}), one can show that the OPE between $O_2 (z)$ and itself is given by
\bea
O_{2}(z) \, O_{2}(w) & = & \frac{c_{22}}{(z-w)^4}
+ \frac{1}{(z-w)^2} \left[ c_{22}^{o} O_{2}+c_{2 2}^{t} T \right](w)
\nonu \\
& + & \frac{1}{(z-w)} \left[  \frac{1}{2}c_{2 2}^{o}\pa O_{2}+ \frac{1}{2}c_{2 2}^{t} \pa T \right](w)\cdots.
\label{O2O2a}
\eea
The relative coefficient $\frac{1}{2} (= \frac{2-2+2}{2\times 2})$
in the descendant fields
can be understood from the general
formula in (\ref{PhiPhi}).
There are no new primary fields in the right hand side of
(\ref{O2O2a}). The structure constants in (\ref{O2O2a}) are
\bea
c_{2 2} & = & -\frac{c^3 (2 c (-1 + N) + 3 (3 - 4 N) N) (c + 2 c N + 3 N (-3 + 2 N))}{18 (c + 6 (-1 + N) N)^2}, \nonu \\
c_{2 2}^{o} & = & \frac{2 c (c - 4 c N + 6 N^2)}{3 (c + 6 (-1 + N) N)}, \nonu \\
c_{2 2}^{t} & = & -\frac{2 c^2 (2 c (-1 + N) + 3 (3 - 4 N) N) (c + 2 c N + 3 N (-3 + 2 N))}{9 (c + 6 (-1 + N) N)^2}.
\nonu
\eea
The fusion rule can be summarized by
$[O_2] [O_2] = [I]+ [O_2]$.

$\bullet$
The OPE $O_2(z) \, O_\frac{5}{2}(w)$

Now let us move to the OPE between  $O_2 (z)$ and $O_{\frac{5}{2}} (w)$\footnote{
From now on,
when we compute the OPEs, we compute them for $N=2$ case
in order to determine the algebraic structures of the OPEs
firstly.
Then, we will compute them for general $N$. The reason is as follows.
The $SO(2N+1)$ or $SO(2N)$
invariant tensor of rank $2$, Kronecker delta, appears in the
fundamental OPEs of (\ref{KKcoset}) and (\ref{JJcoset}). In any OPE between
any two higher spin currents, the group index structure in the right hand
side of this OPE appears in the fields using the Kronecker delta's. That is,
the coefficients in front of the fields do not have tensorial structures
as one varies the $N$. Therefore,
it is possible to extract the $N$-dependence of these
coefficients from those with several fixed $N$. This feature is different from
the $SU(N)$ coset model where there are tensorial structures between $f$-symbol and $d$-symbol as well as Kronecker delta.        }.
We follow the method used in \cite{Ahn1211}
to find the complete structure of the OPE. The OPE between $O_2 (z)$ and
$O_{\frac{5}{2}} (w)$ for $N=2$ shows that
\bea
O_{2}(z) \, O_{\frac{5}{2}}(w) & = & \frac{1}{(z-w)^3} \, c_{2 \frac{5}{2}}^{g} G(w)+\frac{1}{(z-w)^2} \left[ \frac{1}{3} c_{2 \frac{5}{2}}^{g} \pa G + c_{2 \frac{5}{2}}^{o} O_{\frac{5}{2}}\right](w)
\nonu \\
& + & \frac{1}{(z-w)} \left[ \frac{1}{12} c_{2 \frac{5}{2}}^{g}\pa^2 G + \frac{2}{5}c_{2 \frac{5}{2}}^{o} \pa O_{\frac{5}{2}}
\right.
 \nonu \\
& + & \left.
 c_{2 \frac{5}{2}}^{go}
\left( G O_{2} -\frac{2}{5} \pa O_{\frac{5}{2}} \right)+ c_{2 \frac{5}{2}}^{tg}
\left( T G -\frac{3}{8} \pa^2 G \right)
\right.
 \nonu \\
& + & \left. O_{\frac{7}{2}} \right](w)
+\cdots,
\label{O2O5a}
\eea
where the structure constants for fixed $N=2$ are given by
\bea
c_{2 \frac{5}{2}}^{g}(N=2) & =
& -\frac{12 k^2 (5 + k) (1 + 2 k)}{(2 + k)^2 (3 + k)^2}, \qquad
c_{2 \frac{5}{2}}^{o}(N=2) = -\frac{2 (-4 + k) k}{(2 + k) (3 + k)}, \nonu \\
c_{2 \frac{5}{2}}^{go}(N=2) &=& \frac{6 (-4 + k) k}{(2 + k) (1 + 2 k)},\qquad
c_{2 \frac{5}{2}}^{tg}(N=2) =
-\frac{24 k^2 (5 + k) (1 + 2 k)}{(2 + k)^2 (3 + k) (7 + 5 k)}.
\label{struct}
\eea
Moreover, the summation indices (appearing
in $O_2(z), O_{\frac{5}{2}}(z)$ and $G(w)$)
$a, b =1, \cdots, 4$ because $N=2$.
For the primary field $G(w)$ with the structure constant
$c_{2 \frac{5}{2}}^{g}(N=2)$ with (\ref{struct}) in the right hand side,
the relative coefficients, $\frac{1}{3}$ and $\frac{1}{12}$,
for its descendant fields appearing in the second- and first-order pole
can be read off from (\ref{PhiPhi}). Refer to   \cite{Ahn1211}
for details on how to compute the relative coefficients of descendant fields.
For the second-order pole in (\ref{O2O5a}),
the first term descending from $G(w)$ is fixed.
To check if there are other fields besides $ \pa G(w)$ in the second-order pole, we compute $[\{O_2\,O_{\frac{5}{2}}\}_{-2}-(\frac{1}{3}) c_{2 \frac{5}{2}}^{g}(N=2)
\pa G](w)$ \footnote{Here one uses the notation \cite{BS}
for the $n$-th order pole in the OPE $\Phi_i(z) \, \Phi_j(w)$
as follows: $\{\Phi_i \,\Phi_j\}_{-n}(w)$ which is written in terms of WZW
currents.}.
It turns out that this doesn't vanish implying that
there should be  extra fields
besides $ \pa G(w)$.
The spin of fields in the second-order pole should be  $\frac{5}{2}$ and the only candidate for this extra field is
the primary field $O_{\frac{5}{2}}(w)$ in (\ref{currents})
with $c_{2 \frac{5}{2}}^{o}(N=2)$.
Therefore, the extra field is proportional to $O_{\frac{5}{2}}(w)$.

For the first-order pole,
the first-  and second-term descending from $G(w)$ and
$O_{\frac{5}{2}}(w)$ respectively are completely fixed.
The relative coefficient ${\frac{2}{5}}$ in front of $\pa O_{\frac{5}{2}}(w)$ in the first-order pole is
also determined from (\ref{PhiPhi}).
As we did for the second-order pole, to find if there are other fields
besides $ \pa^2 G(w)$ and $ \pa O_{\frac{5}{2}}(w)$, we
compute $[\{O_2\,O_{\frac{5}{2}}\}_{-1} -
\mbox{first line}](w)$ where the first line contains two terms in
(\ref{O2O5a}).
It doesn't vanish and  there should be extra fields.
To find the extra fields in the first-order pole,
let us consider the OPE
between $T(z)$ and $[\{O_2\,O_{\frac{5}{2}}\}_{-1}-\mbox{first line}](w)$ and
the OPE between $G(z)$ and $[\{O_2\,O_{\frac{5}{2}}\}_{-1}-\mbox{first line}](w)$.
The OPE $T(z)$ with $ \left[ \{O_2\,O_{\frac{5}{2}}\}_{-1} -
\mbox{first line} \right](w)$ is given by
$\frac{1}{(z-w)^4} \left[
\frac{(-27k^2(5+k)(1+2k))}{(2+k)^2(3+k)^2} \right] G(w) + {\cal O}((z-w)^{-2})$.
This means that
a quasi-primary field, $T G(w)$ plus derivative terms,
should be considered to cancel
the fourth-order term  of this OPE
for the primary condition.
Moreover, the OPE between $G(z)$ and $[\{O_2\,O_{\frac{5}{2}}\}_{-1}-\mbox{first line}](w)$ leads to
$\frac{1}{(z-w)^3} \,[ -\frac{36k^2(5+k)(1+2k)}{(2+k)^2(3+k)^2}T+\frac{72k (-4+k)}{5(2+k)(3+k)}O_2](w) + {\cal O}((z-w)^{-2})$.
This means that we should consider
another quasi-primary field,
$G O_2(w)$ plus derivative terms, to remove
the third-order terms of this OPE.
With the help of superprimary condition (primary under the stress tensor  and
the equation (\ref{GO7})), the consistent coefficients,
$c_{2 \frac{5}{2}}^{go}(N=2)$ and
$c_{2 \frac{5}{2}}^{tg}(N=2)$, in the second line of the first-order pole in
(\ref{O2O5a})
are
determined.

Then we subtract the first and second line in the first-order pole from $\{O_2\,O_{\frac{5}{2}}\}_{-1}(w)$.
We are left with
a new primary spin-$\frac{7}{2}$ current $O_{\frac{7}{2}}(z)$ in the first-order pole.
The explicit form of the $O_{\frac{7}{2}}(z)$  can be expressed as
\bea
O_{\frac{7}{2}}(w) & = &  \{ O_{2} \; O_{\frac{5}{2}} \}_{-1}(w)
-\left[ \frac{1}{12} c_{2 \frac{5}{2}}^{g}\pa^2 G + \frac{2}{5}c_{2 \frac{5}{2}}^{o} \pa O_{\frac{5}{2}}+c_{2 \frac{5}{2}}^{go}
\left( G O_{2} -\frac{2}{5} \pa O_{\frac{5}{2}} \right)
\right.
 \nonu \\
& + & \left.  c_{2 \frac{5}{2}}^{tg}
\left( T G -\frac{3}{8} \pa^2 G \right) \right](w).
\label{O7}
\eea
As expected, $O_{\frac{7}{2}}(z)$ obeys the following OPE:
\bea
G(z) \, O_{\frac{7}{2}}(w) = \frac{1}{(z-w)} \,O_{4}(w) +\cdots,
\label{GO7}
\eea
where $O_4(z)$ is a spin-4 primary field in (\ref{currents}) which
will appear in (\ref{O5O5})
when the OPE between $O_{\frac{5}{2}}(z)$ and itself is computed later.

To determine the structure constants in (\ref{O2O5a}) for general $N$,
we compute the OPE between $O_2 (z)$ and $O_{\frac{5}{2}} (z)$  by hand explicitly with the help of (\ref{O2O5a}).
Explicit calculations for the third- and second-order pole show that
the coefficients $c_{2 \frac{5}{2}}^{g} $ and
$c_{2 \frac{5}{2}}^{o}$ for general $N$ are
\bea
c_{2 \frac{5}{2}}^{g} & = & \frac{c^2 (2 c (-1 + N) + 3 (3 - 4 N) N) (c + 2 c N + 3 N (-3 + 2 N))}{3 (c + 6 (-1 + N) N)^2}, \nonu \\
c_{2 \frac{5}{2}}^{o} & = & \frac{c (c - 4 c N + 6 N^2)}{3 (c + 6 (-1 + N) N)}.
\label{strstr}
\eea
For the first-order pole, we find
\bea
\{ O_{2} \; O_{\frac{5}{2}} \}_{-1}(w) & = &
c_1 \, K^a J^{ab} J^{cb} J^{cN'}(w) +c_2 \, K^a J^{ab} \pa J^{bN'}(w)+c_3 \, K^a \pa J^{ab} J^{bN'}(w)
\nonu \\
& + & c_4 \, K^a \pa^2 J^{aN'}(w) +c_5 \,\pa K^a J^{ab} \pa J^{bN'}(w)+c_6 \,\pa K^a \pa J^{aN'}(w)
\nonu \\
&+& c_7 \,\pa^2 K^a J^{aN'}(w), \qquad
N' \equiv 2N+1,
\label{O2O5p1}
\eea
where the coefficients are given by
\bea
c_1 &=& -9D c^2 (c - 3 N)^2,
\qquad
c_2 = -6D c^2 (c - 3 N) (2 c (-1 + N) + 3 (3 - 4 N) N),
\nonu \\
c_3 & = &-3D c^2 (c - 3 N) (-6 N^2 + c (-1 + 4 N)),
\qquad
c_4  = Dc^2 (2 c (-1 + N) + 3 (3 - 4 N) N)^2,
\nonu \\
c_5 & = & -6D c^2 (c - 3 N) (c + 2 c N + 3 N (-3 + 2 N)),
\nonu \\
c_6  & = & 2Dc^2 (2 c (-1 + N) + 3 (3 - 4 N) N) (c + 2 c N + 3 N (-3 + 2 N)),
\nonu \\
c_7  & = & D c^2 (c + 2 c N + 3 N (-3 + 2 N))^2,
\qquad
D \equiv \frac{-\sqrt{3 N-c}}{9 (c + 6 (-1 + N) N)^2\sqrt{3 (1 - 2 N) N}}.
\label{O2O5p1coeff}
\eea
To fix the remaining undetermined coefficients
$c_{2 \frac{5}{2}}^{go}$ and $c_{2 \frac{5}{2}}^{tg}$, we apply $G(z)$ to the both sides of (\ref{O7}). From the superprimary condition  $(\ref{GO7}$),
we find that
\bea
\{G \; \{ O_{2} \; O_{\frac{5}{2}} \}_{-1}\}_{-3}(w) &=& \{\frac{1}{12} c_{2 \frac{5}{2}}^{g}G\pa^2 G
  +\frac{2}{5}c_{2 \frac{5}{2}}^{o}G \pa O_{\frac{5}{2}}
  +c_{2 \frac{5}{2}}^{go}G\left( G O_{2} -\frac{2}{5} \pa O_{\frac{5}{2}} \right)
 \nonu \\
& + &  c_{2 \frac{5}{2}}^{tg}G
\left( T G -\frac{3}{8} \pa^2 G \right)\}_{-3}(w).
\label{GGO7}
\eea
The first two coefficients in the right hand side
of (\ref{GGO7}) are known from (\ref{strstr}).
From the equation (\ref{GGO7}), we find that
\bea
c_{2 \frac{5}{2}}^{go} &=& \frac{18 c (-6 N^2 + c (-1 + 4 N))}{(6 + 5 c) (c + 6 (-1 + N) N)},\nonu \\
c_{2 \frac{5}{2}}^{tg}  & = & \frac{6 c^2 (2 c (-1 + N) + 3 (3 - 4 N) N) (c + 2 c N + 3 N (-3 + 2 N))}{(21 + 4 c) (c + 6 (-1 + N) N)^2}.
\label{coeffabove}
\eea
Therefore, the spin-$\frac{7}{2}$ current is given by
(\ref{O7}), (\ref{strstr}), (\ref{O2O5p1}), (\ref{O2O5p1coeff}) and
(\ref{coeffabove}). The OPE $O_2(z) \, O_{\frac{5}{2}}(w)$
for general $N$
is given by (\ref{O2O5a}) with (\ref{strstr}) and (\ref{coeffabove}).
The fusion rule can be summarized by
$[O_2] [O_\frac{5}{2}] = [I]+ [O_\frac{5}{2}] +[O_{\frac{7}{2}}]$.

$\bullet$
The OPE $O_\frac{5}{2}(z) \, O_\frac{5}{2}(w)$

The computation of OPEs is getting more and more complicated and difficult as we move to the next higher spin currents.
From now on, we compute the OPEs
by the mathematica package \cite{Thielemans} exclusively
instead of computing by hand.
Our strategy is as follows. First we compute the OPEs
between the higher spin currents
for $N=2$ to find the complete structures of the OPEs. Then we
continue to compute the OPEs for different $N=3,4,5, \cdots, $
until we can find the $N$-dependence of structure constants fully.
Our method will be clear as we compute the simplest OPE
between $O_{\frac{5}{2}} (z)$ and itself.
From the computation of the OPE between $O_{\frac{5}{2}} (z)$
and itself for  $N=2$, we find that
\bea
O_{\frac{5}{2}}(z) \, O_{\frac{5}{2}}(w)
 & = & \frac{c_{\frac{5}{2}\frac{5}{2}}}{(z-w)^5}   +
\frac{1}{(z-w)^3} \, \left(c_{\frac{5}{2}\frac{5}{2}}^{o} O_{2} + c_{\frac{5}{2}\frac{5}{2}}^{t} T \right)+\frac{1}{(z-w)^2} \, \left[\frac{1}{2}c_{\frac{5}{2}\frac{5}{2}}^{o}\pa O_{2}
 + \frac{1}{2}c_{\frac{5}{2}\frac{5}{2}}^{t}\pa T \right](w)\nonu \\
& + & \frac{1}{(z-w)}
\left[ \frac{3}{20}c_{\frac{5}{2}\frac{5}{2}}^{o}\pa^2 O_{2} + \frac{3}{20}c_{\frac{5}{2}\frac{5}{2}}^{t}\pa^2 T +c_{\frac{5}{2}\frac{5}{2}}^{tt}\left( TT-\frac{3}{10}\pa^2 T \right)
\right. \nonu \\
& + &  \left.
c_{\frac{5}{2}\frac{5}{2}}^{to}\left( T O_{2}-\frac{3}{10}\pa^2 O_{2} \right)
+c_{\frac{5}{2}\frac{5}{2}}^{go}\left( G O_{\frac{5}{2}} -\frac{1}{5}\pa^2 O_{2} \right)+c_{\frac{5}{2}\frac{5}{2}}^{gg}\left( G \pa G -\frac{7}{10}\pa^2 T \right)
\right. \nonu \\
& + &  \left. O_{4} \right](w) +\cdots,
\label{O5O5}
\eea
with the structure constants for fixed $N=2$ are
\bea
c_{\frac{5}{2}\frac{5}{2}}(N=2) & = &-\frac{48 k^3 (5 + k) (1 + 2 k)}{(2 + k)^2 (3 + k)^3}, \qquad
c_{\frac{5}{2}\frac{5}{2}}^{o}(N=2) = \frac{8 (-4 + k) k}{(2 + k) (3 + k)}, \nonu \\
c_{\frac{5}{2}\frac{5}{2}}^{t}(N=2)& =&
-\frac{40 k^2 (5 + k) (1 + 2 k)}{(2 + k)^2 (3 + k)^2}, \qquad
c_{\frac{5}{2}\frac{5}{2}}^{tt}(N=2)  =
-\frac{48 k^2 (5 + k) (1 + 2 k)}{(2 + k)^2 (3 + k) (7 + 5 k)}, \nonu \\
c_{\frac{5}{2}\frac{5}{2}}^{to}(N=2) & = &
\frac{12 (-4 + k) k}{(2 + k) (1 + 2 k)},\qquad
c_{\frac{5}{2}\frac{5}{2}}^{go}(N=2) = -\frac{6 (-4 + k) k}{(2 + k) (1 + 2 k)},
\nonu \\
c_{\frac{5}{2}\frac{5}{2}}^{gg}(N=2) & = &
 \frac{12 k^2 (5 + k) (1 + 2 k)}{(2 + k)^2 (3 + k) (7 + 5 k)}.
\label{n2expp}
\eea
Of course, all the summation indices (appearing in the currents)
run as $a, b=1, \cdots, 4$.
As before, the relative coefficients of descendent fields of $O_2(z)$ and $T(z)$ are fixed by (\ref{PhiPhi}).
To check if there are extra fields in the second-order pole, we compute $[\{O_{\frac{5}{2}}\,O_{\frac{5}{2}}\}_{-2}-(\frac{1}{2}c_{\frac{5}{2}\frac{5}{2}}^{o}\pa O_{2}
 + \frac{1}{2}c_{\frac{5}{2}\frac{5}{2}}^{t}\pa T )](w)$.
It turns out that it is identically zero.
There are no extra fields besides $\pa O_2(w)$ and $\pa T(w)$.
The presence of the extra fields
in the first-order pole can be checked by calculating
the following quantity
$[\{O_{\frac{5}{2}}\,O_{\frac{5}{2}}\}_{-1}-
(\frac{3}{20}c_{\frac{5}{2}\frac{5}{2}}^{o}\pa^2 O_{2}
 + \frac{3}{20}c_{\frac{5}{2}\frac{5}{2}}^{t}\pa^2 T )](w)$.
Because this doesn't vanish, we should compute the following two OPEs:
\bea
&& T(z)\,
[\{O_{\frac{5}{2}}\,O_{\frac{5}{2}}\}_{-1}-(\frac{3}{20}c_{\frac{5}{2}\frac{5}{2}}^{o}\pa^2 O_{2}
 + \frac{3}{20}c_{\frac{5}{2}\frac{5}{2}}^{t}\pa^2 T )](w),  \nonu \\
&& G(z) \,
[\{O_{\frac{5}{2}}\,O_{\frac{5}{2}}\}_{-1}-(\frac{3}{20}c_{\frac{5}{2}\frac{5}{2}}^{o}\pa^2 O_{2}
 + \frac{3}{20}c_{\frac{5}{2}\frac{5}{2}}^{t}\pa^2 T )](w).
\label{aboveOPE}
\eea
These OPEs (\ref{aboveOPE}) have higher-order poles with order $n > 2$.
Looking at these higher-order
 poles enables us to find the extra four quasi-primary fields in the first-order pole in (\ref{O5O5}).
Then we compute $[\{O_{\frac{5}{2}}\,O_{\frac{5}{2}}\}_{-1}-(\mbox{first and second line in the first-order pole})](w)$ and it turns out that
this doesn't vanish and there exists
a new primary spin-4 field $O_4(z)$ which is the superpartner of
$O_{\frac{7}{2}}(z)$ given in (\ref{O7}).
As expected, the $O_4(z)$ obeys the following OPEs:
\bea
T(z) \;O_{4}(w)& =& \frac{1}{(z-w)^2} 4\,O_{4}(w) +\frac{1}{(z-w)}\,\pa O_{4}(w) +\cdots,\nonu \\
G(z) \;O_{4}(w) & = &\frac{1}{(z-w)^2} 7\,O_{\frac{7}{2}}(w)+\frac{1}{(z-w)} \,\pa O_{\frac{7}{2}}(w) +\cdots.
\label{TO4}
\eea
Therefore, the two currents $O_{\frac{7}{2}}(z)$ and $O_4(z)$
provide the correct ${\cal N}=1$ supermultiplet.
One can read off the explicit form of $O_{4}(z)$
from (\ref{O5O5}) as  follows:
\bea
O_{4}(w) & = &  \{ O_{\frac{5}{2}} \; O_{\frac{5}{2}} \}_{-1}(w)
-\left[ \frac{3}{20}c_{\frac{5}{2}\frac{5}{2}}^{o}\pa^2 O_{2} + \frac{3}{20}c_{\frac{5}{2}\frac{5}{2}}^{t}\pa^2 T +c_{\frac{5}{2}\frac{5}{2}}^{tt}\left( TT-\frac{3}{10}\pa^2 T \right)
\right.
\label{O4} \\
& + &  \left.
c_{\frac{5}{2}\frac{5}{2}}^{to}\left( T O_{2}-\frac{3}{10}\pa^2 O_{2} \right)
+c_{\frac{5}{2}\frac{5}{2}}^{go}\left( G O_{\frac{5}{2}} -\frac{1}{5}\pa^2 O_{2} \right)+c_{\frac{5}{2}\frac{5}{2}}^{gg}\left( G \pa G -\frac{7}{10}\pa^2 T \right)
 \right](w).
\nonu
\eea

Next we compute $O_{\frac{5}{2}}(z) \, O_{\frac{5}{2}}(w)$ for $N=3$ (where
the summation indices for the various currents
run as $a, b=1, \cdots, 6$) with the help of the package.
To find  the structure constants
$c_{\frac{5}{2}\frac{5}{2}}$, $c_{\frac{5}{2}\frac{5}{2}}^{o}$, and $ c_{\frac{5}{2}\frac{5}{2}}^{t}$,
we calculate $[\{ O_{\frac{5}{2}} \; O_{\frac{5}{2}} \}_{-5}-c_{\frac{5}{2}\frac{5}{2}}](w)$ and
$[\{ O_{\frac{5}{2}} \; O_{\frac{5}{2}} \}_{-3}-c_{\frac{5}{2}\frac{5}{2}}^{o} O_{2} - c_{\frac{5}{2}\frac{5}{2}}^{t} T](w)$ that should vanish.
Then we find the correct values for
$c_{\frac{5}{2}\frac{5}{2}}, c_{\frac{5}{2}\frac{5}{2}}^o$
and $c_{\frac{5}{2}\frac{5}{2}}^{t}$ when $N=3$.
For the remaining four structure constants, we use the equations
(\ref{O4}) and (\ref{TO4}) which should satisfy for any $N$.
Then the final
structure constants for $N=3$ are given by
\bea
c_{\frac{5}{2}\frac{5}{2}} & = &-\frac{162 k^3 (9 + k) (3 + 2 k)}{(4 + k)^2 (5 + k)^3}, \qquad
c_{\frac{5}{2}\frac{5}{2}}^{o}= \frac{12 (-6 + k) k}{(4 + k) (5 + k)}, \qquad
c_{\frac{5}{2}\frac{5}{2}}^{t} = -\frac{90 k^2 (9 + k) (3 + 2 k)}{(4 + k)^2 (5 + k)^2}, \nonu \\
c_{\frac{5}{2}\frac{5}{2}}^{tt} & = & -\frac{324 k^2 (9 + k) (3 + 2 k)}{(4 + k)^2 (5 + k) (35 + 19 k)}, \qquad
c_{\frac{5}{2}\frac{5}{2}}^{to} =\frac{108 (-6 + k) k}{(4 + k) (10 + 17 k)},\nonu \\
c_{\frac{5}{2}\frac{5}{2}}^{go} &= &-\frac{54 (-6 + k) k}{(4 + k) (10 + 17 k)},\qquad
c_{\frac{5}{2}\frac{5}{2}}^{gg} = \frac{81 k^2 (9 + k) (3 + 2 k)}{(4 + k)^2 (5 + k) (35 + 19 k)}.
\label{n3expp}
\eea

We repeat the above procedures
for $N=4$ and $N=5$ (the indices $a, b=1, \cdots, 8$ and
the indices $a, b=1, \cdots, 10$ respectively) and put all the results
together\footnote{
For $c_{\frac{5}{2}\frac{5}{2}}^{o}$, we obtain the following structure constants
for
different values of $N$
\bea
c_{\frac{5}{2}\frac{5}{2}}^{o}(N=2) & = & \frac{8 (-4+k) k}{(2+k) (3+k)},
\qquad
c_{\frac{5}{2}\frac{5}{2}}^{o}(N=3)=\frac{12 (-6+k) k}{(4+k) (5+k)},
\nonu \\
c_{\frac{5}{2}\frac{5}{2}}^{o}(N=4) & = & \frac{16 (-8+k) k}{(6+k) (7+k)},
\qquad
c_{\frac{5}{2}\frac{5}{2}}^{o}(N=5)=\frac{20 (-10+k) k}{(8+k) (9+k)}.
\label{foot1}
\eea
For  $c_{\frac{5}{2}\frac{5}{2}}^{t}$, the following structure constants
for each value of $N$ can be obtained
\bea
c_{\frac{5}{2}\frac{5}{2}}^{t}(N=2) & = &
\frac{(-40 k^2 (5+k) (1+2 k))}{(2+k)^2 (3+k)^2},
\qquad
c_{\frac{5}{2}\frac{5}{2}}^{t}(N=3)=
\frac{(-90 k^2 (9+k) (3+2 k))}{(4+k)^2 (5+k)^2},
\nonu \\
c_{\frac{5}{2}\frac{5}{2}}^{t}(N=4) & = &
\frac{(-160 k^2 (13+k) (5+2 k))}{(6+k)^2 (7+k)^2},
\qquad
c_{\frac{5}{2}\frac{5}{2}}^{t}(N=5)=\frac{(-250 k^2 (17+k) (7+2 k))}{(8+k)^2 (9+k)^2}.
\label{foot2}
\eea}.
Looking into the above results (\ref{foot1}) and (\ref{foot2}),
one can easily figure out
 that, for general $N$,
\bea
c_{\frac{5}{2}\frac{5}{2}}^{o}&=& \frac{4 k (k - 2 N) N}{(-2 + k + 2 N) (-1 + k + 2 N)},\nonu \\
c_{\frac{5}{2}\frac{5}{2}}^{t} &=& -\frac{10 k^2 N^2 (-3 + 2 k + 2 N) (-3 + k + 4 N)}{(-2 + k + 2 N)^2 (-1 + k + 2 N)^2}.
\label{known}
\eea
In this way, one can find all the structure constants for general $N$ in (\ref{O5O5}).
By converting $k$ into the central charge $c$ using
(\ref{ctilde}),
the structure constants of  (\ref{O5O5}) for general $N$ (which contains
(\ref{n2expp}) and (\ref{n3expp})) in terms of
$c$ and $N$  (rather than $k$ and $N$) are
\bea
c_{\frac{5}{2}\frac{5}{2}} & = &\frac{2 c^3 (2 c (-1+N)+3 (3-4 N) N) (c+2 c N+3 N (-3+2 N))}{9 (c+6 (-1+N) N)^2}, \nonu \\
c_{\frac{5}{2}\frac{5}{2}}^{o} & = & \frac{4 c (-6 N^2+c (-1+4 N))}{3 (c+6 (-1+N) N)}, \nonu \\
c_{\frac{5}{2}\frac{5}{2}}^{t} & = & \frac{10 c^2 (2 c (-1+N)+3 (3-4 N) N) (c+2 c N+3 N (-3+2 N))}{9 (c+6 (-1+N) N)^2}, \nonu \\
c_{\frac{5}{2}\frac{5}{2}}^{tt} & = & \frac{12 c^2 (2 c (-1+N)+3 (3-4 N) N) (c+2 c N+3 N (-3+2 N))}{(21+4 c) (c+6 (-1+N) N)^2}, \nonu \\
c_{\frac{5}{2}\frac{5}{2}}^{to} & = &\frac{36 c (-6 N^2+c (-1+4 N))}{(6+5 c) (c+6 (-1+N) N)},
\qquad
c_{\frac{5}{2}\frac{5}{2}}^{go} = -\frac{18 c (-6 N^2+c (-1+4 N))}{(6+5 c) (c+6 (-1+N) N)},\nonu \\
c_{\frac{5}{2}\frac{5}{2}}^{gg} & = & -\frac{3 c^2 (2 c (-1+N)+3 (3-4 N) N) (c+2 c N+3 N (-3+2 N))}{(21+4 c) (c+6 (-1+N) N)^2},
\label{expp}
\eea
where the second and third coefficients of (\ref{expp})
are the same as the ones in (\ref{known}).

Now let us compute the explicit form of $\{ O_{\frac{5}{2}} \;
O_{\frac{5}{2}} \}_{-1}(w)$ for general $N$.
To find all the operators and their coefficients in
$\{ O_{\frac{5}{2}} \; O_{\frac{5}{2}} \}_{-1}(w)$,
we use the same method as we did for the structure constants.
First we find all the operators in $\{ O_{\frac{5}{2}}
\; O_{\frac{5}{2}} \}_{-1}(w)$ for $N=2$.
In this case, there are three hundred twenty terms.
Then we investigate
them and write down the possible various spin-4 fields with coefficients as follows:
\bea
\{ O_{\frac{5}{2}} \; O_{\frac{5}{2}} \}_{-1}(w) & = &
c_1 \, J^{ab} J^{ad} J^{b\, 2N+1}J^{dN'}(w) +c_2 \, J^{ab} J^{aN'} \pa J^{bN'}(w)+c_3 \pa J^{aN'} \pa J^{aN'}(w)\nonu \\
&+ & c_4 \pa J^{ab} \pa J^{ab}(w) + 
c_5 \, \pa^2 J^{ab} J^{ab}(w) + c_6 \pa^2 J^{aN'}J^{bN'}(w)
\nonu \\
&+ & c_7 K^{a} K^{b} J^{ac} \pa J^{bc}(w)+ c_8 K^{a} K^{b} \pa^2 J^{ab}(w) +c_9 \, K^a \pa K^b J^{ac} J^{bc}(w)
\nonu \\
&+ & c_{10} K^a \pa K^b \pa J^{ab}(w)+ c_{11} K^a \pa^2 K^b J^{ab}(w) +c_{12} \pa^3 K^a K^a(w)
\nonu \\
&+ & c_{13}K^a K^b J^{ab} J^{cN'}J^{cN'}(w)+ c_{14} K^a K^b J^{ac} J^{bd} J^{cd}(w)
\nonu \\
&+ & c_{15} \pa K^a K^a J^{bc} J^{bc}(w) +
c_{16} K^a K^b J^{aN'}\pa J^{bN'}(w)+c_{17} K^a \pa K^b J^{aN'} J^{bN'}(w)
\nonu \\
&+ & c_{18} \pa K^a K^a J^{bN'}J^{bN'}(w)+c_{19} \pa K^a \pa K^b J^{ab}(w) +c_{20} \pa^2 K^a \pa K^a(w),
\label{O5O5p1}
\eea
where $N' \equiv   2N+1$ and the dummy indices
run as $a, b, c, d = 1, \cdots, 2N$.
Of course, for $N=2$, the expression (\ref{O5O5p1})
consists  of $320$ independent terms (by expanding them out completely)
as mentioned before. 
With the tensorial structures in (\ref{O5O5p1}), there exist only 
twenty coefficient functions for general $N$.
Then we solve $[\{ O_{\frac{5}{2}} \; O_{\frac{5}{2}} \}_{-1}-
(\mbox{ the right hand side of (\ref{O5O5p1})})](w)=0$
for the coefficients.
The left hand side of (\ref{O5O5p1}) is given by too many
WZW currents
and we want to write it using the tensorial structure to simplify.
We repeat this procedure for $N=3, 4,5$\footnote{
For the coefficient $c_1$ appearing
in the first term of (\ref{O5O5p1}), each coefficient function can be obtained
\bea
c_1(N=2) & = & \frac{-36 k^2}{(2+k)^2 (3+k)^3},
\qquad
c_1(N=3)=\frac{-81 k^2}{(4+k)^2 (5+k)^3},
\nonu \\
c_1(N=4) & = & \frac{-144 k^2}{(6+k)^2 (7+k)^3},
\qquad
c_1(N=5)=\frac{-225 k^2}{(8+k)^2 (9+k)^3}.
\label{inter1}
\eea
Similarly, for the coefficient $c_2$ appearing in the second term of (\ref{O5O5p1}),
the coefficient functions for different $N$-values are given by
\bea
c_2(N=2) & = & \frac{-72 k^2 (4+k)}{(2+k)^2 (3+k)^3},
\qquad
c_2(N=3)  = \frac{-162 k^2 (7+k)}{(4+k)^2 (5+k)^3},
\nonu \\
c_2(N=4) & = & \frac{-288 k^2 (10+k)}{(6+k)^2 (7+k)^3},
\qquad
c_2(N=5)=\frac{-450 k^2 (13+k)}{(8+k)^2 (9+k)^3}.
\label{inter2}
\eea}.
From the above results (\ref{inter1}) and (\ref{inter2}),
one can easily figure out the general forms of  $c_1$ and $c_2$. For general $N$,
 the coefficients $c_1$ and $c_2$ are given by
\bea
c_1(N,k)&=& -\frac{9 k^2 N^2}{(-2+k+2 N)^2 (-1+k+2 N)^3},\nonu \\
c_2(N,k) &=& -\frac{18 k^2 N^2 (-2+k+3 N)}{(-2+k+2 N)^2 (-1+k+2 N)^3}.
\nonu
\eea
In this way one can find all coefficients in (\ref{O5O5p1}) for general $N$.
 We also put the equations of the coefficients for $N=2$ in (\ref{n2expres})
of Appendix $A$.
The general forms of coefficients are presented in
(\ref{O5O5p1coeff}) of Appendix $A$.
All results regarding the OPE $O_{\frac{5}{2}}(z)\,
O_{\frac{5}{2}}(w)$ have been checked up to $N=8$ by mathematica package
\cite{Thielemans}.
The OPE $O_{\frac{5}{2}}(z) \, O_{\frac{5}{2}}(w)$ for general $N$
can be obtained from
(\ref{O5O5}) with (\ref{expp}).
The fusion rule can be summarized by
$[O_{\frac{5}{2}}] [O_\frac{5}{2}] = [I]+ [O_2] +[O_{4}]$.

Then  the ${\cal N}=1$ fusion rule is summarized by
$[\hat{O}_2][\hat{O}_2]= [\hat{I}] +[\hat{O}_2] +[\hat{O}_{\frac{7}{2}}]$.
The explicit OPE for this is given by (\ref{singleO2O2}).
By rescaling the currents
as $O_2(z) \rightarrow N_2 O_2(z)$ and $O_{\frac{5}{2}}(z)
\rightarrow N_{\frac{5}{2}}
O_{\frac{5}{2}}(z)$, where
\bea
N_2^2 &= &-\frac{9 (c + 6 (-1 + N) N)^2}{c^2 (2 c (-1 + N) + 3 (3 - 4 N) N) (c + 2 c N + 3 N (-3 + 2 N))},\nonu \\
N_{\frac{5}{2}}^2 &= &-\frac{1}{5}N_2^2,
\nonu
\eea
one has  the standard normalizations: $O_2(z) \, O_2(w) \rightarrow
\frac{1}{(z-w)^4} \, \frac{c}{2} +\cdots$ and
 $O_{\frac{5}{2}}(z) \, O_{\frac{5}{2}}(w) \rightarrow
\frac{1}{(z-w)^5} \, \frac{2c}{5} +\cdots$.
Therefore all the previous OPEs can be rewritten in terms of these rescaled
currents. For example, in the OPE of (\ref{O2O5a}),
the spin-$\frac{7}{2}$ is simply taken from the first-order pole subtracted by
the descendant fields and quasi-primary fields in the
right hand side. Strictly speaking, one should consider
the structure constant coming from the spin-$2$ current,
the spin-$\frac{5}{2}$ current
and the spin-$\frac{7}{2}$ current. We will see that the normalization of
spin-$\frac{7}{2}$ current is needed also. Then there  should exist
a nontrivial structure constant in front of spin-$\frac{7}{2}$ current 
\footnote{The reason why we do not continue to do in this direction is 
that once we obtain the higher spin current for given spin, then one should 
determine the highest singular term of this 
OPE between this current and itself in order to fix the 
normalization. For lower higher spin 
currents, this can be done by hand (or by package). However, as the spin 
increases, it is hard to obtain the highest singular term for this OPE even 
by the package (it takes too much time). 
Note that in this paper, we have 
considered the normalization of higher spin current
up to the spin $\frac{9}{2}$ (See the end of the section $3$).  }.

\subsection{ The OPEs between the higher spin currents of spins-$(2',
\frac{5}{2})$ and the higher spin currents of
spins-$(\frac{7}{2},4)$}

$\bullet$ The OPE $O_2(z) \, O_{\frac{7}{2}}(w)$

Now we compute the OPE between  $O_2 (z)$ and $O_{\frac{7}{2}} (w)$ in order to
calculate the OPEs between the
two ${\cal N}=1$ multiplets, $(2',\frac{5}{2})$ and
$(\frac{7}{2},4)$, in (\ref{currents}).
To find the structure of the OPE, we compute it for $N=2$ as mentioned before.
The final result is presented first which explains how the result can be
obtained explicitly.
This OPE is described as
\bea
O_{2}(z) \, O_{\frac{7}{2}}(w) & = & \frac{1}{(z-w)^3}  c_{2 \frac{7}{2}}^{o} O_{\frac{5}{2}}(w)
+\frac{1}{(z-w)^2} \left[ \frac{1}{5} c_{2 \frac{7}{2}}^{o} \pa O_{\frac{5}{2}} + c_{2 \frac{7}{2}}^{o'} O_{\frac{7}{2}}
+c_{2 \frac{7}{2}}^{go}\left( G O_{2} -\frac{2}{5} \pa O_{\frac{5}{2}}  \right) \right] (w) \nonu \\
& + & \frac{1}{(z-w)} \left[  \frac{1}{30} c_{2 \frac{7}{2}}^{o} \pa^2 O_{\frac{5}{2}} +\frac{2}{7} c_{2 \frac{7}{2}}^{o'} \pa O_{\frac{7}{2}}
 +\frac{2}{7} c_{2 \frac{7}{2}}^{go} \pa\left( G O_{2}-\frac{2}{5} \pa O_{\frac{5}{2}}  \right)
 \right. \nonu \\
& + & \left.  c_{2 \frac{7}{2}}^{go'} \left( \pa G O_{2}  -\frac{3}{4} G \pa O_{2}+ \frac{1}{8}\pa^2 O_{\frac{5}{2}} \right)
+ c_{2 \frac{7}{2}}^{to} \left( T O_{\frac{5}{2}} - \frac{1}{4}\pa^2 O_{\frac{5}{2}} \right)
 \right. \nonu \\
& + & \left.
O_{\frac{9}{2}}\right](w)
\cdots.
\label{O2O7}
\eea
All the coefficients for $N=2$ in the right hand side of (\ref{O2O7})
are determined.
The relative coefficients of descendent fields of $O_{\frac{5}{2} }(w)$,
$O_{\frac{7}{2} }(w)$, and
$\left( G O_{2} -\frac{2}{5} \pa O_{\frac{7}{2}}  \right)(w)$
 are determined by (\ref{PhiPhi}).
For the second-order pole, we compute $T(z) \,
[\{ O_{2} \; O_{\frac{7}{2}} \}_{-2}-\frac{1}{5} c_{2 \frac{7}{2}}^{o}
\pa O_{\frac{5}{2}}](w)$
and $G(z) \,
[\{ O_{2} \; O_{\frac{7}{2}} \}_{-2}-\frac{1}{5} c_{2 \frac{7}{2}}^{o} \pa O_{\frac{5}{2}}](w)$ to find extra fields in the second-order pole.
These OPEs have higher-order poles with order $n > 2$.
Adding  $\left( G O_{2} -\frac{2}{5} \pa O_{\frac{5}{2}}  \right)(w)$
with the coefficient $c_{2 \frac{7}{2}}^{go}$  to the second-order pole makes
the higher-order poles of the above OPEs disappear. Then we compute $
[\{ O_{2} \; O_{\frac{7}{2}} \}_{-2}-\frac{1}{5} c_{2 \frac{7}{2}}^{o} \pa O_{\frac{5}{2}}
-c_{2 \frac{7}{2}}^{go}\left( G O_{2} -\frac{2}{5} \pa O_{\frac{5}{2}}  \right)](w)$
which doesn't vanish implying that
there exists the primary spin-$\frac{7}{2}$
field $O_{\frac{7}{2}}(z)$ (that has been found in (\ref{O7}))
in the second-order pole).

For the first-order pole, we compute $[\{ O_{2} \; O_{\frac{7}{2}} \}_{-1}-\mbox{first line }](w)$ leading to the nonvanishing quantity.
So let us compute $T(z) \,
[\{ O_{2} \; O_{\frac{7}{2}} \}_{-1}-\mbox{first line }](w)$
and $G(z) \, [\{ O_{2} \; O_{\frac{7}{2}} \}_{-1}-\mbox{first line }](w)$ as before.
It turns out that the OPEs have higher-order poles with order $n >2$.
To remove the higher-order poles, we should add
the quasi-primary fields in the second line of the first-order pole
in (\ref{O2O7}). Then we
compute $[\{ O_{2} \; O_{\frac{7}{2}} \}_{-1}-\mbox{first line }-\mbox{second line }](w)$ which doesn't vanish. This implies that there exists
a new primary spin-$\frac{9}{2}$ field $O_{\frac{9}{2} }(z)$.
As expected, the spin-$\frac{9}{2}$ current
$O_{\frac{9}{2} }(z)$ obeys the following OPEs:
\bea
T(z) \, O_{\frac{9}{2}}(w)& =& \frac{1}{(z-w)^2} \,
\frac{9}{2}\,O_{\frac{9}{2}}(w) +\frac{1}{(z-w)}\,\pa O_{\frac{9}{2}}(w) +\cdots,\nonu \\
G(z) \, O_{\frac{9}{2}}(w) & = &\frac{1}{(z-w)^2} \,8\,O_{4'}(w)+
\frac{1}{(z-w)} \,\pa O_{4'}(w) +\cdots,
\label{GO9}
\eea
where $O_{4'}(z)$ is a new spin-4 primary field that will appear in
(\ref{O44}) in the next OPE.
The explicit form of $O_{\frac{9}{2}}(z)$ can be derived as follows:
\bea
O_{\frac{9}{2}}(w) & = &  \{ O_{2} \; O_{\frac{7}{2}} \}_{-1}(w)
-\left[  \frac{1}{30} c_{2 \frac{5}{2}}^{o} \pa^2 O_{\frac{5}{2}} +\frac{2}{7} c_{2 \frac{5}{2}}^{o'} \pa O_{\frac{7}{2}} +\frac{2}{7} c_{2 \frac{5}{2}}^{go} \pa\left( G O_{2}-\frac{2}{5} \pa O_{\frac{5}{2}}  \right)
 \right. \nonu \\
& + & \left.  c_{2 \frac{5}{2}}^{go'} \left( \pa G O_{2}  -\frac{3}{4} G \pa O_{2}+ \frac{1}{8}\pa^2 O_{\frac{5}{2}} \right)
+ c_{2 \frac{5}{2}}^{to} \left( T O_{\frac{5}{2}} - \frac{1}{4}\pa^2 O_{\frac{5}{2}} \right)\right](w).
\label{O9}
\eea

To find the structure constants in (\ref{O2O7}) for general $N$, we compute the OPE for $N=3,4,5$ as we did before.
To find  the structure constants
$c_{2 \frac{7}{2}}^{o}$, $c_{2 \frac{7}{2}}^{o'}$, and $c_{2 \frac{7}{2}}^{go}$, we solve the following equations:
\bea
\left[\{ O_{2} \; O_{\frac{7}{2}} \}_{-3}-c_{2 \frac{5}{2}}^{o} O_{\frac{5}{2}}
\right](w) & = &
0,\nonu \\
\left[\{ O_{2} \; O_{\frac{7}{2}} \}_{-2}-\frac{1}{5} c_{2 \frac{5}{2}}^{o} \pa O_{\frac{5}{2}} - c_{2 \frac{5}{2}}^{o'} O_{\frac{7}{2}}
-c_{2 \frac{5}{2}}^{go}\left( G O_{2} -\frac{2}{5} \pa O_{\frac{5}{2}}  \right)
\right](w) & = &
0.
\label{O2O7eq1}
\eea
Then from these relations
(\ref{O2O7eq1}), the three structure constants
can be determined for $N=3, 4, 5$ and therefore the first three coefficients
in (\ref{O9}) are determined completely.
To find  the remaining undetermined coefficients,
$c_{2 \frac{7}{2}}^{go'}$ and $c_{2 \frac{7}{2}}^{to}$, we compute
$T(z)\,O_{\frac{9}{2}}(w)$ and $G(z)\, O_{\frac{9}{2}}(w)$ and use
the fact that there should be no higher-order poles with order $n >2$
as in (\ref{GO9}). We repeat this procedure for $N=3, 4, 5$ cases.

Then we put all the results together, and find the general forms of structure constants in (\ref{O2O7}). The structure constants for general $N$ in
(\ref{O2O7}) are given by
\bea
c_{2 \frac{7}{2}}^{o} & = & -\frac{6 c^2 (-3 + 2 c) (1 + N) (c^2 - 3 c (-1 + N) + 18 (1 - 3 N) N) (6 N +
   c (-3 + 2 N))}{(21 + 4 c) (6 + 5 c) (c + 6 (-1 + N) N)^2}, \nonu \\
c_{2 \frac{7}{2}}^{o'}& = & \frac{2 c (21 + 4 c) (c - 4 c N + 6 N^2)}{3 (6 + 5 c) (c + 6 (-1 + N) N)}, \nonu \\
c_{2 \frac{7}{2}}^{go}& = & -\frac{108 c^2 (-3 + 2 c) (1 + N) (c^2 - 3 c (-1 + N) +
   18 (1 - 3 N) N) (6 N + c (-3 + 2 N))}{(21 + 4 c) (6 + 5 c)^2 (c + 6 (-1 + N) N)^2},\nonu \\
c_{2 \frac{7}{2}}^{go'} & = & -\frac{48 c^2 (-45 + 2 c) (3 + c - 9 N) (1 + N) (c + 6 N) (6 N +
   c (-3 + 2 N))}{7 (29 + 2 c) (21 + 4 c) (6 + 5 c) (c + 6 (-1 + N) N)^2},\nonu \\
c_{2 \frac{7}{2}}^{to} & = & -\frac{12 c^2 (-3 + 10 c) (3 + c - 9 N) (1 + N) (c + 6 N) (6 N +
   c (-3 + 2 N))}{(29 + 2 c) (21 + 4 c) (6 + 5 c) (c + 6 (-1 + N) N)^2}.
\label{coeff9half}
\eea

To find the explicit structure of $\{ O_{2} \; O_{\frac{7}{2}} \}_{-1}(w)$
for general $N$, we examine the operators in
$\{ O_{2} \; O_{\frac{7}{2}} \}_{-1}(w)$ for $N=2$ and
write down the possible various spin-$\frac{9}{2}$
fields with undetermined coefficients.
Then we solve $[\{ O_{2} \; O_{\frac{7}{2}} \}_{-1}-(\mbox{the possible various spin-$\frac{9}{2}$ fields})](w)=0$ for these coefficients.
We repeat this procedure for $N=3, 4,5$ cases. The
first-order pole $\{ O_{2} \; O_{\frac{7}{2}} \}_{-1}(w)$, which consists of
thirty-two terms, is presented as follows:
\bea
\{ O_{2} \; O_{\frac{7}{2}} \}_{-1}(w) &=&
c_1 K^a J^{ab} J^{bc} J^{dc} J^{d N'}(w)+ c_2  K^a J^{ab} J^{cd} J^{cd} J^{b N'}(w)+c_3  K^a J^{ab} J^{cd} J^{bd} J^{c N'}(w)\nonu \\
&+&  c_4  K^a J^{bc} J^{bc} \pa  J^{a N'}(w) +c_5 K^a J^{ab} J^{bc} \pa J^{cN'}(w) +c_6 K^a J^{ab} J^{bN'} J^{cN'} J^{c N'}(w)\nonu \\
&+& c_7 K^a J^{ab}  \pa^2 J^{b N'}(w) +c_8 K^a J^{aN'} \pa J^{bN'} J^{bN'} (w)+c_9 K^a K^b K^c J^{ab} J^{cd}  J^{d N'}(w)\nonu \\
&+& c_{10} K^a J^{bc} \pa J^{ac}  J^{b N'}(w)+c_{11} K^a J^{bN'} J^{bN'} \pa  J^{a N'}(w)+c_{12} K^a K^b K^c J^{ab} \pa J^{cN'}(w) \nonu \\
&+& c_{13}K^a K^b\pa K^c J^{ab} J^{cN'}(w)+ c_{14}K^a K^b K^c \pa J^{ab} J^{cN'}(w)+ c_{15}K^a K^b\pa K^c J^{ac} J^{bN'}(w)\nonu \\
&+&  c_{16}K^a \pa J^{ab} J^{bc} J^{cN'}(w) +c_{17}K^a \pa J^{bc} J^{bc} J^{aN'}(w)+ c_{18}K^a \pa J^{ab}\pa J^{bN'}(w)\nonu \\
&+& c_{19}K^a \pa K^b K^b J^{ca}  J^{cN'}(w) +c_{20 }K^a \pa K^b K^b \pa J^{aN'}(w) +c_{22}K^a \pa^2 K^b K^b  J^{aN'}(w)\nonu \\
&+& c_{23}K^a \pa^3 J^{aN'}(w) +c_{24} \pa K^a J^{bc} J^{bc} J^{aN'}(w)  +c_{25} \pa K^a J^{ab} J^{bc} J^{cN'}(w)\nonu \\
&+& c_{26} \pa K^a J^{ab} \pa J^{bN'}(w) +c_{27} \pa K^a J^{aN'} J^{bN'} J^{bN'}(w) +c_{28} \pa K^a \pa J^{ab} J^{bN'} (w)\nonu \\
&+& c_{29} \pa K^a K^a \pa K^b J^{bN'}(w) +c_{30} \pa K^a \pa^2 J^{aN'}(w) +c_{31} \pa^2 K^a J^{ab} J^{bN'} (w)\nonu \\
&+& c_{32} \pa^2 K^a \pa J^{aN'}(w) +c_{33} \pa^3 K^a J^{aN'} (w),
\label{O2O7p1}
\eea
where $N' \equiv 2N+1$ and the summation indices run $a,b,c,d =1, \cdots, 2N$.
The coefficients in (\ref{O2O7p1}) are presented in
(\ref{O2O7p1coeff}) of Appendix $B$.
All the results regarding the OPE between $O_2(z)$ and $O_{\frac{7}{2}}(z)$ are checked up to $N=7$ case by mathematica package \cite{Thielemans}.

Therefore, the spin-$\frac{9}{2}$ current $O_{\frac{9}{2}}(z)$
is summarized by (\ref{O9}), (\ref{coeff9half}), and (\ref{O2O7p1}).
The OPE $O_2(z) \, O_{\frac{7}{2}}(w)$ for general $N$
is given by (\ref{O2O7}) with (\ref{coeff9half}).
The fusion rule is summarized by
$[O_2][O_\frac{7}{2}]= [I] + [O_{\frac{5}{2}}] +[{O}_{\frac{7}{2}}] +
[O_{\frac{9}{2}}]$.

$\bullet$ The OPE $O_2(z) \, O_{4}(w)$

Now we move to the OPE between the $O_2 (z)$ and $O_{4} (z)$.
From the computation of the OPE for $N=2$, we find that
this OPE takes the form
\bea
O_{2}(z) \, O_{4}(w)
 & = & \frac{1}{(z-w)^4} c_{24}^{o} O_{2} \nonu \\
& + & \frac{1}{(z-w)^2}
\left[ c_{24}^{go}\left( G O_{\frac{5}{2}}-\frac{1}{5}\pa^2 O_{2} \right)+c_{24}^{to}\left( T O_{2}-\frac{3}{10}\pa^2 O_{2} \right)+c_{24}^{o'} O_{4}+c_{24}^{o''}\,O_{4^{'}}  \right](w) \nonu \\
& + & \frac{1}{(z-w)}
\left[\frac{1}{4} c_{24}^{go} \pa \left( G O_{\frac{5}{2}}-\frac{1}{5}\pa^2 O_{2} \right)+\frac{1}{4}c_{24}^{to} \pa \left( T O_{2}-\frac{3}{10}\pa^2 O_{2} \right)
+\frac{1}{4}c_{24}^{o'} \pa O_{4}\right. \nonu \\
& + & \frac{1}{4}c_{24}^{o''}\pa O_{4^{'}} +c_{24}^{to'} \left( T \pa O_{2} -\pa T O_{2} -\frac{1}{6} \pa^3 O_{2} \right)+ c_{24}^{go'}\left( G \pa O_{\frac{5}{2}}-\frac{5}{3} \pa G O_{\frac{5}{2}}-\frac{1}{9} \pa^3 O_{2} \right)\nonu \\
& +  &  \left.  c_{24}^{go''} \left( G O_{\frac{7}{2}}-\frac{1}{4} \pa O_{4} \right) \right](w) +\cdots.
\label{O2O4}
\eea
At the moment, the right hand side holds for $N=2$ case only.
We would like to obtain this OPE for general $N$. In particular, the
structure constants for general $N$.
The relative coefficients of descendants fields are found by (\ref{PhiPhi})
once again.
According to (\ref{PhiPhi}), there is no third-order pole because
the coefficient for the descendant field of $O_{2}(w)$ is zero in the OPE.
To find the
fields in the second-order pole, first we compute $T(z) \,
\{ O_{2} \; O_{4} \}_{-2}(w)$ and $G(z) \,
\{ O_{2} \; O_{4} \}_{-2}(w)$ as before.
The two OPEs have higher-order poles with order $n>2$.
We look at the higher-order poles and find that the following
two quasi-primary fields can remove all the higher-order poles in the above
OPEs:
$\left( G O_{\frac{5}{2}}-\frac{1}{5}\pa^2 O_{2} \right)(w)$
and $\left( T O_{2}-\frac{3}{10}\pa^2 O_{2} \right)(w)$
with appropriate coefficients
$c_{24}^{go}$ and $c_{24}^{to}$ respectively.
Then we subtract
$c_{24}^{go}\left( G O_{\frac{5}{2}}-\frac{1}{5}\pa^2 O_{2} \right)(w)$
and $c_{24}^{to}\left( T O_{2}-\frac{3}{10}\pa^2 O_{2} \right)(w)$
from the second-order pole term $\{ O_{2} \; O_{4} \}_{-2}(w)$.
This doesn't vanish implying
that the second-order pole has at least one primary spin-4 field.
So we solve the following equation for $c_{24}^{o'}$:
\bea
[\{ O_{2} \; O_{4} \}_{-2}-c_{24}^{go}\left( G O_{\frac{5}{2}}-\frac{1}{5}\pa^2 O_{2} \right)-c_{24}^{to}\left( T O_{2}-\frac{3}{10}\pa^2 O_{2} \right)-c_{24}^{o'} O_{4}](w)=0.
\label{O2O4eq1}
\eea
It turns out that there is no solution for $(\ref{O2O4eq1})$.
This implies that there should be  a new primary spin-4 field and
that this new spin-4 field would be the superpartner of $O_{\frac{9}{2}}(z)$.
To check this, we consider the following equation for $c_{24}^{o'}$ and $c'$ :
\bea
&& [\{ O_{2} \; O_{4} \}_{-2}-c_{24}^{go} \left(
G O_{\frac{5}{2}}-\frac{1}{5}\pa^2 O_{2} \right)
-c_{24}^{to} \left( T O_{2}-\frac{3}{10}\pa^2 O_{2} \right)-c_{24}^{o'} O_{4}](w)
\nonu \\
&& +\frac{c' c_{24}^{o''}}{8}\{ G \; O_{\frac{9}{2}} \}_{-2}(w)=0.
\label{O2O4eq2}
\eea
See also the equation (\ref{GO9}).
It turns out that there exists
a solution for $(\ref{O2O4eq2})$ when $c'=-1$. So we can be sure about that the new spin-4 field is nothing but $O_{4'}(z)$.
Therefore, the explicit form of $O_{4'}(z)$, from (\ref{O2O4eq2}),
can be written as
\bea
O_{4'}(z)  =  \frac{1}{c_{24}^{o''}}[\{ O_{2} \; O_{4} \}_{-2}(w)-
 c_{24}^{go}( G O_{\frac{5}{2}}-\frac{1}{5}\pa^2 O_{2} )-c_{24}^{to}
( T O_{2}-\frac{3}{10}\pa^2 O_{2}) -c_{24}^{o'} O_{4}](z).
\label{O44}
\eea
As expected, the spin-$4$ current, $O_{4'}(z)$,
obeys the following OPEs:
\bea
T(z) \, O_{4'}(w)& =& \frac{1}{(z-w)^2} \,4\,O_{4'}(w) +
\frac{1}{(z-w)}\,\pa O_{4'}(w) +\cdots,\nonu \\
G(z) \, O_{4'}(w) & = &\frac{1}{(z-w)} \, O_{\frac{9}{2}}(w) +\cdots,
\label{GO44}
\eea
where the spin-$\frac{9}{2}$ current is given by (\ref{O9}).
Therefore, the $O_{4'}(z)$ and $O_{\frac{9}{2}}(z)$  provide the correct
${\cal N}=1$ multiplet.

To check if there are extra fields besides the
descendant fields in the first-order pole, we compute $[\{ O_{2} \; O_{4} \}_{-1}-(\mbox{desc. fields in the 1st-order pole})](w)$.
Because this doesn't vanish, we should compute the following two OPEs:
\bea
&&T(z) \,
[\{ O_{2} \; O_{4} \}_{-1}-(\mbox{ desc. fields in the 1st-order pole})](w),
\nonu \\
&&G(z) \,
[\{ O_{2} \; O_{4} \}_{-1}-(\mbox{ desc. fields in the 1st-order pole})](w).
\label{O2O4eq22}
\eea
Here the descendant fields in the 1st-order pole contain the first four terms
in (\ref{O2O4}).
The above OPEs (\ref{O2O4eq22})
have higher-order poles with order $n>2$ and these higher-order
poles are removed by the following two quasi-primary fields:
$\left( T \pa O_{2} -\pa T O_{2} -\frac{1}{6} \pa^3 O_{2} \right)(w)$
and $\left( G \pa O_{\frac{5}{2}}-\frac{5}{3} \pa G O_{\frac{5}{2}}-\frac{1}{9} \pa^3 O_{2} \right)(w)$
with the coefficients $c_{24}^{to'} $ and $ c_{24}^{go'}$ respectively.
Furthermore, there exists $G O_{\frac{7}{2}}$ term with derivative.
Then we compute the following expression:
\bea
[\{ O_{2} \; O_{4} \}_{-1}-(\mbox{ desc. fields + quasi. fields})](w),
\nonu
\eea
which vanishes and there are no extra fields in the first-order pole.

To find the structure constants in (\ref{O2O4}) for general $N$,
we compute the OPEs for $N=3,4,5$ cases.
We use $[\{ O_{2} \; O_{4} \}_{-4}- c_{24}^{o} O_{2}](w)=0$ to
find the structure constant
$c_{24}^{o}$. To find the next coefficients
$c_{24}^{go}, c_{24}^{to}, c_{24}^{o'}$, and $c_{24}^{o''}$,
the equations (\ref{O44}) and (\ref{GO44}) can be used.
To find the remaining structure constants
$c_{24}^{to'}, c_{24}^{go'}$ and $c_{24}^{go''}$,
we solve the following equation for them:
\bea
\left[\{ O_{2} \; O_{4} \}_{-1} - (\mbox{the 1st-order pole}) \right](w)=0.
\nonu
\eea
It turns out that
the structure constants in (\ref{O2O4}) for general $N$ are given by
\bea
c_{24}^{o} & = & \frac{24 c^2 (-3 + 2 c) (3 + c - 9 N) (1 + N) (c + 6 N) (6 N + c (-3 + 2 N))}{(21 + 4 c) (6 + 5 c) (c + 6 (-1 + N) N)^2}, \nonu \\
c_{24}^{go} & = &\frac{18 c^2 (-3 + 2 c) (156 + 35 c) (3 + c - 9 N) (1 + N) (c + 6 N) (6 N +
   c (-3 + 2 N))}{(29 + 2 c) (21 + 4 c) (6 + 5 c)^2 (c + 6 (-1 + N) N)^2}, \nonu \\
c_{24}^{to} & = &\frac{24 c^2 (-3 + 2 c) (-123 + 40 c) (3 + c - 9 N) (1 + N) (c + 6 N) (6 N +
    c (-3 + 2 N))}{(29 + 2 c) (21 + 4 c) (6 + 5 c)^2 (c + 6 (-1 + N) N)^2},\nonu \\
c_{24}^{o'} & = & -\frac{10 c (21 + 4 c) (-6 N^2 + c (-1 + 4 N))}{21 (6 + 5 c) (c + 6 (-1 + N) N)},\qquad
c_{24}^{o''}=-8, \nonu \\
c_{24}^{to'} & = & -\frac{6 c^2 (-33 + 40 c) (3 + c - 9 N) (1 + N) (c + 6 N) (6 N +
   c (-3 + 2 N))}{(21 + 4 c) (6 + 5 c)^2 (c + 6 (-1 + N) N)^2},\nonu \\
c_{24}^{go'} & = & -\frac{9 c^2 (-24 + 7 c) (3 + c - 9 N) (1 + N) (c + 6 N) (6 N + c (-3 + 2 N))}{2 (21 + 4 c) (6 + 5 c)^2 (c + 6 (-1 + N) N)^2}, \nonu \\
c_{24}^{go''} & = & \frac{18 c (-6 N^2 + c (-1 + 4 N))}{(6 + 5 c) (c + 6 (-1 + N) N)}.
\label{Coefff}
\eea

To find the explicit form of $\{ O_{2} \; O_{4} \}_{-2}(w)$
for general $N$, we examine the operators in
$\{ O_{2} \; O_{4} \}_{-2}(w)$ for $N=2$ and
write down the possible various spin-4 fields with unknown coefficients.
Then we solve $[\{ O_{2} \; O_{4} \}_{-2}-(\mbox{the various spin-4 fields})](w)=0$ for unknown coefficients when $N=2, 3, 4, 5$ cases.
The second-order pole $\{ O_{2} \; O_{4} \}_{-2}(w)$, which consists of
twenty-nine terms, is presented as follows:
\bea
\{ O_{2} \; O_{4} \}_{-2}(w) &=&
c_1 J^{ab} J^{ab} J^{cd} J^{cd}(w) +c_2 J^{ab} J^{ab} J^{cN'} J^{cN'}(w) +c_3 J^{ab} J^{ac} J^{bd} J^{cd}(w)
\nonu \\
&+& c_4 J^{ab} J^{ac} J^{bN'} J^{cN'}(w)+ c_5 J^{ab} J^{aN'} \pa  J^{bN'}(w) +c_6 J^{aN'} J^{aN'} J^{bN'} J^{bN'}(w)
\nonu \\
&+& c_7 \pa J^{ab} \pa J^{ab} (w) + c_8 \pa^2 J^{ab} J^{ab}(w) +c_9 \pa^2 J^{aN'} J^{aN'}(w)
\nonu \\
&+& c_{10} K^a K^b J^{ab} J^{cd} J^{cd}(w)  +c_{11} K^a K^b J^{ab} J^{cN'} J^{cN'}(w) +c_{12} K^a K^b J^{ac} J^{bd} J^{cd}(w)
\nonu \\
&+& c_{13} K^a K^b J^{ac} J^{bN'} J^{cN'}(w) +c_{14} K^a K^b J^{aN'} \pa J^{bN'}(w) +c_{15} K^a K^b K^c K^d J^{ab} J^{cd}(w)
\nonu \\
&+& c_{16} K^a K^b \pa K^c K^c J^{ab}(w) +c_{17} K^a \pa K^b J^{ac} J^{bc}(w) +c_{18} K^a \pa K^b J^{aN'} J^{bN'}(w)
\nonu \\
&+& c_{19} K^a \pa K^b \pa J^{ab}(w) +c_{20} K^a \pa^2 K^b J^{ab} (w) +c_{21} \pa J^{aN'} \pa J^{aN'} (w)
\nonu \\
&+& c_{22} \pa K^a K^a J^{bc} J^{bc} (w) +c_{23} \pa K^a K^a J^{bN'} J^{bN'}(w) +c_{24} K^a K^b \pa ^2 J^{ab}(w)
\nonu \\
&+& c_{25} K^a K^b J^{ac} \pa J^{bc} (w) +c_{26} \pa K^a K^a  \pa K^b K^b (w) 
+c_{27} \pa K^a \pa K^b J^{ab}(w)
\nonu \\
&+& c_{28} \pa^2 K^a \pa K^a (w) + c_{29} \pa^3 K^a K^a (w),
\label{O2O4p2}
\eea
where $N' \equiv 2N+1$. The coefficients in (\ref{O2O4p2}) for general $N$
are presented in
(\ref{O2O4p2coeff}) of Appendix $C$.
All of the results regarding the OPE $O_2(z) \,
O_4(w)$ are checked for $N=6$ again.

Therefore, the spin-$4$ current $O_{4'}(z)$
is summarized by (\ref{O44}) and (\ref{Coefff}).
The OPE $O_2(z) \, O_4(w)$ for general $N$ is obtained
from (\ref{O2O4}) with (\ref{Coefff}).
The fusion rule is summarized by
$[O_2][O_4]= [I] + [O_2] +[{O}_4] +
[O_{4'}]$.

$\bullet$ The OPE $O_{\frac{5}{2}}(z) \, O_{\frac{7}{2}}(w)$

Now we move to the OPE between $O_{\frac{5}{2}} (z)$ and $O_{\frac{7}{2}} (z)$. The computation of the OPE for $N=2$ shows the following OPE
\bea
O_{\frac{5}{2}}(z) \, O_{\frac{7}{2}}(w)
 & = & \frac{1}{(z-w)^4} c_{\frac{5}{2}\frac{7}{2}}^{o} O_{2}(w) +\frac{1}{(z-w)^3} \frac{1}{4}c_{\frac{5}{2}\frac{7}{2}}^{o} \pa O_{2}(w)\nonu \\
& + & \frac{1}{(z-w)^2}
\left[ \frac{1}{20}c_{\frac{5}{2}\frac{7}{2}}^{o} \pa^2 O_{2}+c_{\frac{5}{2}\frac{7}{2}}^{o'} O_{4}+ c_{\frac{5}{2}\frac{7}{2}}^{o''} O_{4^{'}}\right.\nonu \\
&+& c_{\frac{5}{2}\frac{7}{2}}^{to}\left( T O_{2}-\frac{3}{10}\pa^2 O_{2} \right)+c_{\frac{5}{2}\frac{7}{2}}^{go}\left( G O_{\frac{5}{2}}-\frac{1}{5}\pa^2 O_{2} \right)\left. \right](w) \nonu \\
& + & \frac{1}{(z-w)}
\left[ \frac{1}{120}c_{\frac{5}{2}\frac{7}{2}}^{o} \pa^3 O_{2}+\frac{3}{8}c_{\frac{5}{2}\frac{7}{2}}^{o'}\pa O_{4}+\frac{3}{8} c_{\frac{5}{2}\frac{7}{2}}^{o''}\pa O_{4^{'}}\right.\nonu \\
&+&\frac{3}{8}c_{\frac{5}{2}\frac{7}{2}}^{to}\pa \left( T O_{2}-\frac{3}{10}\pa^2 O_{2} \right)+\frac{3}{8}c_{\frac{5}{2}\frac{7}{2}}^{go}\pa \left( G O_{\frac{5}{2}}-\frac{1}{5}\pa^2 O_{2} \right)\nonu \\
&+& c_{\frac{5}{2}\frac{7}{2}}^{to'}\left( T \pa O_{2} -\pa T O_{2} -\frac{1}{6} \pa^3 O_{2} \right) + c_{\frac{5}{2}\frac{7}{2}}^{go'} \left( G \pa O_{\frac{5}{2}}-\frac{5}{3} \pa G O_{\frac{5}{2}}-\frac{1}{9} \pa^3 O_{2} \right)\nonu \\
&+&c_{\frac{5}{2}\frac{7}{2}}^{go''}\left( G O_{\frac{7}{2}}-\frac{1}{4} \pa O_{4^{'}} \right)\left. \right](w) +\cdots.
\label{O5O7}
\eea
The structure of this OPE is very similar to (\ref{O2O4}) except that it has the third-order pole and there are the descendant fields of $O_2(z)$ in the right
hand side of the OPE.
The relative coefficients of descendant fields are fixed by (\ref{PhiPhi}). No new primary fields appear in (\ref{O5O7}). The quasi-primary
and primary fields are obtained
by the same method we used in the previous OPEs.
To find the structure constants for general $N$, we continue to compute the OPE for $N=3,4,5$ cases.
Then we solve the following equations for the structure constants when $N=3,4,5$:
\bea
\left[\{ O_{\frac{5}{2}} \; O_{\frac{7}{2}} \}_{-4}-
c_{\frac{5}{2}\frac{7}{2}}^{o} O_{2} \right](w)
& = & 0, \nonu \\
\left[\{ O_{\frac{5}{2}} \; O_{\frac{7}{2}} \}_{-2}
-\mbox{the 2nd-order pole } \right](w) & = & 0,
\nonu \\
\left[\{ O_{\frac{5}{2}} \; O_{\frac{7}{2}} \}_{-1}-
\mbox{the 1st-order pole} \right](w) & = & 0.
\nonu
\eea
Here the 2nd-order pole and 1st-order pole are given in the right hand side of
(\ref{O5O7})
respectively.
After that,
we put all the results together
and find the general forms of structure constants.
The general expression of structure constants in (\ref{O5O7}) which
holds for general $N$ are given by
\bea
c_{\frac{5}{2}\frac{7}{2}}^{o} & = & -\frac{24 c^2 (-3 + 2 c) (3 + c - 9 N) (1 + N) (c + 6 N) (6 N + c (-3 + 2 N))}{(21 + 4 c) (6 + 5 c) (c + 6 (-1 + N) N)^2}, \nonu \\
c_{\frac{5}{2}\frac{7}{2}}^{o'} & = &-\frac{4 c (21 + 4 c) (-6 N^2 + c (-1 + 4 N))}{21 (6 + 5 c) (c + 6 (-1 + N) N)}, \qquad
c_{\frac{5}{2}\frac{7}{2}}^{o''} = 8,\nonu \\
c_{\frac{5}{2}\frac{7}{2}}^{to} & = & -\frac{48 c^2 (-3 + 2 c) (69 + 29 c) (3 + c - 9 N) (1 + N) (c + 6 N) (6 N +
   c (-3 + 2 N))}{(29 + 2 c) (21 + 4 c) (6 + 5 c)^2 (c + 6 (-1 + N) N)^2},\nonu \\
c_{\frac{5}{2}\frac{7}{2}}^{go} & =  & -\frac{18 c^2 (-3 + 2 c) (-18 + 23 c) (3 + c - 9 N) (1 + N) (c + 6 N) (6 N +
   c (-3 + 2 N))}{(29 + 2 c) (21 + 4 c) (6 + 5 c)^2 (c + 6 (-1 + N) N)^2},\nonu \\
c_{\frac{5}{2}\frac{7}{2}}^{to'} & = & \frac{6 c^2 (-51 + 25 c) (3 + c - 9 N) (1 + N) (c + 6 N) (6 N +
   c (-3 + 2 N))}{(21 + 4 c) (6 + 5 c)^2 (c + 6 (-1 + N) N)^2},\nonu \\
c_{\frac{5}{2}\frac{7}{2}}^{go'} & = & \frac{9 c^2 (-30 + 29 c) (3 + c - 9 N) (1 + N) (c + 6 N) (6 N +
   c (-3 + 2 N))}{4 (21 + 4 c) (6 + 5 c)^2 (c + 6 (-1 + N) N)^2}, \nonu \\
c_{\frac{5}{2}\frac{7}{2}}^{go''} & = & -\frac{18 c (-6 N^2 + c (-1 + 4 N))}{(6 + 5 c) (c + 6 (-1 + N) N)}.
\label{coeffcoeffco}
\eea
All of the results regarding the OPE $O_{\frac{5}{2}}(z) \,
O_{\frac{7}{2}}(w)$ are checked for $N=6$ again.
The OPE $O_{\frac{5}{2}}(z) \, O_{\frac{7}{2}}(w)$ for general $N$ is obtained
from (\ref{O5O7}) with (\ref{coeffcoeffco}).
The fusion rule is summarized by
$[O_{\frac{5}{2}}][O_\frac{7}{2}]= [I] + [O_2] +[{O}_4] +
[O_{4'}]$.

$\bullet$ The OPE $O_{\frac{5}{2}}(z) \, O_{4}(w)$

Now let us move to the OPE between $O_{\frac{5}{2}} (z)$ and $O_4 (z)$. The computation of the OPE for $N=2$ turns out to be
\bea
O_{\frac{5}{2}}(z) \, O_{4}(w)
&=&\frac{1}{(z-w)^4}c_{\frac{5}{2} 4}^{o}O_{\frac{5}{2}}(w)
\label{O5O4}
\\
&+&\frac{1}{(z-w)^3} \left[\frac{1}{5}c_{\frac{5}{2} 4}^{o}\pa O_{\frac{5}{2}}+c_{\frac{5}{2} 4}^{o'}O_{\frac{7}{2}}+c_{\frac{5}{2} 4}^{go} \left( GO_{2}-\frac{2}{5}\pa O_{\frac{5}{2}}\right)\right](w) \nonu \\
& + & \frac{1}{(z-w)^2}
\left[ \frac{1}{30}c_{\frac{5}{2} 4}^{o}\pa^2 O_{\frac{5}{2}}+\frac{2}{7}c_{\frac{5}{2} 4}^{o'} \pa O_{\frac{7}{2}}+ \frac{2}{7}c_{\frac{5}{2} 4}^{go} \pa \left( GO_{2}-\frac{2}{5}\pa O_{\frac{5}{2}}\right)\right. \nonu \\
&+& c_{\frac{5}{2} 4}^{go'}\left( \pa G O_{2} -\frac{3}{4}G \pa O_{2} +\frac{1}{8} \pa^2 O_{\frac{5}{2}}\right)+ c_{\frac{5}{2} 4}^{to}\left( T O_{\frac{5}{2}}-\frac{1}{4} \pa^2 O_{\frac{5}{2}}\right)+c_{\frac{5}{2} 4}^{o''} O_{\frac{9}{2}}\left. \right](w) \nonu \\
& + & \frac{1}{(z-w)}
\left[  \frac{1}{210}c_{\frac{5}{2} 4}^{o} \pa^3 O_{\frac{5}{2}}+\frac{3}{56}c_{\frac{5}{2} 4}^{o'} \pa^2 O_{\frac{7}{2}}+ \frac{3}{56}c_{\frac{5}{2} 4}^{go} \pa^2 \left( GO_{2}-\frac{2}{5}\pa O_{\frac{5}{2}}\right)\right. \nonu \\
& + &\frac{1}{3}c_{\frac{5}{2} 4}^{go'}\pa \left (\pa G O_{2}-\frac{3}{4} G \pa O_{2}+\frac{1}{8} \pa^2 O_{\frac{5}{2}}\right) + \frac{1}{3}c_{\frac{5}{2} 4}^{to}\pa \left( T O_{\frac{5}{2}}-\frac{1}{4} \pa^2 O_{\frac{5}{2}}\right)+\frac{1}{3}c_{\frac{5}{2} 4}^{o''}\pa  O_{\frac{9}{2}}\nonu \\
&+& c_{\frac{5}{2} 4}^{go''} \left(\pa^2 G O_{2}  +\frac{3}{5}  G \pa^2 O_{2} -2 \pa G \pa O_{2} -\frac{2}{35}\pa^3 O_{\frac{5}{2}}\right)
+c_{\frac{5}{2} 4}^{to'} \left( T O_{\frac{7}{2}}-\frac{3}{16} \pa^2 O_{\frac{7}{2}}\right)\nonu \\
&+&c_{\frac{5}{2} 4}^{go'''}\left( G O_{4}-\frac{1}{8}\pa^2 O_{\frac{7}{2}}\right)+ c_{\frac{5}{2} 4}^{to''} \left(T \pa O_{\frac{5}{2}}-\frac{5}{4} \pa T O_{\frac{5}{2}}-\frac{1}{7} \pa^3 O_{\frac{5}{2}}\right)
\left.  \right](w)  +\cdots.
\nonu
\eea
As done before, the relative coefficients of descendant fields are fixed by (\ref{PhiPhi}), and the quasi-primary, and primary fields are found by the same method we used before.
No new primary field is found in this case.
The field contents up to the second-order pole are the same as the one in
(\ref{O2O7}).
To find the general forms of structure constants,
we continue to compute the OPE between $O_{\frac{5}{2}} (z)$ and $O_4 (w)$ using
 the package for $N=3,4,5$ cases.
After computing the OPEs for $N=3,4,5$, we solve the following
equations for structure constants when $N=3, 4, 5$:
\bea
\left[\{ O_{\frac{5}{2}} \; O_{4} \}_{-4}-c_{\frac{5}{2}4}^{o} O_{\frac{5}{2}} \right](w)
& = & 0, \nonu \\
\left[\{ O_{\frac{5}{2}} \; O_{4} \}_{-3}-\frac{1}{5}
c_{\frac{5}{2} 4}^{o}\pa O_{\frac{5}{2}}
-c_{\frac{5}{2} 4}^{o'}O_{\frac{7}{2}}-c_{\frac{5}{2} 4}^{go} \left( GO_{2}-\frac{2}{5}\pa O_{\frac{5}{2}}\right) \right](w) & = & 0, \nonu \\
\left[\{ O_{\frac{5}{2}} \; O_{4} \}_{-2}-
\mbox{the 2nd-order pole}\right](w) & = & 0, \nonu \\
\left[\{ O_{\frac{5}{2}} \; O_{4} \}_{-1}-
\mbox{the 1st-order pole } \right](w) & = & 0.
\nonu
\eea
Here the 2nd-order pole and 1st-order pole are given in the right hand side of
(\ref{O5O4})
respectively.
Then we put all the
results together and find the structure constants for general $N$.
The general expressions of structure constants in (\ref{O5O4}) are given by
\bea
c_{\frac{5}{2} 4}^{o} & = & \frac{42 c^2 (-3 + 2 c) (3 + c - 9 N) (1 + N) (c + 6 N) (6 N + c (-3 + 2 N))}{(21 + 4 c) (6 + 5 c) (c + 6 (-1 + N) N)^2}, \nonu \\
c_{\frac{5}{2} 4}^{o'} & = &\frac{4 c (21 + 4 c) (-6 N^2 + c (-1 + 4 N))}{3 (6 + 5 c) (c + 6 (-1 + N) N)}, \nonu \\
c_{\frac{5}{2} 4}^{go} & = & \frac{216 c^2 (-3 + 2 c) (3 + c - 9 N) (1 + N) (c + 6 N) (6 N +
   c (-3 + 2 N))}{(21 + 4 c) (6 + 5 c)^2 (c + 6 (-1 + N) N)^2},\nonu \\
c_{\frac{5}{2} 4}^{go'} & = & \frac{72 c^2 (-67 + 26 c) (3 + c - 9 N) (1 + N) (c + 6 N) (6 N +
   c (-3 + 2 N))}{7 (29 + 2 c) (21 + 4 c) (6 + 5 c) (c + 6 (-1 + N) N)^2},\nonu \\
c_{\frac{5}{2} 4}^{to} & =  & \frac{72 c^2 (-19 + 14 c) (3 + c - 9 N) (1 + N) (c + 6 N) (6 N +
   c (-3 + 2 N))}{(29 + 2 c) (21 + 4 c) (6 + 5 c) (c + 6 (-1 + N) N)^2},\nonu \\
c_{\frac{5}{2} 4}^{o''} & = & -9,\qquad
c_{\frac{5}{2} 4}^{go''} = \frac{45 c^2 (-3 + 2 c) (3 + c - 9 N) (1 + N) (c + 6 N) (6 N + c (-3 + 2 N))}{(21 + 4 c) (6 + 5 c)^2 (c + 6 (-1 + N) N)^2}, \nonu \\
c_{\frac{5}{2} 4}^{to'} & = & \frac{36 c (-6 N^2 + c (-1 + 4 N))}{(6 + 5 c) (c + 6 (-1 + N) N)},\qquad
c_{\frac{5}{2} 4}^{go'''} = -\frac{18 c (-6 N^2 + c (-1 + 4 N))}{(6 + 5 c) (c + 6 (-1 + N) N)},
\label{corcoeff}
\\
c_{\frac{5}{2} 4}^{to''} & = & -\frac{144 c^2 (-3+2 c) (1+N) (c^2-3 c (-1+N)+18 (1-3 N) N) (6 N+c (-3+2 N))}{(21+4 c) (6+5 c)^2 (c+6 (-1+N) N)^2}.
\nonu
\eea
All of the results regarding the OPE $O_{\frac{5}{2}}(z) \,
O_{4}(w)$ are checked for $N=6$ again.
The OPE $O_{\frac{5}{2}}(z) \, O_4(w)$ for general $N$ is obtained
from (\ref{O5O4}) with (\ref{corcoeff}).
The fusion rule is summarized by
$[O_{\frac{5}{2}}][O_4]= [I] + [O_\frac{5}{2}] +[{O}_\frac{7}{2}]
+[O_{\frac{9}{2}}] $.

The ${\cal N}=1$ fusion rule is summarized by
$[\hat{O}_2][\hat{O}_{\frac{7}{2}}]=
[\hat{O}_2] +[\hat{O}_{\frac{7}{2}}] +[\hat{O}_4]$.
The explicit OPE is given by (\ref{singleO2O7}).
Among four OPEs between this ${\cal N}=1$ supermultiplet,
half of them are quite related to the other because
they have common field contents.
By rescaling the currents
as $O_{\frac{7}{2}}(z) \rightarrow N_{\frac{7}{2}} O_{\frac{7}{2}}(z)$
and $O_{4}(z)
\rightarrow N_{4}
O_{4}(z)$, where
\bea
N_{\frac{7}{2}}^2 &= &3M (21 + 4 c) (6 + 5 c) (c + 6 (-1 + N) N)^4,\qquad
N_{4}^2 =\frac{1}{8}N_{\frac{7}{2}}^2,
\nonu \\
\frac{1}{M} & \equiv &
(14 c^4 (-3 + 2 c) (3 + c - 9 N) (1 + N) (c + 6 N) \nonu \\
& \times &
(-3 c +     2 (3 + c) N) (2 c (-1 + N) + 3 (3 - 4 N) N) (c + 2 c N +
    3 N (-3 + 2 N))),
\nonu
\eea
the standard normalizations arise: $O_{\frac{7}{2}}(z) \, O_{\frac{7}{2}}(w)
\rightarrow
\frac{1}{(z-w)^7} \, \frac{2c}{7} +\cdots$ and
 $O_{4}(z) \, O_4 (w) \rightarrow
\frac{1}{(z-w)^8} \, \frac{c}{4} +\cdots$. We have checked this up to $N=5$.
Therefore all the previous OPEs can be rewritten in terms of these rescaled
currents.

\subsection{ The OPEs between the higher spin currents of spins-$(2',
\frac{5}{2})$ and the higher spin currents of spins-$(4',\frac{9}{2})$}

$\bullet$ The OPE $O_2(z) \, O_{4'}(w)$

Now let us consider the OPE between $O_{2}(z)$ and $O_{4{'}}(z)$.
The final result for $N=2$
is presented first, which explains how this result can be
obtained explicitly
\bea
O_{2}(z) \, O_{4{'}}(w)
 & = & \frac{1}{(z-w)^4} c_{2 4{'}}^{o} O_{2}  +
\frac{1}{(z-w)^2} \left[  c_{2 4{'}}^{o'} O_{4}+c_{2 4{'}}^{o''} O_{4^{'}} + A_{4}
\right.
\label{O2O44}
\\
& + &
c_{2 4{'}}^{to}\left( T O_{2}-\frac{3}{10}\pa^2 O_{2} \right)+c_{2 4{'}}^{go}\left( G O_{\frac{5}{2}}-\frac{1}{5}\pa^2 O_{2} \right)\left. \right](w) \nonu \\
& + & \frac{1}{(z-w)}
\left[\frac{1}{4}c_{2 4{'}}^{o'} \pa O_{4}+\frac{1}{4}c_{2 4{'}}^{o''}\pa O_{4^{'}}+\frac{1}{4}\pa A_{4}
+\frac{1}{4}c_{2 4{'}}^{to}\pa \left( T O_{2}-\frac{3}{10}\pa^2 O_{2} \right)
\right.\nonu \\
&+&\frac{1}{4}c_{2 4{'}}^{go}\pa \left( G O_{\frac{5}{2}}-\frac{1}{5}\pa^2 O_{2} \right)
+ c_{2 4{'}}^{to'} \left( T \pa O_{2} -\pa T O_{2}
-\frac{1}{6} \pa^3 O_{2} \right)+ \nonu \\
& + &  c_{2 4{'}}^{go'}\left( G \pa O_{\frac{5}{2}}-\frac{5}{3} \pa G O_{\frac{5}{2}}-\frac{1}{9} \pa^3 O_{2} \right)
+ c_{2 4{'}}^{go''} \left( G O_{\frac{7}{2}}-\frac{1}{4} \pa O_{4} \right)\left. \right](w) +\cdots.
\nonu
\eea
The structure of this OPE appears similar to (\ref{O2O4})
except that the OPE (\ref{O2O44}) has a composite spin-4 primary field $A_4(z)$.
The relative coefficients of descendant fields
appearing in the first-order pole
are fixed by the formula (\ref{PhiPhi}).
For the second-order pole, as performed before,
we compute $T(z) \, \{ O_{2} \; O_{4'} \}_{-2}(w)$ and
$G (z) \, \{ O_{2} \; O_{4'} \}_{-2}(w)$
to find the quasi-primary fields in the second-order pole.
By subtracting two candidates from  the second-order singular terms,
we want to check if the
following equation holds:
\bea
[\{ O_{2} \; O_{4'} \}_{-2}-c_{2 4{'}}^{o'} O_{4}-c_{2 4{'}}^{o''} O_{4^{'}}-c_{2 4{'}}^{to} ( T O_{2}-\frac{3}{10}\pa^2 O_{2} )
-c_{2 4{'}}^{go} ( G O_{\frac{5}{2}}-\frac{1}{5}\pa^2 O_{2} )](w)=0.
\label{O2O44eq1}
\eea
It turns out that there is no solution for (\ref{O2O44eq1}).
This  implies that there should be another primary spin-4 field $A_4(z)$.
Since we already have found two spin-4 primary fields,
we expect that the new spin-4 field would be a composite field in terms of
known currents (we have determined so far).
Otherwise, the spin contents of (\ref{currents})
will not be correct.
To check this, let us first express $A_4(z)$ as
\bea
A_{4} (z) & = &
\{O_2 \, O_{4'}\}_{-1}(z) \nonu \\
& - & [ c_{2 4{'}}^{o'} O_{4}+c_{2 4{'}}^{o''} O_{4^{'}}
+c_{2 4{'}}^{to}( T O_{2}-\frac{3}{10}\pa^2 O_{2} )+c_{2 4{'}}^{go}
( G O_{\frac{5}{2}}-\frac{1}{5}\pa^2 O_{2})](z).
\label{A41}
\eea
Then we try to express $A_4(z)$ in terms of other possible
spin-4 fields. It turns out that $A_4(z)$ can be written as
\bea
A_{4} (z) &=& c_{4}^{gg}\left( G \pa G -\frac{7}{10} \pa^2 T\right)(z)+c_{4}^{go}\left( G O_{\frac{5}{2}} -\frac{1}{5} \pa^2 O_2\right)(z) \nonu \\
&+& c_{4}^{tt}\left( T T- \frac{3}{10}\pa^2  T\right)(z)+c_{4}^{to}\left( T O_2- \frac{3}{10}\pa^2  O_2 \right)(z)\nonu \\
&+& c_{4}^{oo}( O_2 O_2 -\frac{3}{10}(-2 \widetilde{c} +c)\, \pa^2 O_2+\frac{3}{10}(\widetilde{c}-c)\pa^2 T)(z)+ c_4^{o} O_4 (z),
\label{A42}
\eea
where $\widetilde{c}$ is the central charge of coset $\frac{\widehat{SO}(2N+1)_k }{\widehat{SO}(2N)_{k}}$.
This field (\ref{A42}) containing $O_2 O_2(z)$ corresponds to
the $A^{(4,0)}$ in \cite{CV}.
As expected it obeys the following OPEs:
\bea
T(z) \, A_{4}(w)& =& \frac{1}{(z-w)^2} \, 4\,A_{4}(w) +
\frac{1}{(z-w)}\,\pa A_{4}(w) +\cdots,\nonu \\
G(z) \, A_{4}(w) & = &\frac{1}{(z-w)} \, A_{\frac{9}{2}}(w) +\cdots,
\label{GA4}
\eea
where $A_{\frac{9}{2}} (z)$ is a composite primary spin-$\frac{9}{2}$ field
and the superpartner of $A_4(z)$. It will appear in (\ref{A91}) or
(\ref{A92}) in the next OPE soon.
For the first-order pole, the derivative terms are completely determined and one can find the extra
quasi-primary fields by the same procedure which we have
performed.

We continue to compute the OPE between $O_{2} (z)$ and $O_{4'} (w)$
by the package \cite{Thielemans} for $N=3,4,5$ cases
to find the general forms of structure constants.
To find the structure constant
$c_{2 4{'}}^{o}$, we should solve $[\{O_2 O_{4'}\}_{-4}- c_{2 4{'}}^{o} O_{2}](w)=0$.
To find the other structure constants
$c_{2 4{'}}^{o'}$, $c_{2 4{'}}^{o''}$,$c_{2 4{'}}^{to}$, and $c_{2 4{'}}^{go}$,
we use (\ref{A41}), and (\ref{GA4}).
Moreover, we solve $[\{O_2 O_{4'}\}_{-1}- (\mbox{the 1st-order pole})](w)=0$
to find the remaining structure constants
$c_{2 4{'}}^{to'}$, $c_{2 4{'}}^{go'}$, and $c_{2 4{'}}^{go''}$.
To find the coefficients in (\ref{A42}), we use (\ref{A42}) and (\ref{GA4}).
We do this procedure for $N=3,4,5$ cases also.
Then the general expressions of the coefficients in (\ref{A42}) are given by
\bea
c_{4}^{gg} & = &-336 F (-33+c) c^4 (29+2 c) (6+5 c) (3+c-9 N) (1+N) (c+6 N) \nonu \\
& \times & (-3 c+2 (3+c) N) (2 c (-1+N)+3 (3-4 N) N) (c+2 c N+3 N (-3+2 N)), \nonu \\
c_{4}^{go} & = & 1008  c^3 F (-33+c) (21+4 c) (-7+10 c) (3+c-9 N) (1+N) (c+6 N)\nonu \\
& \times &  (-3 c+2 (3+c) N) (c+6 (-1+N) N) (-6 N^2+c (-1+4 N)),\nonu \\
c_{4}^{tt} &=& 112 c^4 F(29+2 c) (6+5 c) (3+88 c) (3+c-9 N) (1+N) (c+6 N) \nonu \\
& \times & (-3 c+2 (3+c) N) (2 c (-1+N)+3 (3-4 N) N) (c+2 c N+3 N (-3+2 N)), \nonu \\
c_{4}^{to} &=& 672 c^3 F (21+4 c) (-7+10 c) (129+22 c) (3+c-9 N) (1+N) (c+6 N)\nonu \\
&\times & (-3 c+2 (3+c) N) (c+6 (-1+N) N) (-6 N^2+c (-1+4 N)),\nonu \\
c_4^{o} &=& 72 c^2 F (29+2 c) (21+4 c) (6+5 c) (-7+10 c) (3+c-9 N) (1+N) \nonu \\
&\times & (c+6 N) (-3 c+2 (3+c) N) (c+6 (-1+N) N)^2,\nonu \\
c_4^{oo}&=& 504 c^2 F (29+2 c) (21+4 c) (6+5 c) (-7+10 c) (3+c-9 N) (1+N) \nonu \\
& \times & (c+6 N) (-3 c+2 (3+c) N) (c+6 (-1+N) N)^2, \nonu \\
F& \equiv & \frac{1} {21 (29 + 2 c)^2 (21 + 4 c) (6 + 5 c)^3 (c + 6 (-1 + N) N)^4}.
\nonu
\eea
The structure constants in (\ref{O2O44}) for general $N$ are given by
\bea
c_{2 4{'}}^{o} & = & 210 c^3B (-3+2 c) (29+2 c) (6+5 c) (1+N) (6 N+c (-3+2 N)) (108 N^3 (-1+3 N)\nonu \\
&+&c^3 (-1+4 N)-3 c^2 (1-5 N+6 N^2)-18 c N (1-6 N+11 N^2)), \nonu \\
c_{2 4{'}}^{o'} & = &3 c^2 B (29+2 c) (21+4 c) (6+5 c) (c+6 (-1+N) N) (10 c^3 (-6-N+2 N^2)\nonu \\
&-&126 N^2 (25-50 N+12 N^2) -c^2 (42-623 N+496 N^2+60 N^3)\nonu \\
&-& 3 c N (-350+1155 N-1246 N^2+720 N^3)), \nonu \\
c_{2 4{'}}^{o''} & = & 28 c B (29+2 c)^2 (6+5 c) (84+25 c) (c+6 (-1+N) N)^2 (c-4 c N+6 N^2),\nonu \\
c_{2 4{'}}^{to} & = & 840 c^3 B B'(-3+2 c) (69+10 c) (3+c-9 N) (1+N) (c+6 N) (6 N+c (-3+2 N)),\nonu \\
c_{2 4{'}}^{go} & =  & -23940 c^3 B B'(-3+2 c) (3+c-9 N) (1+N) (c+6 N) (6 N+c (-3+2 N)) ,\nonu \\
c_{2 4{'}}^{to'} & = & -210 c^3 B(29+2 c) (-3+10 c) (1+N) (6 N+c (-3+2 N)) (108 N^3 (-1+3 N)\nonu \\
&+& c^3 (-1+4 N)-3 c^2 (1-5 N+6 N^2)-18 c N (1-6 N+11 N^2)),\nonu \\
c_{2 4{'}}^{go'} & = &945 c^3 B (29+2 c) (1+N) (6 N+c (-3+2 N)) (108 N^3 (-1+3 N)+c^3 (-1+4 N)\nonu \\
&-&3 c^2 (1-5 N+6 N^2)-18 c N (1-6 N+11 N^2)), \nonu \\
c_{2 4{'}}^{go''} & = & 63 c^2 B (29+2 c) (6+5 c) (c+6 (-1+N) N) (10 c^3 (-6-N+2 N^2)\nonu \\
&-& 126 N^2 (25-50 N+12 N^2)-c^2 (42-623 N+496 N^2+60 N^3)\nonu \\
&-&3 c N (-350+1155 N-1246 N^2+720 N^3)),\nonu \\
B & \equiv & \frac{1}{147 (29 + 2 c)^2 (6 + 5 c)^2 (c + 6 (-1 + N) N)^3},
\qquad B' \equiv (-6 N^2+c (-1+4 N)).
\label{anscoeff}
\eea
All of the results regarding the OPE $O_{2}(z) \,
O_{4'}(w)$ are checked for $N=6$ again.
The OPE for general $N$ is obtained from (\ref{O2O44}) with
(\ref{anscoeff}).
The fusion rule is summarized by
$[O_2][O_{4'}]= [I] +[{O}_2]
+[O_4]+[O_{4'}] $.

$\bullet$ The OPE $O_2(z) \, O_{\frac{9}{2}}(w)$

Now let us compute the OPE between $O_{2}(z)$ and $O_{\frac{9}{2}}(z)$. The computation of the OPE for $N=2$ leads to the OPE
\bea
O_{2}(z) \, O_{\frac{9}{2}}(w)
&=&\frac{1}{(z-w)^4}c_{2 \frac{9}{2}}^{o} O_{\frac{5}{2}}(w)
\label{O2O9}
\\
&+&\frac{1}{(z-w)^3} \left[c_{2 \frac{9}{2}}^{o'} O_{\frac{7}{2}}+c_{2 \frac{9}{2}}^{go} \left(G O_{2}-\frac{2}{5}\pa O_{\frac{5}{2}}\right)\right](w) \nonu \\
& + & \frac{1}{(z-w)^2}
\left[ \frac{1}{7}c_{2 \frac{9}{2}}^{o'} \pa O_{\frac{7}{2}}+ \frac{1}{7}c_{2 \frac{9}{2}}^{go} \pa \left( G O_{2}-\frac{2}{5}\pa O_{\frac{5}{2}}\right)
+c_{2 \frac{9}{2}}^{to}\left( T O_{\frac{5}{2}}-\frac{1}{4} \pa^2 O_{\frac{5}{2}}\right)
\right. \nonu \\
&+& c_{2 \frac{9}{2}}^{go'}\left(\pa G O_{2}-\frac{3}{4} G \pa O_{2}+\frac{1}{8} \pa^2 O_{\frac{5}{2}}\right)
 +c_{2 \frac{9}{2}}^{o''} O_{\frac{9}{2}}+c_{2 \frac{9}{2}}^{a} A_{\frac{9}{2}}\left. \right](w) \nonu \\
& + & \frac{1}{(z-w)}
\left[  \frac{1}{56}c_{2 \frac{9}{2}}^{o'} \pa^2 O_{\frac{7}{2}}+ \frac{1}{56} c_{2 \frac{9}{2}}^{go} \pa^2 \left( G O_{2}-\frac{2}{5}\pa O_{\frac{5}{2}}\right)
+ \frac{2}{9} c_{2 \frac{9}{2}}^{to} \pa \left( T O_{\frac{5}{2}}-\frac{1}{4} \pa^2 O_{\frac{5}{2}}\right)\right. \nonu \\
& + &\frac{2}{9}c_{2 \frac{9}{2}}^{go'}\pa \left ( \pa G O_{2}-\frac{3}{4} G \pa O_{2}+\frac{1}{8} \pa^2 O_{\frac{5}{2}}\right)
 +\frac{2}{9}c_{2 \frac{9}{2}}^{o''}\pa  O_{\frac{9}{2}}+\frac{2}{9}c_{2 \frac{9}{2}}^{a}\pa A_{\frac{9}{2}}
 \nonu \\
&+& c_{2 \frac{9}{2}}^{go''} \left(\pa^2 G O_{2}  +\frac{3}{5} G \pa^2 O_{2} -2 \pa G \pa O_{2} -\frac{2}{35}\pa^3 O_{\frac{5}{2}}\right)
+ c_{2 \frac{9}{2}}^{to'} \left( T O_{\frac{7}{2}}-\frac{3}{16} \pa^2 O_{\frac{7}{2}}\right)
\nonu \\
&+&c_{2 \frac{9}{2}}^{to''} \left(T \pa O_{\frac{5}{2}}-\frac{5}{4} \pa T O_{\frac{5}{2}}-\frac{1}{7} \pa^3 O_{\frac{5}{2}}\right)
+c_{2 \frac{9}{2}}^{go'''}\left( G O_{4^{'}}-\frac{2}{9}\pa O_{\frac{9}{2}}\right)\nonu \\
&+& c_{2 \frac{9}{2}}^{ga}\left( G A_{4}-\frac{2}{9}\pa A_{\frac{9}{2}}\right)
+c_{2 \frac{9}{2}}^{go''''}\left( G O_{4}-\frac{1}{8} \pa^2 O_{\frac{7}{2}}\right)+ c_{2 \frac{9}{2}}^{q}Q_{\frac{11}{2}}+  O_{\frac{11}{2}}\left.  \right](w)  +\cdots.
\nonu
\eea
The relative coefficients of descendant fields are fixed by (\ref{PhiPhi}).
As predicted by (\ref{PhiPhi}),
there are no descendant fields of $O_{\frac{5}{2}}(z)$ in the OPE.
For the second-order pole, there is a composite primary
spin-$\frac{9}{2}$ field $A_{\frac{9}{2}} (z)$
which is  the superpartner $A_4(z)$ in (\ref{O2O44}).
It is found by the same method we used to find $A_4(z)$.
The explicit form of $A_{\frac{9}{2}}(z)$ is
\bea
A_{\frac{9}{2}} (z) &=& \frac{1}{c_{2 \frac{9}{2}}^{a}}\{O_2 \,
O_{\frac{9}{2}}\}_{-2}(z)-
\frac{1}{c_{2 \frac{9}{2}}^{a}}\left[ \frac{1}{7}c_{2 \frac{9}{2}}^{o'} \pa O_{\frac{7}{2}}+ \frac{1}{7}c_{2 \frac{9}{2}}^{go} \pa \left( G O_{2}G-\frac{2}{5}\pa O_{\frac{5}{2}}\right)
\right.
\label{A91} \\
&+& c_{2 \frac{9}{2}}^{to}\left( T O_{\frac{5}{2}}-\frac{1}{4} \pa^2 O_{\frac{5}{2}}\right)+c_{2 \frac{9}{2}}^{go'}\left(\pa G O_{2}-\frac{3}{4} G \pa O_{2}+\frac{1}{8} \pa^2 O_{\frac{5}{2}}\right)
 +c_{2 \frac{9}{2}}^{o''} O_{\frac{9}{2}}\left. \right](z).
\nonu
\eea
In terms of other spin-$\frac{9}{2}$ fields, it is expressed as
\bea
A_{\frac{9}{2}} (z) &=& c_{\frac{9}{2}}^{go}\left( \pa G O_{2}-\frac{3}{4} G \pa O_{2}+\frac{1}{8} \pa^2 O_{\frac{5}{2}}\right)(z)+
c_{2 \frac{9}{2}}^{to}\left( T O_{\frac{5}{2}}- \frac{1}{4} \pa^2 O_{\frac{5}{2}}\right)(z) \nonu \\
&+& c_{2 \frac{9}{2}}^{tg}\left( T \pa G- \frac{3}{4}\pa T G- \frac{1}{5} \pa^3 G\right)(z)  \nonu \\
&+& c_{\frac{9}{2}}^{oo}(O_2 O_{\frac{5}{2}}- \frac{3}{4}c'_1\, \pa T G- \frac{3}{4}c'_2\, G \pa O_{2}
-\frac{1}{120}(2c'_4-9c'_1)\, \pa^3 G \nonu \\
&-& \frac{1}{40}(4c'_3-9c'_2)\, \pa^2 O_{\frac{5}{2}}-\frac{3}{7} \pa O_{\frac{7}{2}})(z),
\label{A92}
\eea
where the structure constants are
\bea
c'_1  & = & \frac{6 c^2 (2 c (-1 + N) + 3 (3 - 4 N) N) (c + 2 c N + 3 N (-3 + 2 N))}{(21 + 4 c) (c + 6 (-1 + N) N)^2},\nonu \\
c'_2 &=& \frac{18 c (-6 N^2 + c (-1 + 4 N))}{(6 + 5 c) (c + 6 (-1 + N) N)},\qquad
c'_3 = \frac{c (c - 4 c N + 6 N^2)}{3 (c + 6 (-1 + N) N)},\nonu \\
c'_4 & = & \frac{c^2 (2 c (-1 + N) + 3 (3 - 4 N) N) (c + 2 c N + 3 N (-3 + 2 N))}{3 (c + 6 (-1 + N) N)^2}.
\nonu
\eea
The spin-$\frac{9}{2}$ field
$A_{\frac{9}{2}} (z)$ obeys the following OPEs:
\bea
T(z) \, A_{\frac{9}{2}}(w)& =&
\frac{1}{(z-w)^2} \, \frac{9}{2}\,A_{\frac{9}{2}}(w) +
\frac{1}{(z-w)}\,\pa A_{\frac{9}{2}}(w) +\cdots,\nonu \\
G(z) \, A_{\frac{9}{2}}(w) & = &\frac{1}{(z-w)^2} \, 8\,A_{4}(w)
+\frac{1}{(z-w)} \, \pa A_{4}(w)+\cdots.
\label{GA9}
\eea
Then, the fields $A_4(z)$ and $A_{\frac{9}{2}}(z)$
consist of ${\cal N}=1$ multiplet.
For the first-order pole, we compute the following OPEs:
\bea
&&T(z)[\{ O_{2} \; O_{\frac{9}{2}} \}_{-1}-(\mbox{ desc. fields in the
1st pole })](w),
\nonu \\
&&G(z)[\{ O_{2} \; O_{\frac{9}{2}} \}_{-1}-(\mbox{ desc. fields in the
1st pole })](w).
\nonu
\eea
Here the descendant fields in the first-order pole contains the first
six terms in
(\ref{O2O9}).
Then we examine higher-order poles
with order $n > 2$
and add extra spin-$\frac{11}{2}$ quasi-primary fields to the first-order pole, and
compute the OPEs with $T(z)$ and $G(z)$ again
to check whether the higher-order poles
are removed or not.
If there are still higher-order poles,
we can add another quasi-primary field and
compute the OPEs with $T(z)$ and $G(z)$ again.
We continue this procedure until
the higher-order poles with order $n > 2$
are completely removed. In this case, removing the higher-order poles
was very complicated.
When we added the final quasi-primary field, $Q_{\frac{11}{2}}(z)$,
we could successfully remove all higher-order poles.
The $Q_{\frac{11}{2}}(z)$ is given by
\bea
Q_{\frac{11}{2}}(z)&=& O_{2} O_{\frac{7}{2}} (z)+ 315 c^2 H (-117+4 (-36+c) c) \pa^2 GO_{2}(z)\nonu \\
&+& 1260 c^2 H (-36+c (75+2 c)) \pa G\pa O_{2} (z)\nonu \\
& +& 252 c^2 H (6+5 c) (-3+10 c) T \pa O_{\frac{5}{2}}(z)\nonu \\
&+&2 c^2 H (2304+5 c (-654+c (-247+2 c)))\pa^3 O_{\frac{5}{2}}(z)\nonu \\
&-& \frac{c (21+4 c) (c-4 c N+6 N^2)}{28 (6+5 c) (c+6 (-1+N) N)}\pa^2 O_{\frac{7}{2}}(z)-\frac{1}{3}\pa O_{\frac{9}{2}}(z),
\label{Q11}
\eea
where $H \equiv  \frac{(3+c-9 N)(1 + N) (c + 6 N) (-3 c + 2 (3 + c) N)}{35 (29 + 2 c) (21 + 4 c) (6 + 5 c)^2 (c + 6 (-1 + N) N)^2}$.
This quasi-primary field (\ref{Q11}) containing $O_2 O_{\frac{7}{2}}(z)$
is related to  the primary field $A^{(\frac{11}{2},2)}$ of \cite{CV}.
The coefficients are determined by the fact that the third-pole of the OPE
between $T(z)$ and $Q_{\frac{11}{2}}(w)$  must vanish.
As one can see, $Q_{\frac{11}{2}}(w)$
is very different from the other quasi-primary fields in the sense that the
nonderivative term doesn't contain $T(z)$ or $G(z)$ \footnote{We believe that
this should come from the general formula in $(B.4)$ of \cite{Ahn1211}.
See also
the original paper \cite{BFKNRV}.}.
The OPEs $T(z) \, Q_{\frac{11}{2}}(w)$ and $G(z) \,
Q_{\frac{11}{2}}(w)$ are put in the Appendix $G$
and Appendix $H$ respectively. After finding $Q_{\frac{11}{2}}(z)$, we subtract
all the derivative terms and quasi-primary fields with their coefficients in (\ref{O2O9}) from $\{ O_{2} \; O_{\frac{9}{2}} \}_{-1}(w)$. It doesn't vanish meaning there is a new primary spin-$\frac{11}{2}$ field.
It turns
out that the new primary spin-$\frac{11}{2}$ field cannot be expressed in terms of other spin-$\frac{11}{2}$ fields meaning
it is not a composite field that can be written in terms of known currents.
The explicit form of new primary spin-$\frac{11}{2}$, $O_{\frac{11}{2}}(z)$,
is given by
\bea
O_{\frac{11}{2}}(w) &=& \{ O_{2} \; O_{\frac{9}{2}} \}_{-1}(w)
-\left[  \frac{1}{56}c_{2 \frac{9}{2}}^{o'} \pa^2 O_{\frac{7}{2}}+ \frac{1}{56} c_{2 \frac{9}{2}}^{go} \pa^2 \left( G O_{2}-\frac{2}{5}\pa O_{\frac{5}{2}}\right)
+ \frac{2}{9} c_{2 \frac{9}{2}}^{to} \pa \left( T O_{\frac{5}{2}}-\frac{1}{4} \pa^2 O_{\frac{5}{2}}\right)\right. \nonu \\
& + &\frac{2}{9}c_{2 \frac{9}{2}}^{go'}\pa \left ( \pa G O_{2}-\frac{3}{4} G \pa O_{2}+\frac{1}{8} \pa^2 O_{\frac{5}{2}}\right)
 +\frac{2}{9}c_{2 \frac{9}{2}}^{o''}\pa  O_{\frac{9}{2}}+\frac{2}{9}c_{2 \frac{9}{2}}^{a}\pa A_{\frac{9}{2}}
 \nonu \\
&+& c_{2 \frac{9}{2}}^{go''}
\left(\pa^2 G O_{2}  +\frac{3}{5} G \pa^2 O_{2} -2 \pa G \pa O_{2} -\frac{2}{35}\pa^3 O_{\frac{5}{2}}\right)
+ c_{2 \frac{9}{2}}^{to'} \left( T O_{\frac{7}{2}}-\frac{3}{16} \pa^2 O_{\frac{7}{2}}\right)
\nonu \\
&+&c_{2 \frac{9}{2}}^{to''} \left(T O_{\frac{5}{2}}-\frac{5}{4} \pa T O_{\frac{5}{2}}-\frac{1}{7} \pa^3 O_{\frac{5}{2}}\right)
+c_{2 \frac{9}{2}}^{go'''}\left( G O_{4^{'}}-\frac{2}{9}\pa O_{\frac{9}{2}}\right)\nonu \\
&+& c_{2 \frac{9}{2}}^{ga}\left( G A_{4}-\frac{2}{9}\pa A_{\frac{9}{2}}\right)
+c_{2 \frac{9}{2}}^{go''''}\left( G O_{4}-\frac{1}{8} \pa^2 O_{\frac{7}{2}}\right)+ c_{2 \frac{9}{2}}^{q}Q_{\frac{11}{2}}\left.  \right](w).
\label{O11}
\eea
The spin-$\frac{11}{2}$ current
$O_{\frac{11}{2}}(z)$ obeys the following OPEs:
\bea
T(z) \, O_{\frac{11}{2}}(w)& =& \frac{1}{(z-w)^2} \,\frac{11}{2}\,O_{\frac{11}{2}}(w) +\frac{1}{(z-w)}\,\pa O_{\frac{11}{2}}(w) +\cdots,\nonu \\
G(z) \, O_{\frac{11}{2}}(w) & = &\frac{1}{(z-w)} \, \,O_{6}(w) +\cdots,
\label{GO11}
\eea
where $O_{6}(z)$ is a primary spin-6 field and the superpartner of
$O_{\frac{11}{2}}(z)$.
The
$O_{6}(w)$ will appear in (\ref{O6})
in the OPE between $O_{\frac{5}{2}}(z)$ and $O_{\frac{9}{2}}(w)$ soon.
To find the $N$-dependence of the structure constants, we compute the OPE between $O_2(z)$ and $O_{\frac{9}{2}}(w)$ for $N=3,4,5,6$ cases.
In the previous OPEs, it was enough to compute OPEs up to $N=5$ to
find the $N$-dependence of the structure constants. But in this case, we had to compute the OPE between
$O_2(z)$ and $O_{\frac{9}{2}}(w)$ up to $N=6$ to find the $N$-dependence
of the structure constants completely.
To find the structure constants
$c_{2 \frac{9}{2}}^{o}$, $c_{2 \frac{9}{2}}^{o'}$, and $c_{2 \frac{9}{2}}^{go}$,
we should solve the following equations:
\bea
\left[\{O_2 O_{\frac{9}{2}}\}_{-4}-c_{2 \frac{9}{2}}^{o} O_{\frac{5}{2}}
\right](w) & = &
0,
\nonu \\
\left[\{O_2 O_{\frac{9}{2}}\}_{-3}-c_{2 \frac{9}{2}}^{o'} O_{\frac{7}{2}}
-c_{2 \frac{9}{2}}^{go} \left(G O_{2}-\frac{2}{5}\pa O_{\frac{5}{2}}\right)\right](w)
& = & 0.
\nonu
\eea
To find the structure constants
$c_{2 \frac{9}{2}}^{to}$, $c_{2 \frac{9}{2}}^{go'}$,
$c_{2 \frac{9}{2}}^{o''}$ and $c_{2 \frac{9}{2}}^{a}$,
we use the equations (\ref{A91}) and (\ref{GA9}).
To find the coefficients in (\ref{A92}), the equations
(\ref{A91}) and (\ref{A92}) can be used.
To find the remaining structure constants in (\ref{O2O9}),
we use the equations (\ref{O11}) and (\ref{GO11}).
Then the
structure constants in (\ref{O2O9}) for general $N$ are as follows:
\bea
c_{2 \frac{9}{2}}^{o} & = & 2520 c^3 B B'(-3+2 c) (29+2 c) (53+2 c) (20+3 c) (6+5 c) (1+N) (6 N+c (-3+2 N)),
\nonu \\
c_{2 \frac{9}{2}}^{o'} & = & -252 c^2 B (29+2 c) (53+2 c) (20+3 c) (21+4 c) (6+5 c) (4+73 c)(c+6 (-1+N) N) \nonu \\
& \times &  (10 c^3 (-6-N+2 N^2)-126 N^2 (25-50 N+12 N^2)\nonu \\
&-&c^2 (42-623 N+496 N^2+60 N^3)-3 c N (-350+1155 N-1246 N^2+720 N^3)),
\nonu \\
c_{2 \frac{9}{2}}^{go} & = & -75600 c^3B B' (-3+2 c) (29+2 c) (53+2 c) (20+3 c) (1+N) (6 N+c (-3+2 N)),
\nonu \\
c_{2 \frac{9}{2}}^{go'} & = & -5760 c^3B B' (53+2 c)^2 (20+3 c) (6+5 c) (1+N) (6 N+c (-3+2 N)),
\nonu \\
c_{2 \frac{9}{2}}^{to} & = & 20160 c^3 B B'(5+2 c) (53+2 c) (20+3 c) (6+5 c) (1+N) (6 N+c (-3+2 N)),
\nonu \\
c_{2 \frac{9}{2}}^{o''} & = & 252 c B(29+2 c)^2 (53+2 c) (20+3 c) (6+5 c) (84+25 c)\nonu \\
& \times & (4+73 c) (c+6 (-1+N) N)^2 (c-4 c N+6 N^2),
\qquad
c_{2 \frac{9}{2}}^{a} =\frac{3}{4},
\nonu \\
c_{2 \frac{9}{2}}^{go''} & = & 2450 c^3 B (53+2 c) (20+3 c) (23601-27059 c+17768 c^2+100 c^3) (1+N)\nonu \\
& \times &(c^2-3 c (-1+N)+18 (1-3 N) N) (6 N+c (-3+2 N)) (-6 N^2+c (-1+4 N)),
\nonu \\
c_{2 \frac{9}{2}}^{to'} & = & 588 c^2 B (29+2 c) (20+3 c) (6+5 c) (c+6 (-1+N) N) \nonu \\
&\times &(-2268 N^2 (-100+200 N+5447 N^2)+20 c^5 (4402-3023 N+6046 N^2)\nonu \\
&+& c^4 (274408-1893242 N+2911384 N^2-362760 N^3)\nonu \\
&+&216 c N (-350+2135 N+35259 N^2+76222 N^3)\nonu \\
&+&6 c^3 (76222-900143 N+2198091 N^2-1893242 N^3+528240 N^4)\nonu \\
&+& 9 c^2 (-38129+141036 N+908438 N^2-3600572 N^3+1097632 N^4)),
\nonu \\
c_{2 \frac{9}{2}}^{to''} & = &-784 c^3 B (53+2 c) (20+3 c) (2703-180995 c+104480 c^2+16900 c^3)(1+N) \nonu \\
&\times & (c^2-3 c (-1+N)+18 (1-3 N) N) (6 N+c (-3+2 N)) (-6 N^2+c (-1+4 N)),
\nonu \\
c_{2 \frac{9}{2}}^{go'''} & = & 3528 c B(29+2 c) (53+2 c) (6+5 c) (32718+357445 c+113528 c^2+4300 c^3)\nonu \\
&\times& (c+6 (-1+N) N)^2 (-6 N^2+c (-1+4 N)),
\nonu \\
c_{2 \frac{9}{2}}^{ga} & = & -37044 B(29+2 c)^2 (53+2 c) (6+5 c)^2 (4+73 c) (c+6 (-1+N) N)^3,
\nonu \\
c_{2 \frac{9}{2}}^{go''''} & = &-126 c^2 B(29+2 c) (20+3 c) (6+5 c) (c+6 (-1+N) N) \nonu \\
& \times &(100 c^5 (-1334-59 N+118 N^2)-5292 N^2 (800-1600 N+2799 N^2)\nonu \\
&-& 10 c^4 (148336-41789 N-78272 N^2+3540 N^3)\nonu \\
&-& 252 c N (-5600+343485 N-703756 N^2+95568 N^3)\nonu \\
&-&6 c^3 (111496-2344454 N+1931133 N^2-417890 N^3+800400 N^4)\nonu \\
&-&3 c^2 (137151-9852584 N+31095988 N^2-28133448 N^3+17800320 N^4)),
\nonu \\
c_{2 \frac{9}{2}}^{q} & = & 4410 c B (29+2 c) (53+2 c) (20+3 c) (6+5 c)^2 (-147+182 c+40 c^2) \nonu \\
&\times &(c+6 (-1+N) N)^2 (-6 N^2+c (-1+4 N)),
\nonu \\
\frac{1}{B} &\equiv& 1764 (29 + 2 c) (53 + 2 c) (6 + 5 c) (4 + 73 c)\nonu \\
& \times & (c + 6 (-1 + N) N)^2 (20 + 3 c)  (29 + 2 c) (6 + 5 c) (c + 6 (-1 + N) N),
\nonu \\
B' &\equiv&(4 + 73 c) (108 N^3 (-1 + 3 N) + c^3 (-1 + 4 N)\nonu \\
&-& 3 c^2 (1 - 5 N + 6 N^2) - 18 c N (1 - 6 N + 11 N^2)).
\label{O2O9re1}
\eea
The coefficients in (\ref{A92}) are given by
\bea
c_{\frac{9}{2}}^{go} & = &-96 c^3 F (21+4 c) (-7+10 c) (327+16 c) (3+c-9 N) (1+N) (c+6 N)\nonu \\
& \times &  (-3 c+2 (3+c) N) (c+6 (-1+N) N) (-6 N^2+c (-1+4 N)), \nonu \\
c_{2 \frac{9}{2}}^{to} & = & 480 c^3 F(21+4 c) (6+5 c) (-7+10 c) (3+c-9 N) (1+N) (c+6 N)\nonu \\
& \times &  (-3 c+2 (3+c) N) (c+6 (-1+N) N) (-6 N^2+c (-1+4 N)),\nonu \\
c_{2 \frac{9}{2}}^{tg} &=& -32 c^4 F(29+2 c) (6+5 c) (-195+94 c) (3+c-9 N) (1+N) (c+6 N) \nonu \\
& \times & (-3 c+2 (3+c) N) (2 c (-1+N)+3 (3-4 N) N) (c+2 c N+3 N (-3+2 N)), \nonu \\
c_{\frac{9}{2}}^{oo} &=& 144 c^2F(29+2 c) (21+4 c) (6+5 c) (-7+10 c) (3+c-9 N) (1+N)\nonu \\
& \times &  (c+6 N) (-3 c+2 (3+c) N) (c+6 (-1+N) N)^2,\nonu \\
F & \equiv & \frac{1} {3 (29 + 2 c)^2 (21 + 4 c) (6 + 5 c)^3 (c + 6 (-1 + N) N)^4}.
\label{O2O9re2}
\eea
The equations
$(\ref{O2O9re1})$ and $(\ref{O2O9re2})$ are checked for $N=7$ again.

To find the explicit form of $\{ O_{2} \; O_{\frac{9}{2}} \}_{-1}(w)$
for general $N$, we examine the operators in  $\{ O_{2} \; O_{\frac{9}{2}}
\}_{-1}(w)$ for $N=2$ and
write down the possible various spin-$\frac{11}{2}$ fields with unknown
coefficients. Then we subtract them from the first-order pole
$\{ O_{2} \; O_{\frac{9}{2}} \}_{-1}(w)$
and solve it for the coefficients for $N=2,3,4,5$ cases.
The explicit expression for the first-order pole
$\{ O_{2} \; O_{\frac{9}{2}} \}_{-1}(w)$
is presented as follows:
\bea
&&\{ O_{2} \; O_{\frac{9}{2}} \}_{-1}(w) =
c_1 K^a J^{bc} J^{bc} J^{ad} J^{de}J^{eN'}(w)+c_2 K^a J^{bc} J^{bc} J^{ad} \pa J^{dN'}(w)
\nonu \\
&&+c_3 K^a J^{ab} J^{bc} J^{cd} J^{de}J^{eN'}(w)+c_4 K^a J^{bc} J^{bc} \pa J^{ad} J^{dN'}(w)+c_5 K^a J^{bc} J^{bc} \pa^2 J^{aN'}(w)
\nonu \\
&&+c_6 K^a J^{ab} J^{bc} J^{cd} \pa J^{dN'}(w)+c_7 K^a J^{ab} J^{bc} J^{dN'} J^{dN'} J^{cN'}(w)+c_8 K^a J^{ab} J^{bc} \pa^2 J^{cN'}(w)
\nonu \\
&&+c_9 K^a J^{ab} J^{bc} \pa J^{cd} J^{dN'} (w)+c_{10} K^a J^{ab}J^{cN'} J^{cN'} \pa J^{bN'}(w) +c_{11} K^a J^{ab}J^{bN'} \pa J^{cN'} J^{cN'}(w)
\nonu \\
&&+c_{12} K^a J^{ab} \pa J^{cd} J^{cd} J^{bN'}(w)+c_{13} K^a J^{bc}\pa J^{ad} J^{cd}J^{bN'}(w)+c_{15} K^a J^{ab}\pa J^{bc} \pa J^{cN'}(w)
\nonu \\
&&+c_{16} K^a J^{ab}\pa^2 J^{bc} J^{cN'}(w)+c_{17} K^a J^{ab} \pa J^{bc}J^{dc} J^{dN'}(w) +c_{18} K^a J^{ab} \pa^3 J^{bN'}(w)
\nonu \\
&&+c_{19} K^a J^{aN'}\pa J^{bN'} \pa J^{bN'}(w)+c_{20} K^a J^{aN'}\pa ^2 J^{bN'} J^{bN'}(w)+c_{21} K^a K^b K^c J^{ab}J^{de} J^{ce}  J^{dN'}(w)
\nonu \\
&&+c_{23} K^a K^b K^c J^{ab}J^{cd} \pa J^{dN'}(w)+c_{25} K^a K^b K^c J^{ab}\pa^2 J^{cN'}(w)+c_{27} K^a K^b K^c \pa^2 J^{ab}J^{cN'}(w)
\nonu \\
&&+c_{31} K^a K^b \pa K^d J^{ab}J^{cd} J^{cN'}(w)+c_{32} K^a K^b \pa^2 K^c J^{ac}J^{bN'}(w)+c_{33} K^a K^b \pa K^c \pa J^{ab}J^{cN'}(w)
\nonu \\
&&+c_{34} K^a K^b \pa^2 K^c J^{ab}J^{cN'}(w)+c_{35} K^a K^b \pa K^c \pa J^{ac}J^{bN'}(w)+c_{36} K^a K^b \pa K^c J^{bc}\pa J^{aN'}(w)
\nonu \\
&&+c_{37} K^a K^b \pa K^c J^{ac}J^{bd}J^{dN'}(w)+c_{38} K^a K^b\pa K^c J^{ab}\pa J^{cN'}(w)+c_{39} K^a K^b K^c \pa J^{ab}J^{cd}  J^{dN'}(w)
\nonu \\
&&+c_{42} K^a \pa J^{ab} J^{bc} \pa J^{cN'}(w)+c_{43} K^a \pa J^{ab} J^{cN'} J^{cN'} J^{bN'}(w)+c_{44} K^a \pa J^{ab}\pa  J^{cb} J^{cN'}(w)
\nonu \\
&&+c_{45} K^a \pa J^{ab} \pa^2 J^{bN'}(w)+c_{46} K^a \pa J^{aN'} \pa J^{bN'}  J^{bN'}(w)+c_{47} K^a \pa J^{bc} \pa J^{bc} J^{aN'}(w)
\nonu \\
&&+c_{48} K^a \pa J^{bc} J^{bc} \pa J^{aN'}(w)+
c_{49} K^a \pa K^b K^b J^{ac} J^{cd} J^{dN'}(w)
+c_{50} K^a \pa K^b K^b J^{ac} \pa J^{cN'}(w)
\nonu \\
&&+c_{51} K^a \pa K^b K^b \pa J^{ac} J^{cN'}(w) +c_{52} K^a \pa K^b K^b \pa^2 J^{aN'}(w) +c_{53} K^a \pa K^b \pa K^c  J^{ac} J^{bN'}(w)
\nonu \\
&&+c_{54} K^a \pa K^b \pa K^c  J^{bc} J^{aN'}(w)+c_{55} K^a \pa^2 J^{ab}J^{bc} J^{cN'}(w) +c_{56} K^a \pa^2 J^{ab} \pa J^{bN'}(w)
\nonu \\
&&+c_{57} K^a \pa^2 J^{bc}J^{bc} J^{aN'}(w) +c_{58} K^a \pa^2 J^{aN'}J^{bN'} J^{bN'}(w) +c_{59} K^a \pa^2 K^b K^b \pa J^{aN'}(w)
\nonu \\
&&+c_{60} K^a \pa^2 K^b K^b  J^{ac}J^{cN'}(w) +
c_{61} K^a \pa^2 K^b \pa K^b J^{aN'}(w)+c_{62} K^a \pa^3 J^{ab}J^{bN'}(w)
\nonu \\
&&+c_{63} K^a \pa^3 K^b K^b  J^{aN'}(w)+c_{64} K^a \pa^4 J^{aN'}(w)+
c_{65} \pa^2 K^a J^{ba} J^{bc} J^{cN'}(w)
\nonu \\
&&+c_{66} \pa K^a J^{ab} J^{bc} J^{cd} J^{dN'}(w)+c_{67} \pa K^a J^{ab} J^{bc} \pa J^{cN'}(w)+c_{68} \pa K^a J^{ab} J^{cd} J^{cd} J^{bN'}(w)
\nonu \\
&&+c_{69} \pa K^a J^{ab} J^{bN'} J^{cN'} J^{cN'}(w)+c_{70} \pa K^a J^{ab} \pa J^{bc} J^{cN'}(w)+c_{71} \pa K^a J^{ab} \pa^2  J^{bN'}(w)
\nonu \\
&&+c_{72} \pa K^a J^{bc} J^{bc}\pa J^{aN'}(w)+c_{73} \pa K^a J^{aN'} \pa J^{bN'} J^{bN'}(w)+c_{74} \pa K^a K^a \pa K^b J^{bc} J^{cN'}(w)
\nonu \\
&&+c_{75} \pa K^a \pa J^{ab} J^{bc} J^{cN'}(w)+c_{76} \pa K^a K^a \pa K^b  \pa J^{bN'}(w)+c_{77} \pa K^a K^a \pa^2 K^b J^{bN'}(w)
\nonu \\
&&+c_{78} \pa K^a \pa J^{ab} \pa J^{bN'}(w)+c_{79} \pa K^a J^{aN'} \pa J^{bN'} J^{bN'}(w)+c_{80} \pa K^a \pa J^{bc} J^{bc} J^{aN'}(w)
\nonu \\
&&+c_{81} \pa K^a \pa^2 J^{ab} J^{bN'}(w)+c_{82} \pa K^a \pa^2 K^b K^b J^{aN'}(w)+c_{83} \pa K^a \pa^3 J^{aN'}(w)
\nonu \\
&&+c_{84} \pa^2 K^a J^{bc} J^{bc} J^{aN'}(w)+c_{85} \pa^2 K^a J^{ab} \pa J^{bN'}(w)+c_{86} \pa^2 K^a J^{aN'} J^{bN'} J^{bN'}(w)
\nonu \\
&&+c_{87} \pa^2 K^a \pa J^{ab} J^{bN'}(w)+c_{88} \pa^2 K^a \pa^2 J^{aN'}(w)+c_{89} \pa^3 K^a J^{ab} J^{bN'}(w)+c_{90} \pa^3 K^a \pa J^{aN'}(w)
\nonu \\
&&+c_{91} K^a K^b K^c J^{ab} \pa J^{dc} J^{dN'}(w)+c_{92} \pa^4 K^a J^{aN'}(w)+c_{93} K^a K^b K^c \pa J^{ab} \pa J^{cN'}(w),
\label{O2O9p1}
\eea
where $ N' \equiv 2N+1$. The coefficients in $(\ref{O2O9p1})$ are presented in
(\ref{O2O9p1coeff}) of Appendix $D$. This result is checked for $N=6$ again.
The spin-$\frac{11}{2}$ current $O_{\frac{11}{2}}(z)$ is given by
(\ref{O11}), (\ref{O2O9re1}) and (\ref{O2O9p1}).
The OPE for general $N$ is obtained from (\ref{O2O9}) with (\ref{O2O9re1}).
The fusion rule is summarized by
$[O_2][O_{\frac{9}{2}}]= [I] +[O_{\frac{5}{2}}]+[{O}_\frac{7}{2}]
+[O_{\frac{9}{2}}]+[O_{\frac{11}{2}}]$.

$\bullet$ The OPE $O_{\frac{5}{2}}(z) \, O_{4'}(w)$

Now let us compute
the OPE between $O_{\frac{5}{2}}(z)$ and $O_{4'}(z)$. The computation of the OPE for $N=2$ provides
\bea
O_{\frac{5}{2}}(z) \, O_{4^{'}}(w)
&=&\frac{1}{(z-w)^3} \left[c_{\frac{5}{2}4^{'}}^{o} O_{\frac{7}{2}}+c_{\frac{5}{2}4^{'}}^{go} \left( G O_{2}-\frac{2}{5}\pa O_{\frac{5}{2}}\right)\right](w) \nonu \\
& + & \frac{1}{(z-w)^2}
\left[ \frac{2}{7}c_{\frac{5}{2}4^{'}}^{o} \pa O_{\frac{7}{2}}+ \frac{2}{7}c_{\frac{5}{2}4^{'}}^{go} \pa \left(G  O_{2}-\frac{2}{5}\pa O_{\frac{5}{2}}\right)
\right. \nonu \\
&+&c_{\frac{5}{2}4^{'}}^{go'}\left(\pa G O_{2}-\frac{3}{4} G\pa O_{2} +\frac{1}{8} \pa^2 O_{\frac{5}{2}}\right)
 +c_{\frac{5}{2}4^{'}}^{to}\left( T O_{\frac{5}{2}}-\frac{1}{4} \pa^2 O_{\frac{5}{2}}\right)
 \nonu \\
&+& c_{\frac{5}{2}4^{'}}^{o'} O_{\frac{9}{2}}+c_{\frac{5}{2}4^{'}}^{a} A_{\frac{9}{2}}\left. \right](w) \nonu \\
& + & \frac{1}{(z-w)}
\left[  \frac{3}{56}c_{\frac{5}{2}4^{'}}^{o} \pa^2 O_{\frac{7}{2}}
+ \frac{3}{56}c_{\frac{5}{2}4^{'}}^{go} \pa^2 \left( G O_{2}-\frac{2}{5}\pa O_{\frac{5}{2}}\right)\right. \nonu \\
& + &\frac{1}{3}c_{\frac{5}{2}4^{'}}^{go'}\pa \left (\pa G O_{2}-\frac{3}{4}G \pa O_{2} +\frac{1}{8} \pa^2 O_{\frac{5}{2}}\right)
+ \frac{1}{3}c_{\frac{5}{2}4^{'}}^{to}\pa \left( T O_{\frac{5}{2}}-\frac{1}{4} \pa^2 O_{\frac{5}{2}}\right)
\nonu \\
&+& \frac{1}{3}c_{\frac{5}{2}4^{'}}^{o'}\pa  O_{\frac{9}{2}}+\frac{1}{3}c_{\frac{5}{2}4^{'}}^{a}\pa A_{\frac{9}{2}}+ c_{\frac{5}{2}4^{'}}^{go''}\left( \pa^2 GO_{2} +\frac{3}{5} G \pa^2 O_{2}-2 \pa G \pa O_{2} -\frac{2}{35}\pa^3 O_{\frac{5}{2}}\right)
\nonu \\
&+& c_{\frac{5}{2}4^{'}}^{to'}\left( T O_{\frac{7}{2}}-\frac{3}{16} \pa^2 O_{\frac{7}{2}}\right)+
c_{\frac{5}{2}4^{'}}^{to''} \left(T \pa O_{\frac{5}{2}}-\frac{5}{4} \pa T O_{\frac{5}{2}}-\frac{1}{7} \pa^3 O_{\frac{5}{2}}\right)
\nonu \\
&+& c_{\frac{5}{2}4^{'}}^{go'''}\left( G O_{4^{'}}-\frac{2}{9}\pa O_{\frac{9}{2}}\right)+c_{\frac{5}{2}4^{'}}^{ga}\left( G A_{4}-\frac{2}{9}\pa A_{\frac{9}{2}}\right)
+c_{\frac{5}{2}4^{'}}^{go''''}\left( G O_{4}-\frac{1}{8} \pa^2 O_{\frac{7}{2}}\right)\nonu \\
&+&  c_{\frac{5}{2}4^{'}}^{q}Q_{\frac{11}{2}}+ c_{\frac{5}{2}4^{'}}^{o''} O_{\frac{11}{2}}\left.  \right](w)  +\cdots.
\label{O5O44}
\eea
The field contents of (\ref{O5O44}) are almost the same as the ones in
(\ref{O2O9}).
The relative coefficients of descendant fields are determined from (\ref{PhiPhi}). The quasi-primary
and primary fields are found by the same method we have used so far.
The quasi-primary field $Q_{\frac{11}{2}}(w)$ is given in (\ref{Q11}).
From the above OPE, $O_{\frac{11}{2}}(z)$ can be expressed as
\bea
O_{\frac{11}{2}}(w) &=& \frac{1}{c_{\frac{5}{2}4^{'}}^{o''}}\{ O_{\frac{5}{2}}
\; O_{4'} \}_{-1}(w)
-\frac{1}{c_{\frac{5}{2}4^{'}}^{o''}}\left[  \frac{3}{56}c_{\frac{5}{2}4^{'}}^{o} \pa^2 O_{\frac{7}{2}}
+ \frac{3}{56}c_{\frac{5}{2}4^{'}}^{go} \pa^2 \left( G O_{2}-\frac{2}{5}\pa O_{\frac{5}{2}}\right)\right. \nonu \\
& + &\frac{1}{3}c_{\frac{5}{2}4^{'}}^{go'}\pa \left (\pa G O_{2}-\frac{3}{4}G \pa O_{2} +\frac{1}{8} \pa^2 O_{\frac{5}{2}}\right)
+ \frac{1}{3}c_{\frac{5}{2}4^{'}}^{to}\pa \left( T O_{\frac{5}{2}}-\frac{1}{4} \pa^2 O_{\frac{5}{2}}\right)
\nonu \\
&+& \frac{1}{3}c_{\frac{5}{2}4^{'}}^{o'}\pa  O_{\frac{9}{2}}+\frac{1}{3}c_{\frac{5}{2}4^{'}}^{a}\pa A_{\frac{9}{2}}+ c_{\frac{5}{2}4^{'}}^{go''}\left( \pa^2 GO_{2} +\frac{3}{5} G \pa^2 O_{2}-2 \pa G \pa O_{2} -\frac{2}{35}\pa^3 O_{\frac{5}{2}}\right)
\nonu \\
&+& c_{\frac{5}{2}4^{'}}^{to'}\left( T O_{\frac{7}{2}}-\frac{3}{16} \pa^2 O_{\frac{7}{2}}\right)+
c_{\frac{5}{2}4^{'}}^{to''} \left(T \pa O_{\frac{5}{2}}-\frac{5}{4} \pa T O_{\frac{5}{2}}-\frac{1}{7} \pa^3 O_{\frac{5}{2}}\right)
\nonu \\
&+& c_{\frac{5}{2}4^{'}}^{go'''}
\left( G O_{4^{'}}-\frac{2}{9}\pa O_{\frac{9}{2}}\right)+c_{\frac{5}{2}4^{'}}^{ga}\left( G A_{4}-\frac{2}{9}\pa A_{\frac{9}{2}}\right)
+c_{\frac{5}{2}4^{'}}^{go''''}
\left( G O_{4}-\frac{1}{8} \pa^2 O_{\frac{7}{2}}\right)\nonu \\
&+&  c_{\frac{5}{2}4^{'}}^{q}Q_{\frac{11}{2}}\left.  \right](w).
\label{O111-1}
\eea
Note that this is equal to (\ref{O11}) found before.
As done before, we compute the OPE $O_{\frac{5}{2}}(z) \,
O(w)_{4'}$ for $N=3,4,5$ cases
to find the general forms of structure constants.
To find the structure constants
$c_{\frac{5}{2}4^{'}}^{o}$,
$c_{\frac{5}{2}4^{'}}^{go}$, $c_{\frac{5}{2}4^{'}}^{go'}$, $c_{\frac{5}{2}4^{'}}^{to}$,
$c_{\frac{5}{2}4^{'}}^{o'}$, and $c_{\frac{5}{2}4^{'}}^{ga}$,
we use the following equations:
\bea
\left[\{O_{\frac{5}{2}} \, O_{4'}\}_{-3}-c_{\frac{5}{2}4^{'}}^{o} O_{\frac{7}{2}}
-c_{\frac{5}{2}4^{'}}^{go} \left( G O_{2}-\frac{2}{5}\pa O_{\frac{5}{2}}\right)
\right](w)
& = & 0, \nonu \\
\left[\{O_{\frac{5}{2}} \,
O_{4'}\}_{-2}-(\mbox{the 2nd-order pole })\right](w) & = & 0.
\nonu
\eea
Here the 2nd-order pole contains six terms in (\ref{O5O44}).
The remaining structure constants are determined completely from
(\ref{O11}) and
(\ref{O111-1}).
The structure constants in (\ref{O5O44}) for general $N$ are given by
\bea
c_{\frac{5}{2}4^{'}}^{o}  &=& 168 c^2 B(29+2 c) (53+2 c) (20+3 c) (21+4 c) (6+5 c) (4+73 c)\nonu \\
& \times & (c+6 (-1+N) N) (10 c^3 (-6-N+2 N^2)-126 N^2 (25-50 N+12 N^2)\nonu \\
&-& c^2 (42-623 N+496 N^2+60 N^3)-3 c N (-350+1155 N-1246 N^2+720 N^3)),
 \nonu \\
c_{\frac{5}{2}4^{'}}^{go} &=&50400 c^3 B B'(-3+2 c) (29+2 c) (53+2 c) (20+3 c) (4+73 c) (1+N) (6 N+c (-3+2 N)),
 \nonu \\
c_{\frac{5}{2}4^{'}}^{go'} &=&7680 c^3B B' (25+2 c) (53+2 c) (20+3 c) (6+5 c) (4+73 c) (1+N) (6 N+c (-3+2 N)),
 \nonu \\
c_{\frac{5}{2}4^{'}}^{to} &=&-107520 c^3B B' (53+2 c) (20+3 c) (6+5 c) (4+73 c) (1+N) (6 N+c (-3+2 N)),
 \nonu \\
c_{\frac{5}{2}4^{'}}^{o'}&=&56 cB (29+2 c)^2 (53+2 c) (20+3 c) (6+5 c)(84+25 c) (4+73 c) \nonu \\
& \times & (c+6 (-1+N) N)^2 (c-4 c N+6 N^2),
\qquad
c_{\frac{5}{2}4^{'}}^{a}=\frac{1}{4},
 \nonu \\
c_{\frac{5}{2}4^{'}}^{go''}&=&700 c^3B B' (53+2 c) (20+3 c) (-56229+42661 c-28752 c^2+740 c^3)\nonu \\
& \times &  (1+N) (6 N+c (-3+2 N)),
 \nonu \\
c_{\frac{5}{2}4^{'}}^{to'}&=&-56 c^2B (29+2 c) (20+3 c) (6+5 c) (c+6 (-1+N) N) \nonu \\
& \times & (-11340 N^2 (-1200+2400 N+7117 N^2)+20 c^5 (38698-19847 N+39694 N^2)\nonu \\
&+& c^4 (6218392-14192078 N+20287456 N^2-2381640 N^3)\nonu \\
&+&108 c N (-42000+2187745 N-3709300 N^2+2120432 N^3)\nonu \\
&+& 6 c^3 (1060216-14000654 N+22660563 N^2-14192078 N^3+4643760 N^4)\nonu \\
&+& 9 c^2 (-249095-7418600 N+34356780 N^2-56002616 N^3+24873568 N^4)),
 \nonu \\
c_{\frac{5}{2}4^{'}}^{to''} &=&224 c^3B (53+2 c) (20+3 c) (17907-217555 c+116720 c^2+29700 c^3)\nonu \\
& \times & (1+N) (6 N+c (-3+2 N)) (108 N^3 (-1+3 N)+c^3 (-1+4 N)\nonu \\
&-& 3 c^2 (1-5 N+6 N^2)-18 c N (1-6 N+11 N^2)),
 \nonu \\
c_{\frac{5}{2}4^{'}}^{go'''} &=&-2352 cB (29+2 c) (53+2 c) (6+5 c) (32718+357445 c+113528 c^2+4300 c^3)\nonu \\
& \times & (c+6 (-1+N) N)^2 (-6 N^2+c (-1+4 N)),
 \nonu \\
c_{\frac{5}{2}4^{'}}^{ga} &=&24696 B(29+2 c)^2 (53+2 c) (6+5 c)^2 (4+73 c) (c+6 (-1+N) N)^3,
 \nonu \\
c_{\frac{5}{2}4^{'}}^{go''''} &=&-84 c^2B (29+2 c) (20+3 c) (6+5 c) (c+6 (-1+N) N)\nonu \\
& \times & (20 c^5 (4042-143 N+286 N^2)+2268 N^2 (100-200 N+5683 N^2)\nonu \\
&-& 2 c^4 (-25424+52381 N+375988 N^2+8580 N^3)\nonu \\
&-& 216 c N (350-41090 N+120561 N^2+64058 N^3)\nonu \\
&+&9 c^2 (39781-482244 N+1032758 N^2+888388 N^3+203392 N^4)\nonu \\
&+&6 c^3 (-64058+222097 N-493509 N^2-104762 N^3+485040 N^4)),
 \nonu \\
c_{\frac{5}{2}4^{'}}^{q}&=&-2940 cB (29+2 c) (53+2 c) (20+3 c) (6+5 c)^2 (-147+182 c+40 c^2)\nonu \\
& \times & (c+6 (-1+N) N)^2 (-6 N^2+c (-1+4 N)),
\qquad
c_{\frac{5}{2}4^{'}}^{o''}= -1,
 \nonu \\
\frac{1}{B} &\equiv & 1176 (29 + 2 c)^2 (6 + 5 c)^2 (20 + 3 c) (4 + 73 c) (53 + 2 c) (c +   6 (-1 + N) N)^3,
\label{coeffcoeff}
 \\
B' &\equiv & (108 N^3 (-1 + 3 N) + c^3 (-1 + 4 N) - 3 c^2 (1 - 5 N + 6 N^2) -
  18 c N (1 - 6 N + 11 N^2)).
\nonu
\eea
All results regarding the OPE $O_{\frac{5}{2}}(z) \,
O_{4'}(w)$ are checked for $N=6$ again.
The spin-$\frac{11}{2}$ current
$O_\frac{11}{2}(z)$ is given by
(\ref{O111-1}) and (\ref{coeffcoeff}).
Again the OPE $O_{\frac{5}{2}}(z) \, O_{4'}(w)$
is obtained from (\ref{O5O44}) with (\ref{coeffcoeff}).
The fusion rule is summarized by
$[O_{\frac{5}{2}}][O_{4'}]= [I] +[{O}_\frac{7}{2}]
+[O_{\frac{9}{2}}]+[O_{\frac{11}{2}}] $.


$\bullet$ The OPE $O_{\frac{5}{2}}(z) \, O_{\frac{9}{2}}(w)$

Now we compute the OPE between $O_{\frac{5}{2}}(z)$ and $O_{\frac{9}{2}}(w)$,
which is the final and most complicated OPE in this work. From the computation of the OPE for $N=2$, we find
that
\bea
O_{\frac{5}{2}}(z) \, O_{\frac{9}{2}}(w)
 & = & \frac{1}{(z-w)^5} c_{\frac{5}{2}\frac{9}{2}}^{o}  O_{2}(w)
+ \frac{1}{(z-w)^3} \left[c_{\frac{5}{2}\frac{9}{2}}^{go} \left( G O_{\frac{5}{2}}-\frac{1}{5}\pa^2 O_{2}\right)
\right.
 \label{O5O9}
\\
& +&c_{\frac{5}{2}\frac{9}{2}}^{to} \left( T O_{2}-\frac{3}{10}\pa^2 O_{2}\right)
+c_{\frac{5}{2}\frac{9}{2}}^{o'}O_{4}+c_{\frac{5}{2}\frac{9}{2}}^{o''}O_{4^{'}}+c_{\frac{5}{2}\frac{9}{2}}^{a}A_{4} \left.\right](w) \nonu \\
& + & \frac{1}{(z-w)^2}
\left[ \frac{1}{4}c_{\frac{5}{2}\frac{9}{2}}^{go} \pa \left( G O_{\frac{5}{2}}-\frac{1}{5}\pa^2 O_{2}\right)
+\frac{1}{4}c_{\frac{5}{2}\frac{9}{2}}^{to} \pa  \left( T O_{2}-\frac{3}{10}\pa^2 O_{2}\right)
+\frac{1}{4}c_{\frac{5}{2}\frac{9}{2}}^{o'} \pa O_{4} \right. \nonu \\
& + & \frac{1}{4}c_{\frac{5}{2}\frac{9}{2}}^{o''} \pa O_{4^{'}}+\frac{1}{4}c_{\frac{5}{2}\frac{9}{2}}^{a} \pa A_{4}
+c_{\frac{5}{2}\frac{9}{2}}^{to'}\left( T \pa O_{2}-\pa O_{2} T -\frac{1}{6}\pa^3 O_{2}\right)\nonu \\
&+& c_{\frac{5}{2}\frac{9}{2}}^{go'}\left(G \pa O_{\frac{5}{2}}-\frac{5}{3}\pa G O_{\frac{5}{2}}-\frac{1}{9}\pa^3 O_{2} \right)
+c_{\frac{5}{2}\frac{9}{2}}^{go''}\left(G O_{\frac{7}{2}}-\frac{1}{4} \pa O_{4}\right)\left. \right](w) \nonu \\
& + & \frac{1}{(z-w)}
\left[ \frac{1}{24}c_{\frac{5}{2}\frac{9}{2}}^{go} \pa^2 \left( G O_{\frac{5}{2}}-\frac{1}{5}\pa^2 O_{2}\right)
+\frac{1}{24}c_{\frac{5}{2}\frac{9}{2}}^{to} \pa^2  \left( T O_{2}-\frac{3}{10}\pa^2 O_{2}\right) \right. \nonu \\
& + & \frac{1}{24}c_{\frac{5}{2}\frac{9}{2}}^{o'} \pa^2 O_{4} +\frac{1}{24}c_{\frac{5}{2}\frac{9}{2}}^{o''} \pa^2 O_{4^{'}}
+\frac{1}{24}c_{\frac{5}{2}\frac{9}{2}}^{a} \pa^2 A_{4} +\frac{3}{10}c_{\frac{5}{2}\frac{9}{2}}^{to'} \pa\left( T \pa O_{2}-\pa O_{2} T -\frac{1}{6}\pa^3 O_{2}\right)\nonu \\
&+& \frac{3}{10}c_{\frac{5}{2}\frac{9}{2}}^{go'} \pa\left(G \pa O_{\frac{5}{2}}-\frac{5}{3}\pa G O_{\frac{5}{2}}-\frac{1}{9}\pa^3 O_{2} \right)
+\frac{3}{10}c_{\frac{5}{2}\frac{9}{2}}^{go''} \pa\left(G O_{\frac{7}{2}}-\frac{1}{4} \pa O_{4}\right)\nonu \\
&+& c_{\frac{5}{2}\frac{9}{2}}^{go'''} \left( G O_{\frac{9}{2}}-\frac{1}{9} \pa^2 O_{4{'}}\right)+c_{\frac{5}{2}\frac{9}{2}}^{ga}\left( G A_{\frac{9}{2}}-\frac{1}{9}\pa^2 A_{4}\right)\nonu \\
&+& c_{\frac{5}{2}\frac{9}{2}}^{to''}\left( T \pa^2 O_{2}-\frac{5}{2}\pa T \pa O_{2} +\pa^2 T O_{2}-\frac{3}{28} \pa^4 O_{2}\right)\nonu \\
&+& c_{\frac{5}{2}\frac{9}{2}}^{go''''}\left( G \pa^2 O_{\frac{5}{2}}-4\pa G \pa O_{\frac{5}{2}} +\frac{5}{2} \pa^2 G O_{\frac{5}{2}}-\frac{1}{14}\pa^4 O_{2}\right)\nonu \\
&+&c_{\frac{5}{2}\frac{9}{2}}^{go'''''}\left( G \pa O_{\frac{7}{2}}-\frac{7}{3} \pa G O_{\frac{7}{2}}-\frac{1}{9} \pa^2 O_{4}\right)
+ c_{\frac{5}{2}\frac{9}{2}}^{to'''}\left( T O_{4}-\frac{1}{6} \pa^2 O_{4}\right) \nonu \\
&+& c_{\frac{5}{2}\frac{9}{2}}^{to''''}\left( T O_{4^{'}}-\frac{1}{6} \pa^2 O_{4^{'}}\right)+ c_{\frac{5}{2}\frac{9}{2}}^{ta}\left( T A_{4}-\frac{1}{6} \pa^2 A_{4}\right)
+ c_{\frac{5}{2}\frac{9}{2}}^{q} Q_{6} +O_{6}\left.  \right](w)  +\cdots.
\nonu
\eea
The relative coefficients of descendant fields are fixed by (\ref{PhiPhi}).
The fifth-, third- and second-order pole of (\ref{O5O9}) have the same structure to the fourth-, second- and first-order pole of (\ref{O2O44}) respectively.
The spin-6 quasi primary fields in (\ref{O5O9}) are found by the same method we used in  the previous OPEs.
The explicit form of quasi-primary field $Q_{6}(z)$ is given by
\bea
Q_{6}(z)
&=&O_{2}  O_{4}(z) +O_{\frac{5}{2}}  O_{\frac{7}{2}}(z)\nonu \\
&+& 70 c^2 H (1629+c (-2982+c (-911+10 c)))\pa^2 G O_{\frac{5}{2}}(z)\nonu \\
&-& 14 c^2 H(7839+c (-25155-7714 c+80 c^2)) \pa G \pa O_{\frac{5}{2}}(z)\nonu \\
&+& 7 c^2 H(6507+4 c (-4245+2 c (-484+5 c))) G \pa^2 O_{\frac{5}{2}}(z)   \nonu \\
&+& 42 c^2 H(21+4 c) (-13+62 c) T \pa^2 O_{2}(z)\nonu \\
&+&84 c^2 H (21+4 c) (-79+26 c) \pa^2 T O_{2}(z)\nonu \\
&+ & \frac{18 c (-6 N^2+c (-1+4 N))}{7 (6+5 c) (c+6 (-1+N) N)}\, G \pa O_{\frac{7}{2}}(z)\nonu \\
&+& \frac{c (3+2 c) (-6 N^2+c (-1+4 N))}{14 (6+5 c) (c+6 (-1+N) N)} \pa^2 O_{4}(z) -\frac{1}{3} \pa^2 O_{4^{'}}(z),
\label{quasi6}
\eea
where $H \equiv  \frac{(3+c-9 N)(1 + N) (c + 6 N) (-3 c + 2 (3 + c) N)}{35 (29 + 2 c) (21 + 4 c) (6 + 5 c)^2 (c + 6 (-1 + N) N)^2}$.
This quasi-primary field (\ref{quasi6}) containing $O_2 O_4(z)$ and
$O_{\frac{5}{2}} O_{\frac{7}{2}}(z)$ is related to
the primary field $A^{(6,2)}$ in \cite{CV}.
The coefficients in $(\ref{quasi6})$ are determined by the fact that the third-order pole of the OPE between $T(z)$ and $Q_{6}(w)$  should vanish.
The OPEs
$T(z) \, Q_{6}(w)$ and $G(z) \, Q_{6}(w)$ are put in the Appendix $G$
and Appendix $H$ respectively.
The $O_{6}(w)$ is a spin-6 primary field and the superpartner
of spin-$\frac{11}{2}$ primary field $O_{\frac{11}{2}}(z)$.
The $O_{6}(w)$
was found by the same procedure used in finding $O_{\frac{11}{2}}(z)$.
The explicit form of  the $O_{6}(w)$ is expressed as
\bea
O_{6}(w) &=& \{ O_{\frac{5}{2}} \; O_{\frac{9}{2}} \}_{-1}(w)
-\left[ \frac{1}{24}c_{\frac{5}{2}\frac{9}{2}}^{go} \pa^2 \left( G O_{\frac{5}{2}}-\frac{1}{5}\pa^2 O_{2}\right)
+\frac{1}{24}c_{\frac{5}{2}\frac{9}{2}}^{to} \pa^2  \left( T O_{2}-\frac{3}{10}\pa^2 O_{2}\right) \right. \nonu \\
& + & \frac{1}{24}c_{\frac{5}{2}\frac{9}{2}}^{o'} \pa^2 O_{4} +\frac{1}{24}c_{\frac{5}{2}\frac{9}{2}}^{o''} \pa^2 O_{4^{'}}
+\frac{1}{24}c_{\frac{5}{2}\frac{9}{2}}^{a} \pa^2 A_{4} +\frac{3}{10}c_{\frac{5}{2}\frac{9}{2}}^{to'} \pa\left( T \pa O_{2}-\pa O_{2} T -\frac{1}{6}\pa^3 O_{2}\right)\nonu \\
&+& \frac{3}{10}c_{\frac{5}{2}\frac{9}{2}}^{go'} \pa\left(G \pa O_{\frac{5}{2}}-\frac{5}{3}\pa G O_{\frac{5}{2}}-\frac{1}{9}\pa^3 O_{2} \right)
+\frac{3}{10}c_{\frac{5}{2}\frac{9}{2}}^{go''} \pa\left(G O_{\frac{7}{2}}-\frac{1}{4} \pa O_{4}\right)\nonu \\
&+& c_{\frac{5}{2}\frac{9}{2}}^{go'''} \left( G O_{\frac{9}{2}}-\frac{1}{9} \pa^2 O_{4{'}}\right)+c_{\frac{5}{2}\frac{9}{2}}^{ga}\left( G A_{\frac{9}{2}}-\frac{1}{9}\pa^2 A_{4}\right)\nonu \\
&+& c_{\frac{5}{2}\frac{9}{2}}^{to''}\left( T \pa^2 O_{2}-\frac{5}{2}\pa T \pa O_{2} +\pa^2 T O_{2}-\frac{3}{28} \pa^4 O_{2}\right)\nonu \\
&+& c_{\frac{5}{2}\frac{9}{2}}^{go''''}\left( G \pa^2 O_{\frac{5}{2}}-4\pa G \pa O_{\frac{5}{2}} +\frac{5}{2} \pa^2 G O_{\frac{5}{2}}-\frac{1}{14}\pa^4 O_{2}\right)\nonu \\
&+&c_{\frac{5}{2}\frac{9}{2}}^{go'''''}\left( G \pa O_{\frac{7}{2}}-\frac{7}{3} \pa G O_{\frac{7}{2}}-\frac{1}{9} \pa^2 O_{4}\right)
+ c_{\frac{5}{2}\frac{9}{2}}^{to'''}\left( T O_{4}-\frac{1}{6} \pa^2 O_{4}\right) \nonu \\
&+& c_{\frac{5}{2}\frac{9}{2}}^{to''''}\left( T O_{4^{'}}-\frac{1}{6} \pa^2 O_{4^{'}}\right)+ c_{\frac{5}{2}\frac{9}{2}}^{ta}\left( T A_{4}-\frac{1}{6} \pa^2 A_{4}\right)
+ c_{\frac{5}{2}\frac{9}{2}}^{q} Q_{6} \left.  \right](w).
\label{O6}
\eea
As expected, the spin-$6$ current $O_{6}(z)$ obeys the following OPEs:
\bea
T(z) \, O_{6}(w)& =&
\frac{1}{(z-w)^2} \, 6\,O_{6}(w) +
\frac{1}{(z-w)}\,\pa O_{6}(w) +\cdots,\nonu \\
G(z) \, O_{6}(w) & = &\frac{1}{(z-w)} \, 11\,O_{\frac{11}{2}}(w)
+\frac{1}{(z-w)} \, \pa O_{\frac{11}{2}}(w)\cdots.
\nonu
\eea
Therefore, the currents
$O_{\frac{11}{2}}(z)$ and $O_6(z)$ consist of the correct
${\cal N}=1$ supermultiplet.
Finding the general forms of structure constants is much more complicated than the previous OPEs. First we determine the structure constants
$c_{\frac{5}{2}\frac{9}{2}}^{o}$, $c_{\frac{5}{2}\frac{9}{2}}^{go}$, $c_{\frac{5}{2}\frac{9}{2}}^{to}$, $c_{\frac{5}{2}\frac{9}{2}}^{o'}$,
$c_{\frac{5}{2}\frac{9}{2}}^{o''}$, $c_{\frac{5}{2}\frac{9}{2}}^{a}$, $c_{\frac{5}{2}\frac{9}{2}}^{to'}$, $c_{\frac{5}{2}\frac{9}{2}}^{go'}$,
and $c_{\frac{5}{2}\frac{9}{2}}^{go''}$ that appear in
the fifth-, third-, and second-order pole. To find them, we
solve following equations for $N=3,4,5$ cases:
\bea
\left[
\{O_{\frac{5}{2}} O_{\frac{9}{2}}\}_{-5}-c_{\frac{5}{2}\frac{9}{2}}^{o}  O_{2} \right](w) &
= & 0,\nonu \\
\left[\{O_{\frac{5}{2}} O_{\frac{9}{2}}\}_{-3}-(\mbox{the 2nd-order pole})
\right](w) &
= & 0, \nonu \\
\left[\{O_{\frac{5}{2}} O_{\frac{9}{2}}\}_{-2}-(\mbox{the 1st-order pole})
\right](w) & = & 0.
\nonu
\eea
Here the 2nd-order pole and 1st-order pole
are given in the right hand side of (\ref{O5O9}) respectively.
Then we put all results together and find the general forms of them. The general forms of them are checked for $N=6$ again.

To find the remaining structure constants appearing only in the first-order pole,
we use the fact that
the four OPEs (\ref{O2O44}), (\ref{O2O9}), (\ref{O5O44})
and (\ref{O5O9}) should be expressed as a single ${\cal N}=1$ super OPE.
For $N=2$ and $N=3$ cases,
we checked that they are expressed as one single super OPE. The expression of the super OPE is given in (\ref{singleO2O4}) of the Appendix $F$.
Moreover, the following relations are found for $N=2,3$ cases:
\bea
&& \frac{1}{24}c_{\frac{5}{2}\frac{9}{2}}^{go}-\frac{1}{2}c_{\frac{5}{2}\frac{9}{2}}^{go'}+\frac{5}{2}c_{\frac{5}{2}\frac{9}{2}}^{go''''}
+\frac{1}{6}c_{\frac{5}{2}4'}^{to}-\frac{5}{8}c_{\frac{5}{2}4'}^{to''}
-\frac{3}{56}c_{\frac{5}{2} 4'}^{go}+\frac{1}{3}c_{\frac{5}{2} 4'}^{go'}+c_{\frac{5}{2} 4'}^{go''}\nonu \\
&&
+315 c^2 H (-117+4 (-36+c) c)c_{\frac{5}{2} 4'}^{q}\,\nonu \\
&&+70 c^2 H (1629+c (-2982+c (-911+10 c)))c_{\frac{5}{2}\frac{9}{2}}^{q}\,=0,\nonu \\
&&\frac{1}{12}c_{\frac{5}{2}\frac{9}{2}}^{to}-\frac{5}{2}c_{\frac{5}{2}\frac{9}{2}}^{to''}+\frac{1}{3}c_{\frac{5}{2} 4'}^{to}
-\frac{5}{4}c_{\frac{5}{2} 4'}^{to''}+\frac{3}{14}c_{\frac{5}{2} 4'}^{go}+\frac{1}{6}c_{\frac{5}{2} 4'}^{go'}-4c_{\frac{5}{2} 4'}^{go''}\nonu \\
&&+2520 c^2 H (-36+c (75+2 c)) (3+c-9 N)c_{\frac{5}{2} 4'}^{q}=0,\nonu \\
&&\frac{3}{10}c_{\frac{5}{2}\frac{9}{2}}^{go''}+c_{\frac{5}{2}\frac{9}{2}}^{go'''''}
-c_{\frac{5}{2} 4'}^{go''''}+\frac{18 c (-6 N^2+c (-1+4 N))}{7 (6+5 c) (c+6 (-1+N) N)}c_{\frac{5}{2}\frac{9}{2}}^{q}=0,\nonu \\
&&c_{\frac{5}{2}\frac{9}{2}}^{go'''}-c_{\frac{5}{2}4'}^{go'''}=0, \qquad
c_{\frac{5}{2}\frac{9}{2}}^{ga}-c_{\frac{5}{2}4'}^{ga}=0,\qquad
c_{\frac{5}{2}\frac{9}{2}}^{to'''}+2c_{\frac{5}{2}4'}^{go''''}+c_{\frac{5}{2}4'}^{to}=0, \nonu \\
&&c_{\frac{5}{2}\frac{9}{2}}^{to''''}+2c_{\frac{5}{2}4'}^{go'''}=0, \qquad
c_{\frac{5}{2}\frac{9}{2}}^{ta}+2c_{\frac{5}{2}4'}^{ga}=0,\qquad
c_{\frac{5}{2}\frac{9}{2}}^{q}+c_{\frac{5}{2}4'}^{q}=0,
\label{O5O9eq1}
\eea
where $H \equiv  \frac{(3 + c - 9 N)(1 + N) (c + 6 N) (-3 c + 2 (3 + c) N)}{35 (29 + 2 c) (21 + 4 c) (6 + 5 c)^2 (c + 6 (-1 + N) N)^2}$.
Since the four OPEs (\ref{O2O44}), (\ref{O2O9}), (\ref{O5O44})
and (\ref{O5O9}) should be expressed as a single ${\cal N}=1$ super OPE for any $N$, we safely can assume that the equations in (\ref{O5O9eq1})
hold for beyond $N=3$ case.
By solving the equations in (\ref{O5O9eq1}), one can find
the structure constants
$c_{\frac{5}{2}\frac{9}{2}}^{go'''}$, $c_{\frac{5}{2}\frac{9}{2}}^{ga}$,
$c_{\frac{5}{2}\frac{9}{2}}^{to''}$, $c_{\frac{5}{2}\frac{9}{2}}^{go''''}$, $c_{\frac{5}{2}\frac{9}{2}}^{go'''''}$, $c_{\frac{5}{2}\frac{9}{2}}^{to'''}$, $c_{\frac{5}{2}\frac{9}{2}}^{to''''}$,
$c_{\frac{5}{2}\frac{9}{2}}^{ta}$, and $c_{\frac{5}{2}\frac{9}{2}}^{q}$ for general $N$
completely.
The followings are the general forms of structure constants of (\ref{O5O9}):
\bea
c_{\frac{5}{2}\frac{9}{2}}^{o} & = &-8400 c^3BB' (-3+2 c) (29+2 c) (53+2 c) (20+3 c) (6+5 c) (4+73 c) , \nonu \\
c_{\frac{5}{2}\frac{9}{2}}^{go} & = &12600 c^3BB' (-3+2 c) (53+2 c) (20+3 c) (183+10 c) (4+73 c) , \nonu \\
c_{\frac{5}{2}\frac{9}{2}}^{to} & = &-8400 c^3BB'(-3+2 c) (53+2 c) (20+3 c) (573+50 c) (4+73 c), \nonu \\
c_{\frac{5}{2}\frac{9}{2}}^{o'} & = &-270 c^2B (29+2 c) (53+2 c) (20+3 c) (21+4 c) (6+5 c) (4+73 c) (c+6 (-1+N) N)\nonu \\
 & \times &(10 c^3 (-2+N) (3+2 N)-126 N^2 (25+2 N (-25+6 N))\nonu \\
 &+&3 c N (350+N (-1155+2 (623-360 N) N))-c^2 (42+N (-623+496 N+60 N^2))), \nonu \\
c_{\frac{5}{2}\frac{9}{2}}^{o''} & = &-560 c B (29+2 c)^2 (53+2 c) (20+3 c) (6+5 c) (84+25 c) (4+73 c)\nonu \\
& \times &  (c+6 (-1+N) N)^2 (c-4 c N+6 N^2), \qquad
c_{\frac{5}{2}\frac{9}{2}}^{a} = -2, \nonu \\
c_{\frac{5}{2}\frac{9}{2}}^{to'} & = &10500 c^3 BB'(29+2 c) (53+2 c) (20+3 c) (9+10 c) (4+73 c), \nonu \\
c_{\frac{5}{2}\frac{9}{2}}^{go'} & = &-15750 c^3BB' (3+2 c) (29+2 c) (53+2 c) (20+3 c) (4+73 c), \nonu \\
c_{\frac{5}{2}\frac{9}{2}}^{go''} & = &-630 c^2B (29+2 c) (53+2 c) (20+3 c) (6+5 c) (4+73 c)\nonu \\
 & \times & (c+6 (-1+N) N) (10 c^3 (-2+N) (3+2 N)-126 N^2 (25+2 N (-25+6 N))\nonu \\
 &+& 3 c N (350+N (-1155+2 (623-360 N) N))-c^2 (42+N (-623+496 N+60 N^2))), \nonu \\
c_{\frac{5}{2}\frac{9}{2}}^{go'''} & =& -2940 c B(29+2 c) (53+2 c) (6+5 c) (32718+c (357445+4 c (28382+1075 c)))\nonu \\
& \times &  (c+6 (-1+N) N)^2 (-6 N^2+c (-1+4 N)), \nonu \\
c_{\frac{5}{2}\frac{9}{2}}^{ga} & =&30870 B (29+2 c)^2 (53+2 c) (6+5 c)^2 (4+73 c) (c+6 (-1+N) N)^3, \nonu \\
c_{\frac{5}{2}\frac{9}{2}}^{to''} & =& -980 c^3B (-3+2 c) (53+2 c) (20+3 c) (16851+c (33139+3730 c)) \nonu \\
& \times & (3+c-9 N) (1+N) (c+6 N) (-3 c+2 (3+c) N) (-6 N^2+c (-1+4 N)), \nonu \\
c_{\frac{5}{2}\frac{9}{2}}^{go''''} & =& -1470 c^3 BB'(53+2 c) (20+3 c) (-2817+c (74535+2 c (-29859+10 c (-983+10 c)))), \nonu \\
c_{\frac{5}{2}\frac{9}{2}}^{go'''''} & =& -21 c^2B (29+2 c) (20+3 c) (21+4 c) (6+5 c) (c+6 (-1+N) N)\nonu \\
 & \times & (10 c^4 (12976+7229 N (-1+2 N))-756 N^2 (-450+N (900+3709 N))\nonu \\
 &+&c^3 (198962+N (-2104453+4 (717164-108435 N) N))\nonu \\
 &+&18 c N (-6300+N (310765-492478 N+397924 N^2))\nonu \\
 &+&3 c^2 (-25963+N (-492478+N (2293561+2 N (-2104453+778560 N))))), \nonu \\
c_{\frac{5}{2}\frac{9}{2}}^{to'''} & =& 140 c^2 B (29+2 c) (20+3 c) (21+4 c) (6+5 c)^2 \nonu \\
& \times & (c+6 (-1+N) N) (-252 N^2 (-225+450 N+662 N^2)+2 c^3 (12706+5069 N (-1+2 N))\nonu \\
&+&3 c N (-6300+N (328405-594538 N+304944 N^2))\nonu \\
&-&c^2 (4634+N (297269+4 N (-80002+15207 N)))), \nonu \\
c_{\frac{5}{2}\frac{9}{2}}^{to''''} & =&5880 c B(29+2 c) (53+2 c) (6+5 c) \nonu \\
& \times & (32718+c (357445+4 c (28382+1075 c))) (c+6 (-1+N) N)^2 (-6 N^2+c (-1+4 N)), \nonu \\
c_{\frac{5}{2}\frac{9}{2}}^{ta} & =& -61740 B(29+2 c)^2 (53+2 c) (6+5 c)^2 (4+73 c) (c+6 (-1+N) N)^3, \nonu \\
c_{\frac{5}{2}\frac{9}{2}}^{q} & =& 3675 cB (29+2 c) (53+2 c) (20+3 c) (21+4 c) \nonu \\
& \times &(6+5 c)^2 (-7+10 c) (c+6 (-1+N) N)^2 (-6 N^2+c (-1+4 N)), \nonu \\
\frac{1}{B}& \equiv
& 735 (29 + 2 c)^2 (53 + 2 c) (20 + 3 c) (6 + 5 c)^2 (8 + 146 c) (c +
   6 (-1 + N) N)^3,\nonu \\
B'& \equiv
& (1 + N) (3 + c - 9 N) (c + 6 N) (-3 c + 2 (3 + c) N) (-6 N^2 + c (-1 + 4 N)).
\label{Coeffcoeff}
\eea
To find the explicit form of $\{ O_{\frac{5}{2}} \; O_{\frac{9}{2}} \}_{-1}(w)$
for general $N$, we examine
the operators in  $\{ O_{\frac{5}{2}} \; O_{\frac{9}{2}} \}_{-1}(w)$ for $N=2$ and
write down the possible spin-$6$ fields with undetermined
coefficients.
Then we subtract them from the first-order pole
$\{ O_{\frac{5}{2}} \; O_{\frac{9}{2}} \}_{-1}(w)$
and solve it for the coefficients when $N=2,3,4,5,6$ cases.
The explicit expression of $\{ O_{\frac{5}{2}} \; O_{\frac{9}{2}} \}_{-1}(w)$,
which consist of one hundred eighty seven terms, is presented as follows:
\bea
&&\{ O_{\frac{5}{2}} \; O_{\frac{9}{2}} \}_{-1}(w) =
c_1 J^{ab}J^{ab}J^{cd}\pa J^{cN'} J^{dN'}(w)+ c_2 J^{ab}J^{ac}\pa J^{bd}\pa J^{cd}(w)+c_4 J^{ab}J^{ab}\pa J^{cd} \pa J^{cd}(w)
\nonu \\
&&+c_5 J^{ab}J^{ab}\pa^2 J^{cd} J^{cd}(w)+c_6 J^{ab}J^{ab}\pa^2 J^{cN'} J^{cN'}(w)+c_7 J^{ab}J^{ac}J^{bd}J^{ce} J^{dN'}J^{eN'}(w)
\nonu \\
&&+c_9 K^a K^b K^c K^d J^{cd} \pa J^{aN'} J^{bN'}(w)+c_{11}J^{ab}J^{ac}J^{dc}J^{bN'}\pa J^{dN'}(w)
+c_{12}J^{ab}J^{cb}J^{aN'}J^{cN'}J^{dN'}J^{dN'}(w)
\nonu \\
&&+c_{13}J^{ab}J^{cb}\pa J^{ad}J^{cN'}J^{dN'}(w)+c_{15}J^{ab}J^{ab}\pa J^{cN'}\pa J^{cN'}(w)
+c_{16}J^{ab}J^{cb}\pa^2 J^{aN'} J^{cN'}(w)
\nonu \\
&&+c_{17} J^{ab}J^{ac}J^{bd}\pa^2 J^{cd}(w) +c_{18}J^{ab}J^{aN'}\pa J^{bN'} J^{cN'} J^{cN'}(w)+c_{19}J^{ab}J^{aN'}\pa^3 J^{bN'}(w)
\nonu \\
&&+c_{20}J^{ab}\pa J^{ac}J^{bN'}\pa J^{cN'}(w) +c_{21}J^{ab}\pa J^{ac} \pa J^{bd} J^{cd}(w)+c_{22}J^{ab}\pa J^{ac} \pa J^{bN'} J^{cN'}(w)
\nonu \\
&&+c_{23} J^{ab} \pa J^{ac} \pa^2 J^{bc}(w) +c_{24} J^{ab} \pa J^{aN'} \pa^2 J^{bN'}(w)
+c_{25}J^{ab} \pa^2 J^{ac} J^{bN'} J^{cN'}(w)\nonu \\
&&+c_{26}J^{aN'} J^{aN'} \pa J^{bN'} \pa J^{bN'}(w)+c_{27}J^{aN'}J^{aN'} \pa^2 J^{bN'} J^{bN'}(w)
+c_{28}J^{ab}J^{ac}J^{bd}J^{ed}J^{cN'}J^{eN'}(w)\nonu \\
&&+c_{29}K^a K^b J^{ac}J^{bd} J^{cd} J^{ef}J^{ef}(w)+ c_{30}K^a K^b J^{cd} J^{bc} J^{ad} J^{ef} J^{ef}(w)
+c_{31}K^a K^b J^{ad} J^{cd} J^{cb} J^{ef} J^{ef}(w)\nonu \\
&&+c_{32}K^a K^b J^{ae} J^{cd} J^{ce} J^{bf} J^{df}(w)+c_{34}K^a K^b J^{ab} J^{cd} J^{cd} J^{eN'} J^{eN'}(w)
\nonu \\
&&+c_{35}K^a K^b J^{ac} J^{bd} J^{cd} J^{eN'}J^{eN'}(w)+c_{36}K^a K^b J^{ad} J^{ce} J^{bd} J^{cN'} J^{eN'}(w)
\nonu \\
&&+c_{38}K^a K^b J^{ab} J^{cd} J^{de} J^{cN'} J^{eN'}(w)+c_{41}K^a K^b J^{cd} J^{cd} J^{aN'} \pa J^{bN'}(w)+c_{43}K^a K^b J^{ac} J^{cd} \pa J^{be} J^{de}(w)
\nonu \\
&& +c_{44}K^a K^b J^{cb} J^{ad} J^{de} J^{cN'}J^{eN'}(w)+c_{45}K^a K^b J^{ab} J^{cd} J^{cN'}\pa  J^{dN'}(w)\nonu \\
&&+c_{46}K^a K^b J^{ac} J^{bd} \pa J^{cN'} J^{dN'}(w)+c_{47}K^a K^b J^{ac} J^{cd} \pa J^{bN'} J^{dN'}(w)\nonu \\
&&+c_{48}K^a K^b J^{ab} J^{cN'} J^{cN'} J^{dN'} J^{dN'}(w)+c_{49}K^a K^b J^{ab} \pa J^{cN'} \pa J^{cN'}(w)
+c_{52}K^a K^b J^{ab} \pa J^{cd} \pa J^{cd}(w)\nonu \\
&&+c_{53}K^a K^b J^{ab} \pa^2 J^{cN'} J^{cN'}(w) +c_{54} K^a K^b K^c K^d J^{ad} J^{bc} J^{eN'} J^{eN'}(w)
\nonu \\
&&+c_{56}K^a K^b J^{ac} J^{dc} J^{de} J^{bf} J^{ef}(w)+c_{57}K^a K^b J^{ac} \pa J^{bN'} \pa J^{cN'}(w)+c_{58}K^a K^b J^{ab} \pa^2 J^{cd} J^{cd}(w)
\nonu \\
&&+c_{60}K^a K^b J^{ac} \pa^2 J^{bd} J^{cd}(w) +c_{61}K^a K^b J^{cd} \pa^2 J^{ab} J^{cd}(w)+c_{62}K^a K^b J^{ac} J^{bN'} \pa^2 J^{cN'}(w)
\nonu \\
&& + c_{63} K^a K^b J^{ac} J^{bd} J^{cN'} \pa J^{dN'}(w)+c_{64}K^a K^b J^{ac} J^{bd} \pa^2 J^{cd}(w) +c_{65}K^a K^b J^{ac} \pa J^{bc} J^{dN'} J^{dN'}(w)
\nonu \\
&&+c_{66}K^a K^b J^{ac} J^{cd} J^{bN'} \pa J^{dN'}(w) +c_{67} K^a K^b J^{ac} \pa J^{cd} J^{bN'} J^{dN'}(w)
+c_{68}K^a K^b J^{ac} \pa^2 J^{bN'} J^{cN'}(w)
\nonu \\
&&+c_{69}K^a K^b J^{ad} \pa J^{bc} J^{cN'} J^{dN'}(w)+c_{70} K^a K^b
J^{ad} \pa J^{bc} \pa J^{cd}(w)+c_{71}K^a K^b J^{aN'} J^{cN'} J^{cN'} \pa J^{bN'}(w)
\nonu \\
&&+c_{72} K^a K^b J^{cN'} J^{aN'} J^{cN'} \pa J^{bN'}(w) +c_{73} K^a K^b J^{cN'} J^{cN'} J^{aN'} \pa J^{bN'}(w)
\nonu \\
&&+c_{74}K^a K^b K^c \pa K^d J^{ab} J^{cN'} J^{dN'}(w) +c_{77} K^a K^b K^c \pa^2 K^d J^{ab} J^{cd}(w)
+c_{78}K^a K^b \pa^3 K^c K^c J^{ab}(w)
\nonu \\
&&+c_{79}K^a K^b \pa^3 J^{aN'} J^{bN'}(w) +c_{80}K^a K^b \pa^2 K^c \pa K^c J^{ab}(w)+c_{81} K^a K^b \pa^2 K^c K^c \pa J^{ab}(w)
\nonu \\
&&+c_{82} K^a K^b \pa^2 J^{bc} J^{aN'} J^{cN'}(w)+c_{83} K^a K^b \pa^2 J^{ac} \pa J^{bc}(w) +c_{84}K^a K^b \pa^2 J^{ab} J^{cN'} J^{cN'}(w)
\nonu \\
&&+c_{85} K^a K^b \pa K^c K^c \pa^2 J^{ab}(w)+c_{86}K^a K^b \pa K^c K^c \pa J^{aN'} J^{bN'}(w) +c_{87} K^a K^b \pa K^c K^c \pa J^{ad} J^{bd}(w)
\nonu \\
&&+c_{89} K^a K^b \pa K^c K^d \pa J^{ad} J^{bc}(w) +c_{90} K^a K^b \pa K^c K^c J^{ad} J^{be} J^{ed}(w)
\nonu \\
&&+c_{91} K^a K^b \pa K^c K^e J^{ad} J^{bc} J^{ed}(w) +c_{92} K^a K^b \pa K^c K^c J^{ab} J^{dN'} J^{dN'}(w)
\nonu \\
&&+c_{94}K^a K^b \pa K^c K^c J^{ab} J^{de} J^{de}(w) +c_{95} K^a K^b \pa K^c K^e J^{ab} J^{dc} J^{de}(w)
\nonu \\
&&+c_{97}K^a K^b \pa K^c \pa K^d J^{ab} J^{cd}(w)+c_{99} K^a K^b \pa J^{bc} J^{dc} J^{aN'} J^{dN'}(w)+c_{100} K^a K^b \pa J^{ac} \pa J^{bN'} J^{cN'}(w)
\nonu \\
&&+c_{101} K^a K^b \pa J^{ac} \pa J^{bd} J^{dc}(w) +c_{103} K^a \pa K^c J^{ab} J^{bd} J^{ce} J^{de}(w)+c_{104}K^a \pa K^c J^{ab} J^{bd} J^{de} J^{ce}(w)
\nonu \\
&&+c_{105}K^a \pa K^b \pa^2 K^c K^c J^{ab}(w)+c_{106}K^a \pa K^b \pa^2 J^{aN'} J^{bN'}(w)+c_{107} K^a \pa K^b \pa^2 J^{ac} J^{bc}(w)
\nonu \\
&&+c_{108}K^a \pa K^b \pa K^c K^c \pa J^{ab}(w) +c_{109} K^a \pa K^b \pa K^c K^c J^{aN'} J^{bN'}(w) +c_{110} K^a \pa K^b \pa K^c K^c J^{ad} J^{bd}(w)
\nonu \\
&&+ c_{111}K^a \pa K^b \pa K^c K^d J^{ac} J^{bd}(w) +c_{112} K^a \pa K^b \pa J^{aN'} \pa J^{bN'}(w) +c_{114} K^a \pa K^b \pa J^{ac} \pa J^{bc}(w)
\nonu \\
&&+ c_{115} K^a \pa K^b \pa J^{ac} J^{bN'} J^{cN'}(w) +c_{116}K^a \pa K^b \pa J^{ac} J^{bd} J^{dc}(w) +c_{118} K^a \pa K^b \pa J^{ab} J^{cN'} J^{cN'}(w)
\nonu \\
&&+c_{119}K^a \pa K^b \pa J^{cb} J^{aN'} J^{cN'}(w) +c_{120}K^a \pa K^b  J^{ac} J^{db} J^{dN'} J^{cN'}(w)\nonu \\
&& +c_{122} K^a \pa K^b J^{dc} J^{db} J^{aN'} J^{cN'}(w)+c_{123}K^a \pa K^b J^{ac} J^{dc} J^{dN'} J^{bN'}(w)\nonu \\
&&+c_{125} K^a \pa K^b J^{dc} J^{dc} J^{aN'} J^{bN'}(w) +c_{126} K^a \pa K^b \pa J^{ab} J^{cd} J^{cd}(w)+c_{127} K^a \pa K^b K^b \pa^2 K^{c} J^{ac}(w)
\nonu \\
&&+c_{128} K^a \pa K^b J^{aN'} \pa^2 J^{bN'}(w)+c_{129} K^a \pa K^b J^{aN'} J^{bN'} J^{cN'} J^{cN'}(w)+c_{130}K^a \pa K^b J^{bc} \pa J^{aN'} J^{cN'}(w)
\nonu \\
&&+c_{131}K^a \pa K^b J^{bc} J^{aN'} \pa J^{cN'}(w)+c_{132} K^a \pa K^b J^{ac} \pa J^{bN'} J^{cN'}(w)+c_{133} K^a \pa K^b J^{ac} J^{bN'} \pa J^{cN'}(w)
\nonu \\
&&+c_{134}K^a \pa K^b J^{ac} J^{bc} J^{dN'} J^{dN'}(w)+c_{135}K^a \pa K^b J^{ab} \pa J^{cN'} J^{cN'}(w) +c_{136} K^a \pa^4 K^b J^{ab}(w)
\nonu \\
&&+c_{137}K^a \pa^3 K^b \pa J^{ab}(w) +c_{138} K^a \pa^3 K^b J^{aN'} J^{bN'}(w) +c_{141}K^a \pa^3 K^b J^{ac} J^{cb}(w)
\nonu \\
&&+c_{142}K^a \pa^2 K^b \pa^2 J^{ab}(w) +c_{143}K^a \pa^2 K^b \pa J^{aN'} J^{bN'}(w) +c_{144}K^a \pa^2 K^b \pa J^{ac} J^{cb}(w)
\nonu \\
&&+c_{145}K^a \pa^2 K^b J^{aN'} \pa J^{bN'}(w)+c_{146} K^a \pa^2 K^b J^{cb} J^{aN'} J^{cN'}(w)+c_{147}K^a \pa^2 K^b J^{ab} J^{cN'} J^{cN'}(w)
\nonu \\
&&+c_{148} K^a \pa^2 K^b J^{ac} J^{bN'} J^{cN'}(w)+c_{149}K^a \pa^2 K^b J^{ab} J^{cd} J^{cd}(w)+c_{150}K^a \pa^2 K^b J^{ad} J^{cb} J^{cd}(w)
\nonu \\
&&+c_{151}\pa^5 K^a K^a(w) +c_{152}\pa^4 K^a \pa K^a +c_{153} \pa^4 J^{aN'} J^{aN'}(w)+c_{154} \pa^3 K^a K^a \pa K^b K^b(w)
\nonu \\
&&+c_{155}\pa^3 K^a K^a J^{bN'} J^{bN'}(w) +c_{156} \pa^3 K^a \pa^2 K^a(w)+c_{157}\pa^3 K^a \pa K^b J^{ab}(w)+c_{158}\pa^3 K^a K^a J^{bc} J^{bc}(w)
\nonu \\
&&+c_{159}\pa^3 J^{aN'} \pa J^{aN'}(w) +c_{160}\pa^2 K^a \pa K^a J^{bN'} J^{bN'}(w) +c_{161}\pa^2 K^a \pa K^b J^{aN'} J^{bN'}(w)
\nonu \\
&&+c_{163}\pa^2 K^a \pa K^a J^{bc} J^{bc}(w)+c_{164}\pa^2 K^a \pa K^b J^{ac} J^{bc}(w)+c_{165}\pa^2 K^a \pa K^b J^{bc} J^{ac}(w)
\nonu \\
&&+c_{166}\pa^2 K^a \pa K^a \pa K^b K^b(w) +c_{167}\pa^2 K^a K^a \pa^2 K^b K^b(w) +c_{168}\pa^2 K^a K^a \pa J^{bN'} J^{bN'}(w)
\nonu \\
&&+c_{169}\pa^2 K^a \pa^2 K^b J^{ab}(w) +c_{170}\pa K^a K^a J^{bc} \pa J^{bN'} J^{cN'}(w) +c_{171}\pa K^a K^a J^{bc} J^{dc} J^{bN'} J^{dN'}(w)
\nonu \\
&&+c_{173}\pa K^a K^a J^{bc} J^{bc} J^{dN'} J^{dN'}(w) +c_{174}\pa K^a K^a J^{bc} J^{bc} J^{de} J^{de}(w)
\nonu \\
&&+c_{175} \pa K^a \pa K^b \pa^2 J^{ab}(w) +c_{176} \pa K^a \pa K^b \pa J^{aN'} J^{bN'}(w) +c_{178} \pa K^a \pa K^b J^{ab} J^{cN'} J^{cN'}(w)
\nonu \\
&&+c_{179}\pa K^a \pa K^b J^{ac} J^{bN'} J^{cN'}(w)+c_{180}\pa K^a \pa K^c J^{ac} J^{bd} J^{bd}(w) +c_{181} \pa K^a \pa K^b J^{cb} J^{cN'} J^{aN'}(w)
\nonu \\
&&+c_{182} \pa J^{ab} J^{ab} \pa J^{cN'} J^{cN'}(w) +c_{183} \pa J^{ab} J^{aN'} \pa^2 J^{bN'}(w) +c_{184} \pa J^{ab} \pa J^{ab} J^{cN'} J^{cN'}(w)
\nonu \\
&&+c_{185} \pa^3 J^{ab} \pa J^{ab}(w) +c_{186} J^{ab} J^{ca} \pa J^{cN'} \pa J^{bN'}(w)+c_{187} \pa J^{aN'} J^{aN'} \pa J^{bN'} J^{bN'}(w)
\nonu \\
&&+c_{188} \pa^2 J^{ab} J^{aN'} \pa J^{bN'}(w) +c_{189} \pa^2 J^{ab} \pa^2 J^{ab}(w)+c_{190} \pa^2 J^{aN'} \pa^2 J^{aN'}(w)
\nonu \\
&&+c_{191} K^a K^b \pa J^{ab} \pa J^{cN'} J^{cN'}(w)+c_{192} K^a \pa K^b J^{ab} \pa J^{cd} J^{cd}(w) +c_{194}K^a \pa^2 K^b J^{ac} \pa J^{bc}(w)
\nonu \\
&&+c_{196}\pa K^a K^b J^{bc} J^{ac} J^{de} J^{de}(w)+c_{197} \pa K^a K^a J^{bc} J^{bd} J^{ce} J^{de}(w)+c_{198}\pa K^a K^b J^{ac} J^{bd} J^{ce} J^{de}(w)
\nonu \\
&&+c_{199} \pa K^a K^a J^{bN'} J^{bN'} J^{cN'} J^{cN'}(w)+c_{200} \pa K^a K^a \pa J^{bN'} \pa J^{bN'}(w)
\nonu \\
&&+c_{201} \pa K^a K^a \pa K^b K^b J^{cd} J^{cd}(w)+c_{202} \pa K^a K^a \pa K^b K^b J^{cN'} J^{cN'}(w)+c_{203}\pa K^a K^a \pa K^b \pa K^c J^{bc}(w)
\nonu \\
&&+c_{204} \pa K^a K^a \pa^2 J^{bc} J^{bc}(w) +c_{205} \pa K^a K^a \pa^2 J^{bN'} 
J^{bN'}(w)+c_{206} \pa K^a \pa K^b J^{ac} J^{bd} J^{cd}(w)
\nonu \\
&&+c_{207} \pa^2 K^a K^a \pa J^{bc} J^{bc}(w) +c_{209} J^{ab} J^{ac} J^{de} J^{de} J^{bN'} J^{cN'}(w)+c_{211} J^{ab} J^{cb} J^{de} J^{de} J^{cN'} J^{aN'}(w)
\nonu \\
&&+c_{212} J^{ab} J^{ab} J^{cd} J^{ed} J^{eN'} J^{cN'}(w)+c_{213} J^{ac} J^{bd} J^{aN'} J^{cN'} J^{bN'} J^{dN'}(w)
\nonu \\
&&+c_{217} K^a K^b K^c K^d J^{ae} J^{cf} J^{ef} J^{bd}(w)+c_{222} K^a K^b K^c K^d \pa J^{ac} \pa J^{bd}(w)
\nonu \\
&&+c_{224} K^a \pa K^b J^{ac} \pa^2 J^{cb}(w) +c_{225} K^a \pa K^b \pa^3 J^{ab}(w) +c_{227}\pa K^a K^a \pa J^{bc} \pa J^{bc}(w)
\nonu \\
&& + c_{228} K^a K^b K^c K^d J^{ad} \pa^2 J^{bc}(w) +
c_{229} K^a K^b K^c K^d J^{ec} \pa J^{ed} J^{ab}(w),
\label{O5O9p1}
\eea
where $ N' \equiv 2N+1$. The coefficients in $(\ref{O5O9p1})$ are presented in
(\ref{O5O9p1coeff}) of Appendix $E$.
The spin-$6$ current
$O_6(z)$ is given by
(\ref{O6}), (\ref{Coeffcoeff}) and (\ref{O5O9p1}).
So far, the spin-$4$ and the spin-$5$ currents
in different coset model were constructed in \cite{Ahn1111,AK1308}.
One can analyze the three-point functions of the spin-$6$ current
with scalars.
The OPE for general $N$ is obtained from (\ref{O5O9}) with
(\ref{Coeffcoeff}).
The fusion rule is summarized by
$[O_{\frac{5}{2}}][O_{\frac{9}{2}}]= [I] + [O_2] +[{O}_4]
+[O_{4'}] +[O_6]$.

The ${\cal N}=1$ fusion rule is summarized by
$[\hat{O}_2][\hat{O}_{4}]=
[\hat{O}_2] +[\hat{O}_{\frac{7}{2}}] +[\hat{O}_4] +[\hat{O}_{\frac{11}{2}}]$.
The explicit OPE is given by (\ref{singleO2O4}).
As described before,
among four OPEs between this ${\cal N}=1$ supermultiplet,
half of them are quite related to the others because
they have common field contents.
By rescaling the currents
as $O_{4'}(z) \rightarrow N_{4'} O_{4'}(z)$
and $O_{\frac{9}{2}}(z)
\rightarrow N_{\frac{9}{2}}
O_{\frac{9}{2}}(z)$, where
\bea
N_{4'}^2 &= &-189M (29 + 2 c) (6 + 5 c)^2 N (c - 6 N + 6 N^2)^6,\qquad
N_{\frac{9}{2}}^2 =-\frac{1}{9}N_{4'}^2, \nonu \\
\frac{1}{M} & \equiv
& c^6 (-3 + 2 c) (3 + c - 9 N) (1 + N) (c + 6 N)\nonu \\
& \times & (-3 c +   2 (3 + c) N) (2 c (-1 + N) + 3 (3 - 4 N) N) (c + 2 c N +   3 N (-3 + 2 N))\nonu \\
& \times & (-2268 N^3 (25 + 2 N (-25 + 6 N)) +
   5 c^3 (-2 + N) (-420 + N (437 + N (-12 + 7 N)))\nonu \\
& +&9 c N^2 (6300 + N (-12320 + N (9961 + 5 N (-962 + 7 N))))\nonu \\
&-& 6 c^2 N (4326 + N (-7259 + N (3973 + 5 N (-62 + 7 N))))),
\nonu
\eea
the standard normalizations arise: $O_{4'}(z) \, O_{4'}(w)
\rightarrow
\frac{1}{(z-w)^8} \, \frac{c}{4} +\cdots$ and
 $O_{\frac{9}{2}}(z) \, O_{\frac{9}{2}} (w) \rightarrow
\frac{1}{(z-w)^9} \, \frac{2c}{9} +\cdots$.
We have checked this up to $N=5$.
Then all the previous OPEs can be rewritten in terms of these rescaled
currents.

\section{Conclusions and outlook }

We have found the first four higher spin supercurrents with spins
$(2', \frac{5}{2}), (\frac{7}{2},4), (4',\frac{9}{2})$ and $(\frac{11}{2},6)$
in (\ref{currents})
including the super stress tensor with spins $(\frac{3}{2},2)$ in terms of
the WZW currents in the coset model (\ref{coset1}).
Some of the OPEs between these supercurrents are determined.
In the right  hand side of these OPEs, one sees various kinds of
quasi-primary (and primary) fields
with given spins that can be written in terms of the
above higher spin currents
\footnote{
One might ask what is the spin dependence
of the maximal degree of the polynomials appearing in the structure constants 
which can be expressed as a ratio  of two polynomials in $N$.
For given higher spin current written in terms of WZW currents,
the $N$ dependence arises in many places. That is, the overall factor and 
the relative coefficient functions between various independent terms.
As one calculates the particular OPE between the higher spin currents with 
spins $s_1$ and $s_2$, these $N$ dpendences occur in each singular term of 
the OPE. Furthermore, 
each multiple product of WZW currents of spin $s_1$ and those of spin $s_2$ 
can produce the $N$ dependence also by contracting the group indices during the 
OPE calculation. 

For example, 
in the OPE of $O_2(z) \, O_2(w)$ given in (\ref{O2O2a}), the large $N$ behavior
of $c_{22}$
can be analyzed as follows. The maximal degrees of each polynomial in the 
numerator and denominator are given by $4$ and $4$ respectively from $c_{22}$.  The overall factor of $O_2(z)$ contributes to $\frac{1}{N^2}$ and therefore
by considering the other $O_2(w)$, the total contributions in the OPE 
are given by 
$\frac{1}{N^4}$. How does one obtain the extra $N^4$ behavior?
The $O_2(z)$ contains four independent terms. One realizes that the $N^4$ behavior arises in the OPE between the third term and itself (i.e. 
$K^a K^b J^{ab}(z) \,
K^{c} K^d J^{cd}(w)$) and in the OPE between
the last term and itself (i.e. 
$J^{a 2N+1} J^{a 2N+1}(z) \, J^{b 2N+1} J^{b 2N+1}(w)$).
The fourth-order pole terms of these OPEs behave as $N^2$ and the relative 
coefficient functions of third and fourth terms of $O_2(z)$ behave as 
$N$. Therefore, the total contribution is $N^2 \times N \times N = N^4$ 
as above. Note that 
the contribution from the OPE between the first term and itself in 
$O_2(z)$ is given by $N^3$ (the relative coefficient function in this case 
is a constant).

For the structure constants we have found in this paper,  the  
maximal degree of polynomial in the numerator  is the same as 
the one in the denominator. Let us denote the maximal degree of polynomial
of numerator (or denominator)
by $deg(N)$ and then one realizes that 
$deg(N) \leq s_1 +s_2$.  The $c$ dependent 
coefficients appearing in 
the numerator and denominator can be determined by $deg(N)+1$ linear 
equations which can be obtained from the expressions for 
lower $N=2, 3, \cdots, 
deg(N)+2$. Most of the structure constants have their factorized forms 
and therefore, we do not need all the above $deg(N)+1$ linear 
equations to determine the $c$-dependent coefficients. Of course, as the 
spins $s_1$ and $s_2$ increase, the $deg(N)$ becomes large and it will take 
too much time (by package) 
to obtain the complete OPEs for low several $N$ values. The spin $s_3$ of 
higher spin 
current appearing in the right hand side of above OPE is less than 
$s_1+s_2$: $s_3 < s_1+s_2$. 
There is no definite relation between the $s_3$ and $deg(N)$.
In som examples, $deg(N)$ is greater than $s_3$ but in other examples, 
$deg(N)$ is less than or equal to $s_3$. Therefore, the final $c$-dependent 
coefficient functions (i.e. structure constants of the OPEs) 
for general $N$ can be obtained.      }.
 

$\bullet$ So far, the level of the second numerator current of the
coset model (\ref{coset1}) is fixed by $1$.
It would be interesting to study the higher spin currents for
general levels $(k,l)$.
Or what happens when the level $k$ is replaced by $2N$ or $(2N+1)$ in the
coset model?

$\bullet$
According to recent work in \cite{GG1305},
the large ${\cal N}=4$ minimal model holography is an interesting
subject. For example, there exists a particular ${\cal N}=4$ coset theory
related to the orthogonal group as follows:
$W \times SU(2) \times U(1) =
\frac{SO(N+4)}{SO(N) \times SU(2)} \times U(1)$ where
$W$ is a Wolf space.
Simple computation for the central charge in this model
leads to
$c =\frac{6(k+1)(N+1)}{(k+N+2)}$ which is exactly the
same as the central charge
studied in \cite{GG1305}.
The immediate step is how to construct the large ${\cal N}=4$ superconformal
algebra in this particular coset theory.
For the Wolf space itself, the subgroup of $SO(N+4)$
is realized by $SO(N) \times SO(4) = SO(N) \times SU(2) \times SU(2)$.
One can study this coset theory
for fixed $N$ in order to see the structure of
an extended version of large ${\cal N}=4$ superconformal algebra.

$\bullet$
Furthermore, one of the Kazama-Suzuki models has the following coset model
$\frac{SO(N+2)}{SO(N) \times SO(2)}$ where the central charge
is given by $c =\frac{3Nk}{(N+k)}$.
It would be interesting to find the higher spin currents, along the line of
\cite{Ahn1206,Ahn1208}.
First of all, the ${\cal N}=2$ superconformal algebra should be realized
in this coset model from the ${\cal N}=2$ WZW currents with constraints.
After this is done, then the extension of
the ${\cal N}=2$ superconformal algebra can be obtained
by constructing the higher spin currents with spins
greater than $2$.

$\bullet$ In \cite{CHR1306},
the generalization of the coset (\ref{coset1}) is suggested and it has the
following coset
$\frac{SO(2N+M)_k \oplus SO(2NM)_1}{SO(2N)_{k+M} \oplus SO(M)_{k+2N}}$.
When $M=1$, this reduces to (\ref{coset1}).
When $M=2$, this coset looks like the Kazama-Suzuki model above
except the level $1$ factor in the numerator.
One can calculate the central charge and it is given by
$c=\frac{3kMN}{(-2+k+M+2N)}$. For the stringy limit where
$M, N,$ and $k$ are taken to be very large at the same time,
the central charge
$c$ goes to $N^2$ due to the extra degree of freedom by $M$.
It would be interesting to construct the higher spin
currents in this generalized coset model, along the line of
\cite{Ahn1211,Ahn1305}.

\vspace{.7cm}

\centerline{\bf Acknowledgments}

CA would like to thank Y. Hikida
for discussions.
This work was supported by the Mid-career Researcher Program through
the National Research Foundation of Korea (NRF) grant
funded by the Korean government (MEST) (No. 2012-045385/2013-056327).
CA acknowledges warm hospitality from
the School of  Liberal Arts (and Institute of Convergence Fundamental
Studies), Seoul National University of Science and Technology.

\newpage

\appendix

\renewcommand{\thesection}{\large \bf \mbox{Appendix~}\Alph{section}}
\renewcommand{\theequation}{\Alph{section}\mbox{.}\arabic{equation}}

\section{The coefficient functions
 in (\ref{O5O5p1}) related to the spin-$4$ current
}

The equations for coefficients in (\ref{O5O5p1}) for $N=2$ are given by
\bea
&&\frac{36 k^2}{(2+k)^2 (3+k)^3}+c_1=0,\qquad
\frac{72 k^2}{(2+k)^2 (3+k)^3}-2 c_{14}=0,\qquad
\frac{72 k^2}{(2+k)^2 (3+k)^3}+2 c_{13}=0,\nonu \\
&&\frac{36 (-2+k) k^2}{(2+k)^2 (3+k)^3}+c_7 + c_{14}=0,\qquad
\frac{36 k^3}{(2+k)^2 (3+k)^3}+c_{18}=0,\nonu \\
&&-\frac{6 k^2 (7+k (13+4 k)}{(2+k)^2 (3+k)^3}+c_7-2 c_8-c_{14}+\frac{c_{16}}{2}=0, \qquad
\frac{6 k^2 (1+2 k) (7+5 k)}{(2+k)^2 (3+k)^3)}+c_{11}=0,\nonu \\
&&\frac{36 k^2 (4+3 k}{(2+k)^2 (3+k)^3}+c_9=0,\qquad
\frac{36 k^2}{(2+k)^2 (3+k)^2}+c_{17}=0,\qquad
\frac{24 k^2 (1+2 k)^2}{(2+k)^2 (3+k)^3}-2 c_{19}=0,\nonu \\
&&\frac{72 k^2}{(2+k)^2 (3+k)^3}+2 c_{15}=0, \qquad
\frac{6 k^3 (1+2 k)}{(2+k) (3+k)^3)}+c_{12}=0,\qquad
\frac{6 k^3 (1+2 k)^2}{(2+k)^2 (3+k)^3}-c_{20}=0,\nonu \\
&&\frac{6 k^2 (5+k) (11+4 k)}{(2+k)^2 (3+k)^3}-c_6=0,\qquad
\frac{36 k^2}{(2+k)^2 (3+k)^2}+c_{16}=0,\nonu \\
&&54 (2 c_9 - c_{10} + c_{17}) + k (81 (2 c_9 -c_{10} + c_{17}) +    k (84 + 90 c_9 - 45 c_{10} + 45 c_{17}\nonu \\
&&+  k (-12 + 2 (11 + k) c_{9} - (11 + k) c_{10} + (11 + k)c_{17})))=0,\nonu \\
&&\frac{12 k^2 (5+k)^2}{(2+k)^2 (3+k)^3}+c-3=0,\qquad
\frac{72 k^2 (4+k)}{(2+k)^2 (3+k)^3}+c_2=0,\nonu \\
&&-108 (2 c_1-c_2+4 c_5)+k (k (-2 (171+k (67+k (13+k)))c_1+9 (-24+19 c_2-76 c_5)\nonu \\
&&+k (67+k (13+k)) (c_2-4 c_5))-216 (2 c_1-c_2+4 c_5))=0,\nonu \\
&&\frac{72 k^2}{(2+k)^2 (3+k)^3}+2 (c_1+c_4)=0.
\label{n2expres}
\eea
One performs similar analysis for $N=3,4,5$.
Then the  coefficients  in (\ref{O5O5p1}) for general $N$ (in terms of
$N$ and $c$)   are given by
\bea
c_1 &=& 6 c^2 D(c - 3 N)^3,
\qquad
c_2  = 12 c^2 D(c - 3 N)^2 (c (-1 + N) + 3 (2 - 3 N) N),
\nonu \\
c_3 & = & 2c^2D(c - 3 N) (2 c (-1 + N) + 3 (3 - 4 N) N)^2,
\qquad c_4 =0,
\nonu \\
c_5  & = & 3 c^2 D(c - 6 N) (c - 3 N)^2 (-1 + 2 N),
\nonu \\
c_6 & = & -c^2 D(c - 3 N) (2 c (-1 + N) + 3 (3 - 4 N) N) (3 (9 - 10 N) N +
   c (-5 + 2 N)),
\nonu \\
c_7  & = & -6 c^2D (-3 + 2 c) (c - 3 N)^2 N,
\nonu \\
c_8  & = & -c^2 D(c - 3 N) (-5 c^2 + (45 - 2 c) c N + 4 (-9 + c) (3 + c) N^2 +
   6 (27 - 2 c) N^3 - 72 N^4),
\nonu \\
c_9  & = & -6 c^2 D(c - 3 N)^2 (c + 2 c N + 12 (-1 + N) N),
\nonu \\
c_{10}  & = & -2 c^2 D (c - 3 N) (6 c (-1 + N) N (-9 + 2 N)
\nonu \\
& +&   4 c^2 (-1 + N) (1 + 2 N) - 9 N^2 (15 + 2 N (-15 + 8 N))),
\nonu \\
c_{11}  & = &c^2 D(c - 3 N) (c + 2 c N + 3 N (-3 + 2 N)) (2 c (2 + N) +
   3 N (-9 + 8 N)),
\nonu \\
c_{12}  & = & -c^3 D(-1 + 2 N) (c + 6 (-1 + N) N) (c + 2 c N + 3 N (-3 + 2 N)),
\nonu \\
c_{13} & = & 6 c^2 D (c - 3 N)^3,
\qquad
c_{14} = -6 c^2D (c - 3 N)^3,
\qquad
c_{15}  = 6 c^2 D(c - 3 N)^3,
\nonu \\
c_{16}  & = & -18 c^2 D (c - 3 N)^2 N (-1 + 2 N),
\qquad
c_{17} = -18 c^2 D(c - 3 N)^2 N (-1 + 2 N),
\nonu \\
c_{18}  & = & -6 c^3 D(c - 3 N)^2 (-1 + 2 N),
\qquad
c_{19}  = -2 c^2 D (c - 3 N) (c + 2 c N + 3 N (-3 + 2 N))^2,
\nonu \\
c_{20}  & = & c^3 D(-1 + 2 N) (c + 2 c N + 3 N (-3 + 2 N))^2,
\nonu \\
D & \equiv & \frac{1}{18 N (-1 + 2 N) (c + 6 (-1 + N) N)^2}.
\label{O5O5p1coeff}
\eea

\section{The coefficient functions
 in (\ref{O2O7p1}) associated with the spin-$\frac{9}{2}$ current
}

After the analysis for $N=3,4,5$ in (\ref{O2O7p1}),
the coefficients in (\ref{O2O7p1}) for general $N$ are given by
\bea
c_1 &=& D (c-3 N)^2 (18 (3-2 c) c^2+5 c (144+(159-8 c) c) N-3 (1350+c (9+8 c) (81+10 c)) N^2\nonu \\
&+& 2 (2160+c (4401+16 c (27+2 c))) N^3-24 (171+c (123+4 c)) N^4), \nonu \\
c_2  & = & -9D(21+4 c) (c-3 N)^3 (-1+N) (-6 N^2+c (-1+4 N)),
\nonu \\
c_3 & = & D(c-3 N)^2 (-54 N^2 (-33-46 N+8 N^2)+4 c^3 (9+2 N (20+N (-15+22 N)))\nonu \\
&-& 9 c N (-4+N (267+4 N (9+47 N)))-3 c^2 (18+N (7+4 N (39+N (-147+52 N))))),
\nonu \\
c_4  & = & 3D (21+4 c) (c-3 N)^2 (-1+N) (c-4 c N+6 N^2) (-2 c (-1+N)+3 N (-3+4 N)),
\nonu \\
c_5 & = &2D (c-3 N)^2 (-1+N) (108 c^2 (3+c)-2 c (360+c (681+34 c)) N \nonu \\
&-& 3 (-3+2 c) (-54+c (111+40 c)) N^2+ 2 (3+4 c) (414+c (339+22 c)) N^3\nonu \\
&-& 24 (207+8 c (33+7 c)) N^4),
\nonu \\
c_6  & = & 18 D(3+c-9 N) (c-3 N)^2 (-1+N) (-3 c+2 (3+c) N) (2 c (-1+N)+3 (3-4 N) N),
\nonu \\
c_7  & = & -D(3+c-9 N) (c-3 N) (-1+N) (-3 c+2 (3+c) N) (2 c (-1+N)\nonu \\
&+& 3 (3-4 N) N) (3 N (11+2 N)+c (-33+4 N (5+11 N))),
\nonu \\
c_8  & = &-12D (3+c-9 N) (c-3 N) (-1+N) (-3 c+2 (3+c) N) (2 c (-1+N)+3 (3-4 N) N)^2,
\nonu \\
c_9  & = & -18 D(-3+2 c) (c-3 N)^2 (-1+N) (1+N) (c+6 N) (c+2 c N+3 N (-3+2 N)),
\nonu \\
c_{10}  & = &-2D (c-3 N) (2 c^4 (18+(-1+N) N (55+4 N (-3+11 N)))\nonu \\
&+& 162 N^3 (-9+N (81+4 N (-37+14 N))) + 27 c N^2 (117+N (-719 +2 N (585-286 N \nonu \\
&+&  76 N^2)))-3 c^3 (-36+N (319+N (-763+2 N (395+2 N (-89+22 N))))) \nonu \\
&+& 9 c^2 N (-110+N (701+N (-1307+4 N (283+N (-163+28 N)))))),
\nonu \\
c_{11}  & = & -6D (3+c-9 N) (c-3 N) (-1+N) (-3 c+2 (3+c) N) (2 c (-1+N)+3 (3-4 N) N)^2,
\nonu \\
c_{12} & = & 6 D(-3+2 c) (c-3 N) (-1+N) (1+N) (c+6 N) (2 c (-1+N)\nonu \\
&+&  3(3-4 N) N) (c+2 c N+3 N (-3+2 N)),
\nonu \\
c_{13}  & = &6D (-3+2 c) (c-3 N) (-1+N) (1+N) (c+6 N) (c+2 c N+3 N (-3+2 N))^2,
\nonu \\
c_{14}  & = &6D (-3+2 c) (c-3 N) (-1+N) (1+N) (c+6 N) (c+2 c N+3 N (-3+2 N))^2,
\nonu \\
c_{15}  & = &-12D (-3+2 c) (c-3 N) (-1+N) (1+N) (c+6 N) (c+2 c N+3 N (-3+2 N))^2,
\nonu \\
c_{16}  & = & 2D_3 (c-3 N) (1+N) (2 c^4 N (-5+2 N) (-19+22 N)+324 N^3 (-12+N (-7+24 N))\nonu \\
&+& 27 c N^2 (1+N (511-682 N+96 N^2))-3 c^3 (54+N (-239+4 N (117+2 N (-45+7 N))))\nonu \\
&-& 18 c^2 N (-38+N (279+8 N (-30+N (-13+9 N))))),
\nonu \\
c_{17}  & = & 6D (c-3 N)^2 (-1+N) (2 c^3 (7+11 N (-1+2 N))+54 N^2 (9+2 N (-9+11 N))\nonu \\
&+& 3 c^2 (11+N (-97+4 (40-11 N) N))+9 c N (-18+N (89+2 N (-97+28 N)))),
\nonu \\
c_{18}  & = & -2D (c-3 N) (-1+N) (324 N^3 (12+N (-65+74 N+4 N^2))+ 4 c^4 (-18+N (43+N\nonu \\
&-& 76 N^2+44 N^3))+54 c N^2 (-95+N (506+N (-577+2 N (-47+80 N))))\nonu \\
&+& 9 c^2 N (190+N (-1087+4 N (304+N (139+4 N (-61+10 N)))))\nonu \\
&-&24 c^3 (9+N (-70+N (106+N (29+3 N (-49+22 N)))))),
\nonu \\
c_{19}  & = & -18D (-3+2 c) (c-3 N) (-1+N) (1+N) (c+6 N) \nonu \\
&\times &(-3 c+2 (3+c) N) (c+2 c N+3 N (-3+2 N)),
\nonu \\
c_{20} &=& -6D (-3+2 c) (-1+N) (1+N) (c+6 N) (-3 c+2 (3+c) N)\nonu \\
& \times\ & (2 c (-1+N)+3 (3-4 N) N) (c+2 c N+3 N (-3+2 N)),
\nonu \\
c_{22} &=& -6D (-3+2 c) (-1+N) (1+N) (c+6 N) (-3 c+2 (3+c) N) (c+2 c N+3 N (-3+2 N))^2,
\nonu \\
c_{23} &=& 2 c D(3+c-9 N) (-1+N) (1+N) (-1+2 N) (-3 c+2 (3+c) N)\nonu \\
&\times &  (2 c (-1+N)+3 (3-4 N) N)^2, \nonu \\
c_{24} &=& -3D (21+4 c) (c-3 N)^2 (-1+N) (c-4 c N+6 N^2) (c+2 c N+3 N (-3+2 N)),\nonu \\
c_{25} &=& 2D (c-3 N)^2 (-1+N) (-54 N^2 (-9+4 N (29+5 N))\nonu \\
&+& 4 c^3 (-9+2 N (-22+N (-3+22 N)))+9 c N (10+N (273+4 N (-18+67 N)))\nonu \\
&+& c^2 (54+3 N (-13+4 N (-30+N (3+68 N))))),
\nonu \\
c_{26} &=& -4D (c-3 N) (-1+N) (1+N) (-3 c+2 (3+c) N) (2 c^3 (-5+11 N (-1+2 N))\nonu \\
&+& 54 N^2 (-3+4 N (-3+7 N))+9 c N (24+N (-103+2 (107-92 N) N))\nonu \\
&-& 3 c^2 (4+N (-35+8 N+44 N^2))),\nonu \\
c_{27} &=& -6D (3+c-9 N) (c-3 N) (-1+N) (-3 c+2 (3+c) N) (2 c (-1+N)\nonu \\
&+& 3 (3-4 N) N) (c+2 c N+3 N (-3+2 N)),\nonu \\
c_{28} &=& -2D (c-3 N) (-1+N) (324 N^3 (12+N (23-46 N+48 N^2))\nonu \\
&-& 54 c N^2 (7+N (198+N (-163+36 N (5+N))))\nonu \\
&+& 4 c^4 (-3+N (-9+N (-47+4 N (-3+11 N))))\nonu \\
&-& 27 c^2 N (14+N (-71+4 N (-8+N (-5+8 N (-3+2 N)))))\nonu \\
&+& 3 c^3 (6+N (75+2 N (-3+4 N (-6+11 N (-1+2 N)))))),\nonu \\
c_{29} &=& -6 D(-3+2 c) (-1+N) (1+N) (c+6 N) (-3 c+2 (3+c) N) (c+2 c N+3 N (-3+2 N))^2,\nonu \\
c_{30} &=& D(3+c-9 N) (-1+N) (-1+2 N) (-3 c+2 (3+c) N) (2 c (-1+N)\nonu \\
&+& 3 (3-4 N) N) (-45 N^2 (1+2 N)+6 c N (-7+5 N (1+2 N))+c^2 (3+4 N (4+3 N))),\nonu \\
c_{31} &=& -D(-3+2 c) (c-3 N) (-1+N) (1+N) (c+6 N) (1+22 N)\nonu \\
&\times &  (-3 c+2 (3+c) N) (c+2 c N+3 N (-3+2 N)),\nonu \\
c_{32} &=&D (-3+2 c) (-1+N) (1+N) (-1+2 N) (c+6 N) (-3 c+2 (3+c) N) \nonu \\
& \times & (c+2 c N+3 N (-3+2 N)) (3 (7-16 N) N+c (-2+6 N)),\nonu \\
c_{33} &=& D(-3+2 c) (-1+N) (1+N) (-1+2 N) (c+6 N)\nonu \\
&\times & (-3 c+2 (3+c) N) (c+2 c N+3 N (-3+2 N))^2,\nonu \\
D & \equiv & \frac{i c^3\sqrt{3 c - 9 N}}{18(-1 + N)(21 + 4 c) (6 + 5 c) N (-1 + 2 N) (c + 6 (-1 + N) N)^3 \sqrt{N - 2 N^2}}.
\label{O2O7p1coeff}
\eea

\section{The coefficient functions
 in (\ref{O2O4p2}) corresponding to the spin-$4'$ current
}

After analyzing for fixed $N=3,4,5$, one obtains that
the coefficients in (\ref{O2O4p2}) for general $N$ are
given by
\bea
c_1 &=& -27 c^3 D (21+4 c) (c-3 N)^4 (-6 N^2+c (-1+4 N)), \nonu \\
c_2  & = & 18 c^3 D (c-3 N)^3 (8 c^3 (5-9 N+6 N^2)+54 N^2 (11-64 N+86 N^2)\nonu \\
&+&3 c^2  (43+N (-259+2 (179-72 N) N))+9 c N (-64+N (359+2 N (-259+80 N)))),
\nonu \\
c_3 & = &18 c^3 D (21+4 c) (6+5 c) (c-3 N)^4 N (-1+2 N),
\nonu \\
c_4  & = & -12 c^3 D (21+4 c) (c-3 N)^3 N (-1+2 N)\nonu \\
&\times &  (20 c^2 (-1+N)+18 (6-17 N) N-3 c (17-74 N+40 N^2)),
\nonu \\
c_5 & = &-12 c^3 D (3+c-9 N) (c-3 N)^2 (1+N) (432 (9-14 N) N^3+32 c^3 (-1+N)^2 (-9+10 N)\nonu \\
&-& 36 c N^2 (157+N (-377+222 N))+3 c^2 N (803-4 N (596+5 N (-113+32 N)))),
\nonu \\
c_6  & = &-72 c^3 D (3+c-9 N) (c-3 N)^2 (-3 c+2 (3+c) N) (2 c (-1+N)+3 (3-4 N) N)^2,
\nonu \\
c_7  & = & -6 c^3  D (c-3 N)^2 N (-1+2 N) (4 c^4 (-7+N (-19+8 N))+162 N^2 (15+N (-51+88 N))\nonu \\
&+& 6 c^3 (17+N (-46+(87-16 N) N))+27 c N (-30+N (233+N (-659+136 N)))\nonu \\
&-& 9 c^2 (-44+N (307+N (-615+14 N (-1+8 N))))),
\nonu \\
c_8  & = &-12 c^3  D (3+c-9 N) (c-3 N)^2 (1+N) (108 (9-14 N) N^3-9 c N^2 (157+133 N (-3+2 N))\nonu \\
&+& 3 c^2 N (220+N (-695+6 (119-36 N) N))+4 c^3 (-18+N (52+N (-49+18 N)))),
\nonu \\
c_9  & = & 24 c^3  D (3+c-9 N) (c-3 N) (-1+2 N) (-3 c+2 (3+c) N)\nonu \\
&\times &  (2 c (-1+N)+3 (3-4 N) N)^2 (3 N+c (3+4 N)),
\nonu \\
c_{10}  & = &-18 c^3 D  (c-3 N)^3 (8 c^3 (2+3 N+6 N^2)+54 N^2 (11+2 N (10+N))\nonu \\
&+& 3 c^2 (1+N (-43+70 N+48 N^2))+9 c N (20+N (-145-86 N+64 N^2))),
\nonu \\
c_{11} &=& -12 c^3 D  (c-3 N)^2 (-6 N^2+c (-1+4 N)) (8 c^3 (1+N) (-3+2 N)\nonu \\
&-&27 N^2 (-41+82 N+60 N^2) - 3 c^2 (5+2 N) (3+N (-13+8 N))\nonu \\
&-& 9 c N (41+N (1+2 N) (-83+48 N)))\nonu \\
c_{12} & = & 12 c^3  D (-3+2 c) (21+4 c) (c-3 N)^3 N (-1+2 N) (5 c (1+2 N)+6 N (6+5 N)),
\nonu \\
c_{13}  & = &108 c^3  D (c-3 N)^2 N (-1+2 N) (-108 N^2 (2+N) (-2+5 N)+c^3 (1+27 (1-2 N) N)\nonu \\
&+& 18 c N (-8+N (19+2 N (7+N)))+3 c^2 (-5+N (14+N (-59+54 N)))),
\nonu \\
c_{14}  & = &-36 c^3 D  (3+c-9 N) (c-3 N) N (1+N) (-1+2 N) (c+2 c N-12 N^2)\nonu \\
& \times & (-9 c^2+2 c (24+11 c) N - 12 (12+c) N^2),
\nonu \\
c_{15}  & = &-72 c^3 D  (-3+2 c) (c-3 N)^2 (1+N) (c+6 N) (c+2 c N+3 N (-3+2 N))^2,
\nonu \\
c_{16}  & = & 144 c^3  D (-3+2 c) (c-3 N) (1+N) (c+6 N)\nonu \\
& \times &  (-3 c+2 (3+c) N) (c+2 c N+3 N (-3+2 N))^2,
\nonu \\
c_{17}  & = & 12 c^3 D (c-3 N)^2 (c+6 N) (-3 c+2 (3+c) N) (8 (2+5 N) (c+2 c N)^2\nonu \\
&-& 9 N (-36+N (-98+N (21+62 N)))+3 c (-8+N (-115+N (-335+2 N (-19+80 N))))),
\nonu \\
c_{18}  & = & -72 c^3 D (c-3 N) N (-1+2 N) (c+6 N) (-3 c+2 (3+c) N)\nonu \\
&\times &  (c (-1+N)+3 N^2) (c (-1+11 N)-3 (4+N (-8+9 N)))),
\nonu \\
c_{19}  & = & 12 c^3  D (-3+2 c) (c-3 N) (c+6 N) (-3 c+2 (3+c) N) (8 (c+2 c N)^2 (-2+N (-2+3 N))\nonu \\
&+& 9 N^2 (-108+N (2+N (155+2 N-64 N^2)))\nonu \\
&+& 3 c N (88+N (205+N (-111+2 N (-83+16 N))))),
\nonu \\
c_{20} &=& -48 c^3 D  (-3+2 c) (c-3 N) (1+N) (-1+2 N) (c+6 N)\nonu \\
&\times&  (-3 c+2 (3+c) N) (c+2 c N+3 N (-3+2 N))^2\nonu \\
& \times\ & (2 c (-1+N)+3 (3-4 N) N) (c+2 c N+3 N (-3+2 N)),
\nonu \\
c_{21} &=& -144 c^4 D  (3+c-9 N) (c-3 N) (1+N) (-1+2 N) \nonu \\
&\times &(-3 c+2 (3+c) N) (2 c (-1+N)+3 (3-4 N) N)^2,\nonu \\
c_{22} &=& -6 c^3 D (c-3 N)^2 (-1+2 N) (c+6 N) (-3 c+2 (3+c) N)\nonu \\
& \times &  (8 c^2 (2+N)+63 N (6+5 N)-3 c (-1+N (97+88 N))),
\nonu \\
c_{23} &=& 12 c^3 D (c-3 N) (-1+2 N) (c+6 N) (-3 c+2 (3+c) N) \nonu \\
& \times & (8 c^3 (1+N) (-1+4 N)-81 N^2 (2+N) (-2+5 N)+18 c N (-6+N (-2+N (43+24 N)))\nonu \\
&-& 3 c^2 (5+N (-48+N (83+112 N)))),\nonu \\
c_{24} &=& 12 c^3 D  (c-3 N) (1+N) (4 c^5 (-2+N) (1+2 N) (-3+4 N (-2+3 N))\nonu \\
&+& 1944 N^4 (18+N (-31+2 N (5+4 N)))\nonu \\
&+& 324 c N^3 (-43+N (-95+N (285+N (-201+38 N))))\nonu \\
&-& 18 c^3 N (-20+N (-10+N (387+2 N (-165+8 N (-9+5 N)))))\nonu \\
&+& 3 c^4 (-12+N (-130+N (-53+4 N (128+N (-95+16 N)))))\nonu \\
&-& 27 c^2 N^2 (-13+N (-602+N (117+4 N (254+N (-153+32 N)))))),\nonu \\
c_{25} &=& 12 c^3 D  (-3+2 c) (c-3 N)^2 (-54 N^3 (174+N (1+6 N) (-23+16 N))\nonu \\
&+& 6 c^2 N (-1+2 N) (3+2 N) (-16+N (-29+20 N))+9 c N^2 (107+8 N (78+(24-35 N) N))\nonu \\
&+&8 c^3 (1+2 N) (-3+N (-9+N (-11+10 N)))),
\nonu \\
c_{26} &=& -72 c^4 D\, D'  (-3+2 c) (-1+2 N),\qquad
c_{27} = 144 D\, D' c^3 (-3+2 c) (c-3 N) N,\nonu \\
c_{28} &=& -36 c^4 D\, D' (-3+2 c) (1-2 N)^2,\qquad
c_{29} = 4 c^4 D\, D' (-3+2 c) (1-2 N)^2,\nonu \\
D & \equiv &
\frac{1}{54 N^2(-1 + 2 N) (21 + 4 c) (6 + 5 c)  (-1 + 2 N)
(c + 6 (-1 + N) N)^3}, \nonu \\
D' & \equiv&
(1 + N) (c + 6 N) (6 N + c (-3 + 2 N)) (c + 2 c N + 3 N (-3 + 2 N))^2.
\label{O2O4p2coeff}
\eea

\section{The coefficient functions
 in (\ref{O2O9p1}) related to the spin-$\frac{11}{2}$ current
}

Similar analysis leads to the fact that
the coefficients in (\ref{O2O9p1}) for general $N$ are
given by
\bea
c_1 &=& -567D (29+2 c) (c-3 N)^4 N (-6 N^2+c (-1+4 N)), \nonu \\
c_2  & = & 54 D(c-3 N)^3 N (8 c^2 (49+c)-c (1071+2 c (791+60 c)) N\nonu \\
&+& 2 (-1134+c (4557+4 c (187+9 c))) N^2-6 (1071+2 c (791+60 c)) N^3+288 (49+c) N^4),
\nonu \\
c_3 & = &-378 D(29+2 c) (6+5 c) (c-3 N)^4 N^2 (-1+2 N),
\nonu \\
c_4  & = & 27D(c-3 N)^3 N (2 c^3 (-13+36 N (-1+2 N))+252 N^2 (-18+N (36+25 N))\nonu \\
&-&24 c N (-63+N (154+N (119+39 N)))+c^2 (175-4 N (119+2 N (-185+54 N)))),
\nonu \\
c_5 & = &-9 D (c-3 N)^2 N (2 c (-1+N)+3 (3-4 N) N) (4 c^3 (-3+N (-25+8 N))\nonu \\
&+&378 N^2 (-12+N (-5+36 N))-9 c N (35+48 N (-14+N (14+N)))\nonu \\
&+&6 c^2 (63-4 N (42+N (-22+25 N)))),
\nonu \\
c_6  & = &36 (c-3 N)^3 N (504 c^2 (3+c)+8 c (-462+c (-568+5 c)) N\nonu \\
&-&3 (2394+c (-1323+2182 c+248 c^2)) N^2+4 (5418+5 c (47+2 c) (39+8 c)) N^3\nonu \\
&-&12 (2163+25 c (103+10 c)) N^4),
\nonu \\
c_7  & = &-378 D(3+c-9 N) (c-3 N)^3 N (-3 c+2 (3+c) N) (2 c (4+N)+3 N (-13+16 N)),
\nonu \\
c_8  & = &-18D (c-3 N)^2 N (162 N^3 (-98+N (577+2 N (-843+748 N)))\nonu \\
&+&2 c^4 (-492+N (478+N (597+4 N (-187+45 N))))\nonu \\
&+&27 c N^2 (-1131+2 N (1605+4 N (897+N (-2353+1070 N))))\nonu \\
&-&3 c^3 (984+N (-5918+N (2725+2 N (4535+6 N (-743+190 N)))))\nonu \\
&+&9 c^2 N (2286+N (-9376+N (-1147+4 N (5563+N (-4079+840 N)))))),
\nonu \\
c_9  & = & 18D (29+2 c) (c-3 N)^3 N^2 (-1+2 N) (252 (3-8 N) N\nonu \\
&+&5 c^2 (-29+32 N)-6 c (56+5 N (-49+29 N))),
\nonu \\
c_{10}  & = &108 D  (3+c-9 N) (c-3 N)^2 N (-3 c+2 (3+c) N) (2 c (-1+N)\nonu \\
&+&3 (3-4 N) N) (2 c (8+3 N)+N (-49+96 N)),
\nonu \\
c_{11} &=& 72D (3+c-9 N) (c-3 N)^2 N (-3 c+2 (3+c) N) (2 c (-1+N)\nonu \\
&+&3 (3-4 N) N) (c (34+4 N)+3 N (-77+68 N)),\nonu \\
c_{12} & = & 108D (c-3 N)^3 N (23 c^2 (7+4 c)-c (1386+c (1603+36 c)) N\nonu \\
&+&(4158+c (5649+2210 c+72 c^2)) N^2-6 (1386+c (1603+36 c)) N^3+828 (7+4 c) N^4),
\nonu \\
c_{13}  & = &18D(c-3 N)^3 N (-504 N^2 (15+N (-59+13 N))\nonu \\
&+&6 c N (2044+N (-7056+(1699-1102 N) N))+2 c^3 (-504+N (145+54 N+320 N^2))\nonu \\
&+&c^2 (-3024+N (14789+2 N (-4593+2906 N+780 N^2)))),
\nonu \\
c_{15}  & = &-36 D(c-3 N)^2 N (-984 c^3 (3+c)+c^2 (22737+c (20329+1206 c)) N\nonu \\
&+&3 c (-20490+c (-39767+c (-6205+178 c))) N^2\nonu \\
&+&(71442+c (272709+c (108279-14 c (887+84 c)))) N^3\nonu \\
&+&12 (-3+c) (7410+c (9085+c (1537+30 c))) N^4\nonu \\
&-&12 (-14661+c (4038+c (6197+410 c))) N^5+72 (1203+5 c (127+34 c)) N^6),
\nonu \\
c_{16}  & = & -9D (c-3 N)^2 N (108 N^3 (-1974+N (5643-6806 N+4456 N^2))\nonu \\
&+&6 c^4 (384+N (-23+4 N (51-86 N+30 N^2)))\nonu \\
&+&18 c N^2 (9801+N (-36408+N (53177+2 N (-22567+6292 N))))\nonu \\
&+&3 c^2 N (-18370+N (92811+2 N (-61705+4 N (11891-4714 N+780 N^2))))\nonu \\
&+&c^3 (5532+N (-44257+2 N (20115-2 N (8585+12 N (-577+125 N)))))),
\nonu \\
c_{17}  & = & -180 D(29+2 c) (21+4 c) (c-3 N)^3 N^2 (-1+2 N) (-6 N^2+c (-1+4 N)),
\nonu \\
c_{18}  & = & 12D (3+c-9 N) (c-3 N) N (1+N) (-3 c+2 (3+c) N) (2 c (-1+N)\nonu \\
&+&3 (3-4 N) N) (86 c^2-15 c (31+10 c) N+18 (21+c (55+2 c)) N^2-216 (3+2 c) N^3),
\nonu \\
c_{19}  & = & -36D (3+c-9 N) N (-c+3 N) (-3 c+2 (3+c) N) (2 c (-2+N)\nonu \\
&+&3 (21-8 N) N) (2 c (-1+N)+3 (3-4 N) N)^2,
\nonu \\
c_{20} &=&-36 D(3+c-9 N) N (-c+3 N) (-3 c+2 (3+c) N) (2 c (-2+N)\nonu \\
&+&3 (21-8 N) N) (2 c (-1+N)+3 (3-4 N) N)^2,
\nonu \\
c_{21} &=& -378D (c-3 N)^3 N (1+N) (c+6 N) (3 (5-6 c) c+(-117+4 c (30+c)) N+18 (5-6 c) N^2),
\nonu \\
c_{23} &=& -108D (c-3 N)^2 N (1+N) (c+6 N) (4 c^3 (-2+N) (-11+6 N)\nonu \\
&-&9 N^2 (399-862 N+416 N^2)+3 c N (431+2 N (127+12 N (-79+44 N)))\nonu \\
&-&4 c^2 (26+N (237+N (-431+138 N)))),
\nonu \\
c_{25} &=& 6D (-3+2 c) (c-3 N) N (1+N) (c+6 N) (2 c (-1+N)+3 (3-4 N) N) (9 N^2 (525\nonu \\
&+&46 N (-25+12 N))+2 c^2 (69+N (-61+22 N))-3 c N (575+4 N (-230+61 N))),
\nonu \\
c_{27} &=&18D N (1+N) (-c+3 N) (c+6 N) (c+2 c N+3 N (-3+2 N))^2 (9 (21-34 N) N\nonu \\
&+&c^2 (6+4 N)+c (-51+36 N (1+N))),
\nonu \\
c_{31} &=&108 D (c-3 N)^2 N (1+N) (c+6 N) (c+2 c N+3 N (-3+2 N))\nonu \\
&\times& (3 N (-49+2 N)+2 c^2 (-19+6 N)+c (1+76 (4-3 N) N)),
\nonu \\
c_{32} &=&-36 D N (1+N) (-c+3 N) (c+6 N) (c+2 c N+3 N (-3+2 N))^2 \nonu \\
& \times & (9 (21-34 N) N+c^2 (6+4 N)+c (-51+36 N (1+N))),
\nonu \\
c_{33} &=&-12 D N (1+N) (-c+3 N) (c+6 N) (c+2 c N+3 N (-3+2 N))^2 \nonu \\
& \times &(4 c^2 (-15+4 N)+9 N (-35+18 N)+c (27+6 (73-60 N) N)),
\nonu \\
c_{34} &=&-6  D N (1+N) (-c+3 N) (c+6 N) (c+2 c N+3 N (-3+2 N))^2\nonu \\
& \times & (2 c^2 (-51+22 N)-9 N (7+66 N)+c (-99+12 (82-51 N) N)),
\nonu \\
c_{35} &=&-72 D N (1+N) (-c+3 N) (c+6 N) (c+2 c N+3 N (-3+2 N))^2\nonu \\
& \times & (9 (21-34 N) N+c^2 (6+4 N)+c (-51+36 N (1+N))),
\nonu \\
c_{36} &=&24 D (c-3 N) N (1+N) (c+6 N) (c+2 c N+3 N (-3+2 N))\nonu \\
& \times & (4 c^3 (39+8 (-4+N) N)-27 N^2 (357-854 N+480 N^2)\nonu \\
&-&6 c^2 (60+N (227+2 N (-229+64 N)))+9 c N (427+4 N (-64+N (-227+156 N)))),
\nonu \\
c_{37} &=&72 D (c-3 N)^2 N (1+N) (c+6 N) (c+2 c N+3 N (-3+2 N)) \nonu \\
& \times & (8 c^2 (-9+N)+9 N (-77+86 N)+c (129+6 (61-72 N) N)),
\nonu \\
c_{38} &=&12 D  (c-3 N) N (1+N) (c+6 N) (c+2 c N+3 N (-3+2 N)) (4 c^3 (60+N (-67+22 N))\nonu \\
&-&27 N^2 (273+2 N (-245+72 N))+9 c N (245+2 N (341+4 N (-257+120 N)))\nonu \\
&-&12 c^2 (9+N (257+N (-425+134 N)))),
\nonu \\
c_{39} &=&-36 D (c-3 N)^2 N (1+N) (c+6 N) (c+2 c N+3 N (-3+2 N))\nonu \\
& \times & (8 c^2 (-9+N)+9 N (-77+86 N)+c (129+6 (61-72 N) N)),
\nonu \\
c_{42} &=&-36 D (c-3 N)^2 N (504 c^3 (3+c)-c^2 (15771+13 c (847+10 c)) N\nonu \\
&+&3 c (16422+c (27620+c (3593+2 c))) N^2\nonu \\
&-&(48762+c (227997+c (101709+170 c (49+4 c)))) N^3\nonu \\
&+&6 (36036+c (44577+2 c (6320+c (557+30 c)))) N^4\nonu \\
&-&12 (19467+c (13098+125 c (17+2 c))) N^5+288 (210+(217-20 c) c) N^6),
\nonu \\
c_{43} &=&54 D (3+c-9 N) (c-3 N)^2 N (-3 c+2 (3+c) N) (c^2 (-8+6 N (9+4 N))\nonu \\
&-&6 N^2 (-126+N (125+48 N))+c N (-125+4 N (-67+81 N))),
\nonu \\
c_{44} &=&-18 D (c-3 N)^2 N (1872 c^3 (3+c)-c^2 (47424+c (37951+1558 c)) N\nonu \\
&+&2 c (62820+c (109767+c (17813+14 c))) N^2\nonu \\
&-&2 (37422+c (266247+2 c (35124+c (3017+158 c)))) N^3\nonu \\
&+&8 (42633+c (36792+c (-3081+c (-167+70 c)))) N^4\nonu \\
&+&144 (-1251+c (1326+5 c (35+2 c))) N^5-144 (942+c (533+50 c)) N^6),
\nonu \\
c_{45} &=&6 D (c-3 N) N (-324 c N^3 (1608+N (-10316+N (17687-4518 N-9268 N^2+4248 N^3)))\nonu \\
&-&972 N^4 (-294+N (2040+N (-3707+2 N (725+396 N))))\nonu \\
&+&c^5 (912+2 N (-1174+N (415+4 N (388+N (-409+108 N)))))\nonu \\
&-&27 c^2 N^2 (-8639+2 N (27044+N (-40858+N (-3123+4 N (9903+N (-4495+72 N))))))\nonu \\
&+&3 c^4 (912+N (-8842+N (15705+2 N (333-4 N (3006+N (-2633+702 N))))))\nonu \\
&+&9 c^3 N (-4586+N (31236+N (-45873+2 N (-6287+4 N (8373+N (-5471+1098 N))))))),
\nonu \\
c_{46} &=&24 D (3+c-9 N) N (-c+3 N) (-3 c+2 (3+c) N) (2 c (-1+N)\nonu \\
&+&3 (3-4 N) N)^2 (c (26+8 N)+3 N (-35+52 N)),
\nonu \\
c_{47} &=&54 D (c-3 N)^2 N (c-4 c N+6 N^2) (c^2 (65+14 c)-c (1161+c (283+6 c)) N\nonu \\
&+&(3483+c (2391+4 c (104+3 c))) N^2-6 (1161+c (283+6 c)) N^3+36 (65+14 c) N^4),
\nonu \\
c_{48} &=&-108 D (c-3 N)^2 N (-4 c^3 (16+13 c)+c^2 (633+2 c (545+22 c)) N\nonu \\
&-&c (4608+c (5193+2 c (965+24 c))) N^2+(12474+c (12231+4 c (3645+c (205+4 c)))) N^3\nonu \\
&-&6 (4608+c (5193+2 c (965+24 c))) N^4+36 (633+2 c (545+22 c)) N^5-864 (16+13 c) N^6),
\nonu \\
c_{49} &=&-378 D (c-3 N)^2 N (1+N) (c+6 N) (-3 c+2 (3+c) N) (3 (5-6 c) c\nonu \\
&+&(-117+4 c (30+c)) N+18 (5-6 c) N^2),
\nonu \\
c_{50} &=&-108 D N (1+N) (-c+3 N) (c+6 N) (-3 c+2 (3+c) N) (4 c^3 (-2+N) (-11+6 N)\nonu \\
&-&9 N^2 (399-862 N+416 N^2)+3 c N (431+2 N (127+12 N (-79+44 N)))\nonu \\
&-&4 c^2 (26+N (237+N (-431+138 N)))),
\nonu \\
c_{51} &=&-54 D N (1+N) (-c+3 N) (c+6 N) (-3 c+2 (3+c) N) (2 c^3 (25+6 N (-13+4 N))\nonu \\
&-& 18 N^2 (126+N (-379+206 N))+3 c N (379+8 N (-94+75 (-1+N) N))\nonu \\
&-&c^2 (103+4 N (75+26 N (-17+9 N)))),
\nonu \\
c_{52} &=&-6 D (-3+2 c) N (1+N) (c+6 N) (-3 c+2 (3+c) N) (2 c (-1+N)\nonu \\
&+&3 (3-4 N) N) (9 N^2 (525+46 N (-25+12 N))+2 c^2 (69+N (-61+22 N))\nonu \\
&-&3 c N (575+4 N (-230+61 N))),
\nonu \\
c_{53} &=&24 D N (1+N) (-c+3 N) (c+6 N) (c+2 c N+3 N (-3+2 N))^2 (4 c^2 (-15+4 N)\nonu \\
&+&9 N (-35+18 N)+c (27+6 (73-60 N) N)),
\nonu \\
c_{54} &=&36 D N (1+N) (-c+3 N) (c+6 N) (c+2 c N+3 N (-3+2 N))^2 (9 (21-34 N) N\nonu \\
&+&c^2 (6+4 N)+c (-51+36 N (1+N))),
\nonu \\
c_{55} &=&-9 D (c-3 N)^2 N (108 N^3 (1050+N (47-2978 N+2288 N^2))\nonu \\
&+&c^4 (864+2 N (-1145+4 N (-39-46 N+90 N^2)))\nonu \\
&+&18 c N^2 (-79+N (-12724+N (17249+2 N (-5367+404 N))))\nonu \\
&+&3 c^2 N (-6786+N (32719+2 N (-14725+4 N (4801-2654 N+300 N^2))))\nonu \\
&+&3 c^3 (1324+N (-5703+10 N (861+2 N (-443+4 N (37+7 N)))))),
\nonu \\
c_{56} &=&36 D (c-3 N) N (-324 N^4 (-378+N (1666+N (-2077+458 N+316 N^2)))\nonu \\
&+&2 c^5 (24+N (-53+N (13+2 N (43-86 N+36 N^2))))\nonu \\
&+&54 c N^3 (-1477+N (7198+N (-10925+8 N (821+N (-240+59 N)))))\nonu \\
&-&9 c^2 N^2 (-2552+N (16442+N (-30725+2 N (7517+2 N (3527-3530 N+552 N^2)))))\nonu \\
&+&3 c^3 N (-1061+N (9239+2 N (-8552+N (135+2 N (5345+6 N (-555+58 N))))))\nonu \\
&+&c^4 (144+N (-2087+N (3699+2 N (691-6 N (577+2 N (-233+72 N))))))),
\nonu \\
c_{57} &=&9 D (c-3 N)^2 N (540 N^3 (42+N (-1+6 N) (-67+26 N))\nonu \\
&+&c^4 (-924+2 N (-41+2 N (81+88 N)))\nonu \\
&+&c^3 (-2634+N (14555+8 N (-33+N (-616+117 N))))\nonu \\
&-&3 c^2 N (-4790+N (14051+8 N (1667+3 N (-482+251 N))))\nonu \\
&+&18 c N^2 (-1303+N (-580+N (12787+2 N (-2707+504 N))))),
\nonu \\
c_{58} &=&-3 D (3+c-9 N) (c-3 N) (-3 c+2 (3+c) N) (2 c (-1+N)\nonu \\
&+&3 (3-4 N) N)^2 (c (40+22 N)+3 N (-7+80 N)),
\nonu \\
c_{59} &=&-12 D (-3+2 c) N (1+N) (c+6 N) (-3 c+2 (3+c) N) \nonu \\
& \times & (c+2 c N+3 N (-3+2 N)) (2 c^2 (-2+N) (-23+8 N)\nonu \\
&-&3 c N (-7+2 N) (-51+64 N)+27 N^2 (105+8 N (-28+13 N))),
\nonu \\
c_{60} &=&-36 D  N (1+N) (-c+3 N) (c+6 N) (-3 c+2 (3+c) N) (c+2 c N+3 N (-3+2 N))\nonu \\
& \times &  (c^2 (-86+8 N)+9 N (-63+44 N)+c (87+6 (89-72 N) N)),
\nonu \\
c_{61} &=&6 D (-3+2 c) N (1+N) (c+6 N) (-3 c+2 (3+c) N)\nonu \\
& \times &  (c+2 c N+3 N (-3+2 N))^2 (c (23+6 N)+3 N (-35+18 N)),
\nonu \\
c_{62} &=&6 D (c-3 N) N (324 N^4 (-210+N (-41+2 N (1650+N (-2579+1032 N))))\nonu \\
&+& 4 c^5 (-252+N (335+N (-113+N (253-250 N+72 N^2))))\nonu \\
&+&54 c N^3 (1807+N (-2474+N (-17042+N (27203+2 N (-5353+732 N)))))\nonu \\
&-&9 c^2 N^2 (9886+N (-41452+N (23217+4 N (2116+N (-865+36 N (1+14 N))))))\nonu \\
&-&2 c^4 (1512+N (-11839+2 N (7862+N (-5173+N (5281+12 N (-256+27 N))))))\nonu \\
&+&3 c^3 N (9620+N (-53049+N (60735+2 N (-16767+4 N (3551+3 N (-793+186 N))))))),
\nonu \\
c_{63} &=&6 D (-3+2 c) N (1+N) (c+6 N) (-3 c+2 (3+c) N)\nonu \\
& \times &  (c+2 c N+3 N (-3+2 N))^2 (c (23+6 N)+3 N (-35+18 N)),
\nonu \\
c_{64} &=&- D (3+c-9 N) N (-1+2 N) (-3 c+2 (3+c) N) (2 c (-1+N)+3 (3-4 N) N)^2 \nonu \\
& \times & (-64 c^2+(435-34 c) c N+3 (-189+4 (-2+c) c) N^2+72 (3-2 c) N^3),
\nonu \\
c_{65} &=&9 D (c-3 N)^2 N (324 N^3 (-98+N (1719+4 N (-282+25 N)))\nonu \\
&+& 2 c^4 (-183+N (-1439+2 N (-423+10 N (43+18 N))))\nonu \\
&+&54 c N^2 (284+N (-6409+4 N (866+N (-1039+456 N))))\nonu \\
&+&3 c^3 (71+N (1797+8 N (105+2 N (-154+15 N (1+13 N)))))\nonu \\
&+&9 c^2 N (-491+N (4339+2 N (2835+2 N (-143+10 N (-65+72 N)))))),
\nonu \\
c_{66} &=&36 D (c-3 N)^3 N (126 N^2 (57+4 N (-101+37 N))\nonu \\
&+& 3 c N (-392+N (8421-8948 N+5492 N^2))+c^3 (-378+8 N (-58+N (33+40 N)))\nonu \\
&+&c^2 (-1+2 N) (-315+2 N (655+N (509+510 N)))),
\nonu \\
c_{67} &=&-108 D  (c-3 N)^2 N (-378 N^3 (41+2 N (32+3 N (-45+14 N)))\nonu \\
&+& 10 c^4 (32+3 N (7+N (-13+4 (-1+N) N)))\nonu \\
&-&9 c N^2 (-2170+N (1841+8 N (973+N (-1173+569 N))))\nonu \\
&-&c^3 (-224+N (2981+N (1373+90 N (-41+2 N (5+2 N)))))\nonu \\
&-&3 c^2 N (1673+N (-4445+N (-3427+12 N (588+5 N (-87+44 N)))))),
\nonu \\
c_{68} &=&54 D (c-3 N)^3 N (c^3 (-34+24 N (2+3 N))-126 N^2 (18+N (-123+62 N))\nonu \\
&+&c^2 (-217+2 N (553+8 N (-1+18 N)))+3 c N (861-2 N (2135+2 N (-553+102 N)))),
\nonu \\
c_{69} &=&108 D (3+c-9 N) (c-3 N)^2 N (-3 c+2 (3+c) N) (2 c^2 (3+2 N) (4+3 N)\nonu \\
&+&3 N^2 (399-734 N+288 N^2)+c N (-367+4 N (40+51 N))),
\nonu \\
c_{70} &=&-36 D (c-3 N)^2 N (378 N^3 (-129+2 N (-4+N (39+22 N)))\nonu \\
&+&c^4 (378+2 N (227+N (-237+20 N (-1+9 N))))\nonu \\
&-&9 c N^2 (-1414+N (-7147+2 N (2511+4 N (-215+334 N))))\nonu \\
&+&c^3 (-315+N (-1367+N (-2961+2 N (4075+6 N (-83+70 N)))))\nonu \\
&-&3 c^2 N (-427+N (5990+N (-7617+4 N (4670+N (-3403+960 N)))))),
\nonu \\
c_{71} &=&6 D  (c-3 N) N (-3 c+2 (3+c) N) (-81 N^3 (2667+N (-8301+4 N (1556+N (59+24 N))))\nonu \\
&+&2 c^4 (520+N (149+2 N (-467+2 N (-89+54 N))))\nonu \\
&+&3 c^3 (488+N (-6019+4 N (695+N (2161-8 N (41+54 N)))))\nonu \\
&+&27 c N^2 (5610+N (-22421+4 N (6233+N (239+16 N (-266+153 N)))))\nonu \\
&-&9 c^2 N (3317+N (-17773+2 N (8661+4 N (1447+N (-1901+306 N)))))),
\nonu \\
c_{72} &=&-18  D (c-3 N)^2 N (c-4 c N+6 N^2) (8 c^3 (6+N-2 N^2)\nonu \\
&+& 189 N^2 (99+2 N (-99+40 N))+6 c^2 (70+N (-91+82 N+8 N^2))\nonu \\
&+&9 c N (-693+4 N (252+N (-91+48 N)))),
\nonu \\
c_{73} &=&-24 D  (3+c-9 N) (c-3 N) N (-3 c+2 (3+c) N) (2 c (-1+N)\nonu \\
&+&3 (3-4 N) N) (9 N^2 (357-574 N+200 N^2)\nonu \\
&+&2 c^2 (25+8 N (3+N))+3 c N (-287+2 N (65+48 N))),
\nonu \\
c_{74} &=&-36 D N (1+N) (-c+3 N) (c+6 N) (-3 c+2 (3+c) N) ((87-86 c) c\nonu \\
&+& 9 (-77+4 c (16+c)) N+18 (43-38 c) N^2) (c+2 c N+3 N (-3+2 N)),
\nonu \\
c_{75} &=&-36D  (c-3 N)^2 N (-54 N^3 (-1323+2 N (554+N (-1857+886 N)))\nonu \\
&+&6 c^4 (-62+N (-173+N (-115+76 N+60 N^2)))\nonu \\
&-&9 c N^2 (2998+N (355+4 N (6321+2 N (-2461+689 N))))\nonu \\
&+&c^3 (264+N (4037+N (-2811+10 N (-173+6 N (7+46 N)))))\nonu \\
&+&3 c^2 N (-313+N (1421+N (22269+4 N (-3424+N (671+60 N)))))),
\nonu \\
c_{76} &=&-12 D (-3+2 c) N (1+N) (c+6 N) (-3 c+2 (3+c) N)\nonu \\
& \times &  (c+2 c N+3 N (-3+2 N)) (9 N^2 (357-854 N+480 N^2)\nonu \\
&+&2 c^2 (46+N (-53+22 N))-3 c N (385+4 N (-176+67 N))),
\nonu \\
c_{77} &=&-6 D (-3+2 c) N (1+N) (c+6 N) (-3 c+2 (3+c) N)\nonu \\
& \times & (c+2 c N+3 N (-3+2 N))^2 (9 (21-34 N) N+c (-23+22 N)),
\nonu \\
c_{78} &=&36 D (c-3 N) N (c-4 c N+6 N^2) (-2 c^4 (48+N (-1+2 N) (-47+9 N (-1+2 N)))\nonu \\
&-&162 N^3 (-683+2 N (673+5 N (-55+6 N)))\nonu \\
&+&27 c N^2 (-2752+N (4015+4 N (221-692 N+310 N^2)))\nonu \\
&+&3 c^3 (-280+N (305+N (225+2 N (-321+2 N (53+18 N)))))\nonu \\
&+&9 c^2 N (1681+N (-1777+N (-1083+4 N (462+N (-305+156 N)))))),
\nonu \\
c_{79} &=&-12D  (3+c-9 N) (c-3 N) N (-3 c+2 (3+c) N) (2 c (-1+N)\nonu \\
&+&3 (3-4 N) N) (c^2 (64+90 N+44 N^2)\nonu \\
&+&9 N^2 (273-602 N+256 N^2)+3 c N (-301+20 N (10+9 N))),
\nonu \\
c_{80} &=&-108D  (c-3 N)^2 N (8 c^4 (5+N (1+N) (1+2 N))\nonu \\
&+&54 N^3 (-231+2 N (437+N (-573+194 N)))\nonu \\
&+&3 c^2 N (-573+4 N (920+N (-1493+4 N (131+6 N))))\nonu \\
&+&9 c N^2 (874+5 N (-851+4 N (368+3 N (-71+16 N))))\nonu \\
&+&c^3 (97+N (-1065+4 N (262+N (-11+36 N))))),
\nonu \\
c_{81} &=&36 D (c-3 N) N (-324 N^4 (378+N (-1547+2 N (1381+N (-1229+386 N))))\nonu \\
&-&108 c N^3 (-679+N (3004+N (-6917+N (7903-3582 N+520 N^2))))\nonu \\
&+&2 c^5 (-3+N (-39+2 N (13+N (-39+4 N (-1+9 N)))))\nonu \\
&-&3 c^3 N (23+N (2095+N (-3797+4 N (-1920+N (3463+24 N (-61+8 N))))))\nonu \\
&+&c^4 (51+N (403+2 N (519+2 N (-994+3 N (147+4 N (8+9 N))))))\nonu \\
&+&9 c^2 N^2 (-1153+N (6847+N (-18715+4 N (3218+N (1945+16 N (-148+33 N))))))),
\nonu \\
c_{82} &=&-12 D (-3+2 c) N (1+N) (c+6 N) (-3 c+2 (3+c) N)\nonu \\
& \times &  (c+2 c N+3 N (-3+2 N))^2 (3 (49-60 N) N+c (-23+8 N)),
\nonu \\
c_{83} &=&-2 D (3+c-9 N) N (-1+2 N) (-3 c+2 (3+c) N) (2 c (-1+N)\nonu \\
&+&3 (3-4 N) N) (81 N^3 (119+2 N (-89+40 N))\nonu \\
&-&3 c^2 N (-935+4 N (-23+18 N (8+N)))+2 c^3 (-82+N (-153+8 N (-7+3 N)))\nonu \\
&-&18 c N^2 (627+N (-563+16 N (-14+15 N)))),
\nonu \\
c_{84} &=&-9  D (c-3 N)^2 N (c+2 c N+3 N (-3+2 N)) (c^3 (-54+4 N (17+8 N))\nonu \\
&-&378 N^2 (12+N (-53+22 N))-9 c N (-371+4 N (441-280 N+54 N^2))\nonu \\
&+&3 c^2 (-77+8 N (70+N (-41+17 N)))),
\nonu \\
c_{85} &=&6 D  (c-3 N) N (1+N) (-3 c+2 (3+c) N) (162 N^3 (735+2 N (-549-214 N+228 N^2))\nonu \\
&+&4 c^4 (-44+N (-413+4 N (31+27 N)))\nonu \\
&+&12 c^3 (-44+N (567+N (607+4 N (-314+27 N))))\nonu \\
&-&27 c N^2 (2421+4 N (-1958+N (3859+28 N (-167+54 N))))\nonu \\
&-&9 c^2 N (-1115+4 N (1608+N (-2007+4 N (19+288 N))))),
\nonu \\
c_{86} &=&-6 D (3+c-9 N) (c-3 N) N (-3 c+2 (3+c) N) (c+2 c N+3 N (-3+2 N))\nonu \\
& \times &  (c^2 (88+78 N+44 N^2)+9 N^2 (525-950 N+352 N^2)+3 c N (-475+4 N (80+39 N))),
\nonu \\
c_{87} &=&3 D (c-3 N) N (1944 N^4 (-294+N (-864+N (3601+4 N (-950+291 N))))\nonu \\
&+& 2 c^5 (-381+N (371+2 N (-1063+2 N (-617+386 N+216 N^2))))\nonu \\
&+&324 c N^3 (690+N (6893+N (-16301+4 N (3438+N (-743+12 N)))))\nonu \\
&+&3 c^4 (549+N (2255+4 N (-510+N (1359+2 N (93-652 N+756 N^2)))))\nonu \\
&-&54 c^2 N^2 (-718+N (13556+N (-9361+2 N (-5691+2 N (7209-4042 N+864 N^2)))))\nonu \\
&+&9 c^3 N (-2213+N (5427+4 N (2025+N (-2275+2 N (999+4 N (-68+9 N))))))),
\nonu \\
c_{88} &=&-9D  N (-1+2 N) (-3 c+2 (3+c) N) (2 c^5 (16+N (-77+2 N (-49+4 N+8 N^2)))\nonu \\
&+&c^4 (-180+N (133+2 N (1537-2 N (55+6 N) (-5+8 N))))\nonu \\
&-&81 N^4 (-525+N (-225+4 N (1117+N (-1499+504 N))))\nonu \\
&+&27 c N^3 (900+N (-10375+2 N (8891+4 N (-331+8 N (-166+63 N)))))\nonu \\
&+&3 c^3 N (1099-2 N (2005+2 N (1244+N (-1223+4 N (-196+87 N)))))\nonu \\
&+&9 c^2 N^2 (-2092+N (12167+2 N (-3459+4 N (-1346+N (883+186 N)))))),
\nonu \\
c_{89} &=&6 D (-3+2 c) (c-3 N) N (1+N) (-1+6 N) (c+6 N) (-3 c+2 (3+c) N)\nonu \\
& \times &  (c+2 c N+3 N (-3+2 N)) (c (23+6 N)+3 N (-35+18 N)),
\nonu \\
c_{90} &=&-2D  (-3+2 c) N (1+N) (-1+2 N) (c+6 N) (-3 c+2 (3+c) N)\nonu \\
& \times &  (4 c (-1+N)+3 (9-14 N) N) (c+2 c N+3 N (-3+2 N))\nonu \\
& \times & (c (23+6 N)+3 N (-35+18 N)),
\nonu \\
c_{91} &=&-54 D (c-3 N)^2 N (1+N) (c+6 N) (-18 N^2 (-49+2 N) (-3+2 N)\nonu  \\
&+& 2 c^3 (101+4 (-19+N) N)+3 c N (377+4 N (314+3 N (-181+76 N)))\nonu \\
&-&c^2 (107+2 N (963+2 N (-587+186 N)))),
\nonu \\
c_{92} &=&-D(-3+2 c) N (1+N) (-1+2 N) (c+6 N) (-3 c+2 (3+c) N)\nonu \\
& \times &  (c+2 c N+3 N (-3+2 N))^2 (c (23+6 N)+3 N (-35+18 N)),
\nonu \\
c_{93} &=&12D (c-3 N) N (1+N) (c+6 N) (c+2 c N+3 N (-3+2 N)) (4 c^3 (39+8 (-4+N) N)\nonu \\
&-&27 N^2 (357-854 N+480 N^2)-6 c^2 (60+N (227+2 N (-229+64 N)))\nonu \\
&+&9 c N (427+4 N (-64+N (-227+156 N)))),
\nonu \\
D & \equiv
& -\frac{ c^4 \sqrt{9N-3c}}{1134N(29 + 2 c) (6 + 5 c) N (-1 + 2 N) (c + 6 (-1 + N) N)^4\sqrt{N - 2 N^2}}.
\label{O2O9p1coeff}
\eea

\section{The coefficient functions
 in (\ref{O5O9p1}) associated with the spin-$6$ current
}

As in previous examples,
the coefficients in $(\ref{O5O9p1})$ for general $N$ are
\bea
c_1 &=&-2520 c^4D (29+2 c) (c-3 N)^4 (-1+N) N (-1+2 N) (c^2 (3-10 N+2 N^2)\nonu \\
&+& 18 N^2 (-1-4 N+9 N^2)-6 c N (2-15 N+16 N^2)), \nonu \\
c_2  & = & -360 c^4 D (c-3 N)^3 (-1+N) N (-1+2 N) (2 c^4 (132-267 N+28 N^2+52 N^3)\nonu \\
&+&108 N^3 (-448+2671 N-4642 N^2+2592 N^3)\nonu \\
&+&c^3 (792-7189 N+12742 N^2-5644 N^3+168 N^4)\nonu \\
&+& 18 c N^2 (2398-14907 N+27696 N^2-20132 N^3+5184 N^4)\nonu \\
&-&3 c^2 N (3487-23223 N+41556 N^2-27500 N^3+7248 N^4)),\nonu \\
c_4 &=& \frac{1}{77}c^4 D (c-3 N)^2 (-561330 N^5 (136519124-190461085 N+99067643 N^2\nonu \\
&-&23286276 N^3+2158900 N^4-567 c N^4 (-250673172795+400721752191 N\nonu \\
&-&253411549210 N^2 + 80285572310 N^3-12942693020 N^4+864045664 N^5)\nonu \\
&-&27 c^2 N^3 (3876057446115-6879443743236 N+4921753433246 N^2\nonu \\
&-&1810191324180 N^3+357244952000 N^4-34749846384 N^5+1137076064 N^6)\nonu \\
&-&9 c^3 N^2 (-4253484112800+8207608700457 N-6410229831502 N^2\nonu \\
&+&2596155612624 N^3-577800178760 N^4+68021343488 N^5-3542911968 N^6\nonu \\
&+&39700096 N^7)+4 c^5 (126017881920-276009625179 N+243262702680 N^2\nonu \\
&-&111677980521 N^3+28803431305 N^4-4158821396 N^5+307777160 N^6-8476504 N^7\nonu \\
&+&960 N^8)-6 c^4 N (1160425676115-2398645828212 N+2003282395938 N^2\nonu \\
&-&870443206969 N^3 + 210688708860 N^4-27974557408 N^5+1818472272 N^6\nonu \\
&-&38133616 N^7+1920 N^8)), \nonu \\
c_5 &=& -90 D c^4 (c-3 N)^4 (-1+N) N (-1+2 N) (2 c^3 (-39-94 N+20 N^2)\nonu \\
&+& 252 N^2 (-25-37 N+75 N^2)+c^2 (525-1442 N+520 N^2-1128 N^3)\nonu \\
&-& 6 c N (259-2499 N+1442 N^2+468 N^3)),\nonu \\
c_6 &=& 90 D c^4 (c-3 N)^3 (-1+N) N (-1+2 N) (c^3 (2142-9850 N+10092 N^2-4832 N^3)\nonu \\
&+&12 c^4 (25-48 N+8 N^3)+378 N^3 (132+343 N-1610 N^2+1224 N^3)\nonu \\
&-&3 c^2 N (5635-31432 N+39184 N^2-20184 N^3+6912 N^4)\nonu \\
&+&9 c N^2 (2401-31892 N+62864 N^2-39400 N^3+7200 N^4)), \nonu \\
c_7 &=& -\frac{8}{231} c^4 D (c-3 N)^3 (1122660 N^4 (3731003-5401448 N+2978485 N^2-768196 N^3\nonu \\
&+& 82236 N^4)+378 c N^3 (-16624591875+27777490776 N-18670540180 N^2+6440887055 N^3\nonu \\
&-&1171822700 N^4+93062044 N^5)+9 c^2 N^2 (387960125940-717486059313 N\nonu \\
&+&536660981687 N^2-207226798130 N^3+43229166720 N^4-4499331832 N^5+161167408 N^6)\nonu \\
&+&3 c^3 N (-284527613520+567704947041 N-453994190967 N^2+185882259358 N^3\nonu \\
&-&41032516240 N^4+4619758344 N^5-207952688 N^6+10112 N^7)+2 c^4 (38801306880\nonu \\
&-& 82132149096 N+68767904655 N^2-29188848039 N^3+6642884380 N^4-772039724 N^5\nonu \\
&+&36579920 N^6-87616 N^7+3840 N^8)), \nonu \\
c_9 &=& -7560 c^4  D(c-3 N)^3 (1-2 N)^2 (-1+N) N^2 (1+N) (c+6 N) (2 c^2 (-9+2 N)\nonu \\
&+& 9 N (-13+10 N)-3 c (-5-40 N+36 N^2)),\nonu \\
c_{11} &=& 5040 c^4 D (c-3 N)^4 (-1+N) N (-1+2 N) (-36 N^2 (19-67 N+88 N^2)\nonu \\
&+&3 c N (-127+126 N+1245 N^2-1066 N^3)+c^3 (48-58 N^2+20 N^3)\nonu \\
&+&c^2 (144-490 N-437 N^2+728 N^3-300 N^4)), \nonu \\
c_{12} &=& 2520 c^4 D (3+c-9 N) (c-3 N)^4 (-1+N) N (-1+2 N) (6 N+c (-3+2 N))\nonu \\
& \times &  (2 c (4+N)+3 N (-13+16 N)), \nonu \\
c_{13} &=& \frac{8}{231} c^4 D (c-3 N)^3 (-1122660 N^4 (3732858-5413293 N+3004199 N^2\nonu \\
&-&791064 N^3+89380 N^4)-756 c N^3 (-8314464240+13904327007 N-9373319260 N^2\nonu \\
&+&3261733685 N^3-606720860 N^4+50693908 N^5)-9 c^2 N^2 (387994080780\nonu \\
&-& 717804244827 N+537525927493 N^2-208223832850 N^3+43752437220 N^4\nonu \\
&-&4610679848 N^5+ 165318752 N^6)+3 c^3 N (284520798720-567743058189 N\nonu \\
&+& 454189530093 N^2-186163376912 N^3+41204518640 N^4-4667744496 N^5\nonu \\
&+&214167472 N^6+10112 N^7)+2 c^4 (-38796952320+82114755264 N-68739047235 N^2\nonu \\
&+&29163318141 N^3-6631094030 N^4+769829836 N^5-36185080 N^6-87616 N^7+3840 N^8)), \nonu \\
c_{15} &=& -180 c^4 D (c-3 N)^3 (-1+N) N (-1+2 N) (8 c^4 (81-27 N-56 N^2+22 N^3)\nonu \\
&+& 1134 N^3 (42-45 N-88 N^2+144 N^3)+36 c^2 N (-462+1824 N-157 N^2-666 N^3+64 N^4)\nonu \\
&-&12 c^3 (-231+1017 N-3 N^2-514 N^3+168 N^4)\nonu \\
&-& 27 c N^2 (-413+3654 N-2664 N^2+928 N^3+192 N^4)),\nonu \\
c_{16} &=& 180 c^4 D (c-3 N)^3 (-1+N) N (-1+2 N) (378 N^3 (-231+1022 N-1616 N^2+872 N^3)\nonu \\
&+& 4 c^4 (-252+271 N+387 N^2-376 N^3+60 N^4)+9 c N^2 (-448-6321 N+51892 N^2\nonu \\
&-& 86084 N^3 +38464 N^4)-2 c^3 (1512-8747 N+4358 N^2+15634 N^3-15408 N^4\nonu \\
&+& 3480 N^5) +3 c^2 N (6412-23075 N-14300 N^2+81828 N^3-60832 N^4+16320 N^5)), \nonu \\
c_{17} &=& -180 c^4 D (c-3 N)^4 (-1+N) N (-1+2 N) (-252 N^2 (19-96 N+146 N^2)\nonu \\
&+& 2 c^3 (168-5 N-208 N^2+100 N^3)-6 c N (343+371 N-6022 N^2+4624 N^3)\nonu \\
&+& c^2 (1008-3533 N-3540 N^2+6620 N^3-2400 N^4)), \nonu \\
c_{18} &=& 5040 c^4 D (3+c-9 N) (c-3 N)^3 (-1+N) N (-1+2 N) (c (-1+N)\nonu \\
&+& 3 (2-3 N) N) (6 N+c (-3+2 N)) (2 c (4+N)+3 N (-13+16 N)), \nonu \\
c_{19} &=& 40 c^4  D (3+c-9 N) (c-3 N)^2 (-1+N) N (-1+2 N) (810 N^4 (252-479 N\nonu \\
&+& 46 N^2+168 N^3)+27 c N^3 (-12902+30095 N-9982 N^2-17504 N^3+10464 N^4)\nonu \\
&+& c^4 (2424-4574 N+206 N^2+3704 N^3-2000 N^4+240 N^5)\nonu \\
&-& 3 c^3 N (12225-28081 N + 9210 N^2+19744 N^3-15888 N^4+2880 N^5)\nonu \\
&+& 9 c^2 N^2 (20282-49937 N+22270 N^2+28480 N^3-27664 N^4+7680 N^5)), \nonu \\
c_{20} &=& 720 c^4 D (c-3 N)^3 (-1+N) N (-1+2 N) (378 N^3 (65-269 N+170 N^2+156 N^3)\nonu \\
&+&2 c^4 (-168+208 N+93 N^2-208 N^3+60 N^4)+9 c N^2 (-3626+17003 N-14129 N^2\nonu \\
&-& 11028 N^3+8628 N^4)-3 c^3 (336-2654 N+2833 N^2+1652 N^3-2804 N^4+880 N^5)\nonu \\
&+& 3 c^2 N (3759-20438 N+19432 N^2+13778 N^3-18016 N^4+3960 N^5)), \nonu \\
c_{21} &=& -\frac{8}{231} c^4 D  (c-3 N)^3 (561330 N^4 (3731003-5401245 N\nonu \\
&+&2977470 N^2-766572 N^3+81424 N^4)+189 c N^3 (-16624491390+27776385441 N\nonu \\
&-&18666608395 N^2+6435095060 N^3-1168504220 N^4+92608624 N^5)\nonu \\
&+& 9 c^2 N^2 (193979359575-358736302359 N+268306431616 N^2-103572896680 N^3\nonu \\
&+&21582178680 N^4-2239312496 N^5+80167904 N^6)+6 c^3 N (-71131903380\nonu \\
&+& 141926038389 N-113496736413 N^2+46464724582 N^3-10250000170 N^4\nonu \\
&+& 1150414296 N^5 -51364472 N^6+2528 N^7)+8 c^4 (4850163360-10266518637 N\nonu \\
&+& 8595981585 N^2- 3648547533 N^3+830178635 N^4-96271078 N^5+4468540 N^6\nonu \\
&-& 10952 N^7+480 N^8)), \nonu \\
c_{22} &=& 2160 c^4 D  (c-3 N)^3 (1-2 N)^2 (-1+N) N^2 (10 c^4 (2-5 N+2 N^2)\nonu \\
&+& 252 N^2 (63-184 N+101 N^2)+c^3 (818-1103 N+692 N^2-300 N^3)+ 6 c N (-1694 \nonu \\
&+& 6853 N - 6533 N^2+2458 N^3)+c^2 (1722-11167 N+12617 N^2-4770 N^3-120 N^4)), \nonu \\
c_{23} &=& \frac{4}{231} c^4 D (c-3 N)^2 (1683990 N^5 (102382230-141015895 N\nonu \\
&+& 71750931 N^2-16195600 N^3+1385308 N^4)+567 c N^4 (-557957088765+870822849609 N\nonu \\
&-& 531891742640 N^2+159776917720 N^3-23551346320 N^4+1325324816 N^5)\nonu \\
&+& 27 c^2 N^3 (8547376777470-14717811122874 N+10128970212019 N^2-3531905290290 N^3\nonu \\
&+& 641895014440 N^4-53978593416 N^5+1355007376 N^6)+9 c^3 N^2 (-9303024429315\nonu \\
&+& 17351338484853 N-13020185959823 N^2+5008576828836 N^3-1034298518680 N^4\nonu \\
&+& 107748672472 N^5-4675753872 N^6+96793184 N^7)+4 c^5 (-271832863680\nonu \\
&+& 573501628566 N- 485728888725 N^2+212717494344 N^3-51583421780 N^4\nonu \\
&+& 6834595654 N^5 - 454097800 N^6 +  13200536 N^7+960 N^8)-6 c^4 N (-2519584749900\nonu \\
&+& 5023025276328 N-4030065284037 N^2+ 1666868473781 N^3-377003205180 N^4\nonu \\
&+& 45201848972 N^5-2568568848 N^6+70805984 N^7+1920 N^8)), \nonu \\
c_{24} &=& -360 c^4 D (3+c-9 N) (c-3 N)^2 (-1+N) N (-1+2 N) (54 N^4 (980-2123 N\nonu \\
&+& 614 N^2+672 N^3)+27 c N^3 (-3867+10156 N-4550 N^2-6172 N^3+4064 N^4)\nonu \\
&+& 2 c^4 (444-995 N+197 N^2+810 N^3-536 N^4+80 N^5)-c^3 N (12207-31319 N+13282 N^2\nonu \\
&+& 21932 N^3-19792 N^4+4080 N^5)+3 c^2 N^2 (19235-52456 N+27978 N^2+30668 N^3\nonu \\
&-& 31696 N^4+8640 N^5)), \nonu \\
c_{25} &=& 720 c^4 D  (3+c-9 N) (c-3 N)^3 (-1+N) N^2 (-1+2 N) (2 c^3 (51-116 N+49 N^2+6 N^3)\nonu \\
&-& 18 N^2 (-49+700 N-1432 N^2+864 N^3)+3 c^2 (77-806 N+1473 N^2-768 N^3+124 N^4)\nonu \\
&-& 3 c N (301-3581 N+7529 N^2-5782 N^3+1728 N^4)), \nonu \\
c_{26} &=& 120 c^4 D (3+c-9 N) (c-3 N)^2 (-1+N) N (-1+2 N) (2 c (-1+N)\nonu \\
&+& 3 (3-4 N) N)^2 (6 N+c (-3+2 N)) (c (40+22 N)+3 N (-7+80 N)), \nonu \\
c_{27} &=& 60 c^4 D (3+c-9 N) (c-3 N)^2 (-1+N) N (-1+2 N) (2 c (-1+N)\nonu \\
&+& 3 (3-4 N) N) (6 N+c (-3+2 N)) (2 c^2 (172-69 N+2 N^2)\nonu \\
&-& 3 c N (1391-1816 N+276 N^2)+ 9 N^2 (1323-2782 N+1376 N^2)), \nonu \\
c_{28}&=& -\frac{1}{693} c^4 D  (c-3 N)^3 (-1+N) N (1+N) (56133 N^4 (-2980308+3590060 N\nonu \\
&-& 946135 N^2-182630 N^3+8653 N^4)-27 c N^3 (-9365853492+14104771440 N\nonu \\
&-& 7431900457 N^2+1340713637 N^3+12733180 N^4+432772 N^5)-9 c^2 N^2 (15393018348\nonu \\
&-& 26130193080 N+17307715327 N^2-5674703221 N^3+1000138398 N^4-113096996 N^5\nonu \\
&+& 5081264 N^6)+6 c^3 N (5376847824-9259611516 N+5668320348 N^2-1284880711 N^3\nonu \\
&-& 73815653 N^4+91749600 N^5-16045372 N^6+910720 N^7)+8 c^4 (-329760720 \nonu \\
&+& 474667560 N - 6601644 N^2-376267854 N^3+309667435 N^4-118352105 N^5\nonu \\
&+& 24304394 N^6-2586556 N^7+111930 N^8)), \nonu \\
c_{29} &=& -\frac{1}{16} c^4 D  (c-3 N)^2 (-9720 N^5 (-293424+113244 N+8504 N^2\nonu \\
&-& 82673 N^3+42827 N^4)+162 c N^4 (-33943080+21794844 N-3311584 N^2-989489 N^3\nonu \\
&-& 4812139 N^4+2068962 N^5)-27 c^2 N^3 (-149031120+109666764 N-5854968 N^2\nonu \\
&-& 7511485 N^3-4319005 N^4-2045984 N^5+58588 N^6)-9 c^3 N^2 (144114960-28575924 N\nonu \\
&-& 216116504 N^2+209095967 N^3-88408193 N^4+17110896 N^5-737604 N^6+106080 N^7)\nonu \\
&+& 12 c^4 N (13847400+27499116 N-99100876 N^2+92049550 N^3-39372751 N^4 \nonu \\
&+& 8542056 N^5 - 1135527 N^6+43212 N^7+468 N^8)+4 c^5 (-1321920-13946256 N\nonu \\
&+& 31802712 N^2-10045048 N^3-23534512 N^4+25902661 N^5-11624171 N^6+2679564 N^7\nonu \\
&-& 312132 N^8+14496 N^9)), \nonu \\
c_{30} &=& \frac{1}{16} c^4 D  (c-3 N)^2 (-9720 N^5 (-293424+157092 N-108424 N^2\nonu \\
&-& 24209 N^3+42827 N^4)+162 c N^4 (-34381560+27056604 N-14601184 N^2+4171471 N^3\nonu \\
&-& 5054059 N^4+2068962 N^5)-27 c^2 N^3 (-151369680+129141324 N-43110648 N^2\nonu \\
&+& 8777795 N^3-6254365 N^4-2045984 N^5+58588 N^6)-9 c^3 N^2 (146161200-43091124 N\nonu \\
&-& 189908504 N^2+197161247 N^3-85988993 N^4+17110896 N^5-737604 N^6+106080 N^7)\nonu \\
&+& 12 c^4 N (13993560+26516316 N-97205836 N^2+90900430 N^3-39130831 N^4+8542056 N^5\nonu \\
&-& 1135527 N^6+43212 N^7+468 N^8)+4 c^5 (-1321920-13916016 N+31621272 N^2\nonu \\
&-& 9803128 N^3- 23534512 N^4+25902661 N^5-11624171 N^6+2679564 N^7\nonu \\
&-& 312132 N^8+14496 N^9)),\nonu \\
c_{31} &=& \frac{1}{8} c^4 D (c-3 N)^2 (-9720 N^5 (-293424+149784 N-93808 N^2\nonu \\
&-& 24209 N^3+42827 N^4)+162 c N^4 (-34308480+26325804 N-13693984 N^2+4695631 N^3\nonu \\
&-& 5054059 N^4+2068962 N^5)-27 c^2 N^3 (-151077360+127357164 N-42808248 N^2\nonu \\
&+& 13092035 N^3-6496285 N^4-2045984 N^5+58588 N^6)-9 c^3 N^2 (146015040-42738324 N\nonu \\
&-& 187348184 N^2+191314847 N^3-85021313 N^4+17110896 N^5-737604 N^6+106080 N^7)\nonu \\
&+& 12 c^4 N (13993560+26450796 N-96782476 N^2+90164590 N^3-38828431 N^4\nonu \\
&+& 8542056 N^5 - 1135527 N^6+43212 N^7+468 N^8)+4 c^5 (-1321920-13916016 N\nonu \\
&+& 31606152 N^2 - 9712408 N^3-23655472 N^4+25902661 N^5-11624171 N^6+2679564 N^7\nonu \\
&-& 312132 N^8+14496 N^9)),\nonu \\
c_{32} &=& 3780 c^4 D  (29+2 c) (6+5 c) (c-3 N)^4 N^3 (-1+2 N)^3,\nonu \\
c_{34} &=& 3780 c^4 D  (29+2 c) (c-3 N)^5 (-1+N) N (-1+2 N) (-6 N^2+c (-1+4 N)), \nonu \\
c_{35} &=& -2520 c^4 D (c-3 N)^4 (-1+N) N (-1+2 N) (24 c^3 (-1+N^2)+  c^2 (-72+254 N+56 N^2)\nonu \\
&-& 18 N^2 (10-107 N+144 N^2)-3 c N (-107+510 N-508 N^2+288 N^3)), \nonu \\
c_{36} &=& -2 c^4 D  (c-3 N)^3 (-1+N) (486 N^4 (170320-9676 N-1976 N^2-25851 N^3\nonu \\
&+& 249 N^4)+27 c N^3 (-4721220+2364672 N-1195744 N^2-412299 N^3+108591 N^4\nonu \\
&+& 1214 N^5)-54 c^2 N^2 (-838490-1069044 N+2231184 N^2-1423452 N^3+384633 N^4\nonu \\
&-& 50073 N^5+2534 N^6)-12 c^3 N (-331200+7129221 N-13046236 N^2+9734314 N^3\nonu \\
&-& 3873994 N^4+815765 N^5-90053 N^6+4022 N^7)+8 c^4 (-304560+2236392 N\nonu \\
&-& 3888153 N^2+2509841 N^3-501941 N^4-117054 N^5+71651 N^6-11723 N^7+654 N^8)), \nonu \\
c_{38} &=& 2520 c^4 D  (c-3 N)^4 (-1+N) N (-1+2 N) (6 c^3 (-3-4 N+4 N^2)\nonu \\
&+& 18 N^2 (-10-67 N+30 N^2)+c^2 (15-130 N+272 N^2-144 N^3)\nonu \\
&-& 3 c N (67-534 N+260 N^2+216 N^3)), \nonu \\
c_{41} &=& -11340 c^4 D (29+2 c) (c-3 N)^4 (1-2 N)^2 (-1+N) N^2 (-6 N^2+c (-1+4 N)), \nonu \\
c_{43} &=& -2520 c^4 D (c-3 N)^4 (-1+N) N (-1+2 N) (18 N^2 (-11-268 N+4 N^2)\nonu \\
&+& 4 c^3 (-18-19 N+9 N^2+10 N^3)+3 c N (22+901 N-286 N^2+248 N^3)\nonu \\
&+& 10 c^2 (6-17 N-4 N^2+16 N^3+24 N^4)), \nonu \\
c_{44} &=& 45360 c^4 D (29+2 c) (c-3 N)^4 (1-2 N)^2 (-1+N) N^2 (-6 N^2+c (-1+4 N)),\nonu \\
c_{45} &=& -5040 c^4 D  (c-3 N)^3 (-1+N) N (1+N) (-1+2 N) (c^4 (66-80 N+24 N^2)\nonu \\
&-& 54 N^3 (98-237 N+90 N^2)+c^3 (129-1040 N+1184 N^2-360 N^3)+6 c^2 N (-145+616 N\nonu \\
&-& 442 N^2+108 N^3)+27 c N^2 (95-144 N-248 N^2+216 N^3)), \nonu \\
c_{46} &=&-2160 c^4 D (c-3 N)^2 (1-2 N)^2 (-1+N) N^2 (c^4 (222+34 N-212 N^2+96 N^3)\nonu \\
&+& 108 N^3 (-161+220 N+78 N^2+306 N^3)+c^3 (321-2729 N-884 N^2+3168 N^3-792 N^4)\nonu \\
&-& 3 c^2 N (1322-4373 N-1876 N^2+2856 N^3+72 N^4)-18 c N^2 (-920+2117 N-89 N^2\nonu \\
&+& 474 N^3+216 N^4)),\nonu \\
c_{47} &=& -1080 c^4 D  (c-3 N)^3 (1-2 N)^2 (-1+N) N^2 (2 c^3 (12+65 N+38 N^2)\nonu \\
&-& 126 N^2 (-101+28 N+216 N^2)+3 c N (-196-2121 N+1834 N^2+288 N^3)\nonu \\
&+& c^2 (-756+917 N+1352 N^2+780 N^3)), \nonu \\
c_{48} &=& 2520 c^4 D  (3+c-9 N) (c-3 N)^4 (-1+N) N (-1+2 N) (6 N+c (-3+2 N))\nonu \\
& \times &  (2 c (4+N)+3 N (-13+16 N)), \nonu \\
c_{49} &=& -120 c^4 D (c-3 N)^2 (-1+N) N (-1+2 N) (4 c^5 (-390+641 N+271 N^2\nonu \\
&-& 596 N^3+164 N^4)+486 N^4 (-1834+5011 N-1574 N^2-4384 N^3+2208 N^4)\nonu \\
&+& 27 c^2 N^2 (-4189+12254 N-9038 N^2-6416 N^3+568 N^4+1024 N^5)-6 c^4 (366-5057 N\nonu \\
&+& 7909 N^2+3524 N^3-7220 N^4+2048 N^5)+18 c^3 N (1111-8894 N+14104 N^2\nonu \\
&+& 4534 N^3-11752 N^4+3512 N^5)-81 c N^3 (-6103+11194 N+6992 N^2-14192 N^3\nonu \\
&-& 10688 N^4+8832 N^5)), \nonu \\
c_{52} &=& -\frac{1}{(21+4 c) N (-1+2 N)}60 c^4 D  (c-3 N)^3 (3402 N^5 (44351447\nonu \\
&-& 61956177 N+32308358 N^2-7638900 N^3+718952 N^4)+81 c N^4 (-3301821036\nonu \\
&+& 5013869837 N-2980498275 N^2+882161030 N^3-133961484 N^4+8856008 N^5)\nonu \\
&+& 54 c^2 N^3 (3490542895-5645277003 N+3613609675 N^2-1159782701 N^3+190838642 N^4\nonu \\
&-& 13647676 N^5+96008 N^6)-9 c^3 N^2 (7337907348-12435294511 N+8243780357 N^2\nonu \\
&-& 2624168622 N^3+358445964 N^4+6294712 N^5-7469504 N^6+745536 N^7)\nonu \\
&+& 8 c^5 (-99688320+178269984 N-115164108 N^2+25437843 N^3+5387531 N^4\nonu \\
&-& 4228834 N^5+954208 N^6-100208 N^7+4704 N^8)-6 c^4 N (-1917613440\nonu \\
&+& 3358682364 N-2232994015 N^2+641613101 N^3-28782294 N^4-29210940 N^5\nonu \\
&+& 7594600 N^6-787424 N^7+37248 N^8)), \nonu \\
c_{53} &=& 60 c^4 D  (c-3 N)^2 (-1+N) N (-1+2 N) (4 c^5 (-543+734 N+592 N^2\nonu \\
&-& 824 N^3+176 N^4)+486 N^4 (-2016+2921 N+4408 N^2-6604 N^3+624 N^4)\nonu \\
&+& 54 c^2 N^2 (-599-8250 N+23696 N^2-17524 N^3-512 N^4+2000 N^5)\nonu \\
&-& 6 c^4 (465-6287 N+8108 N^2+8072 N^3-11040 N^4+2960 N^5)\nonu \\
&+& 9 c^3 N (1667-12696 N+13892 N^2+31312 N^3-41696 N^4+12160 N^5)\nonu \\
&-& 81 c N^3 (-4013-16312 N+64500 N^2-34560 N^3-30368 N^4+18624 N^5)), \nonu \\
c_{54} &=& -2520 c^4 D  (c-3 N)^4 (-1+N) N (1+N) (-1+2 N) (c+6 N)\nonu \\
& \times &  (2 c^2 (-9+2 N)+9 N (-13+10 N)-3 c (-5-40 N+36 N^2)), \nonu \\
c_{56} &=& -\frac{1}{2} c^4 D (29+2 c) (c-3 N)^4 (-1+N) N^2 (1+N) \nonu \\
& \times & (-1+2 N) (189 N^2 (1356-620 N+157 N^2-20 N^3+N^4)\nonu \\
&+& 3 c N (-156384+288772 N-152842 N^2+40377 N^3-5198 N^4+260 N^5)\nonu \\
&+& 10 c^2 (16200-43956 N+40432 N^2-18553 N^3+4473 N^4-542 N^5+26 N^6)), \nonu \\
c_{57} &=& -2160 c^4 D  (c-3 N)^2 (1-2 N)^2 (-1+N) N^2 (8 c^4 (-3+8 N-2 N^2\nonu \\
&+& 2 N^3)-162 N^3 (315-400 N-390 N^2+528 N^3)-9 c^2 N (-580+2571 N-254 N^2\nonu \\
&-& 1212 N^3+24 N^4)-6 c^3 (150-388 N-135 N^2-32 N^3+44 N^4)\nonu \\
&+& 27 c N^2 (218+1679 N-2160 N^2-732 N^3+1152 N^4)), \nonu \\
c_{58} &=& -180 c^4 D (c-3 N)^3 (-1+N) N (-1+2 N) (-54 N^3 (399-1618 N\nonu \\
&-& 1384 N^2+24 N^3)+2 c^4 (93-349 N-150 N^2+232 N^3)+c^3 (1317-7311 N\nonu \\
&+& 4286 N^2+6392 N^3-3648 N^4)-3 c^2 N (2619-12662 N-3658 N^2+6604 N^3+1488 N^4)\nonu \\
&+& 9 c N^2 (1884-8065 N-9914 N^2+296 N^3+5472 N^4)), \nonu \\
c_{60} &=& -180 c^4 D  (c-3 N)^3 (-1+N) N (-1+2 N) (-756 N^3 (24-281 N\nonu \\
&+& 162 N^2+32 N^3)+6 c^4 (-42-175 N+54 N^2+12 N^3+40 N^4)-36 c N^2 (-504+5495 N\nonu \\
&-& 5115 N^2+2256 N^3+212 N^4)+c^3 (210-881 N+10572 N^2-14824 N^3+6096 N^4\nonu \\
&+& 2160 N^5)+3 c^2 N (-1652+15459 N-20034 N^2+18700 N^3-7800 N^4+3360 N^5)), \nonu \\
c_{61} &=& -15 c^4 D (c-3 N)^2 (-1+N) N (-1+2 N) (1944 N^4 (-1071+390 N\nonu \\
&+& 530 N^2-1858 N^3+900 N^4)+4 c^5 (-1707-551 N+434 N^2+1154 N^3-704 N^4\nonu \\
&+& 120 N^5)-27 c^2 N^2 (2550+66013 N+7580 N^2-84772 N^3+47808 N^4+4256 N^5)\nonu \\
&-& 162 c N^3 (-5295-27170 N+27256 N^2+2840 N^3-23968 N^4+8544 N^5)\nonu \\
&+& 6 c^4 (933+7217 N-6247 N^2+8966 N^3-14160 N^4+5608 N^5+720 N^6)\nonu \\
&+& 9 c^3 N (-2784+11609 N+45634 N^2-63688 N^3+35192 N^4-18832 N^5+6720 N^6)), \nonu \\
c_{62} &=& -540 c^4 D  (c-3 N)^2 (1-2 N)^2 (-1+N) N^2 (6 c^4 (46-47 N-66 N^2+32 N^3)\nonu \\
&+& 54 N^3 (1295-1318 N-948 N^2+1056 N^3)+c^3 (2070-6461 N-2546 N^2+4872 N^3\nonu \\
&-& 2304 N^4)+ 6 c^2 N (-1672+6235 N+4007 N^2-5898 N^3+288 N^4)\nonu \\
&-& 9 c N^2 (884+6361 N+134 N^2-7728 N^3+2304 N^4)), \nonu \\
c_{63} &=& -6480 c^4 D  (c-3 N)^2 (1-2 N)^2 (-1+N) N^2 (4 c^4 (1-3 N-6 N^2+8 N^3)\nonu \\
&+& 54 N^3 (63+9 N-256 N^2+204 N^3)-3 c^2 N (254-1145 N+70 N^2+1148 N^3+24 N^4)\nonu \\
&-& 2 c^3 (-29+99 N+75 N^2-220 N^3+132 N^4)\nonu \\
&-& 9 c N^2 (-135+1325 N-1296 N^2-356 N^3+144 N^4)),\nonu \\
c_{64} &=& 180 c^4 D  (c-3 N)^2 (-1+N) N (-1+2 N) (4 c^5 (-189-182 N+117 N^2\nonu \\
&-& 30 N^3+40 N^4)-162 N^4 (1001+2116 N-2920 N^2+240 N^3+2448 N^4)+c^4 (630\nonu \\
&+& 3904 N+5988 N^2-12024 N^3+8944 N^4-3552 N^5)+18 c^2 N^2 (-69-16735 N+17100 N^2\nonu \\
&-& 22224 N^3+14192 N^4+144 N^5)+81 c N^3 (1306+6393 N-5554 N^2+3880 N^3-1424 N^4\nonu \\
&+&576 N^5)+ 3 c^3 N (-2018+14019 N-19838 N^2+37476 N^3-29624 N^4+5088 N^5)), \nonu \\
c_{65} &=& -2520 c^4 D  (c-3 N)^3 (-1+N) N (-1+2 N) (54 N^3 (166-245 N+372 N^2)\nonu \\
&+& c^3 (60+288 N+172 N^2-56 N^3)+8 c^4 (-9-13 N+2 N^2+6 N^3)+9 c N^2 (-557+746 N\nonu \\
&-& 2216 N^2+720 N^3)-6 c^2 N (-75+92 N-554 N^2-28 N^3+288 N^4)), \nonu \\
c_{66} &=& 6480 c^4 D (c-3 N)^3 (1-2 N)^2 (-1+N) N^2 (126 N^2 (9-18 N+2 N^2)\nonu \\
&+& 10 c^3 (1-N+2 N^2)+c^2 (7-287 N+484 N^2-60 N^3)\nonu \\
&+& 3 c N (-126+511 N-574 N^2+120 N^3)), \nonu \\
c_{67} &=& 6480 c^4 D (c-3 N)^3 (1-2 N)^2 (-1+N) N^2 (126 N^2 (9-18 N+2 N^2)+10 c^3 (1-N+2 N^2)\nonu \\
&+& c^2 (7-287 N+484 N^2-60 N^3)+3 c N (-126+511 N-574 N^2+120 N^3)),\nonu \\
c_{68} &=& 540 c^4 D  (c-3 N)^2 (1-2 N)^2 (-1+N) N^2 (-378 N^3 (245-338 N-256 N^2+240 N^3)\nonu \\
&+& c^4 (540-342 N-500 N^2+352 N^3)+c^3 (378-5763 N+4770 N^2+6456 N^3-3264 N^4)\nonu \\
&-& 6 c^2 N (-168+1663 N-229 N^2-3434 N^3+912 N^4)\nonu \\
&+& 9 c N^2 (1708+7777 N-13834 N^2-12496 N^3+12096 N^4)), \nonu \\
c_{69} &=& -15120 c^4 D (c-3 N)^3 (1-2 N)^2 (-1+N) N^2 (c+6 N) (c^2 (-24-34 N+20 N^2)\nonu \\
&-& 3 N (73-59 N+42 N^2)+c (66+11 N+236 N^2-180 N^3)), \nonu \\
c_{70} &=& -360 c^4 D  (c-3 N)^3 (-1+N) N (-1+2 N) (-378 N^3 (9+566 N-644 N^2+104 N^3)\nonu \\
&+& 2 c^4 (378+147 N-540 N^2+76 N^3+160 N^4)-9 c N^2 (-560-16625 N+20040 N^2\nonu \\
&-& 14036 N^3+ 9232 N^4)+c^3 (-630-225 N-3670 N^2+5584 N^3-1336 N^4+1200 N^5)\nonu \\
&-& 3 c^2 N (-168+9971 N-9996 N^2+8340 N^3-7072 N^4+3360 N^5)), \nonu \\
c_{71} &=& -\frac{24 c^4 D (c-3 N)}{77 (21+4 c) N (-1+2 N)}  (23575860 N^7 (1126813235-1527071265 N\nonu \\
 &+&759680090 N^2-166434912 N^3+13757504 N^4)-561330 c N^6 (102620164907\nonu \\
&-& 123595173143 N+ 48476464330 N^2-5061841190 N^3-885867856 N^4+166616376 N^5)\nonu \\
&-& 93555 c^2 N^5 (-549219821813+ 479770622798 N+64405558815 N^2-201395927508 N^3\nonu \\
&+& 84931075160 N^4-15015570864 N^5+1013139248 N^6)-81 c^3 N^4 (295533212163005\nonu \\
&-& 60387392248366 N- 430899208834271 N^2+429913090406336 N^3-181198035913120 N^4\nonu \\
&+& 39497729492016 N^5-4377838656464 N^6+194983987904 N^7)\nonu \\
&-& 27 c^4 N^3 (-222749015462150- 279226708420137 N+1019026358801235 N^2\nonu \\
&-& 935734392571366 N^3+424069024043788 N^4- 107491950631488 N^5+15337847831760 N^6\nonu \\
&-& 1124081229664 N^7+30718352192 N^8)- 9 c^5 N^2 (79367760102600+434729432049855 N\nonu \\
&-& 1107648815321053 N^2+1024087154548724 N^3-498970966102000 N^4\nonu \\
&+& 141671684535808 N^5- 23817722027152 N^6+2262818278016 N^7-105784835200 N^8\nonu \\
&+& 1738017792 N^9)-6 c^6 N (-2224499448960- 129005374954188 N+295599570320055 N^2\nonu \\
&-& 281055430434627 N^3+146084030570018 N^4-45363379310320 N^5\nonu \\
&+& 8592780382176 N^6-966953944848 N^7+59550909472 N^8-1697215232 N^9\nonu \\
&+& 20373504 N^{10})+8 c^7 (390475572480-7111268919936 N+15701367135828 N^2\nonu \\
&-& 15435645414755 N^3+8491816854263 N^4-2837477110790 N^5+589509643852 N^6\nonu \\
&-& 74913386952 N^7+5475060192 N^8-199216064 N^9+2542080 N^{10}+33792 N^{11})), \nonu \\
c_{72} &=& \frac{24 c^4D (c-3 N)}{77 (21+4 c) N (-1+2 N))} (70727580 N^7 (770560200-1045761679 N+521386199 N^2\nonu \\
&-& 114631208 N^3+9532252 N^4)-561330 c N^6 (212337984624-260083236667 N\nonu \\
&+& 106175428227 N^2-13376117508 N^3-1205357188 N^4+294464496 N^5)\nonu \\
&-& 93555 c^2 N^5 (-1154223354007+ 1090794720138 N-3454441429 N^2\nonu \\
&-& 330987561762 N^3+148023472440 N^4-26524787464 N^5+1793727760 N^6)\nonu \\
&-& 81 c^3 N^4 (639054650848785-260893941509182 N-682629183140027 N^2\nonu \\
&+& 735544886230612 N^3-312833104771100 N^4+67654896396832 N^5-7369356044288 N^6\nonu \\
&+& 319572284288 N^7)-27 c^4 N^3 (-510138272224550-335609082910749 N\nonu \\
&+& 1703514781620605 N^2-1600957803836012 N^3+721306278372156 N^4\nonu \\
&-& 179798187943296 N^5 + 25068646615920 N^6-1789749821568 N^7+48006105664 N^8)\nonu \\
&-& 9 c^5 N^2 (210659897248200+ 666390601225695 N-1873780811938661 N^2\nonu \\
&+& 1740755708001688 N^3-836731301667660 N^4+232477255533376 N^5\nonu \\
&-& 38059018444784 N^6+3509192992512 N^7-158519085120 N^8+2439085824 N^9)\nonu \\
&-& 6 c^6 N (-15902848566720-208656503319036 N+500842481949735 N^2\nonu \\
&-& 473910220538759 N^3+241830961073706 N^4-73243434739540 N^5+13471444419952 N^6\nonu \\
&-& 1464747924576 N^7+86215439264 N^8-2254189824 N^9+20640768 N^{10})\nonu \\
&+& 8 c^7 (246864360960-11739195624672 N+26544109049436 N^2-25812369240635 N^3\nonu \\
&+& 13894173768421 N^4-4516732135790 N^5+908737641454 N^6-111068377304 N^7\nonu \\
&+& 7680580264 N^8-249939168 N^9+1831680 N^{10}+67584 N^{11})), \nonu \\
c_{73} &=& \frac{24 c^4 D(c-3 N) (-1+N) }{77 (21+4 c) N (-1+2 N)} (-11787930 N^7 (222611143-310059419 N\nonu \\
&+& 160857158 N^2-37656004 N^3+3468600 N^4)+280665 c N^6 (20095612135-24578607139 N\nonu \\
&+& 9830633640 N^2-1022607960 N^3-217061616 N^4+43691216 N^5)\nonu \\
&+& 93555 c^2 N^5 (-53841298366+ 49362458293 N+1661028992 N^2-16270753720 N^3\nonu \\
&+& 7132300096 N^4-1291668592 N^5+91656576 N^6)+81 c^3 N^4 (29545065638225\nonu \\
&-& 11254867766623 N-30544289028438 N^2+31426817196548 N^3-12793486806880 N^4\nonu \\
&+& 2637352195888 N^5-271080981472 N^6+10767842752 N^7)+27 c^4 N^3 (-23631732425640\nonu \\
&-& 14264786376501 N+70436561046360 N^2-62300918312088 N^3+26041125407744 N^4\nonu \\
&-& 5905987618224 N^5+725769138880 N^6-42995357312 N^7+822612736 N^8)\nonu \\
&+& 18 c^5 N^2 (5029522790400+13561743637500 N-36500948626907 N^2\nonu \\
&+& 31312837087266 N^3-13588335894280 N^4+3309594596112 N^5-453650045488 N^6\nonu \\
&+&32148525344 N^7-887392000 N^8+186368 N^9)-24 c^6 N (222551435520\nonu \\
&+& 2047213653696 N-4629172979700 N^2+3973818277659 N^3-1787643788006 N^4\nonu \\
&+& 459732168360 N^5-67629598272 N^6+5209647536 N^7-145014304 N^8-2913536 N^9\nonu \\
&+& 133632 N^{10})+64 c^7 (-136080+55639689156 N-116897577588 N^2+101490793455 N^3\nonu \\
&-& 47170882708 N^4+12688396525 N^5-1989595672 N^6+178776112 N^7-11918872 N^8\nonu \\
&+& 1477984 N^9-177600 N^{10}+8448 N^{11})), \nonu \\
c_{74} &=& 7560 c^4 D  (c-3 N)^3 (1-2 N)^2 (-1+N) N^2 (1+N) (c+6 N) (2 c^2 (-9+2 N)\nonu \\
&+& 9 N (-13+10 N)-3 c (-5-40 N+36 N^2)),\nonu \\
c_{77} &=& 60 c^4 D (-3+2 c) (c-3 N)^2 (-1+N) N (1+N) (-1+2 N) (c+6 N) (c+2 c N\nonu \\
&+& 3 N (-3+2 N)) (2 c^2 (138+67 N+2 N^2)+3 c N (-1255+1000 N+268 N^2)\nonu \\
&+& 9 N^2 (1323-2510 N+1104 N^2)), \nonu \\
c_{78} &=& -20 c^4 D  (-3+2 c) (c-3 N) (-1+N) N (1+N) (-1+2 N) (c+6 N) (6 N\nonu \\
&+& c (-3+2 N)) (c+2 c N+3 N (-3+2 N)) (c^2 (253+110 N+48 N^2)\nonu \\
&+& 3 c N (-1123+912 N+108 N^2)+18 N^2 (567-1039 N+450 N^2)),\nonu \\
c_{79} &=& 360 c^4 D (3+c-9 N) (c-3 N) (1-2 N)^2 (-1+N) N^2 (1+N) (108 N^4 (231-664 N\nonu \\
&+& 408 N^2)+c^3 N (2651-4314 N+2220 N^2-648 N^3)+c^4 (-288+518 N-352 N^2+72 N^3)\nonu \\
&-& 36 c N^3 (10-119 N-186 N^2+144 N^3)-3 c^2 N^2 (2455-3644 N+948 N^2+720 N^3)), \nonu \\
c_{80} &=& -60 c^4 D (-3+2 c) (c-3 N) (-1+N) N (1+N) (-1+2 N) (c+6 N) (6 N+c (-3+2 N))\nonu \\
& \times &  (c+2 c N+3 N (-3+2 N))^2 (9 (21-34 N) N+c (-23+22 N)), \nonu \\
c_{81} &=& 60 c^4 D  (-3+2 c) (c-3 N) (-1+N) N (1+N) (-1+2 N) (c+6 N) (c+2 c N\nonu \\
&+& 3 N (-3+2 N)) (-27 N^3 (903-2014 N+960 N^2)+2 c^3 (69-43 N+70 N^2+24 N^3)\nonu \\
&+& 18 c N^2 (766-1221 N+158 N^2+72 N^3)+3 c^2 N (-815+802 N-72 N^2+192 N^3)), \nonu \\
c_{82} &=& 1080 c^4 D  (c-3 N)^2 (1-2 N)^2 (-1+N) N^2 (4 c^4 (3+68 N-46 N^2+24 N^3)\nonu \\
&+& 54 N^3 (-238+41 N-852 N^2+696 N^3)-4 c^3 (198-781 N+836 N^2-534 N^3+108 N^4)\nonu \\
&+& 9 c N^2 (517+1268 N+4504 N^2-3240 N^3+288 N^4)-3 c^2 N (-925+5518 N-1600 N^2\nonu \\
&+& 696 N^3+432 N^4)), \nonu \\
c_{83} &=& \frac{60 c^4 D  (c-3 N)}{(21+4 c) N (-1+2 N)}  (30618 N^7 (1879156195-2552308699 N+ 74066316 N^2\nonu \\
&-& 280658444 N^3+23411216 N^4)+729 c N^6 (-194696510634+285867969601 N\nonu \\
&-& 161778693257 N^2+ 44463727254 N^3-6000700796 N^4+324597784 N^5)\nonu \\
&-& 243 c^2 N^5 (-611248780951+ 941044890863 N-551579169616 N^2+148343739040 N^3\nonu \\
&-& 15163385212 N^4-615248320 N^5+172707152 N^6)-81 c^3 N^4 (1054441730907\nonu \\
&-& 1652051776145 N+910697130696 N^2-150626552798 N^3-46144176868 N^4\nonu \\
&+& 24060216336 N^5-3891142928 N^6+228965536 N^7)-54 c^4 N^3 (-539128068255\nonu \\
&+& 831459452981 N-366140697300 N^2-70575832003 N^3+120519241062 N^4\nonu \\
&-& 46021509048 N^5+8422933568 N^6-744785648 N^7+23445024 N^8)\nonu \\
&-& 9 c^5 N^2 (652627588086-949225519313 N+196210000785 N^2+428583964264 N^3\nonu \\
&-& 378317491640 N^4+143339605976 N^5-29240429248 N^6+3209784288 N^7-164554432 N^8\nonu \\
&+& 2314240 N^9)+8 c^7 (-3759851520+3788666784 N+5088306162 N^2-11129158163 N^3\nonu \\
&+& 8411293753 N^4-3418763020 N^5+816269610 N^6-114568952 N^7+8826872 N^8\nonu \\
&-&302464 N^9+ 768 N^10)-6 c^6 N (-108015768000+138879230190 N+36859111663 N^2\nonu \\
&-&166824261639 N^3+ 130102162300 N^4-50883535168 N^5+11321335464 N^6\nonu \\
&-& 1432639280 N^7+94135072 N^8-2439168 N^9+21504 N^{10})), \nonu \\
c_{84} &=& -60 c^4  D (c-3 N)^2 (-1+N) N (-1+2 N) (4 c^5 (564-N-713 N^2+16 N^3+164 N^4)\nonu ]\\
&+& 486 N^4 (1015-3408 N+7394 N^2-6360 N^3+1920 N^4)-12 c^4 (-81+2043 N-1121 N^2\nonu \\
&-& 1138 N^3+488 N^4+136 N^5)+27 c^2 N^2 (8359-31420 N+60808 N^2-47112 N^3+9424 N^4\nonu \\
&+& 2656 N^5)-9 c^3 N (3683-20105 N+29100 N^2-13536 N^3-3392 N^4+3344 N^5)\nonu \\
&+& 81 c N^3 (-6964+23723 N-51024 N^2+47380 N^3-20240 N^4+3840 N^5)), \nonu \\
c_{85} &=& 60 c^4 D (c-3 N) (-1+N) N (1+N) (-1+2 N) (c+6 N) (c^5 (4050-8436 N\nonu \\
&+& 6112 N^2-2096 N^3+352 N^4)+486 N^4 (-1197+2241 N-344 N^2-1204 N^3+528 N^4)\nonu \\
&-& 3 c^4 (849+22764 N-60892 N^2+55312 N^3-19104 N^4+2304 N^5)+81 c N^3 (9087-8986 N\nonu \\
&-& 22192 N^2+39448 N^3-20464 N^4+3264 N^5)+9 c^3 N (5175+39944 N-139868 N^2\nonu \\
&+& 153512 N^3-63808 N^4+7008 N^5)-27 c^2 N^2 (10843+18486 N-117332 N^2+155432 N^3\nonu \\
&-& 80288 N^4+13952 N^5)), \nonu \\
c_{86} &=& 7560 c^4 D (c-3 N)^2 (1-2 N)^2 (-1+N) N^2 (1+N) (c+6 N) (12 N+c (-5+2 N))\nonu \\
& \times &  (2 c^2 (-9+2 N)+9 N (-13+10 N)-3 c (-5-40 N+36 N^2)), \nonu \\
c_{87} &=& 120 c^4  D (c-3 N)^2 (-1+N) N (1+N) (-1+2 N) (c+6 N)\nonu \\
 & \times & (4 c^4 (-51+952 N-368 N^2+88 N^3)-81 N^3 (-1785+3036 N-1948 N^2+528 N^3)\nonu \\
 &-& 6 c^3 (33+767 N+6704 N^2-6132 N^3+800 N^4)-54 c N^2 (1305+967 N-1036 N^2\nonu \\
 &-& 1636 N^3+816 N^4)+9 c^2 N (907+6580 N+5964 N^2-14048 N^3+3136 N^4)), \nonu \\
 c_{89} &=& -\frac{120 c^4 D (c-3 N)^2 (-1+N) (1+N)}{(21+4 c) N (-1+2 N)} (10206 N^6 (17893080-24439185 N+12259768 N^2\nonu \\
 &-& 2700208 N^3+221584 N^4)+243 c N^5 (-2162240736+4287473215 N-3334762892 N^2\nonu \\
 &+& 1281132480 N^3-246201968 N^4+19254544 N^5)+81 c^2 N^4 (7281651520-17539261449 N\nonu \\
 &+& 16471340390 N^2-7855180400 N^3+2042274960 N^4-280227280 N^5+16294368 N^6)\nonu \\
 &+& 54 c^3 N^3 (-6262982621+17106009721 N-18109557310 N^2+9857352680 N^3\nonu \\
 &-& 3013592448 N^4+ 517744720 N^5-45340768 N^6+1421824 N^7)-18 c^4 N^2 (-5876547021\nonu \\
 &+& 17572662207 N- 20305606131 N^2+12137100162 N^3-4129969160 N^4+812334064 N^5\nonu \\
 &-& 87262896 N^6+4098976 N^7 + 2560 N^8)+16 c^6 (71850240-243701568 N+318950691 N^2\nonu \\
 &-& 216782257 N^3+84751421 N^4-19616924 N^5+2627536 N^6-185400 N^7+5136 N^8)\nonu \\
 &-& 12 c^5 N (1436659200-4606587915 N+5700740241 N^2-3659607901 N^3+1346504472 N^4\nonu \\
 &-& 290728400 N^5+35603168 N^6-2213424 N^7+54592 N^8)), \nonu \\
 c_{90} &=& 2520 c^4 D  (c-3 N)^3 (-1+N) N (1+N) (-1+2 N) (c+6 N)\nonu \\
& \times &  (12 N+c (-5+2 N)) (2 c^2 (-9+2 N)+9 N (-13+10 N)-3 c (-5-40 N+36 N^2)),\nonu \\
c_{91} &=& 120 c^4 D (c-3 N)^2 (-1+N) N (1+N) (-1+2 N) (c+6 N) (4 c^4 (-51+7 N+220 N^2+4 N^3)\nonu \\
&-& 81 N^3 (-693+2196 N-1948 N^2+528 N^3)+6 c^3 (-33+703 N-1790 N^2+420 N^3\nonu \\
&+& 712 N^4)+ 18 c^2 N (191-1246 N+4116 N^2-5176 N^3+1568 N^4)\nonu \\
&-& 27 c N^2 (825-2014 N+2212 N^2-3272 N^3+1632 N^4)), \nonu \\
c_{92} &=& 5040 c^4 D (c-3 N)^3 (-1+N) N (1+N) (-1+2 N) (c+6 N) (6 N+c (-3+2 N))\nonu \\
& \times &  (2 c^2 (-9+2 N)+9 N (-13+10 N)-3 c (-5-40 N+36 N^2)), \nonu \\
c_{94} &=& -2520 c^4 D  (c-3 N)^4 (-1+N) N (1+N) (-1+2 N) (c+6 N)\nonu \\
& \times & (2 c^2 (-9+2 N)+9 N (-13+10 N)-3 c (-5-40 N+36 N^2)), \nonu \\
c_{95} &=& -2520 c^4 D (c-3 N)^3 (-1+N) N (1+N) (-1+2 N) (c+6 N)(c+2 c N\nonu \\
&+& 12 (-1+N) N) (2 c^2 (-9+2 N)+9 N (-13+10 N)-3 c (-5-40 N+36 N^2)), \nonu \\
c_{97} &=& 120 c^4 D (c-3 N)^2 (-1+N) N (1+N) (-1+2 N) (c+6 N) (c+2 c N\nonu \\
&+& 3 N (-3+2 N))^2 (2 c^2 (-51+22 N)-9 N (7+66 N)+c (-99+984 N-612 N^2)),\nonu \\
c_{99} &=& 6480 c^4 D (c-3 N)^3 (1-2 N)^2 (-1+N) N^2 (126 N^2 (9-47 N+2 N^2)\nonu \\
&+& c^3 (24-66 N+20 N^2)+c^2 (210-1141 N+736 N^2-60 N^3)\nonu \\
&+& 3 c N (-329+1729 N-658 N^2+120 N^3)), \nonu \\
c_{100} &=& 120 c^4 D (c-3 N) (-2916 N^6 (38106977-53043766 N+27498349 N^2-6436712 N^3\nonu \\
&+& 595164 N^4)+486 c N^5 (371362536-345913714 N+31665557 N^2+58140419 N^3\nonu \\
&-& 21702822 N^4 +2377144 N^5)+81 c^2 N^4 (-1380185935+21861735 N\nonu \\
&+& 1823118762 N^2-1384713438 N^3+430669664 N^4-62207024 N^5+3473424 N^6)\nonu \\
&+& 27 c^3 N^3 (1140198925+2601110708 N-5765670627 N^2+3917400066 N^3\nonu \\
&-& 1282992992 N^4+218777456 N^5-18042288 N^6+479584 N^7)+2 c^6 (34145280\nonu \\
&-& 318753852 N+508171813 N^2-359262379 N^3+137626932 N^4-30304928 N^5\nonu \\
&+& 3784712 N^6-243456 N^7+5568 N^8)+18 c^4 N^2 (-140155451-2099436457 N\nonu \\
&+& 3687470901 N^2-2496491546 N^3+867354436 N^4-165291200 N^5+16651520 N^6\nonu \\
&-& 745344 N^7+9792 N^8)-3 c^5 N (138290040-2693063119 N+4402083017 N^2\nonu \\
&-& 3036819792 N^3+1111878548 N^4-229862352 N^5+26270960 N^6\nonu \\
&-& 1474208 N^7+26112 N^8)), \nonu \\
c_{101} &=& -360 c^4 D  (c-3 N)^3 (-1+N) N (-1+2 N) (-378 N^3 (-61-242 N\nonu \\
&+& 128 N^2+48 N^3)+2 c^4 (-126-385 N-158 N^2+136 N^3+80 N^4)-6 c^2 N (434-3655 N\nonu \\
&+& 175 N^2-1730 N^3+864 N^4)-9 c N^2 (-84+13153 N-5590 N^2-1880 N^3+2880 N^4)\nonu \\
&+& c^3 (210+1571 N+1742 N^2-4040 N^3+592 N^4+1440 N^5)), \nonu \\
c_{103} &=& -\frac{4}{77} c^4 D  (c-3 N)^3 (-1+2 N) (1309770 N^4 (41+86 N-836 N^2+656 N^3)\nonu \\
&-& 31185 c N^3 (-4326+9933 N-24296 N^2+7500 N^3+6368 N^4)\nonu \\
&-& 18 c^2 N^2 (4877190+618411 N-1430930 N^2+2041840 N^3-3057520 N^4+3873824 N^5)\nonu \\
&-& 6 c^3 N (-2700495-9200196 N+18224908 N^2-16032350 N^3+6579680 N^4+586616 N^5\nonu \\
&+& 19392 N^6)+4 c^4 (-241920-1090971 N-1166052 N^2+5022886 N^3-3836580 N^4\nonu \\
&+& 993616 N^5-132768 N^6+6784 N^7)), \nonu \\
c_{104} &=& -180 c^4 D (c-3 N)^3 N (-1+2 N) (-378 N^3 (-231+322 N-448 N^2+304 N^3)\nonu \\
&+& 4 c^4 (-63-175 N-129 N^2+88 N^3+60 N^4)-9 c N^2 (4326-7301 N+28120 N^2\nonu \\
&-& 22324 N^3+2000 N^4)+6 c^3 (35+751 N-795 N^2-480 N^3-60 N^4+640 N^5)\nonu \\
&+& 6 c^2 N (-210-408 N+12307 N^2-11152 N^3+3052 N^4+1680 N^5)), \nonu \\
c_{105} &=& 240 c^4 D  (-3+2 c) (c-3 N) (-1+N) N (1+N) (-1+2 N) (c+6 N)\nonu \\
&\times &  (6 N+c (-3+2 N)) (c+2 c N+3 N (-3+2 N))^2 (c (23+6 N)+3 N (-35+18 N)), \nonu \\
c_{106} &=& 180 c^4 D (c-3 N) (1-2 N)^2 (-1+N) N^2 (4 c^5 (21-86 N-214 N^2-94 N^3\nonu \\
&+& 148 N^4)+486 N^4 (-917+771 N+1886 N^2-2400 N^3+1056 N^4)-6 c^4 (-663+2499 N\nonu \\
&-& 1887 N^2-1460 N^3+716 N^4+160 N^5)+54 c^2 N^2 (911-11736 N+16411 N^2-2588 N^3\nonu \\
&-& 6148 N^4+3520 N^5)-9 c^3 N (4615-23999 N+25082 N^2+5048 N^3-18008 N^4\nonu \\
&+& 6160 N^5)+81 c N^3 (3781-515 N-7262 N^2+9120 N^3-13696 N^4+6528 N^5)), \nonu \\
c_{107} &=& \frac{4}{77} c^4 D (c-3 N)^2 (1122660 N^5 (34175025-47493304 N+24542619 N^2\nonu \\
&-& 5702504 N^3+517524 N^4)+187110 c N^4 (-367403364+562199399 N-336217961 N^2 \nonu \\
&+& 99285338 N^3-14644828 N^4+870056 N^5)-27 c^2 N^3 (-1775992371990\nonu \\
&+& 2785716629931 N - 1564546209467 N^2+314684799230 N^3+27095663760 N^4\nonu \\
&-& 19614412856 N^5+2176569872 N^6)-27 c^3 N^2 (603184238670-907088343479 N\nonu \\
&+& 352626759469 N^2+107907557168 N^3-125049798260 N^4+39061512544 N^5\nonu \\
&-& 5496174944 N^6+312372992 N^7)+4 c^5 (-43030189440+42053655483 N\nonu \\
&+& 48678116280 N^2-92590179965 N^3+57274065706 N^4-18051115988 N^5\nonu \\
&+& 3117707380 N^6-280385360 N^7+9708784 N^8)-6 c^4 N (-448247199015\nonu \\
&+& 591513419034 N+19044018070 N^2-402862378847 N^3+272830776122 N^4\nonu \\
&-& 83887701924 N^5+13443154720 N^6-1083809248 N^7+38267168 N^8)), \nonu \\
c_{108} &=& -120 c^4 D (c-3 N) (-1+N) N (1+N) (-1+2 N) (c+6 N)\nonu \\
 &\times & (-972 N^4 (-1155+3285 N-3008 N^2+900 N^3)+4 c^5 (-405-144 N+968 N^2\nonu \\
 &-& 640 N^3+176 N^4)+81 c N^3 (-14613+30842 N-13220 N^2-6408 N^3+4128 N^4)\nonu \\
 &-& 18 c^3 N (2823+12888 N-37806 N^2+29892 N^3-8472 N^4+560 N^5)\nonu \\
 &-& 6 c^4 (-363-6378 N+8178 N^2+2316 N^3-5464 N^4+1264 N^5)\nonu \\
 &+& 27 c^2 N^2 (14729-1222 N-64852 N^2+85272 N^3-44160 N^4+8832 N^5)), \nonu \\
 c_{109} &=& -7560 c^4 D  (c-3 N)^2 (1-2 N)^2 (-1+N) N^2 (1+N) (c+6 N)\nonu \\
 & \times &  (6 N+c (-3+2 N)) (2 c^2 (-9+2 N)+9 N (-13+10 N)-3 c (-5-40 N+36 N^2)), \nonu \\
 c_{110} &=& -2520 c^4 D (c-3 N)^2 (-1+N) N (1+N) (-1+2 N) (c+6 N)\nonu \\
 &\times &  (c+2 c N+12 (-1+N) N) (6 N+c (-3+2 N)) (2 c^2 (-9+2 N)\nonu \\
 &+& 9 N (-13+10 N)-3 c (-5-40 N+36 N^2)), \nonu \\
 c_{111} &=& 720 c^4 D (c-3 N)^2 (-1+N) N (1+N) (-1+2 N) (c+6 N) \nonu \\
 & \times &(c+2 c N+3 N (-3+2 N))^2 (9 (21-34 N) N+c^2 (6+4 N)+c (-51+36 N+36 N^2)), \nonu \\
 c_{112} &=& -720 c^4 D (c-3 N) (1-2 N)^2 (-1+N) N^2 (6 N+c (-3+2 N))\nonu \\
& \times & (2 c^4 (8-71 N-41 N^2+74 N^3)-9 c^2 N (-571+2290 N+1044 N^2-1654 N^3+172 N^4)\nonu \\
&-& 162 N^3 (266-215 N-374 N^2-22 N^3+480 N^4)\nonu \\
&-& 3 c^3 (260-441 N-1205 N^2-346 N^3+560 N^4)\nonu \\
&+& 27 c N^2 (89+2063 N-1559 N^2-2490 N^3+2600 N^4)), \nonu \\
c_{114} &=& \frac{8}{77} c^4 D  (c-3 N)^2 (4490640 N^5 (14721197-20535689 N+10681372 N^2\nonu \\
&-& 2511568 N^3+233088 N^4)+1496880 c N^4 (-77099786+114631341 N-66007098 N^2\nonu \\
&+& 18566610 N^3-2575093 N^4+141426 N^5)-54 c^2 N^3 (-1466209709190\nonu \\
&+& 2214447910908 N-1214642811541 N^2+255067120555 N^3+8530456440 N^4\nonu \\
&-& 11454947908 N^5+1340361376 N^6)-27 c^3 N^2 (988407540210-1447959585159 N\nonu \\
&+& 621188559129 N^2+58838623408 N^3-125696666020 N^4+39465602384 N^5\nonu \\
&-& 5181645344 N^6+249269952 N^7)+4 c^5 (-71650535040+80805966003 N\nonu \\
&+& 31308958320 N^2-88319669105 N^3+54396568006 N^4-16201803908 N^5\nonu \\
&+& 2560988140 N^6-204469520 N^7+6936784 N^8)-6 c^4 N (-736080466275\nonu \\
&+& 988543139814 N-165713029790 N^2-371661474497 N^3+267228917552 N^4\nonu \\
&-& 78882116724 N^5+11562371200 N^6-778408768 N^7+12210368 N^8)),\nonu \\
c_{115} &=& -1080 c^4 D (c-3 N)^2 (1-2 N)^2 (-1+N) N^2 (c^4 (324-202 N-44 N^2\nonu \\
&+& 32 N^3)-108 N^3 (189+214 N-1142 N^2+660 N^3)+c^3 (6-3521 N+6340 N^2-3792 N^3\nonu \\
&+& 696 N^4)-18 c N^2 (-1186+1641 N+873 N^2-2926 N^3+1872 N^4)\nonu \\
&-& 3 c^2 N (1658-7695 N+13740 N^2-10504 N^3+3384 N^4)), \nonu \\
c_{116} &=& \frac{8}{77} c^4 D  (c-3 N)^3 (2619540 N^4 (80522-116869 N\nonu \\
&+& 65865 N^2-18662 N^3+2584 N^4)+124740 c N^3 (-2928876+5595281 N-4326337 N^2\nonu \\
&+& 1727974 N^3-367862 N^4+35340 N^5)-9 c^2 N^2 (-23557535550+47188850859 N\nonu \\
&-& 32268867313 N^2+7401593860 N^3+1104317280 N^4-814885264 N^5+122416048 N^6)\nonu \\
&-& 6 c^3 N (8470667835-15987679002 N+6588462910 N^2+3316995473 N^3-3472049160 N^4\nonu \\
&+& 1088840152 N^5-158304880 N^6+13047792 N^7)+4 c^4 (1077874560-1714907817 N\nonu \\
&-& 307364880 N^2+1663675885 N^3-1055778944 N^4+293841412 N^5-37923560 N^6\nonu \\
&+& 1249840 N^7+6784 N^8)), \nonu \\
c_{118} &=& -120 c^4 D (c-3 N)^2 (-1+N) N (-1+2 N) (6 N+c (-3+2 N)) (8 c^4 (-146-97 N\nonu \\
&+& 131 N^2+82 N^3)+6 c^3 (60+1080 N-465 N^2-1010 N^3+232 N^4)-54 c N^2 (794+464 N\nonu \\
&+&4069 N^2 - 6226 N^3+688 N^4)-9 c^2 N (-678+529 N-6098 N^2+2600 N^3+3344 N^4)\nonu \\
&+& 81 N^3 (553+1857 N+1082 N^2-7344 N^3+3840 N^4)), \nonu \\
c_{119} &=& 6480 c^4 D (3+c-9 N) (c-3 N)^2 (1-2 N)^2 (-1+N) N^2 \nonu \\
&\times & (2 c (-2+N)+3 (21-8 N) N) (2 c (-1+N)+3 (3-4 N) N) (6 N+c (-3+2 N)), \nonu \\
c_{120} &=& 15120 c^4 D  (c-3 N)^3 (1-2 N)^2 (-1+N) N^2 (6 N+c (-3+2 N))\nonu \\
&\times & (10 c^2 (1+N)+c (7-59 N+42 N^2)-3 N (73-233 N+216 N^2)), \nonu \\
c_{122} &=& 6480 c^4 D (c-3 N)^3 (1-2 N)^2 (-1+N) N^2 (126 N^2 (9-18 N+2 N^2)\nonu \\
&+& 10 c^3 (1-N+2 N^2)+c^2 (7-287 N+484 N^2-60 N^3)\nonu \\
&+& 3 c N (-126+511 N-574 N^2+120 N^3)), \nonu \\
c_{123} &=& 1080 c^4 D  (c-3 N)^3 (1-2 N)^2 (-1+N) N^2 (2 c^3 (54-103 N\nonu \\
&+& 38 N^2)+126 N^2 (101-376 N+132 N^2)+c^2 (462-4459 N+4376 N^2\nonu \\
&-& 1236 N^3)+3 c N (-2632+12495 N-8918 N^2+1296 N^3)), \nonu \\
c_{125} &=&  -11340 c^4 D (29+2 c) (c-3 N)^4 (1-2 N)^2 (-1+N) N^2 (-6 N^2+c (-1+4 N)), \nonu \\
c_{126} &=& \frac{4}{77} c^4 D (c-3 N)^3 (-1309770 N^4 (622278-901903 N+498045 N^2\nonu \\
&-& 128024 N^3+13284 N^4)+31185 c N^3 (34417063-49030109 N+26353862 N^2\nonu \\
&-& 6501160 N^3+625608 N^4+9936 N^5)+18 c^2 N^2 (-29451109755+41244298176 N\nonu \\
&-& 21578463917 N^2+5089232180 N^3-426586380 N^4-34917536 N^5+7442912 N^6)\nonu \\
&-& 12 c^3 N (-9698153220+13347002913 N-6785560445 N^2+1521690038 N^3\nonu \\
&-& 117136830 N^4-7236368 N^5+1574840 N^6+19392 N^7)+16 c^4 (-598691520\nonu \\
&+& 808526754 N-395608740 N^2+78197555 N^3-164197 N^4-2894344 N^5\nonu \\
&+& 627020 N^6-68080 N^7+3392 N^8)), \nonu \\
c_{127} &=& -60 c^4 D (-3+2 c) (c-3 N) (-1+N) N (1+N) (-1+2 N) (c+6 N)\nonu \\
& \times &  (6 N+c (-3+2 N)) (c+2 c N+3 N (-3+2 N)) (2 c^2 (138+67 N+2 N^2)\nonu \\
&+& 3 c N (-1255+1000 N+268 N^2)+9 N^2 (1323-2510 N+1104 N^2)), \nonu \\
c_{128} &=& -540 c^4 D (c-3 N)^2 (1-2 N)^2 (-1+N) N^2 (12 c^4 (11+32 N\nonu \\
&-& 28 N^2-22 N^3+12 N^4)+54 N^3 (-1617+3571 N+610 N^2-5448 N^3+2784 N^4)\nonu \\
&-& 2 c^3 (423-587 N+1177 N^2-1104 N^3-24 N^4+504 N^5)+9 c N^2 (3347-2973 N\nonu \\
&-& 10386 N^2+14552 N^3-5232 N^4+1152 N^5)-3 c^2 N (-181+2517 N-4608 N^2\nonu \\
&+& 3464 N^3-5016 N^4+3168 N^5)), \nonu \\
c_{129} &=& -7560 c^4 D (3+c-9 N) (c-3 N)^3 (1-2 N)^2 (-1+N) N^2 \nonu \\
&\times &(6 N+c (-3+2 N)) (2 c (4+N)+3 N (-13+16 N)), \nonu \\
c_{130} &=& 1080 c^4 D  (c-3 N)^2 (1-2 N)^2 (-1+N) N^2 (2 c^4 (228+57 N\nonu \\
&-& 170 N^2+16 N^3)+54 N^3 (-483+412 N-2924 N^2+2880 N^3)+c^3 (-288-2039 N\nonu \\
&+& 2938 N^2+120 N^3-1440 N^4)-6 c^2 N (388-561 N+5439 N^2-4778 N^3+1632 N^4)\nonu \\
&+& 9 c N^2 (2162-2211 N+15894 N^2-17200 N^3+5760 N^4)), \nonu \\
c_{131} &=& 6480 c^4 D (29+2 c) (c-3 N)^2 (1-2 N)^2 (-1+N) N^2 (-6 N^2+c (-1+4 N))\nonu \\
& \times & (-18 N^2 (8-9 N+4 N^2)+c^2 (-9-2 N+4 N^2)-3 c N (-23+8 N+4 N^2)), \nonu \\
c_{132} &=& 2160 c^4 D  (c-3 N)^2 (1-2 N)^2 (-1+N) N^2 (-6 N^2+c (-1+4 N))\nonu \\
&\times &  (c^3 (-42-22 N+44 N^2)-189 N^2 (-45+32 N+48 N^2)\nonu \\
&+& 3 c^2 (119+57 N+76 N^2-44 N^3)-9 c N (518-461 N-282 N^2+336 N^3)), \nonu \\
c_{133} &=& -1080 c^4 (c-3 N)^2 (1-2 N)^2 (-1+N) N^2 (c^4 (-432-202 N\nonu \\
&+& 292 N^2+32 N^3)+54 N^3 (483-736 N-1300 N^2+528 N^3)+c^3 (-1296+7439 N\nonu \\
&+& 2510 N^2-6264 N^3+1152 N^4)-9 c N^2 (2486-4803 N-8922 N^2+4208 N^3+1152 N^4)\nonu \\
&-& 6 c^2 N (-1768+6537 N+1953 N^2-5710 N^3+2928 N^4)), \nonu \\
c_{134} &=& -2520 c^4 D (c-3 N)^3 (-1+N) N (-1+2 N) (6 N+c (-3+2 N))\nonu \\
& \times &  (8 c^3 (4+7 N+3 N^2)+2 c^2 (2-93 N+4 N^2+72 N^3)\nonu \\
&-& 18 N^2 (-98+303 N-616 N^2+288 N^3)-3 c N (155-354 N+580 N^2+288 N^3)),\nonu \\
c_{135} &=& -360 c^4 D (c-3 N)^2 (-1+N) N (-1+2 N) (4 c^5 (18+9 N-17 N^2-4 N^3+4 N^4)\nonu \\
&-& 162 N^4 (-2275+9100 N-10352 N^2+2504 N^3+480 N^4)-2 c^4 (30+1105 N-1062 N^2\nonu \\
&-& 800 N^3+440 N^4+48 N^5)+18 c^2 N^2 (2588-13713 N+22182 N^2-15616 N^3+4248 N^4\nonu \\
&+& 432 N^5)-3 c^3 N (626-9181 N+15616 N^2-5224 N^3-3200 N^4+816 N^5)\nonu \\
&+& 27 c N^3 (-9100+39135 N-54852 N^2+36724 N^3-17680 N^4+3456 N^5)), \nonu \\
c_{136} &=& 5 c^4 D (-3+2 c) (c-3 N) (1-2 N)^2 (-1+N) N (1+N) (c+6 N) (6 N+c (-3+2 N))\nonu \\
&\times & (c+2 c N+3 N (-3+2 N)) (c (23+6 N)+3 N (-35+18 N))\nonu \\
& \times & (2 c (5+N)+3 N (-27+32 N)), \nonu \\
c_{137} &=& -20 c^4 D  (-3+2 c) (c-3 N) (-1+N) N (-1+2 N) (c+6 N) (6 N+c (-3+2 N))\nonu \\
 & \times &(2 c^3 (-138-586 N-545 N^2+400 N^3+548 N^4+96 N^5)+3 c^2 N (1988+4567 N\nonu \\
  &-& 2812 N^2-6628 N^3+1520 N^4+1440 N^5)+9 c N^2 (-4529-1968 N+11286 N^2+3000 N^3\nonu \\
  &-& 9928 N^4+1824 N^5)-27 N^3 (-3255+3899 N+3208 N^2-2060 N^3-4336 N^4+2304 N^5)),\nonu  \\
c_{138} &=& -180 c^4 D (c-3 N) (1-2 N)^2 (-1+N) N^2 (c+6 N) (6 N+c (-3+2 N))\nonu \\
 & \times & (2 c^3 (-9+61 N+136 N^2+36 N^3)-c^2 (-107+283 N+600 N^2+2844 N^3)\nonu \\
 &-& 6 c N (319-1317 N+1307 N^2-2400 N^3+684 N^4)+9 N^2 (735-3593 N+5728 N^2\nonu \\
 &-& 5268 N^3+1728 N^4)), \nonu \\
c_{141} &=& 20 c^4 D (c-3 N)^2 (-1+N) N (-1+2 N) (c+6 N) (6 N+c (-3+2 N)) \nonu \\
&&(10 c^3 (23+148 N+286 N^2+176 N^3+24 N^4)-54 N^2 (357+1621 N-3031 N^2\nonu \\
&+& 364 N^3+396 N^4)+3 c^2 (-115-2058 N-7322 N^2-2928 N^3+6168 N^4+2400 N^5)\nonu \\
&+& 9 c N (619+5022 N+4148 N^2-7460 N^3-8192 N^4+5040 N^5)), \nonu \\
c_{142} &=& 30 c^4 D (c-3 N) (-1+N) N^2 (-1+2 N) (4 c^6 (-93+1361 N-836 N^2\nonu \\
&-& 1556 N^3+752 N^4+288 N^5)-2916 N^4 (-623+4630 N-10446 N^2+11948 N^3-8416 N^4 \nonu \\
&+& 2304 N^5)+6 c^5 (-252-143 N-4422 N^2+7200 N^3-2360 N^4-2928 N^5+3072 N^6)\nonu \\
&+& 486 c N^3 (-4077+33908 N-80059 N^2+88604 N^3-47148 N^4+544 N^5+4608 N^6)\nonu \\
&-& 108 c^3 N (1722-14042 N+26152 N^2-7925 N^3-25694 N^4+28676 N^5-9544 N^6\nonu \\
&& +1344 N^7) +9 c^4 (1317-8478 N+10606 N^2+13164 N^3-32936 N^4+28368 N^5\nonu \\
&-& 15104 N^6+4224 N^7) +81 c^2 N^2 (11781-101906 N+233586 N^2-204800 N^3\nonu \\
&-& 11624 N^4+153792 N^5-92672 N^6+18432 N^7)), \nonu \\
c_{143} &=& -180 c^4 D (c-3 N) (1-2 N)^2 (-1+N) N^2 (c^5 (-690+892 N-448 N^2\nonu \\
&-& 808 N^3+592 N^4)+972 N^4 (-595+1746 N+2266 N^2-6072 N^3+1920 N^4)\nonu \\
&-& 18 c^3 N (-1732+9161 N-641 N^2-3470 N^3-580 N^4+1784 N^5)-3 c^4 (483-5070 N\nonu \\
&+&3144 N^2-2776 N^3-1808 N^4+2048 N^5)+162 c N^3 (3797-13138 N-8941 N^2\nonu \\
&+& 33366 N^3-16520 N^4+3840 N^5)+27 c^2 N^2 (-8063+32586 N+9692 N^2-46144 N^3\nonu \\
&+& 12256 N^4+3872 N^5)), \nonu \\
c_{144} &=& -180 c^4 D (c-3 N)^2 (-1+N) N (-1+2 N) (648 N^4 (-434-872 N+1307 N^2\nonu \\
&-& 10 N^3+72 N^4)+2 c^5 (-69-692 N-1332 N^2-152 N^3+672 N^4+160 N^5)\nonu \\
&-& 18 c^2 N^2 (-1816+19833 N+8928 N^2-14758 N^3-648 N^4+6168 N^5)\nonu \\
&-& 54 c N^3 (-69-26298 N+21973 N^2+6836 N^3-14204 N^4+6336 N^5)+c^4 (207+3858 N\nonu \\
&+& 14104 N^2-3812 N^3-18448 N^4+5840 N^5+6720 N^6)+3 c^3 N (-1791+942 N\nonu \\
&+& 11314 N^2+28840 N^3-4360 N^4-22432 N^5+11520 N^6)), \nonu \\
c_{145} &=& -540 c^4 D (c-3 N) (1-2 N)^2 (-1+N) N^2 (-324 N^4 (-945+3146 N\nonu \\
&-& 2722 N^2+1104 N^3)+2 c^5 (171+88 N-272 N^2-12 N^3+72 N^4)\nonu \\
&-& 54 c N^3 (5197-17938 N+14987 N^2-7250 N^3+360 N^4)-c^4 (-267+5432 N+ 2492 N^2\nonu \\
&-&8712 N^3+2400 N^4+288 N^5)-6 c^3 N (1351-7120 N-2569 N^2+8830 N^3-5556 N^4\nonu \\
&+& 1944 N^5)+9 c^2 N^2 (8701-30414 N+8136 N^2+9904 N^3-11184 N^4+2592 N^5)), \nonu \\
c_{146} &=& -540 c^4 D (c-3 N)^2 (1-2 N)^2 (-1+N) N^2 (6 c^4 (-10-89 N+18 N^2+32 N^3)\nonu \\
&+&54 N^3 (-1295+6886 N-5316 N^2+1728 N^3)+c^3 (1062-7027 N+6866 N^2\nonu \\
&-& 1032 N^3+576 N^4)-6 c^2 N (2330-14873 N+12791 N^2-5178 N^3+1152 N^4)\nonu \\
&+& 9 c N^2 (6452-38183 N+35846 N^2-18672 N^3+3456 N^4)), \nonu \\
c_{147} &=& 60 c^4 D (c-3 N)^2 (-1+N) N (-1+2 N) (8 c^5 (129-440 N-379 N^2+278 N^3\nonu \\
&+& 88 N^4)-486 N^4 (1470-12241 N+27122 N^2-19272 N^3+3840 N^4)-18 c^3 N (-2745\nonu \\
&+& 16740 N-12160 N^2-8556 N^3+5840 N^4+688 N^5)+12 c^4 (-225+755 N+2144 N^2\nonu \\
&-& 284 N^3-1212 N^4+776 N^5)-81 c N^3 (-10057+77854 N-156816 N^2+105696 N^3\nonu \\
&-& 35584 N^4\nonu + 7680 N^5)-27 c^2 N^2 (11629-81012 N+120764 N^2-23776 N^3\nonu \\
&-& 24896 N^4+13376 N^5)), \nonu \\
c_{148} &=& 540 c^4 D  (c-3 N)^2 (1-2 N)^2 (-1+N) N^2 (c^4 (-288-578 N-28 N^2+352 N^3)\nonu \\
&-& 378 N^3 (-245+1226 N-1868 N^2+576 N^3)+3 c^3 (-196+1889 N-310 N^2\nonu \\
&+& 88 N^3+32 N^4)-6 c^2 N (-2331+12779 N-15347 N^2+2114 N^3+2592 N^4)\nonu \\
&-& 9 c N^2 (7924-41951 N+68294 N^2-33616 N^3+8064 N^4)), \nonu \\
c_{149} &=& -90 c^4 D  (c-3 N)^3 (-1+N) N (-1+2 N) (16 c^4 (-33-56 N-2 N^2+6 N^3)\nonu \\
&+& 378 N^3 (24+1171 N-1466 N^2+432 N^3)-2 c^3 (-588+1097 N-1830 N^2+2896 N^3\nonu \\
&+& 432 N^4)-9 c N^2 (-6013+58688 N-55304 N^2+11800 N^3+576 N^4)\nonu \\
&-&3 c^2 N (5383-34120 N+15448 N^2-5304 N^3+6048 N^4)), \nonu \\
c_{150} &=& -180 c^4 D (c-3 N)^3 (-1+N) N^2 (-1+2 N) (4 c^4 (-357-101 N+256 N^2+60 N^3)\nonu \\
&+& 378 N^2 (-309+1432 N-1136 N^2+168 N^3)+c^3 (-30+6164 N-7444 N^2+3840 N^3\nonu \\
&+& 5520 N^4)+ 9 c N (3514-26383 N+36616 N^2-45660 N^3+19056 N^4)\nonu \\
&+& 3 c^2 (-714+10713 N-14096 N^2+18716 N^3-11984 N^4+10080 N^5)), \nonu \\
c_{151} &=& -3 c^5 D (-3+2 c) (-1+N) N (1+N) (-1+2 N)^3 (c+6 N) (6 N+c (-3+2 N))\nonu \\
&\times &  (c+2 c N+3 N (-3+2 N)) (c+2 N (-4+5 N)) (c (23+6 N)+3 N (-35+18 N)), \nonu \\
c_{152} &=& 5 c^5 D (-3+2 c) (-1+N) N (1+N) (-1+2 N)^3 (c+6 N) (6 N+c (-3+2 N))+(c+2 c N\nonu \\
&+& 3 N (-3+2 N)) (c (23+6 N)+3 N (-35+18 N)) (c (7+2 N)+3 N (-19+22 N)), \nonu \\
c_{153} &=& -30 c^4 D  (3+c-9 N) (c-3 N) (1-2 N)^2 (-1+N) N (1+N) (2 c (-1+N)\nonu \\
&+& 3 (3-4 N) N) (6 N+c (-3+2 N)) (72 N^3 (-21+23 N)+c^2 N (-1049+1304 N-252 N^2)\nonu \\
&+& c^3 (80-34 N+4 N^2)+3 c N^2 (1121-1962 N+816 N^2)), \nonu \\
c_{154} &=& 20 c^5 D (-3+2 c) (1-2 N)^2 (-1+N) N (1+N) (c+6 N) (6 N+c (-3+2 N))\nonu \\
& \times & (c+2 c N+3 N (-3+2 N)) (c^2 (253+110 N+48 N^2)\nonu \\
&+& 3 c N (-1123+912 N+108 N^2)+18 N^2 (567-1039 N+450 N^2)), \nonu \\
c_{155} &=& -60 c^4 D (c-3 N) (1-2 N)^2 (-1+N) N (c+6 N) (6 N+c (-3+2 N)) \nonu \\
& \times & (2 c^4 (-9+75 N+88 N^2+4 N^3)+54 N^3 (-280+1359 N-1976 N^2+648 N^3)\nonu \\
&+& c^3 (107-103 N-1724 N^2-836 N^3+768 N^4)-9 c N^2 (-1268+4955 N-3400 N^2\nonu \\
&-& 2732 N^3+864 N^4)-3 c^2 N (749-2202 N-1394 N^2+1800 N^3+1560 N^4)), \nonu \\
c_{156} &=& -30 c^5 D (-3+2 c) (-1+N) N (1+N) (-1+2 N)^3 (c+6 N) (6 N+c (-3+2 N))\nonu \\
&\times & (c+2 c N+3 N (-3+2 N)) (2 c (1+N)+N (-17+16 N))\nonu \\
& \times &  (c (23+6 N)+3 N (-35+18 N)), \nonu \\
c_{157} &=& -40 c^4 D (-3+2 c) (c-3 N) (-1+N) N (1+N) (-1+2 N) (c+6 N)\nonu \\
& \times &  (6 N+c (-3+2 N)) (c+2 c N+3 N (-3+2 N)) (c^2 (-23+223 N+262 N^2+48 N^3)\nonu \\
&+& 3 c N (80-1133 N+480 N^2+372 N^3)+9 N^2 (-63+1210 N-1870 N^2+684 N^3)), \nonu \\
c_{158} &=& 30 c^4 D (c-3 N)^2 (1-2 N)^2 (-1+N) N (c+6 N) (6 N+c (-3+2 N))\nonu \\
& \times &  (5 c+6 N (-4+3 N)) (609 N+c^2 (22+24 N)-c (49+266 N+288 N^2)), \nonu \\
c_{159} &=& 40 c^4 D (3+c-9 N) (c-3 N) (1-2 N)^2 (-1+N) N (2 c (-1+N)\nonu \\
&+& 3 (3-4 N) N) (6 N+c (-3+2 N)) (189 N^3 (-9-40 N+56 N^2)\nonu \\
&+& 3 c^2 N (-1293+241 N+1612 N^2-468 N^3)+2 c^3 (172+69 N-115 N^2+24 N^3)\nonu \\
&+& 9 c N^2 (1183-612 N-1774 N^2+1104 N^3)), \nonu \\
c_{160} &=& 60 c^4 D (c-3 N) (1-2 N)^2 (-1+N) N (c+6 N) (6 N+c (-3+2 N))\nonu \\
 &\times & (2 c^4 (-145+247 N+556 N^2+164 N^3)+243 N^3 (-315+1672 N-1748 N^2+528 N^3)\nonu \\
 &+& 3 c^3 (193-17 N-2888 N^2-2652 N^3+80 N^4)-9 c^2 N (931-3480 N-4090 N^2+128 N^3\nonu \\
 &+& 824 N^4)-27 c N^2 (-1591+7453 N-1340 N^2-3028 N^3+1056 N^4)), \nonu \\
 c_{161} &=& 180 c^4 D (c-3 N) (1-2 N)^2 (-1+N) N^2 (c+6 N) (6 N+c (-3+2 N))\nonu \\
  & \times & (c^3 (290+430 N+328 N^2+296 N^3)-18 c N (-407+1458 N-3727 N^2+992 N^3\nonu \\
  &+& 284 N^4)-3 c^2 (193+211 N+2104 N^2+612 N^3+560 N^4)\nonu \\
  &-& 27 N^2 (805-4309 N+10196 N^2-8156 N^3+2112 N^4)), \nonu \\
c_{163} &=& -90 c^4 D (c-3 N)^2 (-1+N) N (-1+2 N) (c+6 N) (6 N+c (-3+2 N))\nonu \\
 & \times & (-63 N^2 (-77+794 N-788 N^2+168 N^3)+2 c^3 (-33+394 N+604 N^2+168 N^3)\nonu \\
 &-& 6 c N (217-3059 N-1042 N^2+20 N^3+936 N^4)\nonu \\
 &+& c^2 (147-1898 N-6796 N^2-4712 N^3+1056 N^4)), \nonu \\
 c_{164} &=& \frac{c^4  D}{480} (c-3 N) (-1+N) N^2 (-1+2 N) (c+6 N) (6 N+c (-3+2 N)) (-81 N^4 (-74647572\nonu \\
 &+& 60720040 N-4617815 N^2-2106280 N^3+103187 N^4)+27 c N^3 (-358514412\nonu \\
 &+& 536854368 N-451661305 N^2+167280650 N^3-11536283 N^4+579982 N^5)\nonu \\
&-& 18 c^2 N^2 (-211413264+117831380 N+147145572 N^2-202135535 N^3+66630674 N^4\nonu \\
&-& 9293505 N^5+486598 N^6)+6 c^3 N (-70018132-219125424 N+440138353 N^2\nonu \\
&-& 168239926 N^3-22348943 N^4+20206034 N^5-3698398 N^6+220316 N^7)\nonu \\
&+& 4 c^4 (691200+45826132 N-22606704 N^2-79113969 N^3+63441646 N^4\nonu \\
&-& 20922557 N^5+3325154 N^6-241166 N^7+5664 N^8)), \nonu \\
c_{165} &=& \frac{1}{24} c^4 D (-3+2 c) (c-3 N) (-1+N) N^2 (-1+2 N) (c+6 N) (6 N+c (-3+2 N))\nonu \\
& \times & (135 N^3 (642420-313100 N-493337 N^2+320150 N^3+599 N^4)-9 c N^2 (7597620\nonu \\
&-& 1404884 N-3834301 N^2+823824 N^3-11651 N^4+848 N^5)-6 c^2 N (-3035232\nonu \\
&-& 249388 N+444178 N^2+191091 N^3+95993 N^4-14606 N^5+798 N^6)+4 c^3 (-408492\nonu \\
&-& 158380 N+15523 N^2-157740 N^3-833 N^4+11805 N^5-2513 N^6+160 N^7)), \nonu \\
c_{166} &=& 60 c^5 D  (-3+2 c) (1-2 N)^2 (-1+N) N (1+N) (c+6 N) (6 N+c (-3+2 N))\nonu \\
& \times &  (c+2 c N+3 N (-3+2 N))^2 (9 (21-34 N) N+c (-23+22 N)), \nonu \\
c_{167} &=& -60 c^5 D (-3+2 c) (1-2 N)^2 (-1+N) N (1+N) (c+6 N)\nonu \\
& \times &  (6 N+c (-3+2 N)) (c+2 c N+3 N (-3+2 N))^2 (c (23+6 N)+3 N (-35+18 N)), \nonu \\
c_{168} &=& 180 c^4 D (c-3 N) (1-2 N)^2 (-1+N) N (c+6 N) (6 N+c (-3+2 N)) \nonu \\
& \times & (-2349 N^4 (25-50 N+16 N^2)+4 c^4 (-6-7 N+N^2+2 N^3)+c^3 (20\nonu \\
&+& 798 N+44 N^2-704 N^3-96 N^4)-3 c^2 N (50+2893 N-1902 N^2-2804 N^3+216 N^4)\nonu \\
&+& 18 c N^2 (25+2075 N-2770 N^2-872 N^3+432 N^4)), \nonu \\
c_{169} &=& 30 c^4 D (-3+2 c) (c-3 N) (-1+N) N (1+N) (-1+2 N) (c+6 N) (6 N+c (-3+2 N))\nonu \\
& \times &  (c+2 c N+3 N (-3+2 N)) (c (23+6 N)+3 N (-35+18 N))\nonu \\
& \times & (2 c (-2+5 N+6 N^2)+3 N (11-46 N+32 N^2)), \nonu \\
c_{170} &=& -2520 c^4 D (c-3 N)^2 (1-2 N)^2 (-1+N) N (-162 N^4 (-301+254 N+468 N^2)\nonu \\
&+& 4 c^5 (33-7 N-28 N^2+12 N^3)+27 c N^3 (-1248+1037 N+1826 N^2+360 N^3)\nonu \\
&+& c^4 (258-2038 N+198 N^2+1684 N^3-720 N^4)+3 c^3 N (-520+2645 N+242 N^2\nonu \\
&-& 1840 N^3+432 N^4)+18 c^2 N^2 (382-406 N-1055 N^2-222 N^3+648 N^4)), \nonu \\
c_{171} &=& 2520 c^4 D (c-3 N)^3 (1-2 N)^2 (-1+N) N (6 N+c (-3+2 N))\nonu \\
& \times & (1566 N^3+6 c^3 (1+2 N)+c^2 (-5+64 N-72 N^2)-3 c N (-77+328 N+108 N^2)), \nonu \\
c_{173} &=& 1260 c^4  D (c-3 N)^4 (-1+N) N (-1+2 N) (6 N+c (-3+2 N))\nonu \\
& \times & (18 c^2+c (77-256 N+36 N^2)-6 N (10-107 N+144 N^2)), \nonu \\
c_{174} &=& 3780 c^4 D (29+2 c) (c-3 N)^5 (-1+N) N (-1+2 N) (-6 N^2+c (-1+4 N)), \nonu \\
c_{175} &=& 120 c^4 D (c-3 N) (-1+N) N (1+N) (-1+2 N) (2916 N^5 (-371-1195 N+4438 N^2\nonu \\
&-& 3840 N^3+1008 N^4)+2 c^6 (447+683 N-638 N^2-800 N^3+296 N^4+144 N^5)\nonu \\
&-& 486 c N^4 (-1647-13112 N+32341 N^2-20622 N^3+3300 N^4+168 N^5)\nonu \\
&+& 9 c^4 N (1447-1210 N-1448 N^2-13232 N^3+24192 N^4-11104 N^5+576 N^6)\nonu \\
&+& 3 c^5 (-279-3961 N-2126 N^2+11772 N^3-4392 N^4-2752 N^5+1536 N^6)\nonu \\
&-& 162 c^2 N^3 (580+22078 N-33593 N^2+682 N^3+18504 N^4-11688 N^5+2736 N^6)\nonu \\
&-& 27 c^3 N^2 (1842-27517 N+21888 N^2+8344 N^3+3584 N^4-14032 N^5+6144 N^6)), \nonu \\
c_{176} &=& 720 c^4 D (c-3 N) (1-2 N)^2 (-1+N) N^2 (1+N) (2 c^5 (57-113 N-140 N^2+148 N^3)\nonu \\
&+& 972 N^4 (-266+1507 N-1694 N^2+528 N^3)-3 c^4 (369-2227 N+552 N^2+148 N^3\nonu \\
&+& 64 N^4)+162 c N^3 (1913-11287 N+13398 N^2-5864 N^3+1056 N^4)\nonu \\
&-& 9 c^3 N (-2207+13530 N-9818 N^2-384 N^3+1928 N^4)\nonu \\
&-& 27 c^2 N^2 (4720-28695 N+31940 N^2-13500 N^3+2336 N^4)), \nonu \\
c_{178} &=& -120 c^4 D (c-3 N)^2 (-1+N) N (-1+2 N) (4 c^5 (-66-547 N-377 N^2+268 N^3\nonu \\
&+& 164 N^4)-486 N^4 (1834-8569 N+15776 N^2-10076 N^3+2112 N^4)\nonu \\
&+& 6 c^4 (-132+1940 N+1979 N^2-1004 N^3-580 N^4+736 N^5)\nonu \\
&-& 9 c^3 N (-2309+15418 N-11528 N^2-9308 N^3+5696 N^4+1328 N^5)\nonu \\
&-& 54 c^2 N^2 (3881-19339 N+29449 N^2-5564 N^3-5228 N^4+2704 N^5)\nonu \\
&-& 81 c N^3 (-9661+46774 N-88192 N^2+59152 N^3-21632 N^4+4224 N^5)), \nonu \\
c_{179} &=& -360 c^4 D (c-3 N)^2 (-1+N) N (-1+2 N) (-162 N^4 (3031-10342 N+24220 N^2\nonu \\
&-& 20832 N^3+5520 N^4)+4 c^5 (-3-269 N-587 N^2-88 N^3+268 N^4+80 N^5)\nonu \\
&+& 54 c^2 N^2 (-666+4433 N-22569 N^2+18028 N^3-6532 N^4+1232 N^5)\nonu \\
&-& 27 c N^3 (-10916+42899 N-149486 N^2+151548 N^3-65784 N^4+13248 N^5)\nonu \\
&+& 2 c^4 (51+1485 N+4206 N^2+1780 N^3-8236 N^4+1744 N^5+3120 N^6)\nonu \\
&+& 3 c^3 N (48-14121 N+44376 N^2-5456 N^3+3704 N^4-15248 N^5+10080 N^6)),\nonu \\
c_{180} &=& 180 c^4 D (c-3 N)^3 (-1+N) N (-1+2 N) (1134 N^3 (-42+311 N-212 N^2+44 N^3)\nonu \\
&+& 2 c^4 (-225-266 N+92 N^2+88 N^3)+3 c^3 (217-682 N+664 N^2-24 N^3+144 N^4)\nonu \\
&-& 18 c^2 N (413-3242 N+1326 N^2-328 N^3+280 N^4)\nonu \\
&+& 27 c N^2 (1449-11242 N+8656 N^2-3736 N^3+432 N^4)),\nonu \\
c_{181} &=& 360 c^4 D  (c-3 N)^2 (-1+N) N (-1+2 N) (162 N^4 (-1141-3470 N+6332 N^2-3312 N^3\nonu \\
&+& 816 N^4)+4 c^5 (-3-269 N-587 N^2-88 N^3+268 N^4+80 N^5)-27 c N^3 (-1976-25963 N\nonu \\
&+& 10294 N^2+6900 N^3-1416 N^4+576 N^5)+54 c^2 N^2 (229-2714 N-7685 N^2+8780 N^3\nonu \\
&-& 4868 N^4+1536 N^5)+2 c^4 (51+1467 N+4422 N^2+1396 N^3-8140 N^4+1648 N^5\nonu \\
&+& 3120 N^6)+3 c^3 N (-954-5241 N+27396 N^2+520 N^3+4904 N^4-15728 N^5+10080 N^6)), \nonu \\
c_{182} &=& -360 c^4 D (c-3 N)^3 (-1+N) N (-1+2 N) (2 c^4 (66-63 N-61 N^2+38 N^3)\nonu \\
&+& 108 N^3 (-161+759 N-1000 N^2+516 N^3)+c^3 (258-2325 N+2173 N^2+682 N^3\nonu \\
&-& 1152 N^4)-9 c^2 N (401-2009 N+1830 N^2-646 N^3+364 N^4)\nonu \\
&+& 9 c N^2 (1721-8070 N+10143 N^2-7102 N^3+3168 N^4)), \nonu \\
c_{183} &=& -60 c^4 D (c-3 N)^2 (-1+N) N (-1+2 N) (-972 N^4 (2828-17242 N+37221 N^2\nonu \\
&-& 34262 N^3+11368 N^4)+4 c^5 (300-851 N-799 N^2+2726 N^3-1616 N^4+240 N^5)\nonu \\
&-& 162 c N^3 (-14610+99629 N-247163 N^2+288230 N^3-158872 N^4+33440 N^5)\nonu \\
&-& 6 c^4 (-600+1793 N+2740 N^2-28862 N^3+47312 N^4-26808 N^5+4800 N^6)\nonu \\
&-& 27 c^2 N^2 (20768-168241 N+478488 N^2-669308 N^3+482128 N^4\nonu \\
&-& 163296 N^5+19200 N^6)+9 c^3 N (2176-40005 N+156810 N^2-308424 N^3\nonu \\
&+& 311816 N^4-148560 N^5+27840 N^6)), \nonu \\
c_{184} &=& \frac{c^4 D }{1386} (c-3 N)^3 (-1+N) N (1+N) (6 N+c (-3+2 N)) (37422 N^3 (-607188+692932 N\nonu \\
&-& 149987 N^2-39508 N^3+1871 N^4)-9 c N^2 (-2612081772+3380071140 N\nonu \\
&-& 1506983309 N^2+333257314 N^3-47771527 N^4+2526914 N^5)\nonu \\
&+& 12 c^2 N (-647089056+828954720 N-362423618 N^2+66918396 N^3\nonu \\
&-& 2067335 N^4-952239 N^5+87172 N^6)+4 c^3 (200737440-197447904 N\nonu \\
&-& 33958188 N^2+119507128 N^3-62470805 N^4+15130604 N^5-1786777 N^6+83022 N^7)),\nonu \\
c_{185} &=& \frac{c^4 D }{8316}  (c-3 N)^2 (-1+N) N (1+N) (224532 N^5 (-26588880+38443204 N-12620894 N^2\nonu \\
&-& 6289681 N^3+3791531 N^4)+648 c N^4 (17184678408-30615132804 N+20871541546 N^2\nonu \\
&-& 5422780976 N^3-586634977 N^4+430249823 N^5)-27 c^2 N^3 (293477898888\nonu \\
&-& 570089038140 N+464818497844 N^2-198262765875 N^3+40075198927 N^4\nonu \\
&-& 1854160440 N^5+62099116 N^6)+18 c^3 N^2 (150087148416-292741884780 N\nonu \\
&+& 234838285558 N^2-102080038959 N^3+25006084826 N^4-2325816797 N^5\nonu \\
&-& 298541796 N^6+17507132 N^7)-6 c^4 N (73702541412-133222095612 N\nonu \\
&+& 83496330261 N^2-18463293380 N^3-2322038013 N^4+2259981346 N^5\nonu \\
&-& 484741590 N^6+16823104 N^7+686712 N^8)+4 c^5 (6956292960-10476722844 N\nonu \\
&+& 1961540964 N^2+4035022653 N^3-2852313640 N^4+784818761 N^5-89044664 N^6\nonu \\
&+& 1565122 N^7+164760 N^8+7128 N^9)), \nonu \\
c_{186} &=& 360 c^4 D (c-3 N)^3 (-1+N) N (-1+2 N) (378 N^3 (1-40 N-188 N^2+288 N^3)\nonu \\
&+& 4 c^4 (-252+183 N+237 N^2-218 N^3+40 N^4)+9 c N^2 (-5726+23163 N-2790 N^2\nonu \\
&-& 34784 N^3+19296 N^4)-2 c^3 (1512-9441 N+5101 N^2+8612 N^3-8816 N^4+2040 N^5)\nonu \\
&+& 6 c^2 N (4221-18845 N+7614 N^2+19674 N^3-16964 N^4+4320 N^5)), \nonu \\
c_{187} &=& -720 c^4 D (3+c-9 N) (c-3 N)^2 (-1+N) N (-1+2 N) \nonu \\
& \times & (2 c (-2+N)+3 (21-8 N) N) (2 c (-1+N)+3 (3-4 N) N)^2 (6 N+c (-3+2 N)),\nonu \\
c_{188} &=& -120 c^4 D (c-3 N)^2 (-1+N) N (-1+2 N) (486 N^4 (-1295+5687 N-5326 N^2\nonu \\
&-& 4272 N^3+5600 N^4)+8 c^5 (120-224 N-61 N^2+389 N^3-284 N^4+60 N^5)\nonu \\
&+& 81 c N^3 (6814-35827 N+50937 N^2-1208 N^3-31356 N^4+10048 N^5)\nonu \\
&-& 12 c^4 (-240+891 N+587 N^2-3627 N^3+3883 N^4-2064 N^5+540 N^6)\nonu \\
&-& 18 c^3 N (-71+5988 N-24366 N^2+29040 N^3-9198 N^4-2600 N^5+1320 N^6)\nonu \\
&+& 27 c^2 N^2 (-5490+40460 N-89523 N^2+60140 N^3+12916 N^4-26112 N^5+7680 N^6)), \nonu \\
c_{189} &=& 60 c^4 D (c-3 N)^2 (-1+N) N (-1+2 N) (-972 N^4 (-49+26 N+1399 N^2-3624 N^3\nonu \\
&+& 2360 N^4)+2 c^5 (264-1097 N+521 N^2+1514 N^3-1580 N^4+408 N^5)-81 c N^3 (2817\nonu \\
&-& 13242 N-351 N^2+62072 N^3-74612 N^4+22880 N^5)-3 c^4 (-528+6839 N\nonu \\
&-& 16337 N^2-1712 N^3+30532 N^4-24352 N^5+5664 N^6)-27 c^2 N^2 (-5424+31449 N\nonu \\
&-& 28979 N^2-79268 N^3+141412 N^4-68240 N^5+9024 N^6)\nonu \\
&+& 9 c^3 N (-3268+24055 N-35987 N^2-36966 N^3+103744 N^4-63528 N^5+11856 N^6)), \nonu \\
c_{190} &=& -60 c^4 D (3+c-9 N) (c-3 N) (1-2 N)^2 (-1+N) N (2 c (-1+N)+3 (3-4 N) N)\nonu \\
& \times & (6 N+c (-3+2 N)) (27 N^3 (385-922 N+520 N^2)+2 c^3 (52+30 N-85 N^2+18 N^3)\nonu \\
&-& 3 c^2 N (166+321 N-892 N^2+276 N^3)+9 c N^2 (-251+967 N-1242 N^2+528 N^3)), \nonu \\
c_{191} &=& -360 c^4 D (c-3 N)^2 (-1+N) N (-1+2 N) (2 c^5 (18+9 N-17 N^2-4 N^3+4 N^4)\nonu \\
&+& 324 N^4 (553-2617 N+4040 N^2-2648 N^3+1104 N^4)-54 c N^3 (3373-17661 N\nonu \\
&+& 29226 N^2 - 17530 N^3+4084 N^4)+9 c^2 N^2 (5351-30960 N+47373 N^2-13600 N^3\nonu \\
&-& 6636 N^4+144 N^5)+c^4 (-30-1861 N+2865 N^2+1004 N^3-2588 N^4+1104 N^5)\nonu \\
&-& 3 c^3 N (1207-11075 N+16928 N^2+1786 N^3-10888 N^4+3576 N^5)), \nonu \\
c_{192} &=& \frac{2}{231} c^4 D  (c-3 N)^3 (-1+N) N (1+N) (62370 N^4 (96660-151576 N+41393 N^2\nonu \\
&-& 287 N^3+10 N^4)+297 c N^3 (-17933616+21185516 N-7639052 N^2+4597707 N^3\nonu \\
&-& 916441 N^4+44326 N^5)-9 c^2 N^2 (-73634364-395450084 N+689581619 N^2\nonu \\
&-& 415320060 N^3+116512759 N^4-15689206 N^5+803136 N^6)+12 c^3 N (26768034\nonu \\
&-&191115792 N+225419346 N^2-99259402 N^3+12886529 N^4+1006414 N^5\nonu \\
&-& 455393 N^6+32994 N^7)+4 c^4 (-12559320+46007100 N+21686916 N^2-129153896 N^3\nonu \\
&+& 126289823 N^4-58085430 N^5+13827169 N^6-1652374 N^7+78312 N^8)), \nonu \\
c_{194} &=& -60 c^4 D (c-3 N)^2 (-1+N) N (-1+2 N) (486 N^4 (-4039+5878 N+6412 N^2\nonu \\
&-& 13384 N^3+4128 N^4)+c^5 (-690+304 N-4408 N^2-2632 N^3+2080 N^4+480 N^5)\nonu \\
&-& 81 c N^3 (-8762-8157 N+30918 N^2-620 N^3-21624 N^4+4992 N^5)+27 c^2 N^2 (-1074\nonu \\
&-& 3191 N-54886 N^2+94696 N^3-48664 N^4+9584 N^5)+3 c^4 (345+4190 N-1888 N^2\nonu \\
&+& 5592 N^3-14008 N^4-1120 N^5+4320 N^6)+9 c^3 N (-1271-9066 N+38796 N^2\nonu \\
&-& 19268 N^3 +14432 N^4-22384 N^5+8640 N^6)), \nonu \\
c_{196} &=& 1260 c^4 D (29+2 c) (c-3 N)^4 (-1+N) N (-1+2 N) (c^2 (-3+16 N+4 N^2)\nonu \\
&-& 36 N^2 (1-8 N+6 N^2)+12 c N (4-21 N+14 N^2)), \nonu \\
c_{197} &=& -2520 c^4 D (29+2 c) (6+5 c) (c-3 N)^5 (1-2 N)^2 (-1+N) N^2,\nonu \\
c_{198} &=& -\frac{1}{231} c^4 D (c-3 N)^3 (-1+N) N (1+N) (-74844 N^4 (-131220-25004 N+87089 N^2\nonu \\
&-& 10444 N^3+499 N^4)+594 c N^3 (-36502812+41875700 N-21987997 N^2+3127102 N^3\nonu \\
&-& 503135 N^4+28062 N^5)-18 c^2 N^2 (-760764852+1137531564 N-672042139 N^2\nonu \\
&+& 153480370 N^3-12194853 N^4+294806 N^5+27064 N^6)-3 c^3 N (1034113860\nonu \\
&-& 1133703084 N-428055073 N^2+1026925922 N^3-603493903 N^4+154666562 N^5\nonu \\
&-& 19225900 N^6+932936 N^7)+2 c^4 (111378240-26026236 N-171512316 N^2\nonu \\
&-& 120145873 N^3 +359911574 N^4-221810639 N^5+61378366 N^6-8038572 N^7\nonu \\
&+& 404856 N^8)), \nonu \\
c_{199} &=& -2520 c^5 D (3+c-9 N) (c-3 N)^3 (1-2 N)^2 (-1+N) N (6 N+c (-3+2 N))\nonu \\
& \times & (2 c (4+N)+3 N (-13+16 N)), \nonu \\
c_{200} &=& 120 c^4 D (c-3 N) (1-2 N)^2 (-1+N) N (6 N+c (-3+2 N)) (4 c^5 (130-127 N-175 N^2\nonu \\
&+& 82 N^3)+1458 N^4 (315-400 N-390 N^2+528 N^3)+c^4 (732-8442 N+8790 N^2\nonu \\
&+& 11604 N^3-6144 N^4)+27 c^2 N^2 (-349+4022 N+1582 N^2-2268 N^3+512 N^4)\nonu \\
&+& 18 c^3 N (11+1234 N-3108 N^2-2374 N^3+1756 N^4)\nonu \\
&-& 81 c N^3 (1543+3116 N-6136 N^2-1736 N^3+4416 N^4)), \nonu \\
c_{201} &=& 2520 c^4 D (c-3 N)^3 (-1+N) N (1+N) (-1+2 N) (c+6 N) \nonu \\
& \times & (6 N+c (-3+2 N)) (2 c^2 (-9+2 N)+9 N (-13+10 N)-3 c (-5-40 N+36 N^2)), \nonu \\
c_{202} &=& -2520 c^5 D (c-3 N)^2 (1-2 N)^2 (-1+N) N (1+N) (c+6 N)\nonu \\
& \times & (6 N+c (-3+2 N)) (2 c^2 (-9+2 N)+9 N (-13+10 N)-3 c (-5-40 N+36 N^2)), \nonu \\
c_{203} &=& -120 c^4 D  (-3+2 c) (c-3 N) (-1+N) N (1+N) (-1+2 N) (c+6 N) \nonu \\
& \times & (6 N+c (-3+2 N)) (c+2 c N+3 N (-3+2 N))^2 (9 (21-34 N) N+c (-23+22 N)), \nonu \\
c_{204} &=& -\frac{1}{924} c^4 D (c-3 N)^2 (-1+N) N (1+N) (449064 N^5 (-346800+561876 N-305066 N^2\nonu \\
&-& 4809 N^3+239 N^4)-14256 c N^4 (-13694220+14680948 N-3159969 N^2-2628911 N^3\nonu \\
&-& 331059 N^4+16711 N^5)+54 c^2 N^3 (-1598560812-349252104 N+3069730897 N^2\nonu \\
&-& 3000141443 N^3+1016895770 N^4-48250564 N^5+2556936 N^6)\nonu \\
&-& 9 c^3 N^2 (-2043595116-3524980196 N+6060217223 N^2-3172587532 N^3+848094755 N^4\nonu \\
&-& 85321378 N^5-7039564 N^6+932968 N^7)-3 c^4 N (881457588+79143996 N\nonu \\
&+& 5622242571 N^2-11829673420 N^3+9181357733 N^4-3683434844 N^5+818015392 N^6\nonu \\
&-& 92510928 N^7+4196592 N^8)+2 c^5 (101010240-264894156 N+1320376932 N^2\nonu \\
&-& 1425911005 N^3-188223486 N^4+1001647877 N^5-591108222 N^6+156677900 N^7\nonu \\
&-& 19855944 N^8+975744 N^9)), \nonu \\
c_{205} &=& -60 c^4 D (c-3 N) (1-2 N)^2 (-1+N) N (6 N+c (-3+2 N)) (4 c^5 (181-124 N-280 N^2\nonu \\
&+& 88 N^3)+486 N^4 (1778-2997 N-362 N^2+1368 N^3)+c^4 (930-12306 N+7620 N^2\nonu \\
&+& 19464 N^3-8880 N^4)+54 c^2 N^2 (-15+2922 N-1596 N^2-2156 N^3+1000 N^4)\nonu \\
&+& 9 c^3 N (-121+4038 N-4368 N^2-9488 N^3+6080 N^4)\nonu \\
&-& 81 c N^3 (3235+2910 N-9120 N^2-7616 N^3+9312 N^4)), \nonu \\
c_{206} &=& 360 c^4 D (c-3 N)^3 (-1+N) N (-1+2 N) (378 N^3 (1+812 N-848 N^2+168 N^3)\nonu \\
&+& 2 c^4 (-252-553 N-60 N^2+276 N^3+80 N^4)+9 c N^2 (238-24549 N+25218 N^2\nonu \\
&-& 18596 N^3+7944 N^4)+6 c^2 N (-588+6427 N-2202 N^2+2286 N^3-2744 N^4+2520 N^5)\nonu \\
&+& c^3 (420+2451 N+1720 N^2-6400 N^3+2224 N^4+3120 N^5)), \nonu \\
c_{207} &=& 90 c^4 D (c-3 N)^2 (-1+N) N (-1+2 N) (-1134 N^4 (-89+1676 N-2120 N^2+552 N^3)\nonu \\
&+& 4 c^5 (-87-527 N-135 N^2+344 N^3+84 N^4)-27 c N^3 (1736-62359 N+84448 N^2\nonu \\
&-& 56156 N^3+19824 N^4)-18 c^2 N^2 (-1441+29569 N-32196 N^2+25472 N^3-8000 N^4\nonu \\
&+& 2736 N^5)+2 c^4 (237+3067 N+3200 N^2-6272 N^3+1920 N^4+3504 N^5)\nonu \\
&+& 3 c^3 N (-2350+11817 N-3652 N^2+19176 N^3-21232 N^4+10128 N^5)), \nonu \\
c_{209} &=& -\frac{1}{73920}c^4 D (c-3 N)^2 (-1+N) N (1+N) (-74844 N^5 (87191280-151499604 N\nonu \\
&+& 99076816 N^2-3053239 N^3+150361 N^4)-2268 c N^4 (-5251847220+9540938976 N\nonu \\
&-& 6743912991 N^2+1614623849 N^3-91488776 N^4+4246058 N^5)\nonu \\
&-& 27 c^2 N^3 (322378473600-604361868364 N+429225569228 N^2-146324932805 N^3\nonu \\
&+& 19595489495 N^4-1816562176 N^5+69272812 N^6)+9 c^3 N^2 (338102138160\nonu \\
&-& 570663367708 N+185004066576 N^2+146449169219 N^3-128894426409 N^4\nonu \\
&+& 38009007844 N^5-5065672580 N^6+256284752 N^7)+12 c^4 N (-38233622520\nonu \\
&+& 18612637176 N+156888130066 N^2-276285197976 N^3+202532062819 N^4\nonu \\
&-& 78889034550 N^5+16932604945 N^6-1887843948 N^7+85219812 N^8)\nonu \\
&+& 4 c^5 (3925687680+31647829104 N-151476259488 N^2+252772338680 N^3\nonu \\
&-& 219033364088 N^4+110406245391 N^5-33454034437 N^6+5993082520 N^7\nonu \\
&-& 583403652 N^8+23728320 N^9)), \nonu \\
c_{211} &=& \frac{1}{221760}c^4 D (c-3 N)^2 (-1+N) N (1+N) (-224532 N^5 (80503920-182002324 N\nonu \\
&+& 126595536 N^2-6282759 N^3+302441 N^4)-324 c N^4 (-93647595060+209806060656 N\nonu \\
&-& 157954346191 N^2+38602856309 N^3-2242283636 N^4+97002538 N^5)\nonu \\
&-& 27 c^2 N^3 (760610867040-1654445631492 N+1202422400204 N^2-383889918975 N^3\nonu \\
&+& 39104123405 N^4-2995368288 N^5+91595876 N^6)+9 c^3 N^2 (744393478320\nonu \\
&-& 1368728592564 N+338339537608 N^2+522142332057 N^3-396199108387 N^4\nonu \\
&+& 110689329292 N^5-14388388140 N^6+717883376 N^7)+12 c^4 N (-76415256360\nonu \\
&+& 11129524488 N+463017483078 N^2-782459460928 N^3+569222124417 N^4\nonu \\
&-& 221488661770 N^5+47645924715 N^6-5327534404 N^7+241156236 N^8)\nonu \\
&+& 4 c^5 (4434963840+95896703952 N-423902631744 N^2+708943133160 N^3\nonu \\
&-& 621553546064 N^4+317496132733 N^5-97441426231 N^6+17655992840 N^7\nonu \\
&-& 1735707756 N^8+71184960 N^9)), \nonu \\
c_{212} &=& -\frac{1}{5544}c^4 D (c-3 N)^3 (-1+N) N (1+N) (1197504 N^4 (-14364-22996 N+37256 N^2\nonu \\
&-& 4361 N^3+205 N^4)+108 c N^3 (392042412-198823224 N-209029297 N^2+101228507 N^3\nonu \\
&-& 8327054 N^4+279736 N^5)-9 c^2 N^2 (3660757020-5955019668 N+4312057037 N^2\nonu \\
&-& 2275299554 N^3+650524029 N^4-79158796 N^5+3775972 N^6)+6 c^3 N (1582427124\nonu \\
&-& 3113065980 N+2254779303 N^2-687259790 N^3-3455725 N^4+55271496 N^5\nonu \\
&-& 10690244 N^6+637136 N^7)+4 c^4 (-199519200+216902988 N+484404120 N^2\nonu \\
&-& 1095979227 N^3+853473115 N^4-338643053 N^5+72799325 N^6-8081368 N^7\nonu \\
&+& 362580 N^8)), \nonu \\
c_{213} &=& -\frac{1}{693} c^4 D (c-3 N)^3 (-1+N) N (1+N) (74844 N^4 (-4488300+5444996 N-1452661 N^2\nonu \\
&-& 270074 N^3+12799 N^4)-54 c N^3 (-9343618884+14040819684 N-7372647175 N^2\nonu \\
&+& 1363735460 N^3+10350349 N^4+535006 N^5)-9 c^2 N^2 (30564243036-51492494508 N\nonu \\
&+& 33716391953 N^2-10992015188 N^3+1906862607 N^4-214388440 N^5+9581860 N^6)\nonu \\
&+& 3 c^3 N (21302925804-36365264676 N+21949080981 N^2-4872965938 N^3\nonu \\
&-& 286148441 N^4+346836258 N^5-60380548 N^6+3417160 N^7)+2 c^4 (-2616274080\nonu \\
&+& 3753067284 N-65521188 N^2-2943492237 N^3+2423775860 N^4-930730759 N^5\nonu \\
&+& 192244348 N^6-20579588 N^7+895440 N^8)), \nonu \\
c_{217} &=& -2520 c^4 D (c-3 N)^4 (-1+N) N (1+N) (-1+2 N) (c+6 N) (2 c^2 (-9+2 N)\nonu \\
&+& 9 N (-13+10 N)-3 c (-5-40 N+36 N^2)), \nonu \\
c_{222} &=& 180 c^4 D (c-3 N)^2 (-1+N) N (1+N) (-1+2 N) (c+6 N) (2 c^4 (-1005\nonu \\
&+& 238 N+20 N^2+8 N^3)-81 N^3 (1631-1570 N+68 N^2+136 N^3)\nonu \\
&+& 27 c N^2 (3697+2420 N-6384 N^2+1648 N^3+48 N^4)+3 c^3 (543+8924 N-6128 N^2\nonu \\
&+& 336 N^3+80 N^4)+9 c^2 N (-2593-11032 N+12488 N^2-2400 N^3+112 N^4)), \nonu \\
c_{224} &=& -60 c^4 D (c-3 N)^2 (-1+N) N (-1+2 N) (486 N^4 (1477+1548 N-220 N^2-4208 N^3\nonu \\
&+& 4000 N^4)+4 c^5 (645+1142 N-473 N^2-1208 N^3+92 N^4+240 N^5)\nonu \\
&-& 81 c N^3 (8994+4367 N+22048 N^2-32580 N^3+10448 N^4+5440 N^5)\nonu \\
&+& 6 c^4 (-435-4060 N+293 N^2+5364 N^3-1624 N^4-3952 N^5+1680 N^6)\nonu \\
&-& 18 c^3 N (-670+1568 N+10795 N^2-5164 N^3-3744 N^4-1744 N^5+2640 N^6)\nonu \\
&-& 27 c^2 N^2 (-5492-9593 N-43976 N^2+57104 N^3-11488 N^4-22640 N^5+13440 N^6)), \nonu \\
c_{225} &=& \frac{1}{1386} c^4 D (c-3 N) (-1+N) N (1+N) (-336798 N^6 (4933140-11038276 N\nonu \\
&+& 11161651 N^2- 5659196 N^3+1096241 N^4)+5346 c N^5 (606646044-1310657020 N\nonu \\
&+& 1249237259 N^2-569121554 N^3+64807285 N^4+23030706 N^5)\nonu \\
&+& 162 c^2 N^4 (-15673645008+30675598004 N-24226318876 N^2+7119528285 N^3\nonu \\
&+& 2302136133 N^4-2363991614 N^5+506456716 N^6)+27 c^3 N^3 (37403594280\nonu \\
&-& 58150782908 N+19427533212 N^2+18490762297 N^3-20200878807 N^4\nonu \\
&+& 8032889410 N^5-1247247204 N^6+10757160 N^7)+9 c^4 N^2 (-23664485052\nonu \\
&+& 20339565872 N+28450790965 N^2-46804352397 N^3+24914128446 N^4\nonu \\
&-& 5854823294 N^5+622956532 N^6-62247480 N^7+1579888 N^8)\nonu \\
&-& 6 c^5 N (-3747503160-698891988 N+12509712878 N^2-10071672503 N^3\nonu \\
&-& 593500925 N^4+3970254092 N^5-2050015150 N^6+503428020 N^7-58478792 N^8\nonu \\
&+& 2682768 N^9)+8 c^6 (-116948880-179637660 N+567497940 N^2+1719495 N^3\nonu \\
&-& 567589066 N^4+306983868 N^5+45911870 N^6-85516605 N^7+27629066 N^8\nonu \\
&-& 3792228 N^9+194040 N^{10})), \nonu \\
c_{227} &=& -\frac{1}{462} c^4 D (c-3 N)^2 (-1+N) N (1+N) (37422 N^5 (-4672044+6466180 N-2200505 N^2\nonu \\
&-& 199150 N^3+9479 N^4)-1782 c N^4 (-172116972+269287380 N-146145943 N^2\nonu \\
&+& 28822148 N^3-1013081 N^4+62788 N^5)-27 c^2 N^3 (7619374512-12147629084 N\nonu \\
&+& 7120738810 N^2-1921448945 N^3+255507703 N^4-30176576 N^5+1141660 N^6)\nonu \\
&+& 18 c^3 N^2 (3635356104-5183846676 N+2002631726 N^2+210112431 N^3-394333572 N^4\nonu \\
&+& 157317619 N^5-21091124 N^6+1058972 N^7)+3 c^4 N (-3258017136+3258582972 N\nonu \\
&+& 1227685558 N^2-2019835167 N^3+248855333 N^4+336647108 N^5-144974828 N^6\nonu \\
&+& 22561552 N^7-1243552 N^8)+2 c^5 (264012480+6100848 N-758803548 N^2\nonu \\
&+& 771818618 N^3-369977641 N^4+175652999 N^5-86585032 N^6+24472412 N^7\nonu \\
&-& 3369904 N^8+177408 N^9)), \nonu \\
c_{228} &=& 60 c^4 D (c-3 N)^2 (-1+N) N (1+N) (-1+2 N) (c+6 N) (2 c^4 (-429-476 N-32 N^2\nonu \\
&+& 88 N^3)-81 N^3 (-483+4206 N-4376 N^2+1104 N^3)-3 c^3 (-177-3454 N-3796 N^2\nonu \\
&+& 1176 N^3+800 N^4)+9 c^2 N (-565-4396 N-8568 N^2+6752 N^3+1232 N^4)\nonu \\
&+& 27 c N^2 (81+4598 N+5572 N^2-10040 N^3+2208 N^4)), \nonu \\
c_{229} &=& 2520 c^4 D (c-3 N)^3 (-1+N) N (1+N) (-1+2 N) (c+6 N) (-15 N+2 c (2+N)) \nonu \\
&\times & (2 c^2 (-9+2 N)+9 N (-13+10 N)-3 c (-5-40 N+36 N^2)),\nonu \\
D &=& \frac{1}{7560(-1 + N) N (-1 + 2 N)(29 + 2 c) (6 + 5 c) N^2 (-1 + 2 N)^2 (c -
   6 N + 6 N^2)^4}.
\label{O5O9p1coeff}
\eea

\section{ The ${\cal N}=1$ superspace description }

The ${\cal N}=1$
superconformal algebra is described as the ${\cal N}=1$ super OPE
\bea
\hat{T}(Z_1) \, \hat{T}(Z_2) &
= & \frac{1}{z_{12}^3} \,\frac{c}{6} +\frac{\theta_{12}}{z_{12}^2}
\,\frac{3}{2} \hat{T}(Z_2) +
\frac{1}{z_{12}} \,\frac{1}{2} D \hat{T}(Z_2)
+\frac{\theta_{12}}{z_{12}} \,\pa \hat{T}(Z_2) +\cdots,
\label{tt}
\eea
where
$z_{12} = z_1 -z_2-\theta_1 \theta_2,
\theta_{12} = \theta_1 -\theta_2, D = \pa_{\theta} + \theta
\pa_z$ and $\pa =\pa_z$.
The ${\cal N}=1$ super stress energy tensor is
\bea
\hat{T}(Z) = \frac{1}{2} G(z) + \theta \, T(z), \qquad
Z=(z, \theta).
\label{that}
\eea
Of course the single OPE (\ref{tt}) consists of
(\ref{cosettt}), (\ref{tg}) and (\ref{gg}).
The primary superfields that we consider are
given by
\bea
\hat{O}_{2}(Z) & = &  O_{2} (z) +\theta \, O_{\frac{5}{2}} (z),
\nonu \\
\hat{O}_{\frac{7}{2}}(Z) & = &  O_{\frac{7}{2}} (z) +\theta \, O_4 (z),
\nonu \\
\hat{O}_{4}(Z) & = &  O_{4'} (z) +\theta \, O_{\frac{9}{2}} (z).
\label{superfields}
\eea

The three OPEs (\ref{O2O2a}), (\ref{O2O5a})
and (\ref{O5O5}), together with (\ref{that}) and (\ref{superfields}),
can be expressed as a single ${\cal N}=1$ super OPE between $\hat{O_{2}}(Z)$ and itself as follows:
\bea
\hat{O}_{2}(Z_1) \, \hat{O}_{2}(Z_2) &= &
\frac{1}{z_{12}^4} \, c_{22} +
\frac{\theta_{12}}{z_{12}^3}\, 3 c_{22}^{t}\, \hat{T}(Z_2)+
\frac{1}{z_{12}^2} \, \left[ c_{22}^{o}\,\hat{O}_{2}+
 c_{22}^{t} D \hat{T} \right](Z_2)
\nonu \\
&+& \frac{\theta_{12}}{z_{12}^2} \, \left[ \frac{1}{2}\,c_{22}^{o}\,D \hat{O}_{2}
 +2\,c_{22}^{t} \pa \hat{T} \right](Z_2)
 +\frac{1}{z_{12}} \, \left[\frac{1}{2}\, c_{22}^{o}\,\pa \hat{O}_{2}
 + \frac{1}{2}c_{22}^{t}\,D \pa \hat{T} \right](Z_2)
\nonu \\
&+& \frac{\theta_{12}}{z_{12}} \left[  \frac{1}{10}\,(3 c_{22}^{o}+4 c_{2\frac{5}{2}}^{go})\,D \pa \hat{O}_{2}
 + \frac{1}{4}\,(3 c_{22}^{t}+c_{2\frac{5}{2}}^{tg})\, \pa^2 \hat{T}-\hat{O}_{\frac{7}{2}}
  \right. \nonu \\
& - &  \left.
 2c_{2\frac{5}{2}}^{go}\hat{T} \hat{O}_2-2c_{2\frac{5}{2}}^{tg}\hat{T} D \hat{T}\right](Z_2) +\cdots,
\label{singleO2O2}
\eea
where the identity $\frac{1}{(z_1-z_2)^n} = \frac{1}{z_{12}^n} -
n \frac{\theta_1 \theta_2}{z_{12}^{n+1}}$ ($n=1, \cdots, 6$) is used.
All the coefficient functions are given in section $3$.


The four OPEs (\ref{O2O7}), (\ref{O2O4}), (\ref{O5O7})
and (\ref{O5O4}) can be expressed as a single ${\cal N}=1$ super OPE as follows:
\bea
\hat{O}_{2}(Z_1) \, \hat{O}_{\frac{7}{2}}(Z_2) &= &
\frac{\theta_{12}}{z_{12}^4} \, 4\,c_{2 \frac{7}{2}}^{o}\hat{O}_{2}(Z_2) +
\frac{1}{z_{12}^3}\, c_{2 \frac{7}{2}}^{o}\, D \hat{O}_{2}(Z_2)+
\frac{\theta_{12}}{z_{12}^3} \, c_{2 \frac{7}{2}}^{o}\, \pa \hat{O}_{2}(Z_2)
\nonu \\
&+&
\frac{1}{z_{12}^2} \, \left[c_{2\frac{7}{2}}^{o'}\,\hat{O}_{\frac{7}{2}}
 +(\frac{1}{5}\,c_{2 \frac{7}{2}}^{o}-\frac{2}{5}\,c_{2\frac{7}{2}}^{go})\,D \pa \hat{O}_{2}+2c_{2\frac{7}{2}}^{go}\,\hat{T}\hat{O}_{2}\right](Z_2)
\nonu \\
&+& \frac{\theta_{12}}{z_{12}^2} \, \left[\frac{2}{7}\,c_{2\frac{7}{2}}^{o'}\,D  \hat{O}_{\frac{7}{2}}+8 \hat{O}_4
  +(\frac{1}{5}\,c_{2 \frac{7}{2}}^{o}-\frac{3}{10}\,c_{\frac{5}{2}\frac{7}{2}}^{to}
 -\frac{1}{5}\,c_{\frac{5}{2}\frac{7}{2}}^{go})\pa^2 \hat{O}_{2}
   \right. \nonu \\
& + &  \left.
2 c_{\frac{5}{2}\frac{7}{2}}^{go}\,\hat{T}D\hat{O}_{2}+ c_{\frac{5}{2}\frac{7}{2}}^{to}\,D \hat{T} \hat{O}_{2}\right](Z_2)
\nonu \\
&+& \frac{1}{z_{12}}  \, \left[\frac{2}{7}\, c_{2\frac{7}{2}}^{o'}\,\pa \hat{O_{\frac{7}{2}}}+D \hat{O}_4
+ (\frac{1}{30}c_{2 \frac{7}{2}}^{o}-\frac{4}{35}c_{2\frac{7}{2}}^{go}+\frac{1}{8}c_{2\frac{7}{2}}^{go'}-\frac{1}{4}c_{2\frac{7}{2}}^{to})\, D \pa^2\hat{O}_2
 \right. \nonu \\
& + &  \left.
 c_{2 \frac{7}{2}}^{to}\, D\hat{T} D \hat{O}_2+
(\frac{4}{7}c_{2 \frac{7}{2}}^{go}+2c_{2 \frac{7}{2}}^{go'})\, \pa \hat{T} \hat{O}_2
 + (\frac{4}{7}c_{2 \frac{7}{2}}^{go}-\frac{3}{2}c_{2 \frac{7}{2}}^{go'})\, \hat{T}\pa \hat{O}_2 \right](Z_2)
\nonu \\
&+& \frac{\theta_{12}}{z_{12}} \left[ 3\pa \hat{O}_4+(\frac{1}{30}c_{2 \frac{7}{2}}^{o}-\frac{9}{80}c_{\frac{5}{2} \frac{7}{2}}^{to}
-\frac{3}{40}c_{\frac{5}{2} \frac{7}{2}}^{go}-\frac{1}{6}c_{\frac{5}{2} \frac{7}{2}}^{to'}-\frac{1}{9}c_{\frac{5}{2} \frac{7}{2}}^{go'})\, \pa^3\hat{O}_2
 \right. \nonu \\
& + &  \left. (-\frac{1}{4}c_{\frac{5}{2} \frac{7}{2}}^{go''}+\frac{3}{28}c_{2\frac{7}{2}}^{o'})\,D \pa \hat{O}_{\frac{7}{2}}
+2 c_{\frac{5}{2} \frac{7}{2}}^{go''}\,\hat{T} \hat{O}_{\frac{7}{2}}
+ (\frac{3}{8}c_{\frac{5}{2} \frac{7}{2}}^{to}-c_{\frac{5}{2} \frac{7}{2}}^{to'})\,D \pa \hat{T} \hat{O}_{2}
 \right. \nonu \\
& + &  \left.
(\frac{3}{4}c_{\frac{5}{2} \frac{7}{2}}^{go}-\frac{10}{3}c_{\frac{5}{2} \frac{7}{2}}^{go'})\,\pa \hat{T}D \hat{O}_{2}
 + (\frac{3}{8}c_{\frac{5}{2} \frac{7}{2}}^{to}+c_{\frac{5}{2} \frac{7}{2}}^{to'})\,D \hat{T}\pa \hat{O}_{2}
  \right. \nonu \\
& + &  \left.
(\frac{3}{4}c_{\frac{5}{2} \frac{7}{2}}^{go}+2 c_{\frac{5}{2} \frac{7}{2}}^{go'})\, \hat{T} D \pa \hat{O}_{2}\right](Z_2)
+ \cdots.
\label{singleO2O7}
\eea
All the coefficient functions in this expression are in section $3$.

The four OPEs (\ref{O2O44}), (\ref{O2O9}), (\ref{O5O44})
and (\ref{O5O9}) can be expressed as a single ${\cal N}=1$ super OPE as follows:
\bea
\hat{O}_{2}(Z_1) \, \hat{O}_{4}(Z_2) &= & \frac{1}{z_{12}^4}c_{2 4'}^{o}\hat{O}_{2}(Z_2)\nonu \\
&+&
\frac{\theta_{12}}{z_{12}^3} \, \left[7 c_{2 4'}^{o'}\,\hat{O}_{\frac{7}{2}}
 -\frac{2}{5}c_{\frac{5}{2} 4'}^{go}\,D \pa \hat{O}_{2}+2 c_{\frac{5}{2} 4'}^{go}\,\hat{T}\hat{O}_{2}\right](Z_2)
\nonu \\
&+&
\frac{1}{z_{12}^2} \, \left[ c_{2 4'}^{o''}\,\hat{O_4}
 +c_{2 4'}^{o'}\,D  \hat{O}_{\frac{7}{2}}+\hat{A}_{4}-\frac{1}{5}(\frac{3}{2} c_{2 4'}^{to}+c_{2 4'}^{go})\pa^2 \hat{O}_2
 \right. \nonu \\
& + &  \left. c_{2 4'}^{to}D\hat{T} \hat{O}_2+2 c_{2 4'}^{go}\hat{T}D \hat{O}_2\right](Z_2)
\nonu \\
&+&
\frac{\theta_{12}}{z_{12}^2} \, \left[ \frac{1}{4}c_{2 4'}^{o''}\,D\hat{O}_4
 +2c_{2 4'}^{o'}\,\pa  \hat{O}_{\frac{7}{2}}+\frac{1}{4}\, D \hat{A}_{4}
 +(-\frac{4}{35} c_{\frac{5}{2} 4'}^{go}+\frac{1}{8} c_{\frac{5}{2} 4'}^{go'}-\frac{1}{8} c_{\frac{5}{2} 4'}^{to})D \pa^2 \hat{O}_2
 \right. \nonu \\
& + &  \left.
c_{\frac{5}{2} 4'}^{to} D\hat{T}D \hat{O}_2+2(\frac{2}{7}c_{\frac{5}{2} 4'}^{go}+c_{\frac{5}{2} 4'}^{go'})\pa \hat{T} \hat{O}_2
+2(\frac{2}{7}c_{\frac{5}{2} 4'}^{go}-\frac{3}{4}c_{\frac{5}{2} 4'}^{go'}) \hat{T} \pa \hat{O}_2\right](Z_2)
\nonu \\
&+&
\frac{1}{z_{12}} \, \left[ \frac{1}{4}c_{2 4'}^{o''}\,\pa \hat{O}_4
 +\frac{1}{4}(c_{2 4'}^{o'}-c_{2 4'}^{go''})\,D \pa  \hat{O}_{\frac{7}{2}}+\frac{1}{4}\,\pa \hat{A}_{4}
  \right. \nonu \\
& - & \left. (\frac{3}{40}c_{2 4'}^{to}+\frac{1}{20}c_{2 4'}^{go}+\frac{1}{6}c_{2 4'}^{to'}+\frac{1}{9}c_{2 4'}^{go'}) \pa^3 \hat{O}_2
 \right. \nonu \\
& + &  \left.  (\frac{1}{4}c_{2 4'}^{to}-c_{2 4'}^{to'})D \pa \hat{T} \hat{O}_2
+(\frac{1}{4}c_{2 4'}^{to}+c_{2 4'}^{to'}) D\hat{T}\pa \hat{O}_2
 \right. \nonu \\
& + &  \left.
2(\frac{1}{4}c_{2 4'}^{go}-\frac{5}{3}c_{2 4'}^{go'})\pa \hat{T} D\hat{O}_2
+2(\frac{1}{4}c_{2 4'}^{go}+c_{2 4'}^{go'}) \hat{T} D \pa \hat{O}_2
+2c_{2 4'}^{go''}\hat{T} \hat{O}_{\frac{7}{2}} \right](Z_2)
\nonu \\
&+&
\frac{\theta_{12}}{z_{12}} \, \left[-\hat{O_\frac{11}{2}}+(\frac{1}{12}-\frac{2}{9}c_{\frac{5}{2} 4'}^{ga})\, D \pa \hat{A}_4
 \right. \nonu \\
& + &  \left.
(\frac{3}{8}c_{2 4'}^{o'}-\frac{3}{16}c_{\frac{5}{2} 4'}^{to'}-\frac{1}{8}c_{\frac{5}{2} 4'}^{go''''}
-\frac{c(21+4c)(c-4cN+6N^2)}{28(6+5c)(c-6(-1+N)N)} c_{\frac{5}{2} 4'}^{q})\, \pa^2 \hat{O}_{\frac{7}{2}}
 \right. \nonu \\
& + &  \left. (\frac{1}{12} c_{2 4'}^{o''}-\frac{2}{9}c_{\frac{5}{2} 4'}^{go'''}-\frac{1}{3}c_{\frac{5}{2} 4'}^{q})\, D \pa \hat{O}_4
 \right. \nonu \\
& + &  \left.
(\frac{1}{24}c_{\frac{5}{2} 4'}^{go'}-\frac{3}{140}c_{\frac{5}{2} 4'}^{go}-\frac{1}{12}c_{\frac{5}{2} 4'}^{to}
-\frac{2}{35}c_{\frac{5}{2} 4'}^{go''}-\frac{1}{7}c_{\frac{5}{2} 4'}^{to''}
 \right. \nonu \\
& + &  \left.
2c^2 H (2304+5c(-654+c(-247+2c)))c_{\frac{5}{2} 4'}^{q})\, D \pa^3 \hat{O}_{2 }
 \right.
\nonu
 \\
& + &  \left.   c_{\frac{5}{2} 4'}^{q}\, \hat{O}_2 \hat{O}_{\frac{7}{2} }
+ 2 c_{\frac{5}{2} 4'}^{go''''}\, \hat{T} D\hat{O}_{\frac{7}{2} }+ c_{\frac{5}{2} 4'}^{to'}\, D  \hat{T} \hat{O}_{\frac{7}{2} }
+ 2 c_{\frac{5}{2} 4'}^{go'''}\,  \hat{T} \hat{O}_{4 }+2 c_{\frac{5}{2} 4'}^{a}\,\hat{T} \hat{A}_{4 }\right](Z_2)
\nonu \\
& + & \cdots, \label{singleO2O4}
\eea
where
$H \equiv
\frac{(3+c-9 N)(1 + N) (c + 6 N) (-3 c + 2 (3 + c) N)}{
35 (29 + 2 c) (21 + 4 c) (6 + 5 c)^2 (c + 6 (-1 + N) N)^2}$ and one introduces
$\hat{O}_{\frac{11}{2}}(Z) =  O_{\frac{11}{2}} (z) +\theta \, O_{6} (z)$, and
$\hat{A}_{4}(Z) = A_{4} (z) +\theta \, A_{\frac{9}{2}} (z)$.
All the coefficient functions can be found in section $3$.

Following \cite{CV}, the equations 
(\ref{singleO2O2}), (\ref{singleO2O7}), and (\ref{singleO2O4}) can be 
summarized as follows:
\bea
\hat{O}_2 \times \hat{O}_2 &=& n_{2} I +C_{22}^{2} \hat{O}_2 + C_{22}^{\frac{7}{2}} \hat{O}_{\frac{7}{2}},\nonu \\
\hat{O}_2 \times \hat{O}_{\frac{7}{2}} &=& C_{2\frac{7}{2}}^{2} \hat{O}_2 + C_{2\frac{7}{2}}^{\frac{7}{2}} \hat{O}_{\frac{7}{2}}
+ C_{2\frac{7}{2}}^{4} \hat{O}_{4},\nonu \\
\hat{O}_2 \times \hat{O}_4 &=& C_{24}^{2} \hat{O}_2 +C_{24}^{\frac{7}{2}} \hat{O}_{\frac{7}{2}}+C_{24}^{4} \hat{O}_{4}
+A_{24}^{4} \hat{A}'_{4}+C_{24}^{\frac{11}{2}} \hat{O}_{\frac{11}{2}},
\eea
where the structure constants in \cite{CV} are related to our structure 
constants as follows
\bea
\hat{A}'_{4}(Z)& =& \frac{(29 + 2 c) (6 + 5 c)^2 (c + 6 (-1 + N) N)^2}{24 c^3 (-7 + 10 c) (3 + c - 9 N) (1 + N) (c + 6 N) (-3 c +
   2 (3 + c) N)}\hat{A}_4 (Z),\nonu \\
n_{2} &=& c_{22},
\qquad
C_{22}^{2} = c_{2 2}^{o},
\qquad
C_{22}^{\frac{7}{2}}= 1,
\nonu \\
C_{2\frac{7}{2}}^{2} &=& c_{2 \frac{7}{2}}^{o},
\qquad
C_{2\frac{7}{2}}^{\frac{7}{2}} =c_{2 \frac{7}{2}}^{o'},
\qquad
C_{2\frac{7}{2}}^{4} = 1,
\nonu \\
C_{24}^{2} &=& c_{2 4'}^{o},
\qquad
C_{24}^{\frac{7}{2}} =   c_{2 4'}^{o'},
\qquad
C_{24}^{4} =  c_{2 4'}^{o''},
\qquad
C_{24}^{\frac{11}{2}} = 1,
\nonu \\
A_{24}^{4}&=& \frac{24 c^3 (-7 + 10 c) (3 + c - 9 N) (1 + N) (c + 6 N) (-3 c +   2 (3 + c) N)}{(29 + 2 c) (6 + 5 c)^2 (c + 6 (-1 + N) N)^2}.
\eea

As in \cite{CV}, the following relations corresponding first few terms in
$(2.23)$ of \cite{CV} are satisfied:
\bea
&&C_{2 \frac{7}{2}}^{2}= \frac{3(c-15)}{2(5c+6)}(C_{2 2}^{2})^2+\frac{12(5c+6)}{c(4c+21)}n_{2},
\qquad
C_{2 \frac{7}{2}}^{\frac{7}{2}}=\frac{(4c+21)}{(5c+6)}C_{2 2}^{2},\nonu \\
&&C_{2 4}^{2}=\frac{15(c-15)(4c+21)}{28(2c+29)(5c+6)}(C_{2 2}^{2})^3
+\frac{30(5c+6)}{7c(2c+29)}n_{2}\,C_{2 2}^{2},\nonu \\
&&C_{2 4}^{\frac{7}{2}}=-\frac{(4c+21)(50c^2 -145c+483)}{98(2c-3)(2c+29)(5c+6)}(C_{2 2}^{2})^2
-\frac{2(4c+21)(14c-25)}{7c(2c-3)(2c+29)}n_{2},\nonu \\
&&C_{2 4}^{4}=\frac{2(25c+84)}{7(5c+6)}C_{2 2}^{2}, \nonu \\
&&A_{2 4}^{4}=-\frac{6c(c-15)(4c+21)(10c-7)}{(2c-3)(2c+29)(5c+6)^2}(C_{2 2}^{2})^2
-\frac{48(10c-7)}{(2c-3)(2c+29)}n_{2}.
\nonu
\eea
Note that in our case, the structure constant $c_{22}^{o}=C_{22}^2$ 
is not equal to $1$.

\section{The OPEs between the stress energy tensor $T(z)$ and the
quasi-primary (or primary) fields}

To find whether a conformal field is
a quasi-primary or not,
the OPE between $T(z)$ and the conformal field should be computed
and the vanishing of third-order pole of the OPE should be checked.
We list the OPEs between the stress tensor and
the quasi-primary fields (with
four primary fields) appeared in the section $3$ here:
\bea
&& T(z) \, \left(G O_2 -\frac{2}{5} \pa O_{\frac{5}{2}}\right)(w) = {\cal O}((z-w)^{-2}),
\nonu \\
&& T(z) \, \left(TG -\frac{3}{8} \pa^2 G \right)(w) = \frac{1}{(z-w)^4} \frac{(21+4c)}{8}
G(w) + {\cal O}((z-w)^{-2}),
\nonu \\
&& T(z) \, \left(TT-\frac{3}{10} \pa^2 T\right)(w) =
 \frac{1}{(z-w)^4} \frac{(22+5c)}{5}
T(w) + {\cal O}((z-w)^{-2}),
\nonu \\
&& T(z) \, \left(TO_{2}-\frac{3}{10} \pa^2 O_2 \right)(w)  = \frac{1}{(z-w)^4} \frac{(44+5c)}{10}
O_2(w) + {\cal O}((z-w)^{-2}),
\nonu \\
&& T(z) \, \left( GO_{\frac{5}{2}}-\frac{1}{5} \pa^2 O_2 \right)(w)   =  \frac{1}{(z-w)^4} \frac{38}{5}
O_2(w) + {\cal O}((z-w)^{-2}),
\nonu \\
&& T(z) \, \left( G \pa G-\frac{7}{10} \pa^2 T \right)(w)  = \frac{1}{(z-w)^4} (-\frac{17}{5})
T(w) + {\cal O}((z-w)^{-2}),
\nonu \\
&& T(z) \;  \left(\pa G O_2 -\frac{3}{4}G \pa O_2 +\frac{1}{8} \pa^2 O_{\frac{5}{2}} \right)(w)  =
\frac{1}{(z-w)^4} \frac{7}{2}
O_{\frac{5}{2}}(w) + {\cal O}((z-w)^{-2}),
\nonu \\
&& T(z) \, \left(TO_{\frac{5}{2}}-\frac{1}{4} \pa^2 O_{\frac{5}{2}}\right)(w)  =
\frac{1}{(z-w)^4} \frac{(25+2c)}{4}
O_{\frac{5}{2}}(w) + {\cal O}((z-w)^{-2}),
\nonu \\
&& T(z) \, \left( T \pa O_2-\pa T O_2-\frac{1}{6}\pa^3 O_2 \right)(w)
= \frac{1}{(z-w)^5} (-2c)O_2(w) \nonu \\
&& +\frac{1}{(z-w)^4}(-\frac{1}{4}) (-2c)\pa O_2(w)+ {\cal O}((z-w)^{-2}),
\nonu \\
&& T(z) \, \left( G \pa O_{\frac{5}{2}}-\frac{5}{3}\pa G O_{\frac{5}{2}}-\frac{1}{9} \pa^3 O_2\right)(w)  = \frac{1}{(z-w)^5} (-8)O_2(w) \nonu \\
&& +\frac{1}{(z-w)^4}(-\frac{1}{4}) (-8)\pa O_2(w)+ {\cal O}((z-w)^{-2}),
\nonu \\
&& T(z) \, \left( G O_{\frac{7}{2}}-\frac{1}{4} \pa O_4 \right) (w)
= {\cal O}((z-w)^{-2}),
\nonu \\
&& T(z) \, \left( \pa^2 G O_2 +\frac{3}{5} G \pa^2 O_2 -2 \pa G \pa O_2 -\frac{2}{35} \pa^3 O_{\frac{5}{2}} \right)(w)  = \frac{1}{(z-w)^5} \frac{76}{35}O_{\frac{5}{2}}(w) \nonu \\
&& +\frac{1}{(z-w)^4}\left[(-\frac{1}{5})\frac{76}{35}\pa O_{\frac{5}{2}}+\frac{81}{5} (G O_2-\frac{2}{5} \pa O_{\frac{5}{2}})\right](w)+ {\cal O}((z-w)^{-2}),
\nonu \\
&& T(z) \, \left( T O_{\frac{7}{2}}-\frac{3}{16} \pa^2 O_{\frac{7}{2}} \right)(w)  =
 \frac{1}{(z-w)^4} \frac{(161+8c)}{16}O_{\frac{7}{2}}(w) +{\cal O}((z-w)^{-2}),
\nonu \\
&& T(z) \, \left( T \pa O_{\frac{5}{2}}-\frac{5}{4} \pa T O_{\frac{5}{2}}-\frac{1}{7} \pa^3 O_{\frac{5}{2}} \right)(w)  =
\frac{1}{(z-w)^5} (-\frac{5}{28})(13+14c)O_{\frac{5}{2}}(w) \nonu \\
&& +\frac{1}{(z-w)^4}(-\frac{1}{5})(-\frac{5}{28})(13+14c) \pa O_{\frac{5}{2}}(w)+ {\cal O}((z-w)^{-2}),
\nonu \\
&& T(z) \, \left( G O_{4'} -\frac{2}{9} \pa O_{\frac{9}{2}} \right)(w)
 = {\cal O}((z-w)^{-2}),
\nonu \\
&& T(z) \, \left( G A_{4} -\frac{2}{9} \pa A_{\frac{9}{2}}\right)(w)  =
 {\cal O}((z-w)^{-2}),
\nonu \\
&& T(z) \, \left( G O_{4} -\frac{1}{8} \pa^2 O_{\frac{7}{2}} \right)(w)  =
 \frac{1}{(z-w)^4} \frac{119}{8} O_{\frac{7}{2}} (w) +{\cal O}((z-w)^{-2}),
\nonu \\
&& T(z) \, \left( G O_{\frac{9}{2}} -\frac{1}{9} \pa^2 O_{4'}\right)(w)  =
 \frac{1}{(z-w)^4} \frac{52}{3} O_{4'} (w) +{\cal O}((z-w)^{-2}),
\nonu \\
&& T(z) \, \left( G A_{\frac{9}{2}} -\frac{1}{9} \pa^2 A_{4} \right)(w)  =
 \frac{1}{(z-w)^4} \frac{52}{3} A_{4} (w) +{\cal O}((z-w)^{-2}),
\nonu \\
&& T(z) \, \left( T \pa^2 O_2 -\frac{5}{2} \pa T \pa O_2 +\pa^2 T O_2 -\frac{3}{28} \pa^4 O_2 \right)(w)
 = \frac{1}{(z-w)^6}\frac{(58+70c)}{7} O_2(w)\nonu \\
&&+\frac{1}{(z-w)^5}(-\frac{1}{2})\frac{(58+70c)}{7}\pa  O_2(w)
+\frac{1}{(z-w)^4}\left[(\frac{1}{20})\frac{(58+70c)}{7} \pa^2 O_2+ 24(TO_2 -\frac{3}{10} \pa^2 O_2)\right](w)\nonu \\
&&  +{\cal O}((z-w)^{-2}), \nonu \\
&& T(z) \, \left( G \pa O_{\frac{7}{2}}-\frac{7}{3} \pa G O_{\frac{7}{2}}-\frac{1}{9} \pa^2 O_4 \right)(w) =
\frac{1}{(z-w)^4} (-\frac{25}{3}) O_{4} (w) +{\cal O}((z-w)^{-2}),
\nonu \\
&& T(z) \, \left( TO_4 -\frac{1}{6} \pa^2 O_4 \right)(w)
= \frac{1}{(z-w)^4} (\frac{24+c}{2}) O_{4} (w) +{\cal O}((z-w)^{-2}),
\nonu \\
&& T(z) \, \left( TO_{4'} -\frac{1}{6} \pa^2 O_{4'} \right)(w)
= \frac{1}{(z-w)^4} (\frac{24+c}{2}) O_{4'} (w) +{\cal O}((z-w)^{-2}),
\nonu \\
&& T(z) \, \left( TA_4 -\frac{1}{6} \pa^2 A_4 \right)(w)
=  \frac{1}{(z-w)^4} (\frac{24+c}{2}) A_{4} (w) +{\cal O}((z-w)^{-2}),
\nonu \\
&& T(z) \, \left( G \pa^2 O_{\frac{5}{2}}-4 \pa G \pa O_{\frac{5}{2}}+\frac{5}{2} \pa^2 G O_{\frac{5}{2}}-\frac{1}{14} \pa^4 O_2 \right)(w)
= \frac{1}{(z-w)^6}(\frac{230}{7}) O_2(w)\nonu \\
&&+\frac{1}{(z-w)^5}(-\frac{1}{2})(\frac{230}{7})\pa  O_2(w)
+\frac{1}{(z-w)^4}\left[(\frac{1}{20})(\frac{230}{7})\pa^2  O_2+ \frac{75}{2}(G O_{\frac{5}{2}} -\frac{1}{5} \pa^2 O_2)\right](w)
\nonu \\
&&  +{\cal O}((z-w)^{-2}),
\nonu \\
&&T(z) \, ( O_2 O_2 -\frac{3}{10}(-2 \widetilde{c} +c)\, \pa^2 O_2+\frac{3}{10}(\widetilde{c}-c)\pa^2 T)(w)
\nonu \\
&&=\frac{1}{(z-w)^4} \left[ c_{t 4}^{t} T + c_{t 4}^{o} O_2\right](w)+{\cal O}((z-w)^{-2}),
\label{Tquasi1}
\\
&&T(z)\, (O_2 O_{\frac{5}{2}}- \frac{3}{4}c'_1\, \pa T G- \frac{3}{4}c'_2\, G \pa O_{2}
-\frac{1}{120}(2c'_4-9c'_1)\, \pa^3 G - \frac{1}{40}(4c'_3-9c'_2)\, \pa^2 O_{\frac{5}{2}}-\frac{3}{7} \pa O_{\frac{7}{2}})(w)
\nonu \\
&&=\frac{1}{(z-w)^5} c_{t \frac{9}{2}}^{g} G(w)
+\frac{1}{(z-w)^4} \left[-\frac{1}{3}c_{t \frac{9}{2}}^{g} \pa G+ c_{t \frac{9}{2}}^{o} O_{\frac{5}{2}}\right](w)+{\cal O}((z-w)^{-2}),
\label{Tquasi2}
\\
&&T(z) \, Q_{\frac{11}{2}}(w)= \frac{1}{(z-w)^5}c_{t \frac{11}{2}}^{o} O_{\frac{5}{2}}(w)\nonu \\
&&\frac{1}{(z-w)^4}\left [-\frac{1}{5}c_{t \frac{11}{2}}^{o}\pa O_{\frac{5}{2}} +c_{t \frac{11}{2}}^{o'}O_{\frac{7}{2}}
+c_{t \frac{11}{2}}^{go}( G O_2 -\frac{2}{5} \pa O_{\frac{5}{2}})\right ](w)+{\cal O}((z-w)^{-2}),
\nonu \\
&&T(z) \,
Q_{6}(w)=\frac{1}{(z-w)^6}c_{t 6}^{o} O_{2}(w)+\frac{1}{(z-w)^5}(-\frac{1}{2}) c_{t 6}^{o}\pa  O_{2}(w)
\nonu \\
&& +\frac{1}{(z-w)^4}\left [\frac{1}{20}c_{t 6}^{o}\pa^2 O_{2} +c_{t 6}^{o'}O_{4}+c_{t 6}^{o''}O_{4'}
+c_{t 6}^{to}( T O_2 -\frac{3}{10} \pa^2 O_2)+c_{t 6}^{go}( G O_{\frac{5}{2}} -\frac{1}{5} \pa^2 O_2)\right ](w)
\nonu \\
&&+{\cal O}((z-w)^{-2}).
\label{Tquasi3}
\eea
The two quasi-primary fields in (\ref{Tquasi1})
and (\ref{Tquasi2}) come from (\ref{A42}) and (\ref{A92}) respectively.
The general forms of structure constants in (\ref{Tquasi2}) are given by
\bea
c_{t 4}^{t} &=& -44 c^2H' (21+4 c) (6+5 c) (2 c (-1+N)+3 (3-4 N) N) (c+2 c N+3 N (-3+2 N)),\nonu \\
c_{t 4}^{o} &=&-132 cH' (21+4 c) (6+5 c) (c+6 (-1+N) N) (-6 N^2+c (-1+4 N)),\nonu \\
c_{t \frac{9}{2}}^{g} &=&-3 c^2H' (6+5 c) (-195+94 c) (2 c (-1+N)+3 (3-4 N) N) (c+2 c N+3 N (-3+2 N)),\nonu \\
c_{t \frac{9}{2}}^{o} &=& -15 c H'(21+4 c) (-132+25 c) (c+6 (-1+N) N) (-6 N^2+c (-1+4 N)),\nonu \\
H'& \equiv &\frac{1}{90 (6 + 5 c) (21 + 4 c) (c + 6 (-1 + N) N)^2}.
\nonu
\eea
The structure constants in (\ref{Tquasi3}) for general $N$ are given by
\bea
c_{t \frac{11}{2}}^{o} &=& -90 c^2H (-1791+c (615+4 c (302+55 c))),\nonu \\
c_{t \frac{11}{2}}^{o'} &=&\frac{23 c (21+4 c) (c-4 c N+6 N^2)}{12 (6+5 c) (c+6 (-1+N) N)},\qquad
c_{t \frac{11}{2}}^{go} = -945 c^2 H(-1041+4 c (316+13 c)),\nonu \\
c_{t 6}^{o} &=& 28 c^2H (25137+c (-115275-20302 c+2120 c^2)),\nonu \\
c_{t 6}^{o'} &=&-\frac{c (1059+176 c) (-6 N^2+c (-1+4 N))}{21 (6+5 c) (c+6 (-1+N) N)},\qquad
c_{t 6}^{o''} = 4,\nonu \\
c_{t 6}^{to} &=& -18648 c^2H (24+c) (-3+2 c),\qquad
c_{t 6}^{go} = 105 c^2H (3267+c (-26439+4 c (-2294+25 c))),\nonu \\
H & \equiv
&  \frac{(3+c-9 N)(1 + N) (c + 6 N) (-3 c + 2 (3 + c) N)}{35 (29 + 2 c) (21 + 4 c) (6 + 5 c)^2 (c + 6 (-1 + N) N)^2}.
\nonu
\eea

\section{The OPEs between $G(z)$ and the quasi-primary (or primary) fields}

To find the correct superpartner in a given superfield, the OPEs between $G(z)$ and quasi-primary (or primary) fields should be computed.
We list the OPEs between $G(z)$ and
the quasi-primary (or primary) fields of the section $3$ here:
\bea
&& G(z) \, \left(GO_2 -\frac{2}{5} \pa O_{\frac{5}{2}}\right)(w) =\frac{1}{(z-w)^3}\frac{2}{15}(6+5c)O_2(w)+ {\cal O}((z-w)^{-2}),
\nonu \\
&& G(z) \, \left(TG -\frac{3}{8} \pa^2 G \right)(w) = \frac{1}{(z-w)^3}\frac{1}{6}(21+4c)T(w)+ {\cal O}((z-w)^{-2}),
\nonu \\
&& G(z) \, \left(TT-\frac{3}{10} \pa^2 T\right)(w) =
 \frac{1}{(z-w)^4} \frac{51}{20}G(w) +\frac{1}{(z-w)^3} (-\frac{1}{3})
\frac{51}{20} \pa G(w) + {\cal O}((z-w)^{-2}),
\nonu \\
&& G(z) \, \left(TO_{2}-\frac{3}{10} \pa^2 O_2 \right)(w)  = \frac{1}{(z-w)^3} \frac{19}{10}O_{\frac{5}{2}}(w) + {\cal O}((z-w)^{-2}),
\nonu \\
&& G(z) \, \left( GO_{\frac{5}{2}}-\frac{1}{5} \pa^2 O_2 \right)(w)   =
\frac{1}{(z-w)^3} \frac{1}{15}(69+10c)O_{\frac{5}{2}}(w) + {\cal O}((z-w)^{-2}),
\nonu \\
&& G(z) \, \left( G \pa G-\frac{7}{10} \pa^2 T \right)(w)  = \frac{1}{(z-w)^4} (-\frac{1}{10})(3+20c) G(w) \nonu \\
&&+\frac{1}{(z-w)^3}(-\frac{1}{3}) (-\frac{1}{10})(3+20c)\pa  G(w) + {\cal O}((z-w)^{-2}),
\nonu \\
&& G(z) \,  \left(\pa G O_2 -\frac{3}{4}G \pa O_2 +\frac{1}{8} \pa^2 O_{\frac{5}{2}} \right)(w)  =
\frac{1}{(z-w)^4} (1+2c)O_2(w) \nonu \\
&&+\frac{1}{(z-w)^3}(-\frac{1}{4})(1+2c) \pa O_2(w) + {\cal O}((z-w)^{-2}),
\nonu \\
&& G(z) \, \left(TO_{\frac{5}{2}}-\frac{1}{4} \pa^2 O_{\frac{5}{2}}\right)(w)  =
\frac{1}{(z-w)^4} 8 O_2(w) +\frac{1}{(z-w)^3}(-\frac{1}{4})8 \pa O_2(w) + {\cal O}((z-w)^{-2}),
\nonu \\
&& G(z) \, \left( T \pa O_2-\pa T O_2-\frac{1}{6}\pa^3 O_2 \right)(w)=
\frac{1}{(z-w)^4}(- \frac{3}{2}) O_{\frac{5}{2}}(w) \nonu \\
&&+\frac{1}{(z-w)^3}\left[(-\frac{1}{5})(- \frac{3}{2})\pa  O_{\frac{5}{2}}-3(G O_2 -\frac{2}{5} \pa O_{\frac{5}{2}}) \right] (w)+ {\cal O}((z-w)^{-2}),
\nonu \\
&& G(z) \, \left( G \pa O_{\frac{5}{2}}-\frac{5}{3}\pa G O_{\frac{5}{2}}-\frac{1}{9} \pa^3 O_2\right)(w)  =
\frac{1}{(z-w)^4}(-\frac{1}{3})(-3+10c) O_{\frac{5}{2}}(w) \nonu \\
&&+\frac{1}{(z-w)^3}\left[(-\frac{1}{5})(-\frac{1}{3})(-3+10c)\pa  O_{\frac{5}{2}}-8(G O_2 -\frac{2}{5} \pa O_{\frac{5}{2}}) \right] (w)+ {\cal O}((z-w)^{-2}),
\nonu \\
&& G(z) \, \left( G O_{\frac{7}{2}}-\frac{1}{4} \pa O_4 \right) (w)
= \frac{1}{(z-w)^3}\frac{1}{6}(21+4c) O_{\frac{7}{2}} (w)+{\cal O}((z-w)^{-2}),
\nonu \\
&& G(z) \, \left( \pa^2 G O_2 +\frac{3}{5} G \pa^2 O_2 -2 \pa G \pa O_2 -\frac{2}{35} \pa^3 O_{\frac{5}{2}} \right)(w)  =  \nonu \\
&& \frac{1}{(z-w)^5} (\frac{32+280c}{35})O_{2}(w)+\frac{1}{(z-w)^4}(-\frac{1}{2})(\frac{32+280c}{35})\pa O_{2}(w)
\nonu \\
&&+\frac{1}{(z-w)^3}\left[(\frac{1}{20})(\frac{32+280c}{35})\pa^2 O_{2}-\frac{6}{5}(G O_{\frac{5}{2}}-\frac{1}{5} \pa^2 O_2) +
4(TO_2 -\frac{3}{10} \pa^2 O_2)\right](w)\nonu \\
&&+ {\cal O}((z-w)^{-2}), \nonu \\
&& G(z) \, \left( T O_{\frac{7}{2}}-\frac{3}{16} \pa^2 O_{\frac{7}{2}} \right)(w)  =
 \frac{1}{(z-w)^3} \frac{17}{8}O_{4}(w) +{\cal O}((z-w)^{-2}),
\nonu \\
&& G(z) \, \left( T \pa O_{\frac{5}{2}}-\frac{5}{4} \pa T O_{\frac{5}{2}}-\frac{1}{7} \pa^3 O_{\frac{5}{2}} \right)(w)  =
\nonu \\
&&\frac{1}{(z-w)^5} (-\frac{19}{7})O_{2}(w)  +\frac{1}{(z-w)^4}(-\frac{1}{2})(-\frac{19}{7})\pa O_{2}(w)\nonu \\
&&+\frac{1}{(z-w)^3}\left[(\frac{1}{20})(-\frac{19}{7})\pa^2 O_{2}(w)-\frac{15}{4} (GO_{\frac{5}{2}}-\frac{1}{5} \pa^2 O_2) +8(TO_2-\frac{3}{10} \pa^2 O_2)\right](w)\nonu \\
&&+ {\cal O}((z-w)^{-2}),
\nonu \\
&& G(z) \, \left( G O_{4'} -\frac{2}{9} \pa O_{\frac{9}{2}} \right)(w)
 =\frac{1}{(z-w)^3}\frac{2}{9}(20+3c) O_{4'}(w) +{\cal O}((z-w)^{-2}),
\nonu \\
&& G(z) \, \left( G A_{4} -\frac{2}{9} \pa A_{\frac{9}{2}}\right)(w)  =
\frac{1}{(z-w)^3}\frac{2}{9}(20+3c) A_{4}(w) +{\cal O}((z-w)^{-2}),
\nonu \\
&& G(z) \, \left( G O_{4} -\frac{1}{8} \pa^2 O_{\frac{7}{2}} \right)(w)  =
\frac{1}{(z-w)^3}\frac{1}{12}(93+8c) O_{4}(w) +{\cal O}((z-w)^{-2}),
\nonu \\
&& G(z) \, \left( G O_{\frac{9}{2}} -\frac{1}{9} \pa^2 O_{4'}\right)(w)  =
\frac{1}{(z-w)^3}\frac{1}{9}(79+6c) O_{\frac{9}{2}}(w) +{\cal O}((z-w)^{-2}),
\nonu \\
&& G(z) \, \left( G A_{\frac{9}{2}} -\frac{1}{9} \pa^2 A_{4} \right)(w)  =
\frac{1}{(z-w)^3}\frac{1}{9}(79+6c) A_{\frac{9}{2}}(w) +{\cal O}((z-w)^{-2}),
\nonu \\
&& G(z) \, \left( T \pa^2 O_2 -\frac{5}{2} \pa T \pa O_2 +\pa^2 T O_2 -\frac{3}{28} \pa^4 O_2 \right)(w)
 = \frac{1}{(z-w)^5}(\frac{69}{14}) O_{\frac{5}{2}}(w)\nonu \\
&&+\frac{1}{(z-w)^4}\left[(-\frac{2}{5})(\frac{69}{14} )\pa O_{\frac{5}{2}}+9(G O_2 -\frac{2}{5} \pa O_{\frac{5}{2}})\right](w)\nonu \\
&&+\frac{1}{(z-w)^3}\left[(\frac{1}{30})(\frac{69}{14} )\pa^2 O_{\frac{5}{2}}-\frac{9}{7}\pa (G O_2 -\frac{2}{5} \pa O_{\frac{5}{2}})
+2(TO_{\frac{5}{2}}-\frac{1}{4} \pa^2 O_{\frac{5}{2}}) \right. \nonu \\
&& \left. + \frac{58}{7}(\pa G O_2 -\frac{3}{4} G \pa O_2 +\frac{1}{8} \pa^2 O_{\frac{5}{2}})\right](w)+{\cal O}((z-w)^{-2}),\nonu \\
&& G(z) \, \left( G \pa O_{\frac{7}{2}}-\frac{7}{3} \pa G O_{\frac{7}{2}}-\frac{1}{9} \pa^2 O_4 \right)(w) =
\frac{1}{(z-w)^4}(-\frac{7}{3})(3+2c) O_{\frac{7}{2}} (w)\nonu \\
&&+\frac{1}{(z-w)^3}(-\frac{1}{7})(-\frac{7}{3})(3+2c) \pa O_{\frac{7}{2}} (w) +{\cal O}((z-w)^{-2}),
\nonu \\
&& G(z) \, \left( TO_4 -\frac{1}{6} \pa^2 O_4 \right)(w)
= \frac{1}{(z-w)^4}(\frac{35}{2}) O_{\frac{7}{2}} (w)+\frac{1}{(z-w)^3}(-\frac{1}{7})(\frac{35}{2})\pa O_{\frac{7}{2}} (w)\nonu \\
&& +{\cal O}((z-w)^{-2}),
\nonu \\
&& G(z) \, \left( TO_{4'} -\frac{1}{6} \pa^2 O_{4'} \right)(w)
= \frac{1}{(z-w)^3} (\frac{13}{6}) O_{\frac{9}{2}} (w) +{\cal O}((z-w)^{-2}),
\nonu \\
&& G(z) \, \left( TA_4 -\frac{1}{6} \pa^2 A_4 \right)(w)
= \frac{1}{(z-w)^3} (\frac{13}{6}) A_{\frac{9}{2}} (w) +{\cal O}((z-w)^{-2}),
\nonu \\
&& G(z) \, \left( G \pa^2 O_{\frac{5}{2}}-4 \pa G \pa O_{\frac{5}{2}}+\frac{5}{2} \pa^2 G O_{\frac{5}{2}}-\frac{1}{14} \pa^4 O_2 \right)(w)
= \frac{1}{(z-w)^5}(\frac{93}{7}+20c) O_{\frac{5}{2}}(w)\nonu \\
&&\frac{1}{(z-w)^4}\left[(-\frac{2}{5})(\frac{93}{7}+20c)\pa O_{\frac{5}{2}} -24(GO_2-\frac{2}{5} \pa O_{\frac{5}{2}}) \right](w)\nonu \\
&&\frac{1}{(z-w)^3}\left[(\frac{1}{30})(\frac{93}{7}+20c) \pa^2 O_{\frac{5}{2}} +\frac{24}{7}\pa (GO_2-\frac{2}{5} \pa O_{\frac{5}{2}})
+10(TO_{\frac{5}{2}}-\frac{1}{4} \pa^2 O_{\frac{5}{2}})  \right. \nonu \\
&& \left.  -\frac{200}{7}( \pa G O_2 - \frac{3}{4} G \pa O_2 +\frac{1}{8} \pa^2 O_{\frac{5}{2}})\right](w) +{\cal O}((z-w)^{-2}),
\nonu \\
&&G(z)\, ( O_2 O_2 -\frac{3}{10}(-2 \widetilde{c} +c)\, \pa^2 O_2+\frac{3}{10}(\widetilde{c}-c)\pa^2 T)(w)
\nonu \\
&&=\frac{1}{(z-w)^4} c_{g4}^{g} G(w) + \frac{1}{(z-w)^3} \left[ -\frac{1}{3} c_{g4}^{g} \pa G + c_{g4}^{o} O_{\frac{5}{2}}\right](w)+{\cal O}((z-w)^{-2}),
\nonu \\
&&G(z)\, (O_2 O_{\frac{5}{2}}- \frac{3}{4}c'_1\, \pa T G- \frac{3}{4}c'_2\, G \pa O_{2}
-\frac{1}{120}(2c'_4-9c'_1)\, \pa^3 G - \frac{1}{40}(4c'_3-9c'_2)\, \pa^2 O_{\frac{5}{2}}-\frac{3}{7} \pa O_{\frac{7}{2}})(w)
\nonu \\
&&=\frac{1}{(z-w)^4} \left[ c_{g \frac{9}{2}}^{t} T + c_{g \frac{9}{2}}^{o} O_2 \right](w) +
\frac{1}{(z-w)^3} \left[ -\frac{1}{4}c_{g \frac{9}{2}}^{t}\pa T -\frac{1}{4} c_{g \frac{9}{2}}^{o} \pa O_2 \right](w)+{\cal O}((z-w)^{-2}),
\label{Gquasi2}
\\
&&G(z) \,
Q_{\frac{11}{2}}(w)=\frac{1}{(z-w)^5} c_{g \frac{11}{2}}^{o} O_2(w)+\frac{1}{(z-w)^4} (-\frac{1}{2})c_{g \frac{11}{2}}^{o}\pa O_2(w)
\nonu \\
&&\frac{1}{(z-w)^3}\left[ \frac{1}{20}c_{g \frac{11}{2}}^{o}\pa^2 O_2+c_{g \frac{11}{2}}^{o'}O_4+c_{g \frac{11}{2}}^{o''} O_{4'}
+ c_{g \frac{11}{2}}^{to}(T O_2 -\frac{3}{10} \pa^2 O_2)+c_{g \frac{11}{2}}^{go}( G O_{\frac{5}{2}}-\frac{1}{5} \pa^2 O_2)\right](w)
\nonu \\
&& +{\cal O}((z-w)^{-2}),
\nonu \\
&&G(z) \,
Q_{6}(w)=\frac{1}{(z-w)^5}c_{g 6}^{o} O_{\frac{5}{2}}(w)
+\frac{1}{(z-w)^4}\left[-\frac{2}{5}c_{g 6}^{o} \pa O_{\frac{5}{2}}+ c_{g 6}^{o'} O_{\frac{7}{2}} +c_{g 6}^{go}( G O_2 -\frac{2}{5} \pa O_{\frac{5}{2}}) \right](w)
\nonu \\
&&+\frac{1}{(z-w)^3}\left[\frac{1}{30}c_{g 6}^{o} \pa^2 O_{\frac{5}{2}}-\frac{1}{7} c_{g 6}^{o'} \pa O_{\frac{7}{2}}
 -\frac{1}{7} c_{g 6}^{go} \pa ( G O_2 -\frac{2}{5} \pa O_{\frac{5}{2}})
+c_{g 6}^{go'} ( \pa G O_2 -\frac{3}{4} G \pa O_2 +\frac{1}{8} \pa^2 O_{\frac{5}{2}})
  \right. \nonu \\
&&  \left. +c_{g 6}^{to} ( T O_{\frac{5}{2}}-\frac{1}{4} \pa^2 O_{\frac{5}{2}}) +c_{g6}^{o''} O_{\frac{9}{2}}  \right](w)  +{\cal O}((z-w)^{-2}).
\label{Gquasi3}
\eea
The general forms of structure constants in(\ref{Gquasi2}) are given by
\bea
c_{g 4}^{g} &=& -3 c^2 H' (21+4 c) (6+5 c) (2 c (-1+N)+3 (3-4 N) N) (c+2 c N+3 N (-3+2 N)),\nonu \\
c_{g 4}^{o} &=&6 c H'(21+4 c) (6+5 c)(-108 (-1+N) N^3+10 c^3 (-1+N) (1+2 N)\nonu \\
&-& 9 c N (-2+N (57-98 N+40 N^2))-3 c^2 (1+N (-49+20 N (2+N)))),\nonu \\
c_{g \frac{9}{2}}^{t} &=&4 c^2 H' (6+5 c) (-195+94 c) (2 c (-1+N)+3 (3-4 N) N) (c+2 c N+3 N (-3+2 N)),\nonu \\
c_{g \frac{9}{2}}^{o} &=& 60 cH' (21+4 c) (3+16 c) (c+6 (-1+N) N) (-6 N^2+c (-1+4 N)),\nonu \\
H'& \equiv &\frac{1}{90 (6 + 5 c) (21 + 4 c) (c + 6 (-1 + N) N)^2}.
\nonu
\eea
The general forms of structure constants in (\ref{Gquasi3}) are given by
\bea
c_{g \frac{11}{2}}^{o} &=& -24 c^2 H(-2493+10 c (75+2 c (739+10 c))),\nonu \\
c_{g \frac{11}{2}}^{o'} &=&\frac{5 c (21+4 c) (c-4 c N+6 N^2)}{42 (6+5 c) (c+6 (-1+N) N)},\qquad
c_{g \frac{11}{2}}^{o''} = \frac{8}{3},\nonu \\
c_{g \frac{11}{2}}^{to} &=& -84 c^2H (1953+100 (-21+c) c),\qquad
c_{g \frac{11}{2}}^{go} = -630 c^2H (-3+2 c) (-18+23 c),\nonu \\
c_{g 6}^{o} &=& 140 c^2H (3078+c (-2709+c (-15777-3794 c+40 c^2))),\nonu \\
c_{g 6}^{o'} &=& \frac{c (201+46 c) (c-4 c N+6 N^2)}{3 (6+5 c) (c+6 (-1+N) N},\qquad
c_{g 6}^{go} = -420 c^2 H(-675+2 c (-1731-724 c+8 c^2)),\nonu \\
c_{g 6}^{go'} &=& 8 c^2H (90207+c (-307725+2 c (-44221+500 c))),\nonu \\
c_{g 6}^{to} &=& 28 c^2 H(7101+c (-31785+4 c (-1829+25 c))),\qquad
c_{g 6}^{o''}= -\frac{8}{3},\nonu \\
H & \equiv&  \frac{(3+c-9 N)(1 + N) (c + 6 N) (-3 c + 2 (3 + c) N)}{35 (29 + 2 c) (21 + 4 c) (6 + 5 c)^2 (c + 6 (-1 + N) N)^2}.
\nonu
\eea




\begin{thebibliography}{99}

\bibitem{GG}
  M.~R.~Gaberdiel and R.~Gopakumar,
  ``An $AdS_3$ Dual for Minimal Model CFTs,''
Phys.\ Rev.\ D {\bf 83}, 066007 (2011)
[arXiv:1011.2986 [hep-th]].

\bibitem{GG1}
  M.~R.~Gaberdiel and R.~Gopakumar,
  ``Triality in Minimal Model Holography,''  JHEP {\bf 1207}, 127
  (2012)
[arXiv:1205.2472 [hep-th]].  

\bibitem{Ahn1106}
  C.~Ahn,
  ``The Large N 't Hooft Limit of Coset Minimal Models,''  JHEP {\bf
  1110}, 125 (2011)  [arXiv:1106.0351 [hep-th]].

\bibitem{GV}
  M.~R.~Gaberdiel and C.~Vollenweider,
  ``Minimal Model Holography for SO(2N),''  JHEP {\bf 1108}, 104 (2011)  [arXiv:1106.2634 [hep-th]].  

\bibitem{CHR}
  T.~Creutzig, Y.~Hikida and P.~B.~R©ªnne,
  ``N=1 supersymmetric higher spin holography on $AdS_3$,''
JHEP {\bf 1302}, 019 (2013)  [arXiv:1209.5404 [hep-th]].  

\bibitem{PV}
  S.~F.~Prokushkin and M.~A.~Vasiliev,
  ``Higher spin gauge interactions for massive matter fields in 3-D AdS space-time,''  Nucl.\ Phys.\ B {\bf 545}, 385 (1999)  [hep-th/9806236].  

\bibitem{CV}
  C.~Candu, and C.~Vollenweider,
  ``The N=1 algebra  $\cal{W}_{\infty}[\mu]$ and its truncations,''
[arXiv:1305.0013v1 [hep-th]].

\bibitem{Ahn1202}
  C.~Ahn,
  ``The Primary Spin-4 Casimir Operators in the Holographic SO(N) Coset Minimal Models,''  JHEP {\bf 1205}, 040 (2012)  [arXiv:1202.0074 [hep-th]].  

\bibitem{Bowcock}
  P.~Bowcock,
  ``Quasi-primary Fields And
Associativity Of Chiral Algebras,''
Nucl.\ Phys.\ B {\bf 356}, 367 (1991).  

\bibitem{BFKNRV}
  R.~Blumenhagen, M.~Flohr, A.~Kliem, W.~Nahm, A.~Recknagel and R.~Varnhagen,
  ``W algebras with two and three generators,''
  Nucl.\ Phys.\  B {\bf 361}, 255 (1991).

\bibitem{BS}
  P.~Bouwknegt and K.~Schoutens,
  ``W symmetry in conformal field theory,''
Phys.\ Rept.\  {\bf 223}, 183 (1993)  [hep-th/9210010].

\bibitem{Nahm1}
  W.~Nahm,
  ``Algebras of two-dimensional chiral fields and their classification,''
In *Islamabad 1989, Proceedings, Mathematical physics* 283-300. (see Conference Index)

\bibitem{Nahm2}
  W.~Nahm,
  ``Chiral algebras of
two-dimensional chiral field theories and their normal ordered products,''
In *Trieste 1989, Proceedings, Recent developments in conformal field theories* 81-84.
(see Conference Index)

\bibitem{Thielemans}
  K.~Thielemans,
  ``A Mathematica package for
computing operator product expansions,''
Int.\ J.\ Mod.\ Phys.\ C {\bf 2}, 787 (1991).

\bibitem{CGKV}
  C.~Candu, M.~R.~Gaberdiel, M.~Kelm and C.~Vollenweider,
  ``Even spin minimal model holography,''  JHEP {\bf 1301}, 185 (2013)  [arXiv:1211.3113 [hep-th]].  

\bibitem{AP1301}
  C.~Ahn and J.~Paeng,
  ``The OPEs of Spin-4 Casimir Currents in the Holographic $SO(N)$ Coset Minimal Models,''  Class.\ Quant.\ Grav.\  {\bf 30}, 175004 (2013)  [arXiv:1301.0208].  

\bibitem{FMS}
  D.~Friedan, E.~J.~Martinec and S.~H.~Shenker,
  ``Conformal Invariance, Supersymmetry and String Theory,''  Nucl.\ Phys.\ B {\bf 271}, 93 (1986).  

\bibitem{Ahn1211}
  C.~Ahn,
  ``The Higher Spin Currents in the N=1 Stringy Coset Minimal Model,''  JHEP {\bf 1304}, 033 (2013)  [arXiv:1211.2589 [hep-th]].  

\bibitem{Ahn1111}
  C.~Ahn,
  ``The Coset Spin-4 Casimir Operator and Its Three-Point Functions with Scalars,''  JHEP {\bf 1202}, 027 (2012)  [arXiv:1111.0091 [hep-th]].  

\bibitem{AK1308}
  C.~Ahn and H.~Kim,
  ``Spin-5 Casimir Operator and Its Three-Point Functions with Two Scalars,''  arXiv:1308.1726 [hep-th].  

\bibitem{GG1305}
  M.~R.~Gaberdiel and R.~Gopakumar,
  ``Large $\mathcal{N}=4$ Holography,''  JHEP {\bf 1309}, 036 (2013)  [arXiv:1305.4181 [hep-th]].  

\bibitem{Ahn1206}
  C.~Ahn,
  ``The Large N 't Hooft Limit of Kazama-Suzuki Model,''  JHEP {\bf 1208}, 047 (2012)  [arXiv:1206.0054 [hep-th]].  

\bibitem{Ahn1208}
  C.~Ahn,
  ``The Operator Product Expansion of the Lowest Higher Spin Current at Finite N,''  JHEP {\bf 1301}, 041 (2013)  [arXiv:1208.0058 [hep-th]].  

\bibitem{CHR1306}
  T.~Creutzig, Y.~Hikida and P.~B.~Ronne,
  ``Extended higher spin holography and Grassmannian models,''  arXiv:1306.0466 [hep-th].  

\bibitem{Ahn1305}
  C.~Ahn,
  ``Higher Spin Currents with Arbitrary N in the ${\cal N} = 1$
Stringy Coset Minimal Model,''  JHEP {\bf 1307}, 141 (2013)  [arXiv:1305.5892 [hep-th]].  



\end{thebibliography}
\end{document}